\documentclass[12pt,english,american]{article}
\usepackage[T1]{fontenc}
\usepackage[latin9]{inputenc}
\usepackage{geometry}
\usepackage{color}
\usepackage{yfonts}
\usepackage{slashed}
\usepackage{babel}
\usepackage{amsthm}
\usepackage{mathrsfs}
\usepackage{amstext}
\usepackage{amssymb}
\usepackage[unicode=true,pdfusetitle,
 bookmarks=true,bookmarksnumbered=false,bookmarksopen=false,
 breaklinks=true,pdfborder={0 0 0},backref=false,colorlinks=true]
 {hyperref}
\hypersetup{
 linkcolor=blue,citecolor=blue, urlcolor=blue}

\makeatletter

\def\qed{$\Box$\medskip}

\newtheorem{theoreme}{Theorem } [section]
\newtheorem{proposition}[theoreme]{Proposition}
\newtheorem{lemma}[theoreme]{Lemma}
\newtheorem{definition}[theoreme]{Definition}
\newtheorem{corollary}[theoreme]{Corollary}
\newtheorem{remark}[theoreme]{Remark}
\newtheorem{example}[theoreme]{Example}
\newtheorem{criterion}[theoreme]{Criterion}

\newcommand{\beq}{\begin{equation}}
\newcommand{\eeq}{\end{equation}}
\newcommand{\beqa}{\begin{eqnarray}}
\newcommand{\eeqa}{\end{eqnarray}}
\newcommand{\ben}{\begin{arabicenumerate}}
\newcommand{\een}{\end{arabicenumerate}}
\newcommand{\bex}{\begin{example}}
\newcommand{\eex}{\end{example}}
\newcommand{\ber}{\begin{remark}}
\newcommand{\eer}{\end{remark}}
\newcommand{\bec}{\begin{corollary}}
\newcommand{\eec}{\end{corollary}}
\newcommand{\bep}{\begin{proposition}}
\newcommand{\eep}{\end{proposition}}
\newcommand{\becr}{\begin{criterion}}
\newcommand{\eecr}{\end{criterion}}

\def\bel{\begin{lem} } 
\def\eel{\end{lem} }
\def\bet{\begin{theoreme}}
\def\eet{\end{theoreme}}
\def\bed{\begin{defn}}
\def\eed{\end{defn} }

\usepackage{enumitem}		% customizable list environments
\theoremstyle{plain}
\newtheorem{thm}{\protect\theoremname}[section]
\theoremstyle{definition}
\newtheorem{defn}[thm]{\protect\definitionname}
\theoremstyle{plain}
\newtheorem{prop}[thm]{\protect\propositionname}
\theoremstyle{remark}

\theoremstyle{plain}
\newtheorem{lem}[thm]{\protect\lemmaname}
\theoremstyle{plain}

\usepackage{txfonts,refstyle,xcolor}

\newref{con}{name = Conjecture\ }
\newref{prop}{name = Proposition\ }
\newref{def}{name = Definition\ }
\newref{sec}{name = Section\ }
\newref{sub}{name = Section\ }
\newref{thm}{name = Theorem\ }
\newref{lem}{name = Lemma\ }
\newref{cor}{name = Corollary\ }
\newref{fig}{name = Figure\ }

\usepackage{bbm}
\newcommand{\charf}{\mathbbm{1}}

\usepackage[all]{xy}

\newcommand{\xyR}[1]{%
     \makeatletter
     \xydef@\xymatrixrowsep@{#1}
     \makeatother
}

\newcommand{\xyC}[1]{%
     \makeatletter
     \xydef@\xymatrixcolsep@{#1}
     \makeatother
}

\newcommand{\ncol}[1]{\color{normalcolor}}

\makeatother

\addto\captionsamerican{\renewcommand{\corollaryname}{Corollary}}
\addto\captionsamerican{\renewcommand{\definitionname}{Definition}}
\addto\captionsamerican{\renewcommand{\lemmaname}{Lemma}}
\addto\captionsamerican{\renewcommand{\propositionname}{Proposition}}
\addto\captionsamerican{\renewcommand{\remarkname}{Remark}}
\addto\captionsamerican{\renewcommand{\theoremname}{Theorem}}
\addto\captionsenglish{\renewcommand{\corollaryname}{Corollary}}
\addto\captionsenglish{\renewcommand{\definitionname}{Definition}}
\addto\captionsenglish{\renewcommand{\lemmaname}{Lemma}}
\addto\captionsenglish{\renewcommand{\propositionname}{Proposition}}
\addto\captionsenglish{\renewcommand{\remarkname}{Remark}}
\addto\captionsenglish{\renewcommand{\theoremname}{Theorem}}
\providecommand{\corollaryname}{Corollary}
\providecommand{\definitionname}{Definition}
\providecommand{\lemmaname}{Lemma}
\providecommand{\propositionname}{Proposition}
\providecommand{\remarkname}{Remark}
\providecommand{\theoremname}{Theorem}
\begin{document}
\title{Bose particles in a box III. A convergent expansion of  the ground state of the Hamiltonian in the mean field limiting regime.} 
% \author{ A. Pizzo}
  \author{A. Pizzo \footnote{email: pizzo@mat.uniroma2.it}\\
 Dipartimento di Matematica, Universit\`a di Roma ``Tor Vergata",\\
 Via della Ricerca Scientifica 1, I-00133 Roma, Italy}
 % \author{ A. Pizzo}

\date{11/05/2016}

\maketitle

\abstract{In this paper we consider an interacting Bose gas at zero temperature, constrained to a finite box  and in the mean field limiting regime. The $N$ gas particles interact through a pair potential of positive type and with an ultraviolet cut-off. The (nonzero) Fourier components of the potential are assumed to be sufficiently large with respect to the corresponding kinetic energies of the modes like in the companion papers \cite{Pi1}-\cite{Pi2}. Using the multi-scale technique in the occupation numbers of particle states introduced in \cite{Pi1}-\cite{Pi2}, we provide a convergent expansion of the ground state of the Hamiltonian  in terms of the bare operators. In the limit $N\to \infty$ the expansion is  up to any desired precision. 
% for a pair potential of positive type, with an ultraviolet cut-off,  and \emph{high} with respect to the kinetic energy of the corresponding modes; this notion is made precise in {\color{red}...}
%This procedure also provides an algorithm to expand the ground state vector in terms of the bare quantities up to any precision.}}
\\

\noindent
{\bf{Summary of contents}}
\begin{itemize}
\item In Section \ref{introduction},  the model of an interacting Bose gas in a finite box and at zero temperature  is defined along with the notation used throughout the paper. The model is analyzed  at fixed total number of particles. In this context, we define the second quantized Hamiltonian and the associated  \emph{particle number preserving} Bogoliubov Hamiltonian (from now on Bogoliubov Hamiltonian).
\item In Section \ref{multiscale-HBog},  we review the main ideas and results of the multi-scale analysis in the occupation numbers of particle states carried out for the Bogoliubov Hamiltonian in \cite{Pi1} and \cite{Pi2}. The Feshbach flows implemented in \cite{Pi1}, \cite{Pi2} are described informally in Sections \ref{informal-0} and \ref{informal}, respectively.
\item In Section \ref{groundstateH},  in the mean field limiting regime (i.e.,  at fixed box volume $|\Lambda|$, for a number of particles, $N$,  sufficiently large, and for a coupling constant  inversely proportional to the particle density) the ground state vector of the Hamiltonian is constructed  as a byproduct of subsequent Feshbach flows.  Each flow is associated with a couple of components, $\phi_{\pm \bold{j}}$,  of the Fourier expansion of the pair potential. The Fourier expansion consists of only a finite number of components because of the u.v. cut-off. This construction provides a convergent expansion of the ground state vector of the Hamiltonian in terms of the bare operators applied to the vector with all the particles in the zero mode.
% up to any precision in terms of the bare quantities.}
%\item In Section \ref{new-proj}, ....
%\item
%Some of the proofs are deferred to the Appendix in Section \ref{appendix}.
\end{itemize}
\setcounter{equation}{0}

\section{Interacting Bose gas in a box}\label{introduction}
\setcounter{equation}{0}
%\[
%a(x)=\frac{\sum_{\boldsymbol{n}}a_{\boldsymbol{n}}e^{ik_{\boldsymbol{n}}\cdot x}}{|\Lambda_{L'}|^{\frac{1}{2}}}
%\]
%where $|\Lambda_{L'}|=L'^{3}$and $k_{\boldsymbol{n}}=\frac{2\pi\boldsymbol{n}}{L'}$
%and $\boldsymbol{n}\in Z^{3}$. Commutation relations:

%\[
%[a_{\boldsymbol{n}},a_{\boldsymbol{n'}}^{*}]=\delta_{\boldsymbol{n},\boldsymbol{n}'}
%\]
%Hamiltonian
%\[
%H'=\int\frac{1}{2m}(\nabla a^{*})(\nabla a)(x)dx+\frac{\lambda}{2}\int\int(a^{*}(x)a(x)-\rho)\phi(x-y)(a^{*}(x)a(x)-\rho)dxdy
%\]
%where the integration is over $\Lambda_{L'}$ , and 
%\[
%\phi(x-y)=(2\pi)^{-\frac{3}{2}}\sum_{\boldsymbol{n}}\tilde{\phi}(k_{\boldsymbol{n}})\frac{e^{ik_{\boldsymbol{n}}\cdot(x-y)}}{|\Lambda_{L'}|^{\frac{1}{2}}}
%\]
%\\
We study the Hamiltonian describing a gas of (spinless) nonrelativistic Bose particles that, at zero temperature, are constrained to a $d-dimensional$ box of side $L$ with $d\geq 1$. The particles interact through a pair potential with a coupling constant proportional to the inverse of the particle density  $\rho$.  

The rigorous description of this system has many intriguing mathematical aspects not completely clarified yet. In spite of remarkable contributions also in recent years, some important problems are still open to date, in particular in connection to the thermodynamic limit and the exact structure of the ground state vector. We shall briefly mention the results closer to our present work  and give references to the reader for the details. 

Some of the results have been concerned with the low energy spectrum of the Hamiltonian that in the mean field limit was predicted by Bogoliubov \cite{Bo1}, \cite{Bo2}. 
Starting from the Hamiltonian of the system he defined an approximated one, the Bogoliubov Hamiltonian. For a finite box and a large class of pair potentials, upon a  unitary transformation the Bogoliubov Hamiltonian describes\footnote{In the canonical ensemble approach the diagonalization of the (particle preserving) Bogoliubov Hamiltonian is exact only in the mean field limit (see \cite{Se1}.}  a system of non-interacting bosons with a new energy dispersion law, which is in fact the correct description of the energy spectrum of the Bose particles system in the mean field limit.

The expression predicted by Bogoliubov for the ground state energy has been rigorously proven for certain systems in \cite{LS1}, \cite{LS2}, \cite{ESY},\cite{YY}. Concerning  the excitation spectrum, in Bogoliubov theory it consists of elementary excitations whose energy is linear in the momentum for small momenta. 
After some important results restricted to one-dimensional models (see \cite{G}, \cite{LL}, \cite{L}), this conjecture was proven by Seiringer in \cite{Se1} (see also \cite{GS}) for the low-energy spectrum of an interacting Bose gas in a finite box and in the mean field limiting regime,  where  the pair potential is of positive type.  In  \cite{LNSS}  it has been extended to a more general class of potentials and the limiting behavior of the low energy eigenstates has been studied.  Later,  the result of \cite{Se1} has been proven to be valid in a sort of diagonal limit where the particle density and the box volume diverge according to a prescribed asymptotics; see  \cite{DN}. Recently, Bogoliubov's prediction on the energy spectrum in the mean field limiting regime has been shown to be valid also for the high energy eigenvalues (see \cite{NS}).

\noindent
These results are based on energy estimates starting from the spectrum of the corresponding Bogoliubov Hamiltonian. 

A different approach to studying a gas of Bose particles is based on renormalization group.  In this respect, we mention the paper by Benfatto, \cite{Be}, where he provided \emph{an order by order control} of the Schwinger  functions of this system in three dimensions and with an ultraviolet cut-off. His analysis holds at zero temperature in the infinite volume limit  and at finite particle density. Thus, it contains  a fully consistent treatment of the infrared divergences at a perturbative level. This program has been later developed in  \cite{CDPS1}, \cite{CDPS2}, and, more recently, in  \cite{C} and \cite{CG} by making use of \emph{Ward identities} to deal also with two-dimensional systems where some partial control of the renormalization flow has been provided; see \cite{C} for a detailed review of previous related results.

\noindent
Within the renormalization group approach, we also mention some results towards a rigorous construction of the functional integral for this system contained in  \cite{BFKT1}, \cite{BFKT2}, and \cite{BFKT}.
%Somehow related to the content of this paper and of the companion papers,
%The literature on the Bose gas in general is vast and goes beyond the problems and the model addressed in this paper. However, it is surely .... 
%We also mention  the {\color{red}progress in the control} of the dynamical properties of Bose gases (for references and for an update of the state of the art we refer the reader to the introduction of \cite{DFPP}).
\\

In this paper we study  a gas of (spinless) nonrelativistic Bose  particles that, at zero temperature, are constrained to a $d-dimensional$  box, $d\geq 1$,  and interact through a pair potential of positive type and with an ultraviolet cut-off.  We consider the number of particles fixed but we use the formalism of second quantization. We use units such that the particle mass is set equal to $\frac{1}{2}$ and $\hbar$ equal to $1$. The Hamiltonian corresponding to the pair potential $\phi(x-y)$ and to the coupling constant $\lambda>0$ is
\begin{equation}\label{initial-ham}
\mathscr{H}:=\int (\nabla a^{*})(\nabla a)(x)dx+\frac{\lambda}{2}\int\int a^{*}(x)a^*(y)\phi(x-y)a(x)a(y)dxdy\,,
\end{equation}
where 
reference to the integration domain $\Lambda:=\{x\in \mathbb{R}^d\,|\,|x_i|\leq \frac{L}{2}\,,\,i=1,2,\dots, d\}$ is omitted, periodic boundary conditions are assumed, and $dx$ is the Lebesgue measure in $d$ dimensions. Here the operators $a^{*}(x)\:,a(x)$ are the usual operator-valued distributions on $$\mathcal{F}:= \Gamma\left(L^{2}\left(\Lambda,\mathbb{C};dx\right)\right)\,$$  that satisfy the canonical
commutation relations
\[
[a^{\$}(x),a^{\$}(y)]=0,\quad\quad[a(x),a^{*}(y)]=\delta(x-y)\charf,
\]
with $a^{\$}:=a$ or $a^{*}$. In terms of the field modes they read
\[
a(x)=\sum_{\bold{j}\in \mathbb{Z}^d}\frac{a_{\bold{j}}e^{ik_{\bold{j}}\cdot x}}{|\Lambda|^{\frac{1}{2}}},\quad a^{*}(x)=\sum_{\bold{j}\in \mathbb{Z}^d}\frac{a_{\bold{j}}^*e^{-ik_{\bold{j}}\cdot x}}{|\Lambda|^{\frac{1}{2}}},\]
where $k_{\bold{j}}:=\frac{2\pi}{L} \bold{j}$, $\bold{j}=( j_1, \dots, j_d)$, $j_1, \dots, j_d \in \mathbb{Z}$, and $|\Lambda|=L^d$,  with CCR

\[
[a^{\$}_{\bold{j}},a^{\$}_{\bold{j}'}]=0,\quad\quad[a_{\bold{j}},a^{*}_{\bold{j}'}]=\delta_{\bold{j}\,,\,\bold{j}'}\,\,,\,\,\text{with}\quad\,a^{\$}_{\bold{j}}=a_{\bold{j}}\,\,\text{or}\,\,a^{*}_{\bold{j}}\,.
\]
The unique (up to a phase) vacuum vector of $\mathcal{F}$ is denoted by $\Omega$ ($\|\Omega\|=1$).

Given any function $\varphi\in L^{2}\left(\Lambda,\mathbb{C};dz\right)$, we express it in terms of its Fourier components $\varphi_{\bold{j}}$, i.e.,
\begin{equation}
\varphi(z)=\frac{1}{|\Lambda|}\sum_{\bold{j}\in \mathbb{Z}^d}\varphi_{\bold{j}}e^{ik_{\bold{j}}\cdot z}\,,
\end{equation}
and the Parseval identity reads
\begin{equation}
\int dz |\varphi|^2(z)=\frac{1}{|\Lambda|}\sum_{\bold{j}\in \mathbb{Z}^d}|\varphi_{\bold{j}}|^2<\infty\,.
\end{equation}

\begin{definition}\label{def-pot}
The potential $\phi(x-y)$  is a bounded, real-valued function that is periodic, i.e., $\phi(z)=\phi(z+\bold{j}L)$  for $\bold{j}\in \mathbb{Z}^{d}$, and satisfies  the following conditions:
\begin{enumerate}
%\item
%$\phi(x-y)\geq 0$
%\item
%$\phi(x-y)=\phi(y-x)$, hence we write $\phi(x-y)=\frac{1}{|\Lambda|}\sum_{\bold{j}\in \mathbb{Z}^3}\phi_{\bold{j}}e^{ik_{\bold{j}}(x-y)}$
\item  $\phi(z)$ is an even function, in consequence $\phi_{\bold{j}}=\phi_{-\bold{j}}$. 
\item 
$\phi(z)$ is of positive type, i.e., the Fourier components $\phi_{\bold{j}}$ are nonnegative.
\item
The pair interaction has a fixed but arbitrarily large ultraviolet cutoff (i.e., the nonzero Fourier components $\phi_{\bold{j}}$  form a finite set $\{\phi_{\bold{0}}, \phi_{\pm\bold{j}_1},\dots, \phi_{\pm\bold{j}_M}\}$)  with the requirement below to be satisfied:

 (\underline{Strong Interaction Potential Assumption})
The ratio $\epsilon_{\bold{j}}$ between the kinetic energy of the modes $\pm\bold{j}\neq \bold{0}$ and the corresponding Fourier component, $\phi_{\bold{j}}(\neq 0)$,  of the potential, i.e., $\frac{k_{\bold{j}}^2}{\phi_{\bold{j}}}=:\epsilon_{\bold{j}}$, is required to be small enough to ensure the estimates used in \cite{Pi1}. 
%Notice that $\epsilon_{\bold{j}}$ small corresponds either to a low energy mode  $\frac{2\pi\bold{j}}{L}$  or/and to a large potential $\phi_{\bold{j}}$.
%\item [ii)] (\underline{Non-connected Frequencies Assumption}) For any $l,l'$ in the set $\{1,\dots, M\}$, the modes
%\begin{equation}
%\bold{j}_{l}\pm \bold{j}_{l'}\quad \text{and}\quad -(\bold{j}_{l}\pm \bold{j}_{l'})
%\end{equation}
%do not belong to the set $\{\pm \bold{j}_1,\dots,\pm \bold{j}_M\}$.
\end{enumerate}

\end{definition}
%\begin{remark}
%The class of potentials chosen in Definition 1.1 is just to use only one method, i.e., the multi-scale analysis in the number of zero-mode particles using the Feshbach map. This method is quite effective with high potentials, because of the artificial gap induced by the chosen projection in the resolvents of the expansion: the higher is the potential with respect to the kinetic energy,  the higher is the gap. In fact,  the gap is the difference between the $\inf$ of the free Hamiltonian and the value of the spectral parameter $z\in \mathbb{R}, z<0$,  entering the Feshbach map. The procedure is expected to work for $z$ less or close to the Bogoliubov energy.
%\end{remark}
%\begin{remark}
%As it will be explained in the last section, in order to have an explicit expression (i.e., in terms of bare quantities) for the ground state --  we call this expression main term -- up to a remainder which is estimated in norm less than some fraction, say $1/16$, of the main term, we have to decompose the expansion into two steps (at least), by considering ``high" and ``low" components of the potential separately. Hopefully, at least in the case in which we have a bunch of high frequencies and another bunch of very low frequencies, the two sets of frequencies being well separated, this method should work.
%\end{remark}

We restrict $\mathscr{H}$ to the Fock subspace $\mathcal{F}^{N}$ of vectors with $N$ particles 
\begin{equation}
\mathscr{H}\upharpoonright_{\mathcal{F}^N}=\Big(\int (\nabla a^{*})(\nabla a)(x)dx+\frac{\lambda}{2}\int\int a^{*}(x)a^{*}(y) \phi(x-y)a(y)a(x)dxdy\Big)\upharpoonright_{\mathcal{F}^N}
\end{equation}
From now on, we shall study the Hamiltonian
\begin{equation}
H:=\int (\nabla a^{*})(\nabla a)(x)dx+\frac{\lambda}{2}\int\int a^{*}(x)a^{*}(y)\phi(x-y)a(y)a(x)dxdy+c_N\charf
\end{equation}
where $c_N:=\frac{\lambda\phi_{\bold{0}}}{2|\Lambda|}N-\frac{\lambda\phi_{\bold{0}}}{2|\Lambda|}N^2$ with $\bold{0}=(0,\dots,0)$, and it is always meant to be restricted to the subspace $\mathcal{F}^N$. Notice that
\begin{equation}
\mathscr{H}\upharpoonright_{\mathcal{F}^N}=(H-c_N\charf)\upharpoonright_{\mathcal{F}^N}\,.
\end{equation}

In the present paper we proceed with the study  of the Hamiltonian $H$ that we started in the companion papers \cite{Pi1}, \cite{Pi2}. 
 Here, we deal with the complete Hamiltonian of the system  in the limiting regime where the box size is fixed,  the particle density is large,  and  the coupling constant scales like the inverse of the particle density. In this regime we provide the construction of the ground state of the Hamiltonian $H$ under the \emph{Strong Interaction  Potential Assumption} (see 3. in Definition \ref{def-pot}) already used in the companion papers \cite{Pi1} and \cite{Pi2}.

For this result,  we have to finally control the so called ``cubic" and ``quartic" (in the nonzero modes)  terms in  the second quantized Hamiltonian of the system (see (\ref{cub-qua-in})-(\ref{cub-qua-fin})),  terms that are neglected in the corresponding Bogoliubov Hamiltonian. 

Like in the case of the Bogoliubov Hamiltonian (see \cite{Pi2}), with each couple  of Fourier components $\{\phi_{\bold{j}},\phi_{-\bold{j}}\}$ of the pair potential (see Definition \ref{def-pot}) we associate a Feshbach flow. The new terms in the interaction are controlled thanks to a refined choice of the projections associated with the Feshbach flows. At the first step of the Feshbach flow corresponding to the couple of modes $\{\bold{j},-\bold{j}\}$,  a new (perpendicular) projection projects out in one single step the subspace of vectors with a number of particles in the modes $\{\bold{j},-\bold{j}\}$ larger or equal to a minimum number that is chosen to be $\mathcal{O}( N^{\frac{1}{16}})$. 
In spite of this modification, the flow can be still controlled and the new terms of the interaction Hamiltonian, i.e., the  ``cubic" and ``quartic" ones, turn out to irrelevant. To show this we make use of the \emph{short range property of the potential in the particle states numbers} by which we mean the following:

\emph{Consider a vector $\psi\in \mathcal{F}^N$ obtained as product of single particle states of the type $a^{*}_{\bold{j}}\Omega$, and containing $N_{\bold{j}_*}$ particles in the modes $\pm \bold{j}_*$. Then, it is necessary to apply the Hamiltonian to $\psi$ at least $r$ times  in order to get a vector with a nonzero component in the subspace of vectors with $N_{\bold{j}_*}+2r$ or $N_{\bold{j}_*}-2r$ particles in the modes $\pm \bold{j}_*$.
}

This simple but crucial property would not be enough without a refined control of a three-modes Bogoliubov Hamiltonian (see \cite{Pi1} and \cite{Pi2}) that will be combined with the semigroup property of the Feshbach map. In this respect, the key result is Theorem \ref{theorem-junction} in Section \ref{rig-cos-H}.

\subsection{The Hamiltonian $H$ and the Hamiltonian $H^{Bog}$}\label{hamiltonians}
\setcounter{equation}{0}
%In the next formulae of this section, we use the notation that will be necessary for a general potential, that means for a potential where more than three modes have corresponding components different from zero (and positive).

Using the definitions
\begin{equation}
\label{b in the strip}
a_{+}(x):=\sum_{\bold{j}\in\mathbb{Z}^d\setminus \{\bold{0}\} }\frac{a_{\bold{j}}}{|\Lambda|^{\frac{1}{2}}}e^{ik_{\bold{j}}\cdot x}\quad\text{and}\quad
a_{\bold{0}}(x):=\frac{a_{\bold{0}}}{|\Lambda|^{\frac{1}{2}}}\,,
\end{equation}
 the Hamiltonian $H$ reads
\begin{eqnarray}
H& = & \sum_{\bold{j}\in  \mathbb{Z}^d} k^2_{\bold{j}}a_{\bold{j}}^{*}a_{\bold{j}}\\
% &  & +\frac{\lambda}{2}\int \int b^*|_{n,l+1}^{n,l}(x)b^*|_{n,l}^{0,0}(y)\phi(x-y)b|_{n,l+1}^{n,l}(x)b|_{n,l}^{0,0}(y)dxdy\\
& &+\frac{\lambda}{2}\int\int a^{*}_{+}(x)a^*_{+}(y)\phi(x-y)a_{+}(x)a_{+}(y)dxdy \\
& &+\lambda\int \int \{ a^*_{+}(x)a^*_{+}(y)\phi(x-y)a_{+}(x)a_{\bold{0}}(y)
 %&  & +\frac{\lambda}{2}\int \int a^*_{+}(x)a^*_{+}(y)\phi(x-y)a_{\bold{0}}(x)a_{+}(y)dxdy
  +h.c.\}dxdy\\
% &  & +\frac{\lambda}{2}\int \int a^*_{\bold{0}}(x)a^*_{+}(y)\phi(x-y)a_{+}(x)a_{+}(y)dxdy\\
 &  & +\frac{\lambda}{2}\int \int \{ a^*_{\bold{0}}(x)a^*_{\bold{0}}(y)\phi(x-y)a_{+}(x)a_{+}(y)
 + h.c.\}dxdy\\
 &  & +\lambda\int \int a^*_{\bold{0}}(x)a^*_{+}(y)\phi(x-y)a_{\bold{0}}(x)a_{+}(y)dxdy\\
 &  & +\lambda\int \int a^*_{\bold{0}}(x)a^*_{+}(y)\phi(x-y)a_{\bold{0}}(y)a_{+}(x)dxdy\\
& & +\frac{\lambda}{2}\int\int a^{*}_{\bold{0}}(x)a^*_{\bold{0}}(y)\phi(x-y)a_{\bold{0}}(x)a_{\bold{0}}(y)dxdy  \\
& &+c_N\charf\,.
\end{eqnarray}
Because of the implicit restriction to $\mathcal{F}^N$ and due to the choice of the constant $c_N$, it turns out that
\begin{eqnarray}
H& = & \sum_{\bold{j}\in  \mathbb{Z}^d}k^2_{\bold{j}}a_{\bold{j}}^{*}a_{\bold{j}}\\
% &  & +\frac{\lambda}{2}\int \int b^*|_{n,l+1}^{n,l}(x)b^*|_{n,l}^{0,0}(y)\phi(x-y)b|_{n,l+1}^{n,l}(x)b|_{n,l}^{0,0}(y)dxdy\\
%& &+ \frac{\lambda\phi_{\bold{0}}}{2|\Lambda|^2}\int\int a^{*}_{+}(x)a_{+}(y)dxdy
& &+\frac{\lambda}{2}\int\int a^{*}_{+}(x)a^*_{+}(y)\phi_{(\neq 0)}(x-y)a_{+}(x)a_{+}(y)dxdy  \\
& &+\lambda\int \int \{a^*_{+}(x)a^*_{+}(y)\phi_{(\neq 0)}(x-y)a_{+}(x)a_{\bold{0}}(y)
 %&  & +\frac{\lambda}{2}\int \int a^*_{+}(x)a^*_{+}(y)\phi(x-y)a_{\bold{0}}(x)a_{+}(y)dxdy\\
 + a^*_{+}(x)a^*_{\bold{0}}(y)\phi_{(\neq 0)}(x-y)a_{+}(x)a_{+}(y)\}dxdy\\
% &  & +\frac{\lambda}{2}\int \int a^*_{\bold{0}}(x)a^*_{+}(y)\phi(x-y)a_{+}(x)a_{+}(y)dxdy\\
 &  & +\frac{\lambda}{2}\int \int \{a^*_{\bold{0}}(x)a^*_{\bold{0}}(y)\phi_{(\neq 0)}(x-y)a_{+}(x)a_{+}(y)
 + a^*_{+}(x)a^*_{+}(y)\phi_{(\neq 0)}(x-y)a_{\bold{0}}(x)a_{\bold{0}}(y)\}dxdy\quad\quad\quad\quad\\
% &  & +\lambda\int \int a^*_{\bold{0}}(x)a^*_{+}(y)\phi(x-y)a_{\bold{0}}(x)a_{+}(y)dxdy\\
 &  & +\lambda\int \int a^*_{\bold{0}}(x)a^*_{+}(y)\phi_{(\neq 0)}(x-y)a_{\bold{0}}(y)a_{+}(x)dxdy
%& & +\frac{\lambda}{2}\int\int a^{*}_{\bold{0}}(x)a^*_{\bold{0}}(y)\phi(x-y)a_{\bold{0}}(x)a_{\bold{0}}(y)dxdy  \\
%& &+c_N
\end{eqnarray}
where $\phi_{(\neq 0)}(x-y):=\phi (x-y)-\phi_{(0)}(x-y)$ with $\phi_{(0)}(x-y):=\frac{\phi_{\bold{0}}}{|\Lambda|}$.

Next, we define the \emph{particle number preserving} Bogoliubov Hamiltonian
 \begin{eqnarray}
H^{Bog} &:=  &\sum_{\bold{j}\in  \mathbb{Z}^d}k^2_{\bold{j}}a_{\bold{j}}^{*}a_{\bold{j}}\\
& &+ \frac{\lambda}{2}\int \int a^*_{\bold{0}}(x)a^*_{\bold{0}}(y)\phi_{(\neq 0)}(x-y)a_{+}(x)a_{+}(y)dxdy\label{quartic-two-one}\\
& &+\frac{\lambda}{2}\int \int a^*_{+}(x)a^*_{+}(y)\phi_{(\neq 0)}(x-y)a_{\bold{0}}(x)a_{\bold{0}}(y)dxdy\label{quartic-two-two}\\
  &  & +\lambda\int \int a^*_{\bold{0}}(x)a^*_{+}(y)\phi_{(\neq 0)}(x-y)a_{\bold{0}}(y)a_{+}(x)dxdy\,
 \end{eqnarray}
that we can express in terms of the field modes
\begin{eqnarray}
H^{Bog}
&=&\sum_{\bold{j}\in\mathbb{Z}^d\setminus \{\bold{0}\}} (k^2_{\bold{j}}+\lambda\frac{\phi_{\bold{j}}}{|\Lambda|}a^*_{\bold{0}}a_{\bold{0}})a_{\bold{j}}^{*}a_{\bold{j}}+\frac{\lambda}{2}\sum_{\bold{j}\in \mathbb{Z}^d\setminus \{\bold{0}\}}\frac{\phi_{\bold{j}}}{|\Lambda|}\,\Big\{a^*_{\bold{0}}a^*_{\bold{0}}a_{\bold{j}}a_{-\bold{j}}+a^*_{\bold{j}}a^*_{-\bold{j}}a_{\bold{0}}a_{\bold{0}}\Big\}\,.
\end{eqnarray}
Hence, the Hamiltonian $H$ corresponds to
\begin{equation}\label{complete-H}
H=H^{Bog}+V\,
\end{equation}
with
\begin{eqnarray}
V& :=&\lambda\int \int a^*_{+}(x)a^*_{\bold{0}}(y)\phi_{(\neq 0)}(x-y)a_{+}(x)a_{+}(y)dxdy \label{cub-qua-in}\\
& &+\lambda\int \int a^*_{+}(x)a^*_{+}(y)\phi_{(\neq 0)}(x-y)a_{+}(x)a_{\bold{0}}(y)dxdy \\
& & +\frac{\lambda}{2}\int\int a^{*}_{+}(x)a^*_{+}(y)\phi_{(\neq 0)}(x-y)a_{+}(x)a_{+}(y)dxdy\,.\label{cub-qua-fin}
\end{eqnarray}

Following the convention of  \cite{Pi1}, we set 
\begin{equation}\label{definitions}
\lambda=\frac{1}{\rho}\quad,\quad N=:\rho |\Lambda|\,\,\, \text{and}\,\,\text{even},
\end{equation}
where $\rho>0$ is the particle density.

Assuming that
\begin{equation}
\phi(z)=\frac{1}{|\Lambda|}\phi_{\bold{0}}+\frac{1}{|\Lambda|}\sum_{m=1}^{M}\phi_{\bold{j}_m}(e^{ik_{\bold{j}_m}\cdot z}+ e^{-ik_{\bold{j}_m}\cdot z})\,
\end{equation}
with  $M<\infty$ and $\bold{j}_m\neq \bold{0}$, we define
\begin{eqnarray}
V_{\bold{j}_1,\dots \bold{j}_m}& :=&\sum_{l=1}^{m}V_{\bold{j}_l}\\
&:= &\frac{1}{N}\sum_{l=1}^{m}\sum_{\bold{j}\in \mathbb{Z}^d\setminus \{-\bold{j}_l,\bold{0}\}}a^*_{\bold{j}+\bold{j}_l}\,a^*_{\bold{0}}\,\phi_{\bold{j}_l}\,a_{\bold{j}}a_{\bold{j}_l}+h.c. \label{pair-1}\\
& &+\frac{1}{N}\sum_{l=1}^{m}\sum_{\bold{j}\in \mathbb{Z}^d\setminus \{\bold{j}_l,\bold{0}\}}a^*_{\bold{j}-\bold{j}_l}\,a^*_{\bold{0}}\,\phi_{\bold{j}_l}\,a_{\bold{j}}a_{-\bold{j}_l}+h.c. \label{pair-1-bis}\\
& &+\frac{1}{N}\sum_{l=1}^{m}\sum_{\bold{j}\in \mathbb{Z}^d\setminus\{ -\bold{j}_l,\bold{0}\}}\,\sum_{\bold{j}'\in \mathbb{Z}^d\setminus \{\bold{j}_l,\bold{0}\}}a^*_{\bold{j}+\bold{j}_l}\,a^*_{\bold{j}'-\bold{j}_l}\,\phi_{\bold{j}_l}\,a_{\bold{j}}a_{\bold{j}'} \,.\label{pair-2}
\end{eqnarray}
Consequently, we can write
\begin{eqnarray}
V_{\bold{j}_1,\dots \bold{j}_M}& \equiv&\frac{1}{\rho}\int \int a^*_{+}(x)a^*_{\bold{0}}(y)\phi_{(\neq 0)}(x-y)a_{+}(x)a_{+}(y)dxdy \\
& &+\frac{1}{\rho}\int \int a^*_{+}(x)a^*_{+}(y)\phi_{(\neq 0)}(x-y)a_{+}(x)a_{\bold{0}}(y)dxdy \\
& & +\frac{1}{2\rho}\int\int a^{*}_{+}(x)a^*_{+}(y)\phi_{(\neq 0)}(x-y)a_{+}(x)a_{+}(y)dxdy\,.
%&=&\frac{1}{N}\sum_{l=1}^{m}\sum_{\bold{j}\in \mathbb{Z}^d}a^*_{\bold{j}+\bold{j}_l}\,a^*_{\bold{0}}\,\phi_{\bold{j}_l}\,a_{\bold{j}}a_{\bold{j}_l}+h.c. \label{def-v1...m}\\
%& &+\frac{1}{N}\sum_{l=1}^{m}\sum_{\bold{j}\in \mathbb{Z}^d}\,\sum_{\bold{j}'\in \mathbb{Z}^d}a^*_{\bold{j}+\bold{j}_l}\,a^*_{\bold{j}'-\bold{j}_l}\,\phi_{\bold{j}_l}\,a_{\bold{j}}a_{\bold{j}'} \label{def-v1...m-bis}\,,
\end{eqnarray}

{\bf{Notation}}

\begin{enumerate}
\item
The symbol $\charf$ stands for the identity operator. If helpful we specify the Hilbert space where it acts. For $c-$number operators, e.g., $z\charf$, we may omit the symbol $\charf$.
\item
The symbol $\langle\,\,,\,\,\rangle$ stands for the scalar product in $\mathcal{F}^N$.
\item
The symbol $\mathcal{O}(\alpha)$ stands for a quantity bounded in absolute value by a constant times $\alpha$ ($\alpha>0$). The symbol $o(\alpha)$ stands for a quantity such that $o(\alpha)/\alpha \to 0$ as $\alpha \to 0$. Throughout the paper the implicit multiplicative constants are always independent of $N$. 
%The symbol $\mathcal{O}(\alpha)$ stands for a quantity bounded in absolute value by a constant times $\alpha$ ($\alpha>0$). Throughout the paper the implicit multiplicative constant in the symbol $\mathcal{O}(\alpha)$ is always independent of $N$. 
%In certain cases we shall use explicit constants if the same quantity will be used in later proofs.
\item
In some cases we shall use explicit constants, e.g., $C^{\#}_I$, if the same quantity will be used in later proofs. Unless otherwise specified or unless it is obvious from the context, the explicit constants may depend on the size of the box and on the details of the potential, in particular on the number, $M$,  of couples of nonzero frequency components ($\neq 0$) in the Fourier expansion of the pair potential.
% and we also mention the dependence on parameters of the system (eg., $\Delta_0:\min \{(k_{\bold{j}})^2\,;\,\bold{j}\in \mathbb{Z}^d\setminus \{\bold{0}\}\}$).
%\item
%The symbol $o(\alpha)$ stands for a quantity such that $o(\alpha)/\alpha \to 0$ as $\alpha \to 0$. Throughout the paper the implicit multiplicative constant in the symbol $o(\alpha)$ is always independent of $N$. 

\item
The symbol $|\psi \rangle \langle \psi|$, with $\|\psi\|=1$,  stands for the one-dimensional projection onto the state $\psi$.
\item
The word mode for the wavelength $\frac{2\pi}{L}\bold{j}$ (or simply for $\bold{j}$) refers to the field mode associated with it.
\item
Theorems and lemmas from the companions papers \cite{Pi1} and \cite{Pi2} are underlined, quoted in italic, and with the numbering that they have in the corresponding paper; e.g., \emph{\underline{Theorem 3.1} of \cite{Pi1}}.
\end{enumerate}

%\begin{remark}
%The Hamiltonian $\mathscr{H}\upharpoonright_{\mathcal{F}^N}$ coincides with the Hamiltonian in Eq. (1) of Seiringer's paper for 
%\begin{itemize}
%\item
%$\lambda=\frac{1}{N-1}$ and $m=\frac{1}{2}$
%\item
% $\Lambda\equiv$ the box of side $L=1$
% \item
%$ \phi(x-y)\equiv v(x-y)$.
%\end{itemize}
%Notice also that for $|\Lambda|=1$ and $\lambda=\frac{1}{N-1}$
%\begin{equation}
%-c_N=\frac{N^2\phi_{\bold{0}}}{2(N-1)}-\frac{N\phi_{\bold{0}}}{2(N-1)}
%\end{equation}
%\end{remark}

\section{Multi-scale analysis in the particle states occupation numbers  for the Bogoliubov  Hamiltonian: Review of results}\label{multiscale-HBog}
\setcounter{equation}{0}
%Since the momentum operator $ \sum_{\bold{j}\in  \mathbb{Z}^3}k_{\bold{j}}a_{\bold{j}}^{*}a_{\bold{j}}$ commutes with $H$ and $H^{Bog}$, it is convenient to consider the decomposition of $\mathcal{F}^N$ into sectors $\mathcal{F}^N_{P}$ where $P$ is the sum of a collection of $k_{\bold{j}}$. Then, we consider the Hamiltonians $H_P$, $H^{Bog}_P$, and the interaction $\Delta H_P$ at any fixed total momentum $P$. 
This section serves the purpose of collecting  formulae and algorithms derived in \cite{Pi1} and \cite{Pi2}. We shall refer to them in Section  \ref{groundstateH}  where the Feshbach flow associated with the Hamiltonian $H$ is defined. For the details of the strategy the reader is encouraged to consult the companion papers \cite{Pi1}, \cite{Pi2}. %where motivations and tools are described in full detail.
\\

\subsection{The Feshbach flow associated with a three-modes system  Hamiltonian $H^{Bog}_{\bold{j}^*}$}\label{informal-0}
We started our study in \cite{Pi1} from the  \emph{three-modes Bogoliubov Hamiltonian}
\begin{eqnarray}
H^{Bog}_{\bold{j}_{*}}
&:=&\sum_{\bold{j}\in\mathbb{Z}^d\setminus \{\bold{0}\,;\, \pm\bold{j}_{*} \}} k^2_{\bold{j}}a_{\bold{j}}^{*}a_{\bold{j}}+\hat{H}^{Bog}_{\bold{j}_{*}}
%+\phi_{\bold{j}_{*}}\frac{a^*_{\bold{0}}a_{\bold{0}}}{N}a_{\bold{j}_{*}}^{*}a_{\bold{j}_{*}}+\phi_{\bold{j}_{*}}\frac{a^*_{\bold{0}}a_{\bold{0}}}{N}a_{-\bold{j}_{*}}^{*}a_{-\bold{j}_{*}}+\frac{\phi_{\bold{j}}}{N}\,\Big\{a^*_{\bold{0}}a^*_{\bold{0}}a_{\bold{j}_{*}}a_{-\bold{j}_{*}}+a^*_{\bold{j}_{*}}a^*_{-\bold{j}_{*}}a_{\bold{0}}a_{\bold{0}}\Big\}\,\quad
\end{eqnarray}
where (see the definitions in (\ref{H0j})-(\ref{interaction-terms}))
\begin{equation}\label{check-HBogj}
\hat{H}^{Bog}_{\bold{j}_{*}}:=\hat{H}^0_{\bold{j} _{*}} +W_{\bold{j}_{*}}+W^*_{\bold{j}_{*}}\,
\end{equation}
involves the three modes $\bold{0},\pm\bold{j}_{*}$ only. Therefore, $H^{Bog}_{\bold{j}_{*}}$ is the sum of:
\begin{itemize}
\item
The operator
 \begin{equation}\label{H0j}
\hat{H}^0_{\bold{j}_{*}}:=(k^2_{\bold{j} _{*}}+\phi_{\bold{j} _{*}}\frac{a^*_{\bold{0}}a_{\bold{0}}}{N})a_{\bold{j} _{*}}^{*}a_{\bold{j} _{*}}+(k^2_{\bold{j} _{*}}+\phi_{\bold{j} _{*}}\frac{a^*_{\bold{0}}a_{\bold{0}}}{N})a_{-\bold{j} _{*}}^{*}a_{-\bold{j} _{*}}
\end{equation}
commuting with each number operator $a^*_{\bold{j}}a_{\bold{j}}$;
\item
The interaction terms
\begin{equation}\label{interaction-terms}
\phi_{\bold{j} _{*}}\frac{a^*_{\bold{0}}a^*_{\bold{0}}a_{\bold{j} _{*}}a_{-\bold{j} _{*}}}{N}=:W_{\bold{j} _{*}}\,\quad,\quad
\phi_{\bold{j} _{*}}\frac{a_{\bold{0}}a_{\bold{0}}a^*_{\bold{j} _{*}}a^*_{-\bold{j} _{*}}}{N}=:W^*_{\bold{j} _{*}}\,
\end{equation} 
changing the number of particles in the three modes $\bold{j}=\bold{0},\pm\bold{j} _{*} $;
\item
The kinetic energy $\sum_{\bold{j}\in\mathbb{Z}^d\setminus \{\bold{0};\pm\bold{j}_{*} \}} k^2_{\bold{j}}a_{\bold{j}}^{*}a_{\bold{j}}$ of the noninteracting modes.
\end{itemize}
The  identity $H^{Bog}=\frac{1}{2}\sum_{\bold{j}\in\mathbb{Z}^d\setminus \{\bold{0}\}}\hat{H}^{Bog}_{\bold{j}}$ follows from the previous definitions.
%\begin{equation}
%H_0=\frac{1}{2}\sum_{\bold{j}\in\mathbb{Z}^3\setminus \{\bold{0}\}}\hat{H}^{0}_{\bold{j}}\,\quad,\quad
%H^{Bog}=\frac{1}{2}\sum_{\bold{j}\in\mathbb{Z}^d\setminus \{\bold{0}\}}\hat{H}^{Bog}_{\bold{j}}\,.
%\end{equation}

%Now, we focus on a three-modes system where $\phi_{\bold{j}}\neq 0$ only for $\bold{j}\equiv \pm \bold{j}_{*}$ and $\bold{j}=0$, and we construct the ground state of the corresponding Bogoliubov Hamiltonian:
%Notice however that because of the selection rules on the particle momentum, the operators $V$, $V^*$, and $U$ are identically zero if the Hilbert space contains only the modes $ \{\bold{0};\bold{j};-\bold{j} \}$.  
%The treatment of the Hamiltonian $H$ in Section \ref{multiscale-H} is very similar but the algebra is slightly more involved. 
%\begin{eqnarray}
%H^{Bog}
%&=&\sum_{\bold{j}\in\mathbb{Z}^3\setminus \{\bold{0}\}} (\frac{k^2_{\bold{j}}}{2m}+\lambda\frac{\phi_{\bold{j}}}{|\Lambda|}a^*_{\bold{0}}a_{\bold{0}})a_{\bold{j}}^{*}a_{\bold{j}}+\frac{\lambda}{2}\sum_{\bold{j}\in \mathbb{Z}^3\setminus \{\bold{0}\}}\frac{\phi_{\bold{j}}}{|\Lambda|}\,\Big\{a^*_{\bold{0}}a^*_{\bold{0}}a_{\bold{j}}a_{-\bold{j}}+a^*_{\bold{j}}a^*_{-\bold{j}}a_{\bold{0}}a_{\bold{0}}\Big\}\,.
%\end{eqnarray}

\begin{remark}
We stress that $H^{Bog}_{\bold{j}_{*}}$ contains the kinetic energy corresponding to all the modes whereas $\hat{H}^{Bog}_{\bold{j}_{*}}$ contains the kinetic energy associated with the interacting modes only.
\end{remark} The multi-scale analysis in the occupation numbers of particle states relies on a novel application of Feshbach map and yields a trivial effective Hamiltonian (i.e., a multiple of a one dimensional orthogonal projection) in a neighborhood of the ground state energy. 
\\

For the Hamiltonian $H^{Bog}_{\bold{j}_*}$ applied to $\mathcal{F}^{N}$, we define:
\begin{itemize}
\item
$Q^{(0,1)}_{\bold{j}_*}:=$
the projection (in $\mathcal{F}^{N}$) onto the subspace generated by vectors with $N-0=N$ or $N-1$ particles in the modes $\bold{j}_{*}$ and $-\bold{j}_{*}$, i.e., the operator $a^*_{\bold{j}_*}a_{\bold{j}_*}+a^*_{-\bold{j}_*}a_{-\bold{j}_*}$ has eigenvalues $N$ and $N-1$ when restricted to $Q^{(0,1)}_{\bold{j}_*}\mathcal{F}^N$;
\item
$Q^{(>1)}_{\bold{j}_*}:=$ the projection onto the orthogonal complement of $Q^{(0,1)}_{\bold{j}_*}\mathcal{F}^{N}$ in $\mathcal{F}^{N}$.
\end{itemize}
 Therefore, we have $$Q^{(0,1)}_{\bold{j}_*}+Q^{(>1)}_{\bold{j}_*}=\charf_{\mathcal{F}^{N}}\,.$$ 
 
Analogously, starting from $i=2$ up to $i=N-2$ with $i$ even, we define 
$Q^{(i, i+1)}_{\bold{j}_*}$
the projection onto the subspace of $Q^{(>1)}_{\bold{j}_*}\mathcal{F}^{N}$ spanned by the vectors with $N-i$ or $N-i-1$ particles in the modes $\bold{j}_{*}$ and $-\bold{j}_{*}$.  Analogously, $Q^{(>i+1)}_{\bold{j}_*}$ is the projection onto the orthogonal complement of $Q^{(i, i+1)}_{\bold{j}_*}Q^{(>i-1)}_{\bold{j}_*}\mathcal{F}^{N}$ in $Q^{(>i-1)}_{\bold{j}_*}\mathcal{F}^{N}$, i.e.,
\begin{equation}
Q^{(>i+1)}_{\bold{j}_*}+Q^{(i,i+1)}_{\bold{j}_*}=Q^{(>i-1)}_{\bold{j}_*}\,.
\end{equation}
%\begin{remark}
%Notice that if we denote by $\mathcal{F}^{N}_{\{\bold{0};\bold{j}_{*};-\bold{j}_{*}\}}$ the subspace of vectors that contain only particles in the modes $\{\bold{0};\bold{j}_{*};-\bold{j}_{*}\}$, then $Q^{(>N-1)}\mathcal{F}^{N}_{\{\bold{0};\bold{j}_{*};-\bold{j}_{*}\}}\equiv Q^{(N)}\mathcal{F}^{N}_{\{\bold{0};\bold{j}_{*};-\bold{j}_{*}\}}$ where the R-H-S is the one-dimensional subspace generated by the state $$\eta:=\frac{1}{\sqrt{N!}}a_{\bold{0}}^*\dots a_{\bold{0}}^*\Omega$$ with all the $N$ particles in the zero-mode state. 
%\end{remark}

We recall that given the (separable) Hilbert space $\mathcal{H}$,  the projections $\mathscr{P}$, $\overline{\mathscr{P}}$ ($\mathscr{P}=\mathscr{P}^2$, $\overline{\mathscr{P}}=\overline{\mathscr{P}}^2$) where $\mathscr{P}+\overline{\mathscr{P}}=\charf_{\mathcal{H}}$, and a closed operator $K-z $, $z$ in a subset of $ \mathbb{C}$,  acting on $\mathcal{H}$, the Feshbach map associated with the couple $\mathscr{P}\,,\,\overline{\mathscr{P}}$ maps $K-z$ to the operator $\mathscr{F}(K-z)$ acting on $\mathscr{P}\mathcal{H}$ where  (formally)
\begin{equation}
\mathscr{F}(K-z):=\mathscr{P}(K-z)\mathscr{P}-\mathscr{P}K\overline{\mathscr{P}}\frac{1}{\overline{\mathscr{P}}(K-z)\overline{\mathscr{P}}}\overline{\mathscr{P}}K\mathscr{P}\,.
\end{equation}

We iterate the Feshbach map starting from $i=0$ up to $i=N-2$ with $i$ even, using the projections $\mathscr{P}^{(i)}$ and $\overline{{\mathscr{P}}^{(i)}}$  for the i-th  step of the iteration where
\begin{equation}\label{projections}
\mathscr{P}^{(i)}:= Q^{(>i+1)}_{\bold{j}_*}\quad,\quad \overline{{\mathscr{P}}^{(i)}}:= Q^{(i,i+1)}_{\bold{j}_*}\,.
\end{equation}
We denote by  $\mathscr{F}^{(i)}$ the Feshbach map at the i-th step. We start applying $\mathscr{F}^{(0)}$ to $H^{Bog}_{\bold{j}_*}-z$ where $z\in \mathbb{R}$ ranges in the interval $(-\infty, z_{max})$ with $z_{max}$ larger but very close to
\begin{equation}
E^{Bog}_{\bold{j}_*}:=-\Big[k^2_{\bold{j}_*}+\phi_{\bold{j}_*}-\sqrt{(k^2_{\bold{j}_*})^2+2\phi_{\bold{j}_*}k^2_{\bold{j}_*}}\Big]\,. 
\end{equation}
More precisely, for $0\leq i \leq N-2$, we consider \begin{equation}\label{range-z}
z\leq E^{Bog}_{\bold{j}_*}+ (\delta-1)\phi_{\bold{j}^*}\sqrt{\epsilon_{\bold{j}_*}^2+2\epsilon_{\bold{j}_*}}\quad,\quad \epsilon_{\bold{j}_*}:=\frac{k^2_{\bold{j}_*}}{\phi_{\bold{j}_*}}\,,
\end{equation} with $\delta=  1+\sqrt{\epsilon_{\bold{j}_*}}$, $\frac{1}{N}\leq \epsilon^{\nu}_{\bold{j}_*}$ for some $\nu >1$, and $\epsilon_{\bold{j}_*}$ sufficiently small; see point 3. in Definition \ref{def-pot}.

As a result of the flow (for details see \cite{Pi1} or \emph{\underline{Section 2.1} of \cite{Pi2}}), for $i=2,4,6,\dots,N-2$ we obtain  the Feshbach Hamiltonians 
\begin{eqnarray}\label{KappaBog-i}
\mathscr{K}^{Bog\,(i)}_{\bold{j}_*}(z)&=&Q^{(>i+1)}_{\bold{j}_*}(H^{Bog}_{\bold{j}_*}-z)Q^{(>i+1)}_{\bold{j}_*}\\
%& &-\sum_{l_{i-1}=0}^{\infty}Q^{(>i)}_{\bold{j}_*}W_{\bold{j}_*}\,Q^{(i-1)}_{\bold{j}_*}R^{Bog}_{\bold{j}_*\,;\,i-1,i-1}(z)\,\Big[\Gamma^{Bog\,}_{\bold{j}_*\,;\,i-1,i-1}(z)R^{Bog}_{\bold{j}_*\,;\,i-1,i-1}(z)\Big]^{l_{i-1}}\,Q^{(i-1)}_{\bold{j}_*}W^*_{\bold{j}_*}Q^{(>i)}_{\bold{j}_*} \nonumber\\
& &-\sum_{l_i=0}^{\infty}Q^{(>i+1)}_{\bold{j}_*}W_{\bold{j}_*}\,R^{Bog}_{\bold{j}_*\,;\,i,i}(z)\Big[\Gamma^{Bog\,}_{\bold{j}_*\,;\,i,i}(z)R^{Bog}_{\bold{j}_*\,;\,i,i}(z)\Big]^{l_i}W^*_{\bold{j}_*}Q^{(>i+1)}_{\bold{j}_*} \nonumber
\end{eqnarray}
where  we have used the notation:
\begin{itemize}
\item
$$W_{\bold{j}_*\,;\,i,i'}:=Q^{(i,i+1)}_{\bold{j}_*}W_{\bold{j}_*}Q^{(i',i'+1)}_{\bold{j}_*}\quad,\quad W^*_{\bold{j}_*\,;\,i,i'}:=Q^{(i,i+1)}_{\bold{j}_*}W^*_{\bold{j}_*}Q^{(i',i'+1)}_{\bold{j}_*}$$ and  $$
R^{Bog}_{\bold{j}_*\,;\,i,i}(z):=Q^{(i,i+1)}_{\bold{j}_*}\frac{1}{Q^{(i,i+1)}_{\bold{j}_*}(H^{Bog}_{\bold{j}_*}-z)Q^{(i,i+1)}_{\bold{j}_*}}Q^{(i,i+1)}_{\bold{j}_*}\,;$$
\item \begin{equation}\label{GammaBog-2}
\Gamma^{Bog\,}_{\bold{j}_*\,;\,2,2}(z):=W_{\bold{j}_*\,;\,2,0}\,R_{\bold{j}_*\,;\,0,0}^{Bog}(z)W_{\bold{j}_*\,;\,0,2}^*\,;
\end{equation}
\item
 for $i\geq 4$,
%\begin{equation}
%\Gamma^{Bog\,}_{i-1,i-1}(z):=W_{i-1,i-3}\,R^{Bog}_{i-3,i-3}(z) \sum_{l_{i-3}=0}^{\infty}\Big[\Gamma^{Bog\,}_{i-3,i-3}(z)R^{Bog}_{i-3,i-3}(z)\Big]^{l_{i-3}}W^*_{i-3,i-1}
%\end{equation}
\begin{eqnarray}\label{GammaBog-i}
\Gamma^{Bog\,}_{\bold{j}_*\,;\,i,i}(z)&:=&W_{\bold{j}_*\,;\,i,i-2}\,R^{Bog}_{\bold{j}_*\,;\,i-2,i-2}(z) \sum_{l_{i-2}=0}^{\infty}\Big[\Gamma^{Bog}_{\bold{j}_*\,;\,i-2,i-2}(z)R^{Bog}_{\bold{j}_*\,;\,i-2,i-2}(z)\Big]^{l_{i-2}}W^*_{\bold{j}_*\,;\,i-2,i}\\
&=&W_{\bold{j}_*\,;\,i,i-2}\,(R^{Bog}_{\bold{j}_*\,;\,i-2,i-2}(z))^{\frac{1}{2}} \sum_{l_{i-2}=0}^{\infty}\Big[(R^{Bog}_{\bold{j}_*\,;\,i-2,i-2}(z))^{\frac{1}{2}}\Gamma^{Bog}_{\bold{j}_*\,;\,i-2,i-2}(z)(R^{Bog}_{\bold{j}_*\,;\,i-2,i-2}(z))^{\frac{1}{2}}\Big]^{l_{i-2}}\times \quad\quad\quad\\
& &\quad\quad\quad \times (R^{Bog}_{\bold{j}_*\,;\,i-2,i-2}(z))^{\frac{1}{2}}W^*_{\bold{j}_*\,;\,i-2,i}\,.\nonumber
\end{eqnarray}
\end{itemize}

For the last implementation of the Feshbach flow a new couple of projections is considered: $\mathscr{P}_{\eta}:=|\eta \rangle \langle \eta |$  (where $\eta$ is the normalized state with all the particles in the zero mode)  and $\overline{\mathscr{P}_{\eta}}$ such that
\begin{equation}
\mathscr{P}_{\eta}+\overline{\mathscr{P}_{\eta}}=\charf_{Q^{(>N-1)}_{\bold{j}_*}\mathcal{F}^N}\,.
\end{equation}
We notice that for $i=N-2$ the projection $Q^{(>i+1\equiv N-1)}_{\bold{j}_*}$ coincides with the projection onto the subspace where less than $N-i-1=N-N+1=1$ particles in the modes $\bold{j}_{*}$ and $-\bold{j}_{*}$ are present, i.e., where no particle in the modes $\bold{j}_{*}$ and $-\bold{j}_{*}$ is present.  

%\noindent
%Including this last step,   in \cite{Pi1} the flow of Feshbach Hamiltonians is defined for $z$ ranging between $-\infty$ and $z_{max}$ where $z_{max}$  is strictly below the expected excitation spectrum of the Hamiltonian $H^{Bog}_{\bold{j}_*}$ and  very close but still above the Bogoliubov energy 
 %if the spectral parameter $z(\in \mathbb{R})$. Indeed,} the estimates are such that %Therefore, for the three-modes system we have
%\begin{equation}
%E^{Bog}_{\bold{j}_{*}}:=-\Big[k^2_{\bold{j}_*}+\phi_{\bold{j}_*}-\sqrt{(k^2_{\bold{j}_*})^2+2\phi_{\bold{j}_*}k^2_{\bold{j}_*}}\Big]\,.
%\end{equation}

Starting from the formal expression
\begin{eqnarray}
& &\mathscr{K}^{Bog\,(N)}_{\bold{j}_*}(z) \\
&:=&\mathscr{F}^{(N)}(\mathscr{K}^{Bog\,(N-2)}_{\bold{j}_*}(z))\\
&=&\mathscr{P}_{\eta}(H^{Bog}_{\bold{j}_*}-z)\mathscr{P}_{\eta}\label{K-last-step-0}\\
& &-\mathscr{P}_{\eta}W_{\bold{j}_*}\,R^{Bog}_{\bold{j}_*\,;\,N-2,N-2}(z)\sum_{l_{N-2}=0}^{\infty}[\Gamma^{Bog}_{\bold{j}_*\,;\,N-2,N-2}(z) R^{Bog}_{\bold{j}_*\,;,N-2,N-2}(z)]^{l_{N-2}}\, W^*_{\bold{j}_*}\mathscr{P}_{\eta}\quad \nonumber\\
& &-\mathscr{P}_{\eta}W_{\bold{j}_*}\,\overline{\mathscr{P}_{\eta}}\,\frac{1}{\overline{\mathscr{P}_{\eta}}\mathscr{K}^{Bog\,(N-2)}_{\bold{j}_*}(z)\overline{\mathscr{P}_{\eta}}}\overline{\mathscr{P}_{\eta}}W^*_{\bold{j}_*}\mathscr{P}_{\eta}\,, \nonumber
\end{eqnarray}
the argument implemented  in \cite{Pi1} shows that
the expression on the R-H-S of (\ref{K-last-step-0}) is well defined  for 
$z$ such that
\begin{equation}\label{range-z}
z<\min\,\Big\{ z_{*}+\frac{\Delta_0}{2}\,;\,E^{Bog}_{\bold{j}_*}+ \sqrt{\epsilon_{\bold{j}_*}}\phi_{\bold{j}^*}\sqrt{\epsilon_{\bold{j}_*}^2+2\epsilon_{\bold{j}_*}}\Big\}\,
\end{equation}
where $z_*$ is the unique solution of $f_{\bold{j}_*}(z)=0$ with
\begin{equation}\label{fp-function}
f_{\bold{j}_*}(z):=-z-\langle \eta\,,\,W_{\bold{j}_*}\,R^{Bog}_{\bold{j}_*\,;\,N-2,N-2}(z)\sum_{l_{N-2}=0}^{\infty}[\Gamma^{Bog}_{\bold{j}_*\,;\,N-2,N-2}(z) R^{Bog}_{\bold{j}_*\,;\,N-2,N-2}(z)]^{l_{N-2}}\,W^*_{\bold{j}_*}\eta\rangle,
\end{equation}
Indeed, for $z$ in the range in (\ref{range-z}) the operator
\begin{equation}
\overline{\mathscr{P}_{\eta}}\mathscr{K}^{Bog\,(N-2)}_{\bold{j}_*}(z)\overline{\mathscr{P}_{\eta}}
\end{equation}
is bounded invertible in $\overline{\mathscr{P}_{\eta}}\mathcal{F}^N$. 
%for $1\gg \mu >0$. 
%$\Delta_0=\min\, \Big\{(k_{\bold{j}})^2\,|\,\bold{j}\in \mathbb{Z}^d\setminus\{\bold{0}\}\Big\}$,  and $(1-\frac{\phi_{\bold{j}_{*}}}{\Delta_0}\frac{ N^{\mu}  }{N})$ is assumed larger than $\frac{1}{2}$.
%But this also implies that  the expression in (\ref{absent}) is identically zero because $\overline{\mathscr{P}_{\eta}}W^*_{\bold{j}_*}\mathscr{P}_{\eta}=0$. 
Finally,
%Let $\eta$ be the (normalized) vector where all the $N$ particles are in the zero-mode state. We want to construct the Feshbach Hamiltonian obtained from $\mathscr{K}^{Bog\,(N-1)}(z)$ using the projection $\mathscr{P}_{\eta}:=|\eta \rangle \langle \eta |$ and the projection $\overline{\mathscr{P}_{\eta}}$ such that
%\begin{equation}
%\mathscr{P}_{\eta}+\overline{\mathscr{P}_{\eta}}=\charf_{Q^{(>N-1)}\mathcal{F}^N}
%\end{equation}
the identitites \begin{equation}\label{domain-z}
\mathscr{P}_{\eta}(H^{Bog}_{\bold{j}_*}-z)\mathscr{P}_{\eta}=-z\mathscr{P}_{\eta}\quad,\quad \overline{\mathscr{P}_{\eta}}W^*_{\bold{j}_*}\mathscr{P}_{\eta}=\mathscr{P}_{\eta}W_{\bold{j}_*}\overline{\mathscr{P}_{\eta}}=0
\end{equation} 
imply
\begin{equation}
\mathscr{K}^{Bog\,(N)}_{\bold{j}_*}(z)=f_{\bold{j}_*}(z)|\eta \rangle \langle \eta |\,.
\end{equation}
%where \footnote{In \cite{Pi1} the fixed point problem $f_{\bold{j}_*}(z)=0$ is actually solved first and then it is deduced that  the inverse $\overline{\mathscr{P}_{\eta}}\mathscr{K}^{Bog\,(N-2)}_{\bold{j}_*}(z)\overline{\mathscr{P}_{\eta}}$ (in the space $\overline{\mathscr{P}_{\eta}}\mathcal{F}^{(N)}$) is well defined in the given domain for $z$ (see (\ref{domain-z})). }
%The latter trivially implies $\mathscr{K}^{Bog\,(N-2)}_{\bold{j}_*}(z)=f_{\bold{j}_*}(z)|\eta \rangle \langle \eta |$.} 

The ground state energy, $z_*$,  and the (non-normalized) ground state vector of the Hamiltonian $H^{Bog}_{\bold{j}_*}$ are then obtained exploiting Feshbach theory:
\begin{eqnarray}
\psi^{Bog}_{\bold{j}_*}&:=&\eta \label{gs-1}\\
& &-\frac{1}{Q^{(N-2,N-1)}_{\bold{j}_*}\mathscr{K}^{Bog\,(N-4)}_{\bold{j}_*}(z_*)Q^{(N-2,N-1)}_{\bold{j}_*}}Q^{(N-2, N-1)}_{\bold{j}_*}W^*_{\bold{j}_*}\eta \label{gs-1-2}\\
& &-\sum_{j=2}^{N/2}\prod^{2}_{r=j}\Big[-\frac{1}{Q^{(N-2r,N-2r+1)}_{\bold{j}_*}\mathscr{K}^{Bog\,(N-2r-2)}_{\bold{j}_*}(z_*)Q^{(N-2r,N-2r+1)}_{\bold{j}_*}}W^*_{\bold{j}_*\,;\,N-2r,N-2r+2}\Big]\times \label{gs-2}\quad\quad\quad\\
& &\quad\quad\quad\quad\quad \quad\quad\times \frac{1}{Q^{(N-2,N-1)}_{\bold{j}_*}\mathscr{K}^{Bog\,(N-4)}_{\bold{j}_*}(z_*)Q^{(N-2,N-1)}_{\bold{j}_*}}Q^{(N-2, N-1)}_{\bold{j}_*}W^*_{\bold{j}_*}\eta \nonumber
\end{eqnarray}
where $\mathscr{K}^{Bog\,(-2)}_{\bold{j}_*}(z_*):=H^{Bog}_{\bold{j}_*}-z_*$.
\noindent
%By the expansion of 
%\begin{equation}
%R^{Bog}_{N-2,N-2}(z)\sum_{l_{N-2}=0}^{\infty}[\Gamma^{Bog}_{N-2,N-2}(z) R^{Bog}_{N-2,N-2}(z)]^{l_{N-2}}\
%\end{equation}
%it will be proven (--to be done, but see also my comment in Remark \ref{expansions}--) that for $z<\xi^{Bog} E^{Bog}$ there exists a unique $z_*$ such that
%\begin{eqnarray}
%z_*&=&-\langle \eta\,,\,W\,Q^{(N-2)}\,R^{Bog}_{N-2,N-2}(z_*)\sum_{l_{N-2}=0}^{\infty}[\Gamma^{Bog}_{N-2,N-2}(z_*)R^{Bog}_{N-2,N-2}(z_*)]^{l_{N-2}}\, Q^{(N-2)}W^*\eta\rangle \quad \label{fixed-point}
%&&-\langle \eta\,,\,\mathscr{V}^{(N-1)}(z_*)\,R_{N-1,N-1}(z_*)\sum_{l_{N-1}=0}^{\infty}[\Gamma_{N-1,N-1}(z_*) R_{N-1,N-1}(z_*)]^{l_{N-1}}\,(\mathscr{V}^{(N-1)}(z_*) )^*\eta\rangle\quad\quad
%\end{eqnarray}
%and that (as we already know from Seiringer's paper)  $z_*=E^{Bog}+o(1)$.
%Then, the condition in (\ref{condition}) is fulfilled for $\xi$ sufficiently close to $1$ and $\mathscr{F}_{\eta}(\mathscr{K}^{Bog\,(N-1)}(z))(z_*)$ is well defined.
%Moreover, because of its uniqueness we can conclude that $z_*$  is the ground state energy of $\mathscr{F}_{\eta}(\mathscr{K}^{Bog\,(N-1)}(z))(z_*)$.
The norm of the sum in (\ref{gs-2})  is bounded by a multiple of
\begin{eqnarray}\label{convergent-series}
& &\sum_{j=2}^{\infty}c_j:=\sum_{j=2}^{\infty}\Big\{\prod^{2}_{l=j}\frac{1}{\Big[1+\sqrt{\eta a_{\epsilon_{\bold{j}_*}}}-\frac{b_{\epsilon_{\bold{j}_*}}/ \sqrt{\eta a_{\epsilon_{\bold{j}_*}}}}{2l-\epsilon^{\Theta}_{\bold{j}_*}}\Big]\Big[1+a_{\epsilon_{\bold{j}_*}}-\frac{2b_{\epsilon_{\bold{j}_*}}}{2l+1}-\frac{1-c_{\epsilon_{\bold{j}_*}}}{(2l+1)^2}]^{\frac{1}{2}}}\Big\}
\end{eqnarray}
which is convergent for $\epsilon_{\bold{j}_*}>0$ because 
\begin{equation}
\frac{c_{j}}{c_{j-1}}=\frac{1}{\Big[1+\sqrt{\eta a_{\epsilon_{\bold{j}_*}}}-\frac{b_{\epsilon_{\bold{j}_*}}/ \sqrt{\eta a_{\epsilon_{\bold{j}_*}}}}{2j-\epsilon^{\Theta}_{\bold{j}_*}}\Big]\Big[1+a_{\epsilon_{\bold{j}_*}}-\frac{2b_{\epsilon_{\bold{j}_*}}}{2j+1}-\frac{1-c_{\epsilon_{\bold{j}_*}}}{(2j+1)^2}]^{\frac{1}{2}}}<1
\end{equation}
for $j$ sufficiently large, where $a_{\epsilon_{\bold{j}_*}}, b_{\epsilon_{\bold{j}_*}}, c_{\epsilon_{\bold{j}_*}}$, and $0<\Theta\leq \frac{1}{4}$ are those defined in \emph{\underline{Lemma 3.6} of \cite{Pi1}}. (The series in (\ref{convergent-series}) diverges  in the limit $\epsilon_{\bold{j}_*} \to 0$.) Hence, for any $\epsilon_{\bold{j}_*}>0$ fulfilling the \emph{Strong Interaction Potential Assumption} (see point 3.) in Definition \ref{def-pot}) we have derived an expansion of $\psi^{Bog}_{\bold{j}_*}$ controlled by the parameter $\theta_{\epsilon_{\bold{j}_*}}:=\frac{1}{1+\sqrt{\epsilon_{\bold{j}_*}}+o(\sqrt{\epsilon_{\bold{j}_*}})}$.
\\

The ground state energy $z_*$ approaches $E^{Bog}_{\bold{j}_*}$ as $N\to \infty$, more precisely (see {\emph{\underline{Lemma 5.5.} of \cite{Pi1}}}) in the mean field limiting regime the estimate $|z_*-E^{Bog}_{\bold{j}_*}|\leq \mathcal{O}(\frac{1}{N^{\beta}})$ holds for any $0<\beta <1$. Starting from the formula in (\ref{gs-1})-(\ref{gs-2}) and from the definitions in (\ref{GammaBog-2})-(\ref{GammaBog-i}), in \emph{\underline{Section 4.4} of \cite{Pi1}} we show how to expand the ground state $\psi^{Bog}_{\bold{j}_*}$ in terms of the bare operators $\frac{1}{\hat{H}^0_{\bold{j}_*}-z}|_{z\equiv E^{Bog}_{\bold{j}_*}}$ and $W_{\bold{j}_*}^*+W_{\bold{j}_*}$ applied to the vector $\eta$, up to any desired precision provided $N$ is sufficiently large.
\begin{remark}\label{same-gs}
We observe that $\psi^{Bog}_{{\bold{j}_*}}$ is also eigenvector of $\hat{H}^{Bog}_{{\bold{j}_*}}$  with the same eigenvalue (see the definition in (\ref{check-HBogj})).
\end{remark}

\subsection{The Feshbach flows associated with the intermediate Hamiltonians $H^{Bog}_{\bold{j}_1,\,\dots, \bold{j}_m}$}\label{informal}

\noindent
In the paper \cite{Pi2} we have shown how the ground state of the Hamiltonian
\begin{equation}
H^{Bog}_{\bold{j}_1,\,\dots, \bold{j}_m}:=\sum_{\bold{j}\in\mathbb{Z}^d\setminus \{\pm\bold{j}_{1}, \dots,\pm\bold{j}_{m} \}} k^2_{\bold{j}}a_{\bold{j}}^{*}a_{\bold{j}}+\sum_{l=1}^{m}\hat{H}^{Bog}_{\bold{j}_l}\,,
\end{equation}
with $1\leq m \leq M$, can be constructed by means of an inductive procedure. At each step of this procedure we exploit the Feshbach map where the (Feshbach) projections are associated with a three-modes system.  In the following we shall outline the procedure.
\\

We start from $H^{Bog}_{\bold{j}_1}$ and using the results of Section \ref{informal-0} we construct 
\begin{eqnarray}
\mathscr{K}_{\bold{j}_1}^{Bog\,(N)}(z)
&:= &\mathscr{P}_{\eta}(H^{Bog}_{\bold{j}_1}-z)\mathscr{P}_{\eta}\\
& &-\mathscr{P}_{\eta}W_{\bold{j}_1}\sum_{l_{N-2}=0}^{\infty}R^{Bog}_{\bold{j}_1;\,N-2,N-2}(z)\,\Big[\Gamma^{Bog\,}_{\bold{j}_1\,;\,N-2,N-2}(z)R^{Bog}_{\bold{j}_1;\,N-2,N-2}(z)\Big]^{l_{N-2}}W_{\bold{j}_1}^*\mathscr{P}_{\eta}\,. \nonumber
%& &-\sum_{l_i=0}^{\infty}Q^{(>i)}W\,Q^{(i)}R^{Bog}_{i,i}(z)\Big[\Gamma^{Bog\,}_{i,i}(z)R^{Bog}_{i,i}(z)\Big]^{l_i}Q^{(i)}W^*Q^{(>i)} \nonumber
\end{eqnarray}
Next, we determine the ground state energy, $z^{Bog}_{\bold{j}_1}$, of $H^{Bog}_{\bold{j}_1}$ by imposing
\begin{eqnarray}
z^{Bog}_{\bold{j}_1}=\langle \eta\,, \,W_{\bold{j}_1}\sum_{l_{N-2}=0}^{\infty}R^{Bog}_{\bold{j}_1;\,N-2,N-2}(z^{Bog}_{\bold{j}_1})\,\Big[\Gamma^{Bog\,}_{\bold{j}_1;\,N-2,N-2}(z^{Bog}_{\bold{j}_1})R^{Bog}_{\bold{j}_1;\,N-2,N-2}(z^{Bog}_{\bold{j}_1})\Big]^{l_{N-2}}W_{\bold{j}_1}^*\eta \rangle\,.\quad\quad
%&=&\langle \eta\,, \,W_{\bold{j}_1}\,Q_{\bold{j}_1}^{(N-2,N-1)}[R^{Bog}_{\bold{j}_1;\,N-2,N-2}(z_{\bold{j}_1})]^{\frac{1}{2}}\,\sum_{l_{N-2}=0}^{\infty}\Big[[R^{Bog}_{\bold{j}_1;\,N-2,N-2}(z_{\bold{j}_1})]^{\frac{1}{2}}\Gamma^{Bog\,}_{\bold{j}_1;\,N-2,N-2}(z_{\bold{j}_1})[R^{Bog}_{\bold{j}_1;\,N-2,N-2}(z_{\bold{j}_1})]^{\frac{1}{2}}\Big]^{l_{N-2}}\times \nonumber\\
%& &\quad\quad\quad\quad \times [R^{Bog}_{\bold{j}_1;\,N-2,N-2}(z_{\bold{j}_1})]^{\frac{1}{2}}\,Q_{\bold{j}_1}^{(N-2, N-1)}W_{\bold{j}_1}^*\eta \rangle
%& &-\sum_{l_i=0}^{\infty}Q^{(>i)}W\,Q^{(i)}R^{Bog}_{i,i}(z)\Big[\Gamma^{Bog\,}_{i,i}(z)R^{Bog}_{i,i}(z)\Big]^{l_i}Q^{(i)}W^*Q^{(>i)} \nonumber
\end{eqnarray}
Hence, the (non-normalized) ground state vector, $\psi^{Bog}_{\bold{j}_1}$, of $H^{Bog}_{\bold{j}_1}$ is given in (\ref{gs-1})-(\ref{gs-2}) with $\bold{j}_*, z_*$ replaced with $\bold{j}_1$ and $z^{Bog}_{\bold{j}_1}$, respectively.
%\begin{eqnarray}
%\psi^{Bog}_{\bold{j}_1}
%&:=&
%\Big[Q^{(>0)}-\frac{1}{Q^{(0)}(H^{Bog}-z_*)Q^{(0)}}Q^{(0)}(H^{Bog}-z_*)Q^{(>0)}\Big]\times\\
%& &\quad\quad\quad\times \Big\{ \prod_{i=0}^{N-2}\Big[Q^{(>i+1)}-\frac{1}{Q^{(i+1)}\mathscr{K}^{Bog\,(i)}(z_*)Q^{(i+1)}}Q^{(i+1)}\mathscr{K}^{Bog\,(i)}(z_*)Q^{(>i+1)}\Big]\Big\}\eta \quad \\
%&= &\eta \\
%& &-\frac{1}{Q^{(N-2)}\mathscr{K}^{Bog\,(N-3)}(z_*)Q^{(N-2)}}Q^{(N-2)}W^*\eta \\
%& &+\frac{1}{Q^{(N-4)}\mathscr{K}^{Bog\,(N-5)}(z_*)Q^{(N-4)}}W^*_{N-4,N-2}\frac{1}{Q^{(N-2)}\mathscr{K}^{Bog\,(N-3)}(z_*)Q^{(N-2)}}Q^{(N-2)}W^*\eta \quad \nonumber \\
%& &-\frac{1}{Q^{(N-6)}\mathscr{K}^{Bog\,(N-7)}(z_*)Q^{(N-6)}}W^*_{N-6,N-4}\frac{1}{Q^{(N-4)}\mathscr{K}^{Bog\,(N-5)}(z_*)Q^{(N-4)}}Q^{(N-4)}W^*Q^{(N-2)}\times \quad \nonumber \\
%& &\quad \times\frac{1}{Q^{(N-2)}\mathscr{K}^{Bog\,(N-3)}(z_*)Q^{(N-2)}}Q^{(N-2)}W^*\eta \\
%& &+\dots \\
%\eta \\
%& &-\frac{1}{Q_{\bold{j}_1}^{(N-2,N-1)}\mathscr{K}_{\bold{j}_1}^{Bog\,(N-4)}(z^{Bog}_{\bold{j}_1})Q_{\bold{j}_1}^{(N-2, N-1)}}Q_{\bold{j}_1}^{(N-2,N-1)}W_{\bold{j}_1}^*\eta\\
%& &-\sum_{j=2}^{N/2}\prod^{2}_{i=j}\Big[-\frac{1}{Q_{\bold{j}_1}^{(N-2i, N-2i+1)}\mathscr{K}_{\bold{j}_1}^{Bog\,(N-2i-2)}(z^{Bog}_{\bold{j}_1})Q_{\bold{j}_1}^{(N-2i, N-2i+1)}}W^*_{\bold{j}_1;\, N-2i,N-2i+2}\Big]\times \quad\quad\\
%& &\quad\quad\quad\quad\quad\quad \times \frac{1}{Q_{\bold{j}_1}^{(N-2,N-1)}\mathscr{K}_{\bold{j}_1}^{Bog\,(N-4)}(z^{Bog}_{\bold{j}_1})Q_{\bold{j}_1}^{(N-2, N-1)}}Q_{\bold{j}_1}^{(N-2, N-1)}W_{\bold{j}_1}^*\eta\nonumber
%\end{eqnarray}

In the next step, we consider the intermediate  Hamiltonian
\begin{equation}\label{ham-12}
H^{Bog}_{\bold{j}_1,\, \bold{j}_2}:=\sum_{\bold{j}\in\mathbb{Z}^d\setminus \{ \pm\bold{j}_{1}\,;\, \pm\bold{j}_{2} \}} k^2_{\bold{j}}a_{\bold{j}}^{*}a_{\bold{j}}+\hat{H}^{Bog}_{\bold{j}_1,\, \bold{j}_2}:=\sum_{\bold{j}\in\mathbb{Z}^d\setminus \{\pm\bold{j}_{1}\,;\, \pm\bold{j}_{2} \}} k^2_{\bold{j}}a_{\bold{j}}^{*}a_{\bold{j}}+\sum_{l=1}^{2}\hat{H}^{Bog}_{\bold{j}_l}\end{equation}
and construct the Feshbach Hamiltonians
\begin{eqnarray}
& &\mathscr{K}_{\bold{j}_1,\,\bold{j}_2}^{Bog\,(i)}(z^{Bog}_{\bold{j}_1}+z)\\
&=&Q_{\bold{j}_2}^{(>i+1)}(H^{Bog} _{\bold{j}_1,\, \bold{j}_2}-z^{Bog}_{\bold{j}_1}-z)Q^{(>i+1)}_{\bold{j}_2}\quad\quad\quad\\
& &-\sum_{l_i=0}^{\infty}Q^{(>i+1)}_{\bold{j}_2}W_{\bold{j}_2}\,R^{Bog}_{\bold{j}_1,\bold{j}_2\,;\,i,i}(z^{Bog}_{\bold{j}_1}+z)\Big[\Gamma^{Bog\,}_{\bold{j}_1,\bold{j}_2\,;\,i,i}(z^{Bog}_{\bold{j}_1}+z)R^{Bog}_{\bold{j}_1,\bold{j}_2\,;\,i,i}(z^{Bog}_{\bold{j}_1}+z)\Big]^{l_i}W^*_{\bold{j}_2}Q^{(>i+1)}_{\bold{j}_2} \nonumber
%&=&Q_{\bold{j}_m}^{(>i)}(H^{Bog} _{\bold{j}_1,\, \bold{j}_2}-z_{\bold{j}_1}-z)Q^{(>i)}_{\bold{j}_m}\quad\quad\quad\\
%& &-Q^{(>i+1)}_{\bold{j}_2}\Gamma^{Bog}_{\bold{j}_1,\bold{j}_2\,;\,N,N}(z_{\bold{j}_1}+z))Q^{(>i+1)}_{\bold{j}_2} \nonumber
%&=&\mathscr{P}_{\psi^{Bog}_{\bold{j}_1}}(H^{Bog}_{\bold{j}_1,\, \bold{j}_2}-z_{\bold{j}_1}-z)\mathscr{P}_{\psi^{Bog}_{\bold{j}_1}}\\
%& &-\mathscr{P}_{\psi^{Bog}_{\bold{j}_1}}W_{\bold{j}_2}\,Q_{\bold{j}_2}^{(N-2, N-1)}\sum_{l_{N-2}=0}^{\infty}R^{Bog}_{\bold{j}_1,\bold{j}_2;\,N-2,N-2}(z_{\bold{j}_1}+z)\,\Big[\Gamma^{Bog\,}_{\bold{j}_1,\bold{j}_2\,;\,N-2,N-2}(z_{\bold{j}_1}+z)R^{Bog}_{\bold{j}_1,\bold{j}_2;\,N-2,N-2}(z_{\bold{j}_1}+z)\Big]^{l_{N-2}}\times \nonumber \\
%& &\quad\quad\quad \times Q_{\bold{j}_2}^{(N-2, N-1)}W_{\bold{j}_2}^*\mathscr{P}_{\psi^{Bog}_{\bold{j}_1}} \nonumber\\
%& &-\sum_{l_i=0}^{\infty}Q^{(>i)}W\,Q^{(i)}R^{Bog}_{i,i}(z)\Big[\Gamma^{Bog\,}_{i,i}(z)R^{Bog}_{i,i}(z)\Big]^{l_i}Q^{(i)}W^*Q^{(>i)} \nonumber
%&=&\mathscr{P}_{\psi^{Bog}_{\bold{j}_1}}(H^{Bog}_{\bold{j}_2}-z)\mathscr{P}_{\psi^{Bog}_{\bold{j}_1}}
\end{eqnarray}
for $0\leq i \leq N-2$ and even, where we use the definitions:
\begin{itemize}
\item
\begin{equation}\label{rBog-ii}
R^{Bog}_{\bold{j}_1,\,\bold{j}_2\,;\,i,i}(z^{Bog}_{\bold{j}_1}+z):=Q^{(i, i+1)}_{\bold{j}_2}{\frac{1}{Q^{(i, i+1)}_{\bold{j}_2}(H^{Bog}_{\bold{j}_1,\, \bold{j}_2}-z^{Bog}_{\bold{j}_1}-z)Q^{(i,i+1)}_{\bold{j}_2}}Q^{(i, i+1)}_{\bold{j}_2}}\,;\end{equation}
\item
\begin{equation}
\Gamma^{Bog\,}_{\bold{j}_1,\,\bold{j}_2\,;\,2,2}(z^{Bog}_{\bold{j}_1}+z):=W_{\bold{j}_*\,;\,2,0}R^{Bog}_{\bold{j}_1,\,\bold{j}_2\,;\,0,0}(z^{Bog}_{\bold{j}_1}+z)W_{\bold{j}_*\,;\,0,2}^*
\end{equation}
%\begin{equation}
%\Gamma^{Bog\,}_{\bold{j}_1,\,\bold{j}_2\,;\,3,3}(z_{\bold{j}_1}+z):=W_{\bold{j}_*\,;\,3,1}R^{Bog}_{\bold{j}_1,\,\bold{j}_2\,;\,1,1}(z_{\bold{j}_1}+z)W_{\bold{j}_*\,;\,1,3}^*\label{gamma-12}
%\end{equation}
and,  for $i\geq 4$ and even,
\begin{eqnarray}
& &\Gamma^{Bog\,}_{\bold{j}_1,\,\bold{j}_2\,;\,i,i}(z^{Bog}_{\bold{j}_1}+z)\\
&:=&W_{\bold{j}_2\,;\,i,i-2}\,R^{Bog}_{\bold{j}_1,\,\bold{j}_2\,;\,i-2,i-2}(z^{Bog}_{\bold{j}_1}+z) \sum_{l_{i-2}=0}^{\infty}\Big[\Gamma^{Bog}_{\bold{j}_1,\,\bold{j}_2\,;\,i-2,i-2}(z^{Bog}_{\bold{j}_1}+z)R^{Bog}_{\bold{j}_1,\,\bold{j}_2\,;\,\,i-2,i-2}(z^{Bog}_{\bold{j}_1}+z)\Big]^{l_{i-2}}W^*_{\bold{j}_2\,;\,i-2,i}\,.\quad\quad\quad\label{gamma-12}
\end{eqnarray}
%\item
%%\check{R}^{Bog}_{\bold{j}_1,\,\bold{j}_2\,;\,i,i}(z_{\bold{j}_1}+z):=\sum_{l_i=0}^{\infty}R^{Bog}_{\bold{j}_1,\bold{j}_2\,;\,i,i}(z_{\bold{j}_1}+z)\Big[\Gamma^{Bog\,}_{\bold{j}_1,\bold{j}_2\,;\,i,i}(z_{\bold{j}_1}+z)R^{Bog}_{\bold{j}_1,\bold{j}_2\,;\,i,i}(z_{\bold{j}_1}+z)\Big]^{l_i}
%\end{equation}
\end{itemize}
In the last implementation of the Feshbach map we make use of the projections
\begin{equation}\label{projection-step-1}
\mathscr{P}_{\psi^{Bog}_{\bold{j}_1}}:=|\frac{ \psi^{Bog}_{\bold{j}_1} }{\| \psi^{Bog}_{\bold{j}_1} \|}\rangle \langle\frac{ \psi^{Bog}_{\bold{j}_1} }{\| \psi^{Bog}_{\bold{j}_1} \|}|\quad,\quad \overline{\mathscr{P}_{\psi^{Bog}_{\bold{j}_1}}}:=\charf_{Q^{(>N-1)}_{\bold{j}_2}\mathcal{F}^N}-\mathscr{P}_{\psi^{Bog}_{\bold{j}_1}}
\end{equation}
where $\charf_{Q^{(>N-1)}_{\bold{j}_2}\mathcal{F}^N}$ is the projection onto the subspace of states of  $\mathcal{F}^N$ with no particles in the modes $\pm \bold{j}_2$,
and we define
\begin{eqnarray}
& &\Gamma^{Bog\,}_{\bold{j}_1,\,\bold{j}_2\,;\,N,N}(z^{Bog}_{\bold{j}_1}+z)\\
&:=&W_{\bold{j}_2}\,R^{Bog}_{\bold{j}_1,\,\bold{j}_2\,;\,N-2,N-2}(z^{Bog}_{\bold{j}_1}+z) \sum_{l_{N-2}=0}^{\infty}\Big[\Gamma^{Bog}_{\bold{j}_1,\,\bold{j}_2\,;\,N-2,N-2}(z^{Bog}_{\bold{j}_1}+z)R^{Bog}_{\bold{j}_1,\,\bold{j}_2\,;\,\,N-2,N-2}(z^{Bog}_{\bold{j}_1}+z)\Big]^{l_{N-2}}W^*_{\bold{j}_2}\,.\quad\quad\quad\label{gamma-NN}
\end{eqnarray}
For the derivation of $\mathscr{K}_{\bold{j}_1,\,\bold{j}_2}^{Bog\,(N)}(z^{Bog}_{\bold{j}_1}+z)$, we point out that (see Remark \ref{same-gs})
$$(\hat{H}^{Bog}_{\bold{j}_1}-z^{Bog}_{\bold{j}_1})\mathscr{P}_{\psi^{Bog}_{\bold{j}_1}}=0\quad,\quad \sum_{\bold{j}\in \mathbb{Z}^d\setminus \{\bold{0},\pm \bold{j}_1\}}a^*_{\bold{j}}a_{\bold{j}}\mathscr{P}_{\psi^{Bog}_{\bold{j}_1}}=0$$
and
$$\mathscr{P}_{\psi^{Bog}_{\bold{j}_1}}(H^{Bog}_{\bold{j}_1,\, \bold{j}_2}-z^{Bog}_{\bold{j}_1})\mathscr{P}_{\psi^{Bog}_{\bold{j}_1}}=\mathscr{P}_{\psi^{Bog}_{\bold{j}_1}}(\hat{H}^{Bog}_{\bold{j}_2})\mathscr{P}_{\psi^{Bog}_{\bold{j}_1}}=\mathscr{P}_{\psi^{Bog}_{\bold{j}_1}}(W_{\bold{j}_2}+W^*_{\bold{j}_2})\mathscr{P}_{\psi^{Bog}_{\bold{j}_1}}=0\,,$$
 $$\mathscr{P}_{\psi^{Bog}_{\bold{j}_1}}(H^{Bog}_{\bold{j}_1,\, \bold{j}_2}-z^{Bog}_{\bold{j}_1}-z)\overline{\mathscr{P}_{\psi^{Bog}_{\bold{j}_1}}}=\mathscr{P}_{\psi^{Bog}_{\bold{j}_1}}(\hat{H}^{Bog}_{\bold{j}_2}-z)\overline{\mathscr{P}_{\psi^{Bog}_{\bold{j}_1}}}=\mathscr{P}_{\psi^{Bog}_{\bold{j}_1}}(W_{\bold{j}_2}+W^*_{\bold{j}_2})\overline{\mathscr{P}_{\psi^{Bog}_{\bold{j}_1}}}=0\,. $$ These identities follow from  the definitions of $\mathscr{P}_{\psi^{Bog}_{\bold{j}_1}}$, $\overline{\mathscr{P}_{\psi^{Bog}_{\bold{j}_1}}}$,  and $H^{Bog}_{\bold{j}_1,\, \bold{j}_2}$ combined with the fact that $Q^{(>N-1)}_{\bold{j}_2}$ is the projection onto the subspace of $\mathcal{F}^N$ of states with no particles in the modes $\pm \bold{j}_2$.
 % and $\psi^{Bog}_{\bold{j}_1}\in \charf_{Q^{(>N-2)}_{\bold{j}_2}\mathcal{F}^N}$.
Formally, we get
\begin{eqnarray}
& &\mathscr{K}_{\bold{j}_1,\,\bold{j}_2}^{Bog\,(N)}(z^{Bog}_{\bold{j}_1}+z)\\
&:=&\mathscr{P}_{\psi^{Bog}_{\bold{j}_1}}(H^{Bog}_{\bold{j}_1,\, \bold{j}_2}-z^{Bog}_{\bold{j}_1}-z)\mathscr{P}_{\psi^{Bog}_{\bold{j}_1}}\\
& &-\mathscr{P}_{\psi^{Bog}_{\bold{j}_1}}\Gamma^{Bog}_{\bold{j}_1,\bold{j}_2;\,N,N}(z^{Bog}_{\bold{j}_1}+z)\mathscr{P}_{\psi^{Bog}_{\bold{j}_1}} \\
& &-\mathscr{P}_{\psi^{Bog}_{\bold{j}_1}}\Gamma^{Bog}_{\bold{j}_1,\bold{j}_2;\,N,N}(z^{Bog}_{\bold{j}_1}+z)\overline{\mathscr{P}_{\psi^{Bog}_{\bold{j}_1}}}\times\\
& &\quad\quad\times \frac{1}{\overline{\mathscr{P}_{\psi^{Bog}_{\bold{j}_1}}}\mathscr{K}_{\bold{j}_1,\,\bold{j}_2}^{Bog\,(N-2)}(z^{Bog}_{\bold{j}_1}+z)\overline{\mathscr{P}_{\psi^{Bog}_{\bold{j}_1}}}}\overline{\mathscr{P}_{\psi^{Bog}_{\bold{j}_1}}}\Gamma^{Bog}_{\bold{j}_1,\bold{j}_2;\,N,N}(z^{Bog}_{\bold{j}_1}+z)\mathscr{P}_{\psi^{Bog}_{\bold{j}_1}}\nonumber \\
%& &-\sum_{l_i=0}^{\infty}Q^{(>i)}W\,Q^{(i)}R^{Bog}_{i,i}(z)\Big[\Gamma^{Bog\,}_{i,i}(z)R^{Bog}_{i,i}(z)\Big]^{l_i}Q^{(i)}W^*Q^{(>i)} \nonumber
%&=&\mathscr{P}_{\psi^{Bog}_{\bold{j}_1}}(\hat{H}^{Bog}_{\bold{j}_2}-z)\mathscr{P}_{\psi^{Bog}_{\bold{j}_1}}\\
%& &-\mathscr{P}_{\psi^{Bog}_{\bold{j}_1}}\Gamma^{Bog}_{\bold{j}_1,\bold{j}_2;\,N,N}(z_{\bold{j}_1}+z)\mathscr{P}_{\psi^{Bog}_{\bold{j}_1}} \nonumber\\
%& &-\mathscr{P}_{\psi^{Bog}_{\bold{j}_1}}{\color{red}\Gamma^{Bog}_{\bold{j}_1,\bold{j}_2;\,N,N}(z_{\bold{j}_1}+z)}\overline{\mathscr{P}_{\psi^{Bog}_{\bold{j}_1}}}\times\\
%& &\quad\quad\times \frac{1}{\overline{\mathscr{P}_{\psi^{Bog}_{\bold{j}_1}}}\mathscr{K}_{\bold{j}_1,\,\bold{j}_2}^{Bog\,(N-2)}(z_{\bold{j}_1}+z)\overline{\mathscr{P}_{\psi^{Bog}_{\bold{j}_1}}}}\overline{\mathscr{P}_{\psi^{Bog}_{\bold{j}_1}}}\Gamma^{Bog}_{\bold{j}_1,\bold{j}_2;\,N,N}(z_{\bold{j}_1}+z)\mathscr{P}_{\psi^{Bog}_{\bold{j}_1}}\nonumber \\
&=&-z\mathscr{P}_{\psi^{Bog}_{\bold{j}_1}}\\
& &-\mathscr{P}_{\psi^{Bog}_{\bold{j}_1}}\Gamma^{Bog}_{\bold{j}_1,\bold{j}_2;\,N,N}(z^{Bog}_{\bold{j}_1}+z)\mathscr{P}_{\psi^{Bog}_{\bold{j}_1}}\\
& &-\mathscr{P}_{\psi^{Bog}_{\bold{j}_1}}\Gamma^{Bog}_{\bold{j}_1,\bold{j}_2;\,N,N}(z^{Bog}_{\bold{j}_1}+z)\overline{\mathscr{P}_{\psi^{Bog}_{\bold{j}_1}}}\times\label{small-term}\\
& &\quad\quad\times \frac{1}{\overline{\mathscr{P}_{\psi^{Bog}_{\bold{j}_1}}}\mathscr{K}_{\bold{j}_1,\,\bold{j}_2}^{Bog\,(N-2)}(z^{Bog}_{\bold{j}_1}+z)\overline{\mathscr{P}_{\psi^{Bog}_{\bold{j}_1}}}}\overline{\mathscr{P}_{\psi^{Bog}_{\bold{j}_1}}}\Gamma^{Bog}_{\bold{j}_1,\bold{j}_2;\,N,N}(z^{Bog}_{\bold{j}_1}+z)\mathscr{P}_{\psi^{Bog}_{\bold{j}_1}}\,.\nonumber
\end{eqnarray}

We determine  the ground state energy, $z^{Bog}_{\bold{j}_1,\bold{j}_2}:=z^{Bog}_{\bold{j}_1}+z^{(2)}$,  of $H^{Bog}_{\bold{j}_1,\, \bold{j}_2}$   by imposing
\begin{eqnarray}
& &z^{(2)}\\
&=&-\langle \, \frac{\psi^{Bog}_{\bold{j}_1}}{\|\psi^{Bog}_{\bold{j}_1}\|}, \Gamma^{Bog}_{\bold{j}_1,\bold{j}_2;\,N,N}(z^{Bog}_{\bold{j}_1}+z^{(2)})\frac{\psi^{Bog}_{\bold{j}_1}}{\|\psi^{Bog}_{\bold{j}_1}\|}\rangle \\
& &-\langle \, \frac{\psi^{Bog}_{\bold{j}_1}}{\|\psi^{Bog}_{\bold{j}_1}\|},\Gamma^{Bog}_{\bold{j}_1,\bold{j}_2;\,N,N}(z^{Bog}_{\bold{j}_1}+z^{(2)})\overline{\mathscr{P}_{\psi^{Bog}_{\bold{j}_1}}}\times\\
& &\quad\quad\times \frac{1}{\overline{\mathscr{P}_{\psi^{Bog}_{\bold{j}_1}}}\mathscr{K}_{\bold{j}_1,\,\bold{j}_2}^{Bog\,(N-2)}(z^{Bog}_{\bold{j}_1}+z^{(2)})\overline{\mathscr{P}_{\psi^{Bog}_{\bold{j}_1}}}}\overline{\mathscr{P}_{\psi^{Bog}_{\bold{j}_1}}}\Gamma^{Bog}_{\bold{j}_1,\bold{j}_2;\,N,N}(z^{Bog}_{\bold{j}_1}+z^{(2)})\frac{\psi^{Bog}_{\bold{j}_1}}{\|\psi^{Bog}_{\bold{j}_1}\|}\rangle.\nonumber
%& &-\sum_{l_i=0}^{\infty}Q^{(>i)}W\,Q^{(i)}R^{Bog}_{i,i}(z)\Big[\Gamma^{Bog\,}_{i,i}(z)R^{Bog}_{i,i}(z)\Big]^{l_i}Q^{(i)}W^*Q^{(>i)} \nonumber
\end{eqnarray}
Hence,  the ground state vector of $H^{Bog}_{\bold{j}_1,\, \bold{j}_2}$ is (up to normalization)
\begin{eqnarray}
& &\psi^{Bog}_{\bold{j}_1,\bold{j}_2}\\
&:=&\psi^{Bog}_{\bold{j}_1} \\
%& &-\Big\{\frac{1}{Q^{(N-2, N-1)}_{\bold{j}_2}\mathscr{K}_{\bold{j}_1,\bold{j}_2}^{Bog\,(N-4)}(z^{Bog}_{\bold{j}_1,\bold{j}_2})Q^{(N-2)}_{\bold{j}_2}}Q^{(N-2, N-1)}_{\bold{j}_2}W_{\bold{j}_2}^*\\
%& &-\sum_{j=2}^{N/2}\prod^{2}_{r=j}\Big[-\frac{1}{Q_{\bold{j}_2}^{(N-2r, N-2r+1)}\mathscr{K}_{\bold{j}_1,\bold{j}_2}^{Bog\,(N-2r-2)}(z^{Bog}_{\bold{j}_1,\bold{j}_2})Q_{\bold{j}_2}^{(N-2r, N-2r+1)}}W^*_{\bold{j}_2;N-2r,N-2r+2}\Big] \,\psi^{Bog}_{\bold{j}_1} \quad\quad\quad \\
& &+\Big\{\sum_{j=2}^{N/2}\Big[\prod^{2}_{r=j}\Big(-\frac{1}{Q_{\bold{j}_2}^{(N-2r, N-2r+1)}\mathscr{K}_{\bold{j}_1,\bold{j}_2}^{Bog\,(N-2r-2)}(z^{Bog}_{\bold{j}_1,\bold{j}_2})Q_{\bold{j}_2}^{(N-2r, N-2r+1)}}W^*_{\bold{j}_2;N-2r,N-2r+2}\Big)\Big]+\charf\Big\}\times\quad\quad\quad\\
%& &\quad\quad\quad\quad\quad \times\Big[ \frac{1}{Q_{\bold{j}_2}^{(N-2, N-1)}\mathscr{K}_{\bold{j}_1,\bold{j}_2}^{Bog\,(N-4)}(z^{Bog}_{\bold{j}_1,\bold{j}_2})Q^{(N-2, N-1)}_{\bold{j}_2}}Q^{(N-2, N-1)}_{\bold{j}_2}W_{\bold{j}_2}^*\Big]\times\\
& &\quad\quad\quad\quad\times \Big[Q^{(>N-1)}_{\bold{j}_2}-\frac{1}{Q^{(N-2,N-1)}_{\bold{j}_2}\mathscr{K}^{Bog\,(N-4)}_{\bold{j}_1,\bold{j}_2}(z^{Bog}_{\bold{j}_1,\bold{j}_2})Q^{(N-2,N-1)}_{\bold{j}_2}}Q^{(N-2, N-1)}_{\bold{j}_2}W^*_{\bold{j}_2}\Big]\times\\
& &\quad\quad\quad\quad\quad\quad \times \Big[\mathscr{P}_{\psi^{Bog}_{\bold{j}_1}}-\frac{1}{\overline{\mathscr{P}_{\psi^{Bog}_{\bold{j}_1}}}\mathscr{K}^{Bog\,(N-2)}_{\bold{j}_1,\bold{j}_2}(z^{Bog}_{\bold{j}_1,\bold{j}_2})\overline{\mathscr{P}_{\psi^{Bog}_{\bold{j}_1}}}}
\overline{\mathscr{P}_{\psi^{Bog}_{\bold{j}_1}}}\mathscr{K}^{Bog\,(N-2)}_{\bold{j}_1,\bold{j}_2}(z^{Bog}_{\bold{j}_1,\bold{j}_2})\Big]\,\psi^{Bog}_{\bold{j}_1}\,\nonumber
\end{eqnarray}
where $\mathscr{K}^{Bog\,(-2)}_{\bold{j}_1,\bold{j}_2}(z^{Bog}_{\bold{j}_1,\bold{j}_2}):=H^{Bog}_{\bold{j}_1,\bold{j}_2}-z^{Bog}_{\bold{j}_1,\bold{j}_2}$.
\\

At the $m-th$ step, first we define
\begin{equation}
H^{Bog}_{\bold{j}_1,\,\dots, \bold{j}_m}:=\sum_{\bold{j}\in\mathbb{Z}^d\setminus \{\pm\bold{j}_{1},\dots , \pm\bold{j}_{m} \}} k^2_{\bold{j}}a_{\bold{j}}^{*}a_{\bold{j}}+\sum_{l=1}^{m}\hat{H}^{Bog}_{\bold{j}_l}\,.\end{equation}
(The reader should notice that the kinetic energy of the interacting nonzero modes, $\pm\bold{j}_{1}\,,\,\dots \,,\, \pm\bold{j}_{m}$,  is contained in $\sum_{l=1}^{m}\hat{H}^{Bog}_{\bold{j}_l}$.) Then, we construct
\begin{eqnarray}
& &\mathscr{K}_{\bold{j}_1,\dots,\bold{j}_m}^{Bog\,(N)}(z+z^{Bog}_{\bold{j}_1,\dots,\bold{j}_{m-1}})\label{final-fesh-step-m}\\
&=&-z\mathscr{P}_{\psi^{Bog}_{\bold{j}_1,\dots,\bold{j}_{m-1}}}\\
& &-\mathscr{P}_{\psi^{Bog}_{\bold{j}_1,\dots,\bold{j}_{m-1}}}\Gamma^{Bog}_{\bold{j}_1,\dots,\bold{j}_m ;N,N}(z+z^{Bog}_{\bold{j}_1,\dots,\bold{j}_{m-1}})\mathscr{P}_{\psi^{Bog}_{\bold{j}_1,\dots,\bold{j}_{m-1}}}\\
& &-\mathscr{P}_{\psi^{Bog}_{\bold{j}_1,\dots,\bold{j}_{m-1}}}\Gamma^{Bog}_{\bold{j}_1,\dots,\bold{j}_m;\,N,N}(z+z^{Bog}_{\bold{j}_1,\dots,\bold{j}_{m-1}})\overline{\mathscr{P}_{\psi^{Bog}_{\bold{j}_1,\dots,\bold{j}_{m-1}}}}\times\label{small-term-m}\\
& &\quad\quad\times \frac{1}{\overline{\mathscr{P}_{\psi^{Bog}_{\bold{j}_1,\dots,\bold{j}_{m-1}}}}\mathscr{K}_{\bold{j}_1,\dots,\bold{j}_m}^{Bog\,(N-2)}(z+z^{Bog}_{\bold{j}_1,\dots,\bold{j}_{m-1}})\overline{\mathscr{P}_{\psi^{Bog}_{\bold{j}_1,\dots,\bold{j}_{m-1}}}}}\overline{\mathscr{P}_{\psi^{Bog}_{\bold{j}_1,\dots,\bold{j}_{m-1}}}}\Gamma^{Bog}_{\bold{j}_1,\dots,\bold{j}_m ;N,N}(z+z^{Bog}_{\bold{j}_1,\dots,\bold{j}_{m-1}})\mathscr{P}_{\psi^{Bog}_{\bold{j}_1,\dots,\bold{j}_{m-1}}}\nonumber\\
&=:&f^{Bog}_{\bold{j}_1,\dots,\bold{j}_m}(z+z^{Bog}_{\bold{j}_1,\dots,\bold{j}_{m-1}})\mathscr{P}_{\psi^{Bog}_{\bold{j}_1,\bold{j}_2,\dots,\bold{j}_{m-1}}}\label{def-f}
\end{eqnarray}
with definitions analogous to (\ref{rBog-ii})-(\ref{gamma-NN}):
\begin{itemize}
\item
\begin{equation}\label{final-proj-Bog}
\mathscr{P}_{\psi^{Bog}_{\bold{j}_1,\dots,\bold{j}_{m-1}}}:=|\frac{\psi^{Bog}_{\bold{j}_1,\dots,\bold{j}_{m-1}}}{\| \psi^{Bog}_{\bold{j}_1,\dots,\bold{j}_{m-1}} \|}\rangle \langle\frac{\psi^{Bog}_{\bold{j}_1,\dots,\bold{j}_{m-1}} }{\|\psi^{Bog}_{\bold{j}_1,\dots,\bold{j}_{m-1}}\|}|\quad,\quad \overline{\mathscr{P}_{\psi^{Bog}_{\bold{j}_1,\dots,\bold{j}_{m-1}}}}:=\charf_{Q^{(>N-1)}_{\bold{j}_m}\mathcal{F}^N}-\mathscr{P}_{\psi^{Bog}_{\bold{j}_1,\dots,\bold{j}_{m-1}}}\,;
\end{equation}
\item
\begin{equation}
R^{Bog}_{\bold{j}_1,\dots,\bold{j}_m\,;\,i,i}(z^{Bog}_{\bold{j}_1,\dots ,\bold{j}_{m-1}}+z):=Q^{(i, i+1)}_{\bold{j}_m}\frac{1}{Q^{(i, i+1)}_{\bold{j}_m}(H^{Bog}_{\bold{j}_1, \dots,\bold{j}_m}-z^{Bog}_{\bold{j}_1,\dots ,\bold{j}_{m-1}}-z)Q^{(i, i+1)}_{\bold{j}_m}}Q^{(i, i+1)}_{\bold{j}_m}\,;\,\label{Rbog-m}
\end{equation}
\item
\begin{equation}
\Gamma^{Bog\,}_{\bold{j}_1,\dots,\bold{j}_m\,;\,2,2}(z^{Bog}_{\bold{j}_1,\dots ,\bold{j}_{m-1}}+z):=W_{\bold{j}_m\,;\,2,0}\,R^{Bog}_{\bold{j}_1,\dots,\bold{j}_m\,;\,0,0}(z^{Bog}_{\bold{j}_1,\dots ,\bold{j}_{m-1}}+z)W_{\bold{j}_m\,;\,0,2}^*
\end{equation}
%\begin{equation}
%\Gamma^{Bog\,}_{\bold{j}_1,\dots,\bold{j}_m\,;\,3,3}(z_{\bold{j}_1,\dots ,\bold{j}_{m-1}}+z):=W_{\bold{j}_m\,;\,3,1}\,R^{Bog}_{\bold{j}_1,\dots,\bold{j}_m\,;\,i,i}(z_{\bold{j}_1,\dots ,\bold{j}_{m-1}}+z)W_{\bold{j}_m\,;\,1,3}^*
%\end{equation}
and, for $N-2\geq i\geq 4$ and even,
\begin{eqnarray}
& &\Gamma^{Bog\,}_{\bold{j}_1,\dots,\bold{j}_m\,;\,i,i}(z^{Bog}_{\bold{j}_1,\dots,\bold{j}_{m-1}}+z)\\
&:=&W_{\bold{j}_m\,;\,i,i-2}\,R^{Bog}_{\bold{j}_1,\dots,\bold{j}_m\,;\,i-2,i-2}(z^{Bog}_{\bold{j}_1,\dots,\bold{j}_{m-1}}+z)\times\\
& &\quad\times \sum_{l_{i-2}=0}^{\infty}\Big[\Gamma^{Bog}_{\bold{j}_1,\dots,\bold{j}_m\,;\,i-2,i-2}(z^{Bog}_{\bold{j}_1,\dots,\bold{j}_{m-1}}+z)R^{Bog}_{\bold{j}_1,\dots,\bold{j}_m\,;\,\,i-2,i-2}(z^{Bog}_{\bold{j}_1,\dots,\bold{j}_{m-1}}+z)\Big]^{l_{i-2}}W^*_{\bold{j}_m\,;\,i-2,i}\,\quad\quad\quad \label{Gammabog-m}
\end{eqnarray}
and 
\begin{eqnarray}
& &\Gamma^{Bog\,}_{\bold{j}_1,\dots,\bold{j}_m\,;\,N,N}(z^{Bog}_{\bold{j}_1,\dots,\bold{j}_{m-1}}+z) \label{GammaNNBog}\\
&:=&W_{\bold{j}_m}\,R^{Bog}_{\bold{j}_1,\dots,\bold{j}_m\,;\,N-2,N-2}(z^{Bog}_{\bold{j}_1,\dots,\bold{j}_{m-1}}+z)\times\\
& &\quad\times \sum_{l_{N-2}=0}^{\infty}\Big[\Gamma^{Bog}_{\bold{j}_1,\dots,\bold{j}_m\,;\,N-2,N-2}(z^{Bog}_{\bold{j}_1,\dots,\bold{j}_{m-1}}+z)R^{Bog}_{\bold{j}_1,\dots,\bold{j}_m\,;\,\,N-2,N-2}(z^{Bog}_{\bold{j}_1,\dots,\bold{j}_{m-1}}+z)\Big]^{l_{N-2}}W^*_{\bold{j}_m}\,.\quad\quad\quad \nonumber
\end{eqnarray}
\end{itemize}

\noindent
We compute the ground state energy, $z^{Bog}_{\bold{j}_1,\dots,\bold{j}_m}:=z^{Bog}_{\bold{j}_1,\dots,\bold{j}_{m-1}}+z^{(m)}$, of $H^{Bog}_{\bold{j}_1,\dots, \bold{j}_m}$ by solving the equation (in $z$)
\begin{equation}\label{fixed-p-eq}
f^{Bog}_{\bold{j}_1,\dots,\bold{j}_m}(z+z^{Bog}_{\bold{j}_1,\dots,\bold{j}_{m-1}})=0;
\end{equation}
see (\ref{def-f}).
% that is we determine $z_*$ such that
%\begin{eqnarray}
%z_*&=& \langle  \frac{\psi^{Bog}_{\bold{j}_1,\dots,\bold{j}_{m-1}}}{\|\psi^{Bog}_{\bold{j}_1,\dots,\bold{j}_{m-1}}\|}\,, \,W_{\bold{j}_m}\,Q_{\bold{j}_m}^{(N-2, N-1)}\times \label{fixed-point-m}\\
%& &\times \sum_{l_{N-2}=0}^{\infty}R^{Bog}_{\bold{j}_1,\dots,\bold{j}_m,;\,N-2,N-2}(z_{\bold{j}_1,\dots,\bold{j}_{m-1}}+z_*)\,\Big[\Gamma^{Bog\,}_{\bold{j}_1,\dots,\bold{j}_m\,,;\,N-2,N-2}(z_{\bold{j}_1,\dots,\bold{j}_{m-1}}+z_*)R^{Bog}_{\bold{j}_1,\dots,\bold{j}_m\,;\,N-2,N-2}(z_{\bold{j}_1,\dots,\bold{j}_{m-1}}+z_*)\Big]^{l_{N-2}}\times \nonumber\\
%& &\quad \times Q_{\bold{j}_m}^{(N-2, N-1)}W_{\bold{j}_m}^* \frac{\psi^{Bog}_{\bold{j}_1,\dots,\bold{j}_{m-1}}}{\|\psi^{Bog}_{\bold{j}_1,\dots,\bold{j}_{m-1}}\|} \rangle\,. \nonumber
%& &-\sum_{l_i=0}^{\infty}Q^{(>i)}W\,Q^{(i)}R^{Bog}_{i,i}(z)\Big[\Gamma^{Bog\,}_{i,i}(z)R^{Bog}_{i,i}(z)\Big]^{l_i}Q^{(i)}W^*Q^{(>i)} \nonumber
%\end{eqnarray}
Hence,  the ground state vector of $H^{Bog}_{\bold{j}_1,\dots, \bold{j}_m}$ is (up to normalization)
\begin{eqnarray}
& &\psi^{Bog}_{\bold{j}_1,\dots,\bold{j}_m} \label{gs-Hm-start}\\
&:=&\psi^{Bog}_{\bold{j}_1, \dots, \bold{j}_{m-1}}\\
%& &+\sum_{j=2}^{N/2}\prod^{2}_{r=j}\Big[-\frac{1}{Q_{\bold{j}_m}^{(N-2r, N-2r+1)}\mathscr{K}_{\bold{j}_1,\dots ,\bold{j}_m}^{Bog\,(N-2r-2)}(z^{Bog}_{\bold{j}_1,\dots ,\bold{j}_m})Q_{\bold{j}_m}^{(N-2r, N-2r+1)}}W^*_{\bold{j}_m;N-2r,N-2r+2}\Big]\psi^{Bog}_{\bold{j}_1, \dots, \bold{j}_{m-1}}\quad\quad\quad\label{gs-Hm-fin} \\
& &+\Big\{\sum_{j=2}^{N/2}\Big[\prod^{2}_{r=j}\Big(-\frac{1}{Q_{\bold{j}_m}^{(N-2r, N-2r+1)}\mathscr{K}_{\bold{j}_1,\dots ,\bold{j}_m}^{Bog\,(N-2r-2)}(z^{Bog}_{\bold{j}_1,\dots ,\bold{j}_m})Q_{\bold{j}_m}^{(N-2r, N-2r+1)}}W^*_{\bold{j}_m;N-2r,N-2r+2}\Big)\Big]+\charf\Big\}\times\quad\quad\quad \label{gs-Hm-fin}\\
& &\quad\quad\quad\times \Big[Q^{(>N-1)}_{\bold{j}_m}-\frac{1}{Q^{(N-2,N-1)}_{\bold{j}_m}\mathscr{K}^{Bog\,(N-4)}_{\bold{j}_1,\dots,\bold{j}_m}(z^{Bog}_{\bold{j}_1,\dots,\bold{j}_m})Q^{(N-2,N-1)}_{\bold{j}_m}}Q^{(N-2, N-1)}_{\bold{j}_m}W^*_{\bold{j}_m}\Big]\times \nonumber \\
& &\quad\quad\quad \times \Big[\mathscr{P}_{\psi^{Bog}_{\bold{j}_1\dots,\bold{j}_{m-1}}}-\frac{1}{\overline{\mathscr{P}_{\psi^{Bog}_{\bold{j}_1,\dots,\bold{j}_{m-1}}}}\mathscr{K}^{Bog\,(N-2)}_{\bold{j}_1,\dots,\bold{j}_m}(z^{Bog}_{\bold{j}_1,\dots,\bold{j}_m})\overline{\mathscr{P}_{\psi^{Bog}_{\bold{j}_1, \dots, \bold{j}_{m-1}}}}}
\overline{\mathscr{P}_{\psi^{Bog}_{\bold{j}_1\dots,\bold{j}_{m-1}}}}\mathscr{K}^{Bog\,(N-2)}_{\bold{j}_1,\dots,\bold{j}_m}(z^{Bog}_{\bold{j}_1,\dots,\bold{j}_m})\Big]\psi^{Bog}_{\bold{j}_1, \dots, \bold{j}_{m-1}}\,\nonumber \\
&=:&T_{m}\,\psi^{Bog}_{\bold{j}_1, \dots, \bold{j}_{m-1}}
\end{eqnarray}
where $\mathscr{K}^{Bog\,(-2)}_{\bold{j}_1,\dots,\bold{j}_m}(z^{Bog}_{\bold{j}_1,\dots,\bold{j}_m}):=H^{Bog}_{\bold{j}_1,\dots,\bold{j}_m}-z^{Bog}_{\bold{j}_1,\dots, \bold{j}_m}$. Thus, we have derived the formula
\begin{equation}
\psi^{Bog}_{\bold{j}_1, \dots, \bold{j}_{M}}=T_{M}\dots T_{1}\eta\,.
\end{equation}

Some observations are in order to understand why the procedure that we have described is not a straightforward iteration of the operations implemented for a three-modes system. Indeed, as more couples of interacting modes are considered (i.e., starting from $H^{Bog}_{\bold{j}_1,\bold{j}_2}$) the main task is showing that the interaction terms associated with the couples of modes, $\pm\bold{j}_1,\dots\,\pm\bold{j}_m$,  are to some extent independent. This becomes apparent since:
\begin{itemize}
\item for $m\geq2$  the term in (\ref{small-term-m})  is shown to be vanishing as $N\to \infty$;
\item
 at later steps (i.e., starting from  $m= 2$) the fixed point equation in (\ref{fixed-p-eq}) can be written as a three-modes system fixed point equation plus a small correction that vanishes as $N\to \infty$.
\end{itemize}
The construction implemented in \cite{Pi2} culminates in the theorem below.
\\
 
\noindent
\emph{\underline{Theorem 4.3} of \cite{Pi2}}
\emph{Let $max_{1\leq m \leq M}\epsilon_{\bold{j}_m}$ be sufficiently small and $N$ sufficiently large. 
%Set $\Delta_0\equiv \min\, \Big\{\epsilon_{\bold{j}}\,|\,\bold{j}\in \mathbb{Z}^3 \Big\}\,$.  
Then the following properties hold true for all $1\leq m \leq M$:}
\begin{enumerate}
\item \emph{The Feshbach Hamiltonian $\mathscr{K}_{\bold{j}_1,\dots,\bold{j}_m}^{Bog\,(N)}(z+z^{Bog}_{\bold{j}_1,\dots,\bold{j}_{m-1}})$ in (\ref{final-fesh-step-m})-(\ref{def-f}) is well defined for}
\begin{equation}
 z\leq\min\,\Big\{ z_{m}+\gamma\Delta_{m-1}-\frac{C^{\perp}}{(\ln N)^{\frac{1}{2}}}\,;\,E^{Bog}_{\bold{j}_m}+ \sqrt{\epsilon_{\bold{j}_m}}\phi_{\bold{j}_m}\sqrt{\epsilon_{\bold{j}_m}^2+2\epsilon_{\bold{j}_m}}\Big\}\quad,\quad \gamma=\frac{1}{2},\label{inter-def}
 \end{equation} 
\emph{ where:}
 \begin{itemize}
 \item $z^{Bog}_{\bold{j}_1,\dots,\bold{j}_{m-1}}$ \emph{is the ground state energy of $H^{Bog}_{\bold{j}_1,\dots,\bold{j}_{m-1}}$ and is defined iteratively in point 2. below;}
 \item
   $z_m$ \emph{is the ground state energy of $H^{Bog}_{\bold{j}_m}$;
   \item $\Delta_{m-1}$ (for $m\geq 1$)  is defined iteratively by $\Delta_{0}:=\min\, \Big\{(k_{\bold{j}})^2\,|\,\bold{j}\in \mathbb{Z}^d \setminus \{\bold{0}\}\Big\}$ and}
    \begin{eqnarray}
\Delta_m &:= & \gamma\Delta_{m-1}-\frac{C^{\perp}}{(\ln N)^{\frac{1}{2}}}-(\frac{2}{\gamma})^m\frac{C_{III}}{(\ln N)^{\frac{1}{4}}}\,\quad \quad\quad\quad \label{Deltagap}
%&=&\gamma \Delta_{m-1}-{\color{red}(\frac{2}{\gamma})^m}\frac{C_{III}}{(\ln N)^{\frac{1}{4}}}
 \end{eqnarray}
\emph{ with $C_{III}:=C_{I}+\frac{C_{II}^2}{(1-\gamma)\Delta_{0}}$, where $C_I, C_{II}$ are introduced in \underline{Lemma 4.3} of \cite{Pi2}.}
\end{itemize} 
 \item
\emph{ For $z$ as in (\ref{inter-def}), there exists a unique value $z^{(m)}$ such that $$f^{Bog}_{\bold{j}_1,\dots,\bold{j}_m}(z+z^{Bog}_{\bold{j}_1,\dots,\bold{j}_{m-1}})|_{z=z^{(m)}}=0\,.$$  The inequality $|z^{(m)}-z_m|\leq (\frac{2}{\gamma})^m \frac{C_{III}}{(\ln N)^{\frac{1}{4}}}$ holds true.  }
 
 \noindent
\emph{ The Hamiltonian $H^{Bog}_{\bold{j}_1,\dots,\bold{j}_{m}}$ has nondegenerate ground state energy $z^{Bog}_{\bold{j}_1,\dots,\bold{j}_{m}}:=z^{Bog}_{\bold{j}_1,\dots,\bold{j}_{m-1}}+z^{(m)}$ where $z^{Bog}_{\bold{j}_1,\dots,\bold{j}_{m-1}}|_{m=1}\equiv 0$. The corresponding (non-normalized) eigenvector is given in  (\ref{gs-Hm-start})-(\ref{gs-Hm-fin}).}

\item 
\emph{ The spectral gap of the two operators\footnote{$\mathcal{F}^N_{\pm\bold{j}_{m+1}}$ is the subspace of vectors in  $\mathcal{F}^N$ with at least one particle in the modes $\pm\bold{j}_{m+1}$.}}
\begin{equation}
H^{Bog}_{\bold{j}_1,\dots,\bold{j}_{m}}\quad, \quad \Big(\hat{H}^{Bog}_{\bold{j}_1, \dots,\bold{j}_{m}}+\sum_{\bold{j}\notin \{\pm\bold{j}_1, \dots, \pm\bold{j}_{m+1}\}}(k_{\bold{j}})^2a^*_{\bold{j}}a_{\bold{j}}\Big)\upharpoonright_{\mathcal{F}^N\ominus \mathcal{F}^N_{\pm\bold{j}_{m+1}}}\,
\end{equation}
\emph{ above the (common) ground state energy} $z^{Bog}_{\bold{j}_1,\dots,\bold{j}_{m}}$ \emph{ is larger or equal to $\Delta_{m}$.}

%\item The ground state $\psi^{Bog}_{\bold{j}_1,\bold{j}_2,\dots,\bold{j}_{m}}$ fulfills the expansion:
%\begin{equation}
%\psi^{Bog}_{\bold{j}_1,\bold{j}_2,\dots,\bold{j}_{m}}=\sum_{l=2}^{m} T_{m}\dots T_{l+1}S_{l}\,\psi^{Bog}_{\bold{j}_1,\bold{j}_2,\dots,\bold{j}_{l-1}}+T_{m}\dots T_0\eta
%\end{equation}
%\begin{equation}
%S_n:=\quad \|S_n\|\leq \frac{C^n}{N}
%\end{equation}
\item
\emph{The lower bound }
\begin{equation}
\text{infspec}\,[\sum_{l=1}^{m}\hat{H}^{Bog}_{\bold{j}_l }]-z^{Bog}_{\bold{j}_1,\dots ,\bold{j}_{m}}\geq -\frac{m}{(\ln N)^{\frac{1}{8}}}\,
\end{equation}
\emph{holds true.}
 \item For $\tilde{C}_m:=\frac{\sum_{l=1}^{m}\phi_{\bold{j}_l}}{\Delta_0}$ \emph{the upper bound}
 \begin{equation}\label{control-number}
\langle \frac{\psi^{Bog}_{\bold{j}_1,\dots,\bold{j}_{m}}}{\|\psi^{Bog}_{\bold{j}_1,\dots,\bold{j}_{m}}\|}\,,\,\sum_{\bold{j}\in\mathbb{Z}^d\setminus \{\bold{0}\}} a_{\bold{j}}^{*}a_{\bold{j}}\,\frac{\psi^{Bog}_{\bold{j}_1,\dots,\bold{j}_{m}}}{\|\psi^{Bog}_{\bold{j}_1,\dots,\bold{j}_{m}}\|}\rangle \leq \tilde{C}_m\,
\end{equation}
\emph{holds true.}
 \end{enumerate}
 
Like for the three-modes system, starting from the formula in (\ref{gs-Hm-start})-(\ref{gs-Hm-fin}), in \emph{\underline{Corollary 4.5} of \cite{Pi2}} we show how to expand the ground state vector $\psi^{Bog}_{\bold{j}_1,\dots,\bold{j}_m}$ in terms of the bare operators $\frac{1}{\hat{H}^0_{\bold{j}_l}-z}|_{z\equiv E^{Bog}_{\bold{j}_l}}$ and $W_{\bold{j}_l}^*+W_{\bold{j}_l}$, with $l=1,\dots,m$, applied to the vector $\eta$, up to any desired precision provided $N$ is sufficiently large.

\section{Ground state of $H$: Outline of the construction}\label{groundstateH} 
\setcounter{equation}{0}

Due to the ultraviolet cut-off on the two-body  potential ($\phi_{\bold{j}}$ in Fourier space) it is convenient to write
\begin{equation}
H\equiv H_{\bold{j}_1,\,\dots, \bold{j}_{M}}=H^{Bog}_{\bold{j}_1,\,\dots, \bold{j}_{M}}+V_{\bold{j}_1,\,\dots, \bold{j}_{M}}\,
\end{equation}
where, for $1\leq m \leq M$, $V_{\bold{j}_1,\,\dots, \bold{j}_{m}}$ is defined in (\ref{pair-1})-(\ref{pair-2}).

The strategy to construct the ground state of $H$ consists in three operations:
\begin{enumerate}
\item We define intermediate Hamiltonians
\begin{equation}
H_{\bold{j}_1,\,\dots, \bold{j}_{m}}=H^{Bog}_{\bold{j}_1,\,\dots, \bold{j}_{m}}+V_{\bold{j}_1,\,\dots, \bold{j}_{m}}
\end{equation}
obtained by adding a couple of modes, $\{\bold{j}_m\,,\,-\bold{j}_m\}$ with $1\leq m \leq M$,  at a time to the pair potential (see (\ref{pair-1})-(\ref{pair-2})) so that we obtain $H_{\bold{j}_1,\,\dots, \bold{j}_{M}}$ at the $M-th$ step;
\item At each step, i.e., for each intermediate Hamiltonian,  we use the Feshbach map flow described in Section \ref{new-proj} and
associated with the new couple of modes, $\{\bold{j}_m\,,\,-\bold{j}_m\}$,  that has been added to the pair potential. In comparison with the construction of the ground state of $H^{Bog}$, the new construction requires ``refined" Feshbach projections. Indeed, the term $V_{\bold{j}_1,\dots,\bold{j}_m}$ cannot be controlled using the projections of Section \ref{informal-0} for all index values $i$ of the Feshbach flow. In the new scheme, we start from $i=\bar{i}=N-\lfloor N^{\frac{1}{16}} \rfloor$ where $\lfloor N^{\frac{1}{16}} \rfloor$ is assumed to be even\footnote{The exponent $\frac{1}{16}$ is not optimal.} .  With a new choice of the couple of projections $(\mathscr{P}^{(\bar{i})}\,,\,\overline{\mathscr{P}^{(\bar{i})}})$,  in the subspace $\overline{\mathscr{P}^{(\bar{i})}}\mathcal{F}^N$ the number of particles in the modes $\{\pm \bold{j}_m\}$ can range between $N-\lfloor N^{\frac{1}{16}} \rfloor-1$ and $N$.
% and the number of particles  in nonzero modes different from $\{\pm \bold{j}_m\}$ is constrained by an upper bound of order $N^{\frac{1}{8}}$. }
\item We use the projection onto the ground state of an auxiliary Hamiltonian, $H^{\#}_{\bold{j}_1,\,\dots, \bold{j}_{m-1}}$,  at the $(m-1)-$th step as the final projection of the Feshbach map flow at the $m-$th step. Differently from the case of the Bogoliubov Hamiltonians, for the Hamiltonian $H_{\bold{j}_1,\,\dots, \bold{j}_{m-1}}$ the restriction of the pair potential to the Fourier modes associated with the set $\{\pm \bold{j}_1,\dots, \pm \bold{j}_{m-1}\}$ does not imply that  the field modes associated with $\pm \bold{j}_m$ are absent in the interaction term. The Hamiltonian $H^{\#}_{\bold{j}_1,\,\dots, \bold{j}_{m-1}}$ will be defined starting from  $H_{\bold{j}_1,\,\dots, \bold{j}_{m-1}}$ by omitting these terms. 
\end{enumerate} 

%\begin{remark}The resulting Feshbach Hamiltonian can be split into two parts: a term the form of which is identical to the form of the Feshbach Hamiltonian associated with $H^{Bog}_{\bold{j}_1,\dots,\bold{j}_m}$; an ``extra" term that connects   to a subspace with {\color{red}vectors containing} many particles in the modes $\pm \bold{j}_m$ (see Remark \ref{extra-term}).  The ``extra" term is estimated  only at the very last implementation of the Feshbach map where it is shown to be a small remainder; see Proposition \ref{invertibility-bis}. 
%\end{remark}

%and makes uses of the ground state of the Bogoliubov Hamiltonians at the $H^{Bog}_{\bold{j}_1,\dots, \bold{j}_m}$-th step. 
The features of the new projections and the details of the Feshbach flow are described in Section \ref{new-proj}. The proofs to make the construction rigorous are deferred to Section \ref{rig-cos-H}.

\subsection{The Feshbach flows associated with the intermediate Hamiltonians $H_{\bold{j}_1,\,\dots, \bold{j}_{m}}$: The new projections $\mathfrak{Q}^{(i,i+1)}_{\bold{j}_m}$ and $\mathfrak{Q}^{(>i+1)}_{\bold{j}_m}$}\label{new-proj}

%
%\noindent
%\underline{Non-connected Frequencies Assumption} 
%
%\noindent
%\emph{For any $l,l'$ in the set $\{1,\dots, M\}$, the modes}
%\begin{equation}
%\bold{j}_{l}\pm \bold{j}_{l'}\quad \text{\emph{and}}\quad -(\bold{j}_{l}\pm \bold{j}_{l'})
%\end{equation}
%\emph{do not belong to the set $\{\bold{j}_1,\dots,\bold{j}_M\}$.}
%\\

For the derivation of the Feshbach Hamiltonians associated with the Hamiltonian $H_{\bold{j}_1,\,\dots, \bold{j}_{m}}$, we assume that the ground state of the related Hamiltonian $H^{\#}_{\bold{j}_1,\,\dots, \bold{j}_{m-1}}$ has been already constructed. Here, we define $H^{\#}_{\bold{j}_1,\,\dots, \bold{j}_{m-1}}$ and
%in (\ref{def-xiham-1-zero}).  
introduce new notation:
\begin{definition}
\begin{equation}\label{def-xiham-1-zero}
H^{\#}_{\bold{j}_1,\dots,\bold{j}_{m-1}}|_{m=1}:=T\quad,\quad\text{for}\quad m\geq 2\quad H^{\#}_{\bold{j}_1,\dots,\bold{j}_{m-1}}:=T_{\bold{j}\notin\{\pm\bold{j}_{1};\dots; \pm\bold{j}_{m-1}\}}+\hat{H}^{Bog}_{\bold{j}_1,\dots,\bold{j}_{m-1}}+V^{\#}_{\bold{j}_1,\dots,\bold{j}_{m-1}}
\end{equation}
with
\begin{itemize}
\item
\begin{equation}\label{def-Tsenza}
T_{\bold{j}\notin\{\pm\bold{j}_{1};\dots; \pm\bold{j}_{m-1}\}}:=\sum_{\bold{j}\in \mathbb{Z}^d\setminus\{\pm\bold{j}_{1};\dots; \pm\bold{j}_{m-1}\}}k_{\bold{j}}^2a^*_{\bold{j}}a_{\bold{j}}\,;\quad\quad
\end{equation}
\item
\begin{eqnarray}
& &V^{\#}_{\bold{j}_1,\dots \bold{j}_{m-1}}\\
&:= &\frac{1}{N}\sum_{l=1}^{m-1}\sum_{\bold{j}\in \mathbb{Z}^d\setminus \{-\bold{j}_l\,;\,\bold{0}\,;\,\pm \bold{j}_m\,;\,\pm \bold{j}_m-\bold{j}_l\}}a^*_{\bold{j}+\bold{j}_l}\,a^*_{\bold{0}}\,\phi_{\bold{j}_l}\,a_{\bold{j}}a_{\bold{j}_l}+h.c. \label{def-v1...m-0}\label{pair-1-cap}\\
& &+\frac{1}{N}\sum_{l=1}^{m-1}\sum_{\bold{j}\in \mathbb{Z}^d\setminus \{\bold{j}_l\,;\,\bold{0}\,;\,\pm\bold{j}_m\,;\,\pm \bold{j}_m+\bold{j}_l\}}a^*_{\bold{j}-\bold{j}_l}\,a^*_{\bold{0}}\,\phi_{\bold{j}_l}\,a_{\bold{j}}a_{-\bold{j}_l}+h.c. \label{def-v1...m-0}\label{pair-1-bis-cap}\\
& &+\frac{1}{N}\sum_{l=1}^{m-1}\sum_{\bold{j}\in \mathbb{Z}^d\setminus\{ -\bold{j}_l\,;\,\bold{0}\,;\,\pm\bold{j}_m\,;\,\pm\bold{j}_m-\bold{j}_l\}}\,\sum_{\bold{j}'\in \mathbb{Z}^d\setminus\{ \bold{j}_l\,;\,\bold{0}\,;\,\pm\bold{j}_m\,;\,\pm\bold{j}_m+\bold{j}_l\}}a^*_{\bold{j}+\bold{j}_l}\,a^*_{\bold{j}'-\bold{j}_l}\,\phi_{\bold{j}_l}\,a_{\bold{j}}a_{\bold{j}'}\,,\quad\quad\quad\label{pair-2-cap}
\end{eqnarray}
i.e., $V^{\#}_{\bold{j}_1,\dots \bold{j}_{m-1}}$ corresponds to $V_{\bold{j}_1,\dots \bold{j}_{m-1}}$ without all the summands containing at least one of the operators $a_{\pm \bold{j}_m}\,,\,a^*_{\pm \bold{j}_m}$. 
\end{itemize}
\end{definition}

Then, we step from the Hamiltonian $H_{\bold{j}_1,\,\dots, \bold{j}_{m-1}}$ to $H_{\bold{j}_1,\,\dots, \bold{j}_{m}}$ by adding the term
\begin{equation}
V_{\bold{j}_m}+H^{Bog}_{\bold{j}_1,\,\dots, \bold{j}_{m}}-H^{Bog}_{\bold{j}_1,\,\dots, \bold{j}_{m-1}}\,.
\end{equation}

%\item
%\begin{equation}
%\check{H}^0_{\bold{j}_1,\dots ,\bold{j}_m}:=\sum_{\bold{j}\in\mathbb{Z}^3\setminus \{\bold{0}\,\cup\, \pm\bold{j}_{1}\,\cup\,\dots \,\cup\, \pm\bold{j}_{m} \}} k^2_{\bold{j}}a_{\bold{j}}^{*}a_{\bold{j}}+\sum_{l=1}^{m}\hat{H}^0_{\bold{j}_l}\,
%\end{equation}
%\end{enumerate}
%where $\hat{H}^{0}_{\bold{j}}$ is defined in (\ref{H0j}).

The construction is by induction in the index $m$ ranging from $m=1$ up to $m=M$. In this outline we present the flow associated with $H_{\bold{j}_1,\,\dots, \bold{j}_{m}}$  and defer the details of the induction scheme to the next section.  
%We shall distinguish the case $m=1$ from all the others, i.e., $2\leq m \leq M$.
\\

%The Feshbach map is applied to the Hamiltonian $H_{\bold{j}_1,\,\dots, \bold{j}_{m}}$ using modified spectral projections with respect to the Bogoliubov Hamiltonian $H^{Bog}_{\bold{j}_1,\,\dots, \bold{j}_{m}}$. Indeed, because of the interaction term $V_{\bold{j}_1,\,\dots, \bold{j}_{m-1}}$ the estimates used to control the Feshbach flow derived for $H^{Bog}$ (see \cite{Pi2}) are not available. 

\begin{definition} \label{def-new-Feshbach-proj}

\noindent
For the first implementation of the Feshbach map applied to the Hamiltonian $H_{\bold{j}_1,\,\dots, \bold{j}_{m}}$ we employ the couple $\mathfrak{Q}^{(\bar{i},\bar{i}+1)}_{\bold{j}_m}$, $\mathfrak{Q}^{(>\bar{i}+1)}_{\bold{j}_m}$, $ \bar{i}=N-\lfloor N^{\frac{1}{16}} \rfloor$ assumed to be even,  that is defined here for $N\gg 1$:
% assuming that $N^{\frac{1}{8}}$ is an even number:
\begin{itemize}
\item
$\mathfrak{Q}^{(\bar{i},\bar{i}+1)}_{\bold{j}_m}:=\sum_{j=0\,,\, even}^{\bar{i}}Q^{(j,j+1)}_{\bold{j}_m}$ where $Q^{(j,j+1)}_{\bold{j}_m}$ is defined in Section \ref{informal-0}.
% and where $\bar{i}\equiv N$ for $m=1$ and $\bar{i}\equiv \lfloor N^{\frac{1}{16}}\rfloor$ for $2\leq m\leq M$.
 Therefore, $\mathfrak{Q}^{(\bar{i},\bar{i}+1)}_{\bold{j}_m}$ projects onto the subspace of vectors  with a  number of particles in the modes $\pm \bold{j}_{m}$ that can range from  $N$ to $\lfloor N^{\frac{1}{16}} \rfloor -1$, i.e.,  the operator $a^*_{\bold{j}_m}a_{\bold{j}_m}+a^*_{-\bold{j}_m}a_{-\bold{j}_m}$ has eigenvalues $N,N-1,\dots, \lfloor N^{\frac{1}{16}} \rfloor, \lfloor N^{\frac{1}{16}} \rfloor -1$ when restricted to $\mathfrak{Q}^{(\bar{i},\bar{i}+1)}_{\bold{j}_m}\mathcal{F}^N$.
%(i.e., the eigenspace corresponding to the eigenvalues $\lfloor N^{\frac{1}{16}} \rfloor$ and $\lfloor  N^{\frac{1}{16}} \rfloor -1$ of the operator $[a^*_{\bold{j}_M}a_{\bold{j}_M}+a^*_{-\bold{j}_M}a_{-\bold{j}_M}]_{Q^{(0,1)}_{\bold{j}_M}\mathcal{F}^N}$) 
%and where the operator ${\color{blue}H^{\#}_{\bold{j}_1,\dots, \bold{j}_{m-1}}-T_{\bold{j}=\pm \bold{j}_m}}$ takes values in the interval (\ref{spectral-interval});
% that do not belong to the set $\{\pm \bold{j}_1,\dots,\pm \bold{j}_{m}\}$. 
\item
$\mathfrak{Q}^{(>\bar{i}+1)}_{\bold{j}_m}$ is the projection onto the orthogonal complement of $\mathfrak{Q}^{(\bar{i},\bar{i}+1)}_{\bold{j}_m}\mathcal{F}^{N}$ in $\mathcal{F}^{N}$.
\end{itemize}
 Thus, we can write $$\mathfrak{Q}^{(\bar{i},\bar{i}+1)}_{\bold{j}_m}+\mathfrak{Q}^{(>\bar{i}+1)}_{\bold{j}_m}=\charf_{\mathcal{F}^{N}}\,.$$ 
 
%Starting from $i=2$ up to $\bar{i}=N-\lfloor N^{\frac{1}{16}} \rfloor$ with  $i$ even, and where, for convenience,  $\lfloor N^{\frac{1}{16}} \rfloor$ is supposed to be even as well:
%\begin{itemize}
%\item  $\mathfrak{Q}^{(i, i+1)}_{\bold{j}_m}$
%is the projection onto the subspace of $\mathfrak{Q}^{(>i-1)}_{\bold{j}_m}\mathcal{F}^{N}$ spanned by the vectors with $N -i$ or $N -i-1$ particles in the modes $\pm \bold{j}_{m}$  and where  the operator $T_{\bold{j}\notin \{\pm \bold{j}_1,\dots,\bold{j}_m\}}$ takes values in the interval (\ref{spectral-interval});
% that do not belong to the set $\{\pm \bold{j}_1,\dots,\pm \bold{j}_{m}\}$. 
%\item
%$\mathfrak{Q}^{(>i+1)}_{\bold{j}_m}$ is the projection onto the orthogonal complement of $\mathfrak{Q}^{(i, i+1)}_{\bold{j}_m}\mathfrak{Q}^{(>i-1)}_{m}\mathcal{F}^{N}$ in $\mathfrak{Q}^{(>i-1)}_{\bold{j}_m}\mathcal{F}^{N}$, i.e.,
%\begin{equation}
%\mathfrak{Q}^{(>i+1)}_{\bold{j}_m}+\mathfrak{Q}^{(i,i+1)}_{\bold{j}_m}=\mathfrak{Q}^{(>i-1)}_{\bold{j}_m}\,.
%\end{equation}
%\end{itemize}
Starting from $i=\bar{i}+2$ up to $i=N-2$ with $i$ even:
\begin{itemize}
\item  $\mathfrak{Q}^{(i, i+1)}_{\bold{j}_m}\equiv Q^{(i, i+1)}_{\bold{j}_m}$
is the projection onto the subspace of $\mathfrak{Q}^{(>i-1)}_{\bold{j}_m}\mathcal{F}^{N}$ spanned by the vectors with $N -i$ or $N -i-1$ particles in the modes $\pm \bold{j}_{m}$; %and not more than $\lfloor N^{\frac{1}{16}}\rfloor $ particles for all the other couples of modes.
% that do not belong to the set $\{\pm \bold{j}_1,\dots,\pm \bold{j}_{m}\}$. 
\item
$\mathfrak{Q}^{(>i+1)}_{\bold{j}_m}$ is the projection onto the orthogonal complement of $\mathfrak{Q}^{(i, i+1)}_{\bold{j}_m}\mathfrak{Q}^{(>i-1)}_{m}\mathcal{F}^{N}$ in $\mathfrak{Q}^{(>i-1)}_{\bold{j}_m}\mathcal{F}^{N}$, i.e.,
\begin{equation}
\mathfrak{Q}^{(>i+1)}_{\bold{j}_m}+\mathfrak{Q}^{(i,i+1)}_{\bold{j}_m}=\mathfrak{Q}^{(>i-1)}_{\bold{j}_m}\,.
\end{equation}
\end{itemize}
We shall iterate the Feshbach map starting from $i=\bar{i}$ up to $i=N-2$ with $i$ even, using the projections $\mathcal{P}^{(i)}$ and $\overline{{\mathcal{P}}^{(i)}}$  for the i-th  step of the flow where
\begin{equation}\label{projections}
\mathcal{P}^{(i)}:= \mathfrak{Q}^{(>i+1)}_{\bold{j}_m}\quad,\quad \overline{{\mathcal{P}}^{(i)}}:=\mathfrak{Q}^{(i, i+1)}_{\bold{j}_m}\,.
\end{equation}
\end{definition}
%\begin{remark}
%{\color{red}We could avoid the use of the treshold $\bar{i}$ that means we could consider $\bar{i}\equiv N-2$. At the price of a longer proof of Corollary \ref{main-lemma-H} we get a more agile formula of the ground state vector for the portion that matters in the mean field limit.}
%\end{remark}
\begin{remark}
In this section the derivation of the Feshbach Hamiltonians is only formal. Hence, we do not specify the values of $w$ such that the Feshbach map is well defined. Some expansions will be justified later in Section \ref{rig-cos-H}.  

\end{remark}
\noindent
We denote by  $\mathfrak{F}^{(i)}$ the Feshbach map at the i-th step of the flow (with $ i \geq \bar{i}$ and even) corresponding to the couple of projections $\mathcal{P}^{(i)}$ and $\overline{{\mathcal{P}}^{(i)}}$. We start applying $\mathfrak{F}^{(\bar{i})}$ to $H_{\bold{j}_1,\dots,\bold{j}_{m}}-w$,  and we get
\begin{eqnarray}
\mathscr{K}^{(\bar{i})}_{\bold{j}_1,\,\dots, \bold{j}_{m}}(w)&:=&\mathfrak{F}^{(\bar{i})}(H_{\bold{j}_1,\dots,\bold{j}_{m}}-w)\label{first-fesh-ham-0}\\
&=&\mathfrak{Q}^{(>\bar{i}+1)}_{\bold{j}_m}(H_{\bold{j}_1,\dots,\bold{j}_{m}}-w)\mathfrak{Q}^{(>\bar{i}+1)}_{\bold{j}_m}\\
& &-\mathfrak{Q}^{(>\bar{i}+1)}_{\bold{j}_m}\check{W}_{\bold{j}_1,\,\dots, \bold{j}_{m}}\mathfrak{Q}^{(\bar{i},\bar{i}+1)}_{\bold{j}_m}\frac{1}{\mathfrak{Q}^{(\bar{i},\bar{i}+1)}_{\bold{j}_m}(H_{\bold{j}_1,\dots,\bold{j}_{m}}-w)\mathfrak{Q}^{(\bar{i},\bar{i}+1)}_{\bold{j}_m}}\mathfrak{Q}^{(\bar{i},\bar{i}+1)}_{\bold{j}_m}\check{W}^*_{\bold{j}_1,\,\dots, \bold{j}_{m}}\mathfrak{Q}^{(>\bar{i}+1)}_{\bold{j}_m}\,\quad\quad\quad \label{first-fesh-ham-1}
\end{eqnarray}
where \begin{equation}
\check{W}_{\bold{j}_1,\,\dots, \bold{j}_{m}}\equiv H_{\bold{j}_1,\dots,\bold{j}_{m}}=H^{\#}_{\bold{j}_1,\dots,\bold{j}_{m-1}}+V'_{\bold{j}_1,\,\dots, \bold{j}_{m-1}}+\hat{H}^{Bog}_{\bold{j}_m}+V_{\bold{j}_{m}}-T_{\bold{j}=\pm\bold{j}_m}\,\label{def-Wcheck}
\end{equation} 
with $V'_{\bold{j}_1,\,\dots, \bold{j}_{m-1}}:=V_{\bold{j}_1,\,\dots, \bold{j}_{m-1}}-V^{\#}_{\bold{j}_1,\,\dots, \bold{j}_{m-1}}$ and $T_{\bold{j}=\pm\bold{j}_m}:=\sum_{\bold{j}=\pm \bold{j}_m} k^2_{\bold{j}}\,a_{\bold{j}}^{*}a_{\bold{j}}$. Notice that for $i\geq \bar{i}+2$
\begin{equation}
\mathfrak{Q}^{(i,i+1)}_{\bold{j}_m}\check{W}_{\bold{j}_1,\,\dots, \bold{j}_{m}}\mathfrak{Q}^{(i-2,i-1)}_{\bold{j}_m}=\mathfrak{Q}^{(i,i+1)}_{\bold{j}_m}(V'_{\bold{j}_1,\,\dots, \bold{j}_{m-1}}+W_{\bold{j}_m}+W_{\bold{j}_m}^*+V_{\bold{j}_{m}})\mathfrak{Q}^{(i-2,i-1)}_{\bold{j}_m}\,.
\end{equation}

\noindent
Next,  by iteration we define
\begin{equation}\label{definition-kappa-operators}
\mathscr{K}^{(i+2)}_{\bold{j}_1,\,\dots, \bold{j}_{m}}(w):=\mathfrak{F}^{(i+2)}(\mathscr{K}^{(i)}_{\bold{j}_1,\,\dots, \bold{j}_{m}}(w))\quad i=\bar{i},\dots,N -4\\,\quad \text{with}\,\, i\,\, \text{even}.
\end{equation}
Using the selection rules of $\check{W}_{\bold{j}_1,\,\dots, \bold{j}_{m}}(\equiv\check{W}^*_{\bold{j}_1,\,\dots, \bold{j}_{m}})$ , in Lemma  \ref{formal-fesh} we provide the formal expression of the Feshbach Hamiltonian  for  $\bar{i}+2\leq i\leq N-2$\,.

\begin{lemma}\label{formal-fesh}
Assume that the Neumann expansion
\begin{eqnarray}
& &\mathfrak{Q}^{(i,i+1)}_{\bold{j}_m}\frac{1}{\mathfrak{Q}^{(i,i+1)}_{\bold{j}_m}(H_{\bold{j}_1,\dots,\bold{j}_{m}}-w)\mathfrak{Q}^{(i,i+1)}_{\bold{j}_m}-\Gamma_{\bold{j}_1,\,\dots, \bold{j}_{m}\,;\,i,i}(w)}\mathfrak{Q}^{(i,i+1)}_{\bold{j}_m}\label{ass-nue-1}\\
&=&\sum_{l_i=0}^{\infty}R_{\bold{j}_1,\,\dots, \bold{j}_{m}\,;\,i,i}(w)\Big[\Gamma_{\bold{j}_1,\,\dots, \bold{j}_{m}\,;\,i,i}(w)R_{\bold{j}_1,\,\dots, \bold{j}_{m}\,;\,i,i}(w)\Big]^{l_i}\label{ass-nue-2}\\
&=:&(R_{\bold{j}_1,\,\dots, \bold{j}_{m}\,;\,i,i}(w))^{\frac{1}{2}}\check{\Gamma}_{\bold{j}_1,\,\dots, \bold{j}_{m}\,;\,i,i}(w)(R_{\bold{j}_1,\,\dots, \bold{j}_{m}\,;\,i,i}(w))^{\frac{1}{2}}\label{def-gamma-check}
\end{eqnarray}
holds for $\bar{i}+2\leq i\leq N-2$.  Then, for $\bar{i}+2\leq i\leq N-2$ and even
\begin{eqnarray}
& &\mathscr{K}^{(i)}_{\bold{j}_1,\,\dots, \bold{j}_{m}}(w)\label{fesh-ham-inizio}\\
&:= &\mathfrak{F}^{(i+2)}(\mathscr{K}^{(i)}_{\bold{j}_1,\,\dots, \bold{j}_{m}}(w))\label{fesh-ham-i-in}\\
&=&\mathfrak{Q}^{(>i+1)}_{\bold{j}_m}(H_{\bold{j}_1,\dots,\bold{j}_{m}}-w)\mathfrak{Q}^{(>i+1)}_{\bold{j}_m}\label{fesh-ham-i-in-bis-0}\\
%& &-\mathfrak{Q}^{(>i+1)}_{\bold{j}_M}\check{W}\mathfrak{Q}^{(i,i+1)}_{\bold{j}_M}\,R_{i,i}(w)\sum_{l_{i}=0}^{\infty}[\Gamma_{i,i}(z) R_{i,i}(w)]^{l_{i}}\, \mathfrak{Q}^{(i,i+1)}_{\bold{j}_M}\check{W}^*\mathfrak{Q}^{(>i+1)}_{\bold{j}_M}\\
& &-\mathfrak{Q}^{(>i+1)}_{\bold{j}_m}\check{W}_{\bold{j}_1,\,\dots, \bold{j}_{m}}\,R_{\bold{j}_1,\,\dots, \bold{j}_{m}\,;\,i,i}(w)\sum_{l_i=0}^{\infty}\Big[\Gamma_{\bold{j}_1,\,\dots, \bold{j}_{m}\,;\,i,i}(w)\,R_{\bold{j}_1,\,\dots, \bold{j}_{m}\,;\,i,i}(w)\Big]^{l_i}\times \quad\quad  \label{fesh-ham-i-fin}
\\
& &\quad\quad\quad\times \check{W}^*_{\bold{j}_1,\,\dots, \bold{j}_{m}}\mathfrak{Q}^{(>i+1)}_{\bold{j}_m} \nonumber\,,%& &-\mathscr{V}^{(i)}(w)\,R_{i,i}(w)\sum_{l_i=0}^{\infty}[\Gamma_{i,i}(w) R_{i,i}(w)]^{l_i}\,(\mathscr{V}^{(i)}(w) )^*
\end{eqnarray}
where :
\begin{itemize}
\item
\begin{equation}
\check{W}_{\bold{j}_1,\,\dots, \bold{j}_{m}\,;\,i,i-2}:=\mathfrak{Q}^{(i,i+1)}_{\bold{j}_m}\check{W}_{\bold{j}_1,\,\dots, \bold{j}_{m}}\mathfrak{Q}^{(i-2,i-1)}_{\bold{j}_m}=:\check{W}^*_{\bold{j}_1,\dots,\bold{j}_m\,;\,i,i-2}
\end{equation}
\item
\begin{equation}\label{def-Rii}
R_{\bold{j}_1,\,\dots, \bold{j}_{m}\,;\,i,i}(w):=\mathfrak{Q}^{(i,i+1)}_{\bold{j}_m}\frac{1}{\mathfrak{Q}^{(i,i+1)}_{\bold{j}_m}(H_{\bold{j}_1,\dots,\bold{j}_{m}}-w)\mathfrak{Q}^{(i,i+1)}_{\bold{j}_m}}\mathfrak{Q}^{(i,i+1)}_{\bold{j}_m}
\end{equation}
%\item \begin{equation}\label{GammaBog-2}
%\Gamma_{\bold{j}_1,\,\dots, \bold{j}_{M}\,;\,2,2}(z):=\check{W}_{\bold{j}_M\,;\,2,0}\,R_{\bold{j}_1,\,\dots, \bold{j}_{M}\,;\,0,0}(z)\check{W}_{\bold{j}_M\,;\,0,2}^*
%\end{equation}
\item
\begin{eqnarray}
\Gamma_{_{\bold{j}_1,\,\dots, \bold{j}_{m}}\,;\,\bar{i}+2,\bar{i}+2}(w)&:=&\mathfrak{Q}^{(\bar{i}+2,\bar{i}+3)}_{\bold{j}_m}\check{W}_{\bold{j}_1,\,\dots, \bold{j}_{m}}\mathfrak{Q}^{(\bar{i},\bar{i}+1)}_{\bold{j}_m}\frac{1}{\mathfrak{Q}^{(\bar{i},\bar{i}+1)}_{\bold{j}_m}(H_{\bold{j}_1,\,\dots, \bold{j}_{m}}-w)\mathfrak{Q}^{(\bar{i},\bar{i}+1)}_{\bold{j}_m}}\mathfrak{Q}^{(\bar{i},\bar{i}+1)}_{\bold{j}_m}\check{W}^*_{\bold{j}_1,\,\dots, \bold{j}_{m}}\mathfrak{Q}^{(\bar{i}+2,\bar{i}+3)}_{\bold{j}_m}\,,\quad\quad\quad \label{def-gamma-2}
\end{eqnarray}
for $\bar{i}+4\leq i\leq N-2$,
%\begin{equation}
%\Gamma^{Bog\,}_{i-1,i-1}(z):=W_{i-1,i-3}\,R^{Bog}_{i-3,i-3}(z) \sum_{l_{i-3}=0}^{\infty}\Big[\Gamma^{Bog\,}_{i-3,i-3}(z)R^{Bog}_{i-3,i-3}(z)\Big]^{l_{i-3}}W^*_{i-3,i-1}
%\end{equation}
\begin{eqnarray}\label{GammaH-i}
& &\Gamma_{\bold{j}_1,\,\dots, \bold{j}_{m}\,;\,i,i}(w)\\
&:=&\check{W}_{\bold{j}_1,\dots,\bold{j}_m\,;\,i,i-2}\,R_{\bold{j}_1,\,\dots, \bold{j}_{m}\,;\,i-2,i-2}(w) \sum_{l_{i-2}=0}^{\infty}\Big[\Gamma_{\bold{j}_1,\,\dots, \bold{j}_{m}\,;\,i-2,i-2}(w)R_{\bold{j}_1,\,\dots, \bold{j}_{m}\,;\,i-2,i-2}(w)\Big]^{l_{i-2}}\check{W}^*_{\bold{j}_1,\dots,\bold{j}_m\,;\,i-2,i}\quad\quad\quad\quad\\
&=&\check{W}_{\bold{j}_1,\dots,\bold{j}_m\,;\,i,i-2}\,(R_{\bold{j}_1,\,\dots, \bold{j}_{m}\,;\,i-2,i-2}(w))^{\frac{1}{2}}\times\\
& &\quad\quad\times  \sum_{l_{i-2}=0}^{\infty}\Big[(R_{\bold{j}_1,\,\dots, \bold{j}_{m}\,;\,i-2,i-2}(w))^{\frac{1}{2}}\Gamma_{\bold{j}_1,\,\dots, \bold{j}_{m}\,;\,i-2,i-2}(w)(R_{\bold{j}_1,\,\dots, \bold{j}_{m}\,;\,i-2,i-2}(w))^{\frac{1}{2}}\Big]^{l_{i-2}}\times \nonumber\quad\quad\quad\\
& &\quad\quad\quad \times (R_{\bold{j}_1,\,\dots, \bold{j}_{m}\,;\,i-2,i-2}(w))^{\frac{1}{2}}\check{W}^*_{\bold{j}_1,\dots,\bold{j}_m\,;\,i-2,i}\,.\nonumber
\end{eqnarray}
\end{itemize}

\end{lemma}

\emph{Proof}

We start computing $\mathscr{K}^{(\bar{i}+2)}_{\bold{j}_1,\,\dots, \bold{j}_{m}}(w)$ from  the operator $\mathscr{K}^{(\bar{i})}_{\bold{j}_1,\,\dots, \bold{j}_{m}}(w)$ given in (\ref{first-fesh-ham-0})-(\ref{first-fesh-ham-1}) and from the formula 
%We assume that for $\bar{i}\leq i\leq N-4$ the operator $\mathscr{K}^{(i)}_{\bold{j}_1,\,\dots, \bold{j}_{m}}(w)$ is given by the expression in (\ref{fesh-ham-i-in})-(\ref{fesh-ham-i-fin}) if $i\geq \bar{i}+2$ or by the expression in (\ref{first-fesh-ham-0})-(\ref{first-fesh-ham-1}) if $i=\bar{i}$. Then we prove that the expression given in (\ref{fesh-ham-i-in})-(\ref{fesh-ham-i-fin}) holds true for $i+2$, i.e., 
\begin{equation}
\mathfrak{F}^{(i+2)}(\mathscr{K}^{(i)}_{\bold{j}_1,\,\dots, \bold{j}_{m}}(w))=\mathscr{K}^{(i+2)}_{\bold{j}_1,\,\dots, \bold{j}_{m}}(w)\,.
\end{equation}  
 We compute
\begin{eqnarray}
%& &\mathfrak{Q}^{(>i+3)}_{\bold{j}_M}\mathscr{K}^{(i)}(w)\mathfrak{Q}^{(>i+3)}_{\bold{j}_M}\\
& &\mathfrak{Q}^{(>\bar{i}+3)}_{\bold{j}_m}\mathscr{K}^{(\bar{i})}_{\bold{j}_1,\,\dots, \bold{j}_{m}}(w)\mathfrak{Q}^{(\bar{i}+2,\bar{i}+3)}_{\bold{j}_m}\label{gener-KH-1-bis}\\
&=&\mathfrak{Q}^{(>\bar{i}+3)}_{\bold{j}_m}(H_{\bold{j}_1,\,\dots, \bold{j}_{m}}-w)\mathfrak{Q}^{(\bar{i}+2,\bar{i}+3)}_{\bold{j}_m}\\
%& &-\mathfrak{Q}^{(>i+1)}_{\bold{j}_M}\check{W}\mathfrak{Q}^{(i,i+1)}_{\bold{j}_M}\,R_{i,i}(w)\sum_{l_{i}=0}^{\infty}[\Gamma_{i,i}(z) R_{i,i}(w)]^{l_{i}}\, \mathfrak{Q}^{(i,i+1)}_{\bold{j}_M}\check{W}^*\mathfrak{Q}^{(>i+1)}_{\bold{j}_M}\\
& &-\mathfrak{Q}^{(>\bar{i}+3)}_{\bold{j}_m}\check{W}_{\bold{j}_1,\,\dots, \bold{j}_{m}}\mathfrak{Q}^{(\bar{i},\bar{i}+1)}_{\bold{j}_m}\frac{1}{\mathfrak{Q}^{(\bar{i},\bar{i}+1)}_{\bold{j}_m}(H_{\bold{j}_1,\dots,\bold{j}_{m}}-w)\mathfrak{Q}^{(\bar{i},\bar{i}+1)}_{\bold{j}_m}}\mathfrak{Q}^{(\bar{i},\bar{i}+1)}_{\bold{j}_m}\check{W}^*_{\bold{j}_1,\,\dots, \bold{j}_{m}}\mathfrak{Q}^{(\bar{i}+2,\bar{i}+3)}_{\bold{j}_m} \nonumber\\
%& &-\mathscr{V}^{(i)}(w)\,R_{i,i}(w)\sum_{l_i=0}^{\infty}[\Gamma_{i,i}(w) R_{i,i}(w)]^{l_i}\,(\mathscr{V}^{(i)}(w) )^*\\
&=&\mathfrak{Q}^{(>\bar{i}+3)}_{\bold{j}_m}\check{W}_{\bold{j}_1,\,\dots, \bold{j}_{m}}\mathfrak{Q}^{(\bar{i}+2,\bar{i}+3)}_{\bold{j}_m}
\end{eqnarray}
because $\mathfrak{Q}^{(>\bar{i}+3)}_{\bold{j}_m}\check{W}_{\bold{j}_1,\,\dots, \bold{j}_{m}}\mathfrak{Q}^{(\bar{i},\bar{i}+1)}_{\bold{j}_m}=0$ (see  (\ref{def-Wcheck})). Next, we compute
\begin{eqnarray}
%& &\mathfrak{Q}^{(>i+3)}_{\bold{j}_M}\mathscr{K}^{(i)}(w)\mathfrak{Q}^{(>i+3)}_{\bold{j}_M}\\
& &\mathfrak{Q}^{(\bar{i}+2,\bar{i}+3)}_{\bold{j}_m}\mathscr{K}^{(i)} _{\bold{j}_1,\,\dots, \bold{j}_{m}}(w)\mathfrak{Q}^{(\bar{i}+2,\bar{i}+3)}_{\bold{j}_m}\label{gener-KH-2-bis}\\
&=&\mathfrak{Q}^{(\bar{i}+2,\bar{i}+3)}_{\bold{j}_m}(H_{\bold{j}_1,\,\dots, \bold{j}_{m}}-w)\mathfrak{Q}^{(\bar{i}+2,\bar{i}+3)}_{\bold{j}_m}\\
%& &-\mathfrak{Q}^{(>i+1)}_{\bold{j}_M}\check{W}\mathfrak{Q}^{(i,i+1)}_{\bold{j}_M}\,R_{i,i}(w)\sum_{l_{i}=0}^{\infty}[\Gamma_{i,i}(z) R_{i,i}(w)]^{l_{i}}\, \mathfrak{Q}^{(i,i+1)}_{\bold{j}_M}\check{W}^*\mathfrak{Q}^{(>i+1)}_{\bold{j}_M}\\
& &-\mathfrak{Q}^{(\bar{i}+2,\bar{i}+3)}_{\bold{j}_m}\check{W}_{\bold{j}_1,\,\dots, \bold{j}_{m}}\mathfrak{Q}^{(\bar{i},\bar{i}+1)}_{\bold{j}_m}\frac{1}{\mathfrak{Q}^{(\bar{i},\bar{i}+1)}_{\bold{j}_m}(H_{\bold{j}_1,\dots,\bold{j}_{m}}-w)\mathfrak{Q}^{(\bar{i},\bar{i}+1)}_{\bold{j}_m}}\mathfrak{Q}^{(\bar{i},\bar{i}+1)}_{\bold{j}_m}\check{W}^*_{\bold{j}_1,\,\dots, \bold{j}_{m}}\mathfrak{Q}^{(\bar{i}+2,\bar{i}+3)}_{\bold{j}_m}\quad \\
%& &-\mathfrak{Q}^{(i+2,i+3)}_{\bold{j}_m}\check{W}_{\bold{j}_1,\,\dots, \bold{j}_{m}}\,\sum_{l_i=0}^{\infty}R_{\bold{j}_1,\,\dots, \bold{j}_{m}\,;\,i,i}(w)\Big[\Gamma_{\bold{j}_1,\,\dots, \bold{j}_{m}\,;\,i,i}(z)R_{\bold{j}_1,\,\dots, \bold{j}_{m}\,;\,i,i}(w)\Big]^{l_i}\times \quad\quad\\
%& &\quad\quad\quad\times \check{W}^*_{\bold{j}_1,\,\dots, \bold{j}_{m}}\mathfrak{Q}^{(i+2,i+3)}_{\bold{j}_m} \nonumber\\
%&= &\mathfrak{Q}^{(i+2,i+3)}_{\bold{j}_m}(H_{\bold{j}_1,\,\dots, \bold{j}_{m}}-w)\mathfrak{Q}^{(i+2,i+3)}_{\bold{j}_m}\\
%& &-\mathfrak{Q}^{(i+2,i+3)}_{\bold{j}_m}\,\check{W}_{\bold{j}_1,\,\dots, \bold{j}_{m}}\,\sum_{l_i=0}^{\infty}R_{\bold{j}_1,\,\dots, \bold{j}_{m}\,;\,i,i}(w)\Big[\Gamma_{\bold{j}_1,\,\dots, \bold{j}_{m}\,;\,i,i}(z)R_{\bold{j}_1,\,\dots, \bold{j}_{m}\,;\,i,i}(w)\Big]^{l_i} \,\check{W}^*_{\bold{j}_1,\,\dots, \bold{j}_{m}}\,\mathfrak{Q}^{(i+2,i+3)}_{\bold{j}_m}\quad\nonumber\\
&= &\mathfrak{Q}^{(\bar{i}+2,\bar{i}+3)}_{\bold{j}_m}(H_{\bold{j}_1,\,\dots, \bold{j}_{m}}-w)\mathfrak{Q}^{(\bar{i}+2,\bar{i}+3)}_{\bold{j}_m}-\Gamma_{\bold{j}_1,\,\dots, \bold{j}_{m}\,;\,\bar{i}+2,\bar{i}+2}(w)\,\quad\nonumber
%& &-\mathscr{V}^{(i)}(w)\,R_{i,i}(w)\sum_{l_i=0}^{\infty}[\Gamma_{i,i}(w) R_{i,i}(w)]^{l_i}\,(\mathscr{V}^{(i)}(w) )^*
\end{eqnarray}
where we have used the definition in (\ref{def-gamma-2}). 
Hence, using the assumption in (\ref{ass-nue-1})-(\ref{ass-nue-2})  we get
\begin{eqnarray}
& &\mathfrak{Q}^{(>\bar{i}+3)}_{\bold{j}_m}\mathscr{K}^{(\bar{i})} _{\bold{j}_1,\,\dots, \bold{j}_{m}}(w)\mathfrak{Q}^{(>\bar{i}+3)}_{\bold{j}_m}\\
& &-\mathfrak{Q}^{(>\bar{i}+3)}_{\bold{j}_m}\mathscr{K}^{(\bar{i})} _{\bold{j}_1,\,\dots, \bold{j}_{m}}(w)\mathfrak{Q}^{(\bar{i}+2,\bar{i}+3)}_{\bold{j}_m}\frac{1}{\mathfrak{Q}^{(\bar{i}+2,\bar{i}+3)}_{\bold{j}_M}\mathscr{K}_{\bold{j}_1,\,\dots, \bold{j}_{m}}^{(\bar{i})}(w)\mathfrak{Q}^{(\bar{i}+2,\bar{i}+3)}_{\bold{j}_m}}\mathfrak{Q}^{(\bar{i}+2,\bar{i}+3)}_{\bold{j}_m}\mathscr{K}^{(\bar{i})}(w)\mathfrak{Q}^{(>\bar{i}+3)}_{\bold{j}_m}\\
&=&\mathfrak{Q}^{(>\bar{i}+3)}_{\bold{j}_m}(H_{\bold{j}_1,\,\dots, \bold{j}_{m}}-w)\mathfrak{Q}^{(>\bar{i}+3)}_{\bold{j}_m}\\
%& &-\mathfrak{Q}^{(>i+1)}_{\bold{j}_M}\check{W}\mathfrak{Q}^{(i,i+1)}_{\bold{j}_M}\,R_{i,i}(w)\sum_{l_{i}=0}^{\infty}[\Gamma_{i,i}(z) R_{i,i}(w)]^{l_{i}}\, \mathfrak{Q}^{(i,i+1)}_{\bold{j}_M}\check{W}^*\mathfrak{Q}^{(>i+1)}_{\bold{j}_M}\\
& &-\mathfrak{Q}^{(>\bar{i}+3)}_{\bold{j}_m}\check{W}_{\bold{j}_1,\,\dots, \bold{j}_{m}}\,\sum_{l_{\bar{i}+2}=0}^{\infty}R_{\bold{j}_1,\,\dots, \bold{j}_{m}\,;\,\bar{i}+2,\bar{i}+2}(w)\Big[\Gamma_{\bold{j}_1,\,\dots, \bold{j}_{m}\,;\,\bar{i}+2,\bar{i}+2}(w)R_{\bold{j}_1,\,\dots, \bold{j}_{m}\,;\,\bar{i}+2,\bar{i}+2}(w)\Big]^{l_{\bar{i}+2}}\times \quad\quad\quad\\
& &\quad\quad\quad\times \check{W}^*_{\bold{j}_1,\,\dots, \bold{j}_{m}}\mathfrak{Q}^{(>\bar{i}+3)}_{\bold{j}_m}\,\\\nonumber
&=&\mathscr{K}^{(\bar{i}+2)} _{\bold{j}_1,\,\dots, \bold{j}_{m}}(w)\,.
%& &-\mathscr{V}^{(i)}(w)\,R_{i,i}(w)\sum_{l_i=0}^{\infty}[\Gamma_{i,i}(w) R_{i,i}(w)]^{l_i}\,(\mathscr{V}^{(i)}(w) )^*
\end{eqnarray}

Assuming that the identity in (\ref{fesh-ham-inizio})-(\ref{fesh-ham-i-fin}) holds for $i\geq \bar{i}+2$ we show that it is also valid for $i+2$. To this purpose we repeat the previous computation for $i\geq \bar{i}+2$:
\begin{itemize}
\item
\begin{eqnarray}
%& &\mathfrak{Q}^{(>i+3)}_{\bold{j}_M}\mathscr{K}^{(i)}(w)\mathfrak{Q}^{(>i+3)}_{\bold{j}_M}\\
& &\mathfrak{Q}^{(>i+3)}_{\bold{j}_m}\mathscr{K}^{(i)}_{\bold{j}_1,\,\dots, \bold{j}_{m}}(w)\mathfrak{Q}^{(i+2,i+3)}_{\bold{j}_m}\label{gener-KH-1}\\
&=&\mathfrak{Q}^{(>i+3)}_{\bold{j}_m}(H_{\bold{j}_1,\,\dots, \bold{j}_{m}}-w)\mathfrak{Q}^{(i+2,i+3)}_{\bold{j}_m}\\
%& &-\mathfrak{Q}^{(>i+1)}_{\bold{j}_M}\check{W}\mathfrak{Q}^{(i,i+1)}_{\bold{j}_M}\,R_{i,i}(w)\sum_{l_{i}=0}^{\infty}[\Gamma_{i,i}(z) R_{i,i}(w)]^{l_{i}}\, \mathfrak{Q}^{(i,i+1)}_{\bold{j}_M}\check{W}^*\mathfrak{Q}^{(>i+1)}_{\bold{j}_M}\\
& &-\mathfrak{Q}^{(>i+3)}_{\bold{j}_m}\check{W}_{\bold{j}_1,\,\dots, \bold{j}_{m}}\,\sum_{l_i=0}^{\infty}R_{\bold{j}_1,\,\dots, \bold{j}_{m}\,;\,i,i}(w)\Big[\Gamma_{\bold{j}_1,\,\dots, \bold{j}_{m}\,;\,i,i}(w)R_{\bold{j}_1,\,\dots, \bold{j}_{m}\,;\,i,i}(w)\Big]^{l_i}\times \quad\quad\\
& &\quad\quad\quad\times \check{W}^*_{\bold{j}_1,\,\dots, \bold{j}_{m}}\mathfrak{Q}^{(i+2,i+3)}_{\bold{j}_m} \nonumber\\
%& &-\mathscr{V}^{(i)}(w)\,R_{i,i}(w)\sum_{l_i=0}^{\infty}[\Gamma_{i,i}(w) R_{i,i}(w)]^{l_i}\,(\mathscr{V}^{(i)}(w) )^*\\
&=&\mathfrak{Q}^{(>i+3)}_{\bold{j}_m}\check{W}_{\bold{j}_1,\,\dots, \bold{j}_{m}}\mathfrak{Q}^{(i+2,i+3)}_{\bold{j}_m}
\end{eqnarray}
because $\mathfrak{Q}^{(>i+3)}_{\bold{j}_m}\check{W}_{\bold{j}_1,\,\dots, \bold{j}_{m}}\mathfrak{Q}^{(i,i+1)}_{\bold{j}_m}=0$ (see  (\ref{def-Wcheck}));
\item
likewise 
\begin{eqnarray}
%& &\mathfrak{Q}^{(>i+3)}_{\bold{j}_M}\mathscr{K}^{(i)}(w)\mathfrak{Q}^{(>i+3)}_{\bold{j}_M}\\
& &\mathfrak{Q}^{(i+2,i+3)}_{\bold{j}_m}\mathscr{K}^{(i)} _{\bold{j}_1,\,\dots, \bold{j}_{m}}(w)\mathfrak{Q}^{(i+2,i+3)}_{\bold{j}_m}\label{gener-KH-2}\\
&=&\mathfrak{Q}^{(i+2,i+3)}_{\bold{j}_m}(H_{\bold{j}_1,\,\dots, \bold{j}_{m}}-w)\mathfrak{Q}^{(i+2,i+3)}_{\bold{j}_m}\\
%& &-\mathfrak{Q}^{(>i+1)}_{\bold{j}_M}\check{W}\mathfrak{Q}^{(i,i+1)}_{\bold{j}_M}\,R_{i,i}(w)\sum_{l_{i}=0}^{\infty}[\Gamma_{i,i}(z) R_{i,i}(w)]^{l_{i}}\, \mathfrak{Q}^{(i,i+1)}_{\bold{j}_M}\check{W}^*\mathfrak{Q}^{(>i+1)}_{\bold{j}_M}\\
& &-\mathfrak{Q}^{(i+2,i+3)}_{\bold{j}_m}\check{W}_{\bold{j}_1,\,\dots, \bold{j}_{m}}\,\sum_{l_i=0}^{\infty}R_{\bold{j}_1,\,\dots, \bold{j}_{m}\,;\,i,i}(w)\Big[\Gamma_{\bold{j}_1,\,\dots, \bold{j}_{m}\,;\,i,i}(z)R_{\bold{j}_1,\,\dots, \bold{j}_{m}\,;\,i,i}(w)\Big]^{l_i}\times \quad\quad\\
& &\quad\quad\quad\times \check{W}^*_{\bold{j}_1,\,\dots, \bold{j}_{m}}\mathfrak{Q}^{(i+2,i+3)}_{\bold{j}_m} \nonumber\\
%&= &\mathfrak{Q}^{(i+2,i+3)}_{\bold{j}_m}(H_{\bold{j}_1,\,\dots, \bold{j}_{m}}-w)\mathfrak{Q}^{(i+2,i+3)}_{\bold{j}_m}\\
%& &-\mathfrak{Q}^{(i+2,i+3)}_{\bold{j}_m}\,\check{W}_{\bold{j}_1,\,\dots, \bold{j}_{m}}\,\sum_{l_i=0}^{\infty}R_{\bold{j}_1,\,\dots, \bold{j}_{m}\,;\,i,i}(w)\Big[\Gamma_{\bold{j}_1,\,\dots, \bold{j}_{m}\,;\,i,i}(z)R_{\bold{j}_1,\,\dots, \bold{j}_{m}\,;\,i,i}(w)\Big]^{l_i} \,\check{W}^*_{\bold{j}_1,\,\dots, \bold{j}_{m}}\,\mathfrak{Q}^{(i+2,i+3)}_{\bold{j}_m}\quad\nonumber\\
&= &\mathfrak{Q}^{(i+2,i+3)}_{\bold{j}_m}(H_{\bold{j}_1,\,\dots, \bold{j}_{m}}-w)\mathfrak{Q}^{(i+2,i+3)}_{\bold{j}_m}-\Gamma_{\bold{j}_1,\,\dots, \bold{j}_{m}\,;\,i+2,i+2}(w)\,.\quad\nonumber
%& &-\mathscr{V}^{(i)}(w)\,R_{i,i}(w)\sum_{l_i=0}^{\infty}[\Gamma_{i,i}(w) R_{i,i}(w)]^{l_i}\,(\mathscr{V}^{(i)}(w) )^*
\end{eqnarray}
\end{itemize}
%where the identity in (\ref{ortho-prop}) has been used used  repeatedly.
%  for  (\ref{gener-KH-2}).

\noindent
Hence, using the assumption in (\ref{ass-nue-1})-(\ref{ass-nue-2})  we get
\begin{eqnarray}
& &\mathfrak{Q}^{(>i+3)}_{\bold{j}_m}\mathscr{K}^{(i)} _{\bold{j}_1,\,\dots, \bold{j}_{m}}(w)\mathfrak{Q}^{(>i+3)}_{\bold{j}_m}\\
& &-\mathfrak{Q}^{(>i+3)}_{\bold{j}_m}\mathscr{K}^{(i)} _{\bold{j}_1,\,\dots, \bold{j}_{m}}(w)\mathfrak{Q}^{(i+2,i+3)}_{\bold{j}_m}\frac{1}{\mathfrak{Q}^{(i+2,i+3)}_{\bold{j}_M}\mathscr{K}_{\bold{j}_1,\,\dots, \bold{j}_{m}}^{(i)}(w)\mathfrak{Q}^{(i+2,i+3)}_{\bold{j}_m}}\mathfrak{Q}^{(i+2,i+3)}_{\bold{j}_m}\mathscr{K}^{(i)}(w)\mathfrak{Q}^{(>i+3)}_{\bold{j}_m}\\
&=&\mathfrak{Q}^{(>i+3)}_{\bold{j}_m}(H_{\bold{j}_1,\,\dots, \bold{j}_{m}}-w)\mathfrak{Q}^{(>i+3)}_{\bold{j}_m}\\
%& &-\mathfrak{Q}^{(>i+1)}_{\bold{j}_M}\check{W}\mathfrak{Q}^{(i,i+1)}_{\bold{j}_M}\,R_{i,i}(w)\sum_{l_{i}=0}^{\infty}[\Gamma_{i,i}(z) R_{i,i}(w)]^{l_{i}}\, \mathfrak{Q}^{(i,i+1)}_{\bold{j}_M}\check{W}^*\mathfrak{Q}^{(>i+1)}_{\bold{j}_M}\\
& &-\mathfrak{Q}^{(>i+3)}_{\bold{j}_m}\check{W}_{\bold{j}_1,\,\dots, \bold{j}_{m}}\,\sum_{l_{i+2}=0}^{\infty}R_{\bold{j}_1,\,\dots, \bold{j}_{m}\,;\,i+2,i+2}(w)\Big[\Gamma_{\bold{j}_1,\,\dots, \bold{j}_{m}\,;\,i+2,i+2}(w)R_{\bold{j}_1,\,\dots, \bold{j}_{m}\,;\,i+2,i+2}(w)\Big]^{l_{i+2}}\times \quad\quad\quad\\
& &\quad\quad\quad\times \check{W}^*_{\bold{j}_1,\,\dots, \bold{j}_{m}}\mathfrak{Q}^{(>i+3)}_{\bold{j}_m}\,\\\nonumber
&=&\mathscr{K}^{(i+2)} _{\bold{j}_1,\,\dots, \bold{j}_{m}}(w)\,.
%& &-\mathscr{V}^{(i)}(w)\,R_{i,i}(w)\sum_{l_i=0}^{\infty}[\Gamma_{i,i}(w) R_{i,i}(w)]^{l_i}\,(\mathscr{V}^{(i)}(w) )^*
\end{eqnarray}

\qed

For the last implementation (corresponding to $i=N$)  of the Feshbach map we employ  the projections 
\begin{eqnarray}
\mathcal{P}^{(N)}&:=&\mathscr{P}_{\psi^{\#}_{\bold{j}_1,\dots,\bold{j}_{m-1}}}:=|\frac{\psi^{\#}_{\bold{j}_1,\dots,\bold{j}_{m-1}}}{\|\psi^{\#}_{\bold{j}_1,\dots,\bold{j}_{m-1}}\|} \rangle \langle \frac{\psi^{\#}_{\bold{j}_1,\dots,\bold{j}_{m-1}}}{\|\psi^{\#}_{\bold{j}_1,\dots,\bold{j}_{m-1}}\|}| \,,\\
\overline{\mathcal{P}^{(N)}}&:=&\overline{\mathscr{P}_{\psi^{\#}_{\bold{j}_1,\dots,\bold{j}_{m-1}}}}:=\charf_{\mathfrak{Q}^{(> N-1)}_{\bold{j}_m}\mathcal{F}^N}-|\frac{\psi^{\#}_{\bold{j}_1,\dots,\bold{j}_{m-1}}}{\|\psi^{\#}_{\bold{j}_1,\dots,\bold{j}_{m-1}}\|} \rangle \langle \frac{\psi^{\#}_{\bold{j}_1,\dots,\bold{j}_{m-1}}}{\|\psi^{\#}_{\bold{j}_1,\dots,\bold{j}_{m-1}}\|} |\,
\end{eqnarray}
where $\psi^{\#}_{\bold{j}_1,\dots,\bold{j}_{m-1}}\equiv \eta$  for $m=1$, and for $m\geq2$ is the ground state vector (non-normalized) of the Hamiltonian $H^{\#}_{\bold{j}_1,\dots,\bold{j}_{m-1}}$ (see (\ref{def-xiham-1-zero})). The vector $\psi^{\#}_{\bold{j}_1,\dots,\bold{j}_{m-1}}(\neq 0)$ and the corresponding eigenvalue $z^{\#}_{\bold{j}_1,\dots,\bold{j}_{m-1}}$  will be iteratively constructed in the next section.
\begin{remark}
The auxiliary Hamiltonian $H^{\#}_{\bold{j}_1,\dots,\bold{j}_{m-1}}$ mediates the step from $H_{\bold{j}_1,\dots,\bold{j}_{m-1}}$ to $H_{\bold{j}_1,\dots,\bold{j}_{m}}$. In the analogous construction for the Bogoliubov Hamiltonians (see \cite{Pi2}) the operator $H^{Bog}_{\bold{j}_1,\dots,\bold{j}_{m-1}}$ plays the role of $H^{\#}_{\bold{j}_1,\dots,\bold{j}_{m-1}}$, thus no auxiliary Hamiltonian is necessary.  Indeed,  the Bogoliubov interaction associated with the Fourier modes $\{\pm\bold{j}_l\,|\,l=1,\dots,m-1\}$ of the pair potential contains only the field modes $\{\bold{0};\pm\bold{j}_l\,|\,l=1,\dots,m-1\}$. Hence, it does not contain the operators $a_{\pm \bold{j}_m}\,,\,a^*_{\pm \bold{j}_m}$.
\end{remark}
%Similarly to the Hamiltonian $\hat{H}^{Bog}_{\bold{j}_1,\dots,\bold{j}_{m}}$ associated with $H^{Bog}_{\bold{j}_1,\dots,\bold{j}_{m}}$ we define for $m\geq 2$
%\begin{equation}\label{def-hdiesis}
%\hat{H}^{\#}_{\bold{j}_1,\dots,\bold{j}_{m}}:= T_{\bold{j}\neq\{\pm\bold{j}_{1},\dots, \pm\bold{j}_{m}\}}+\hat{H}^{Bog}_{\bold{j}_1,\dots,\bold{j}_{m-1}}+V^{\#}_{\bold{j}_1,\dots,\bold{j}_{m-1}}\,
%\end{equation}
%%\begin{equation}
%H^{\#}_{\bold{j}_1,\dots,\bold{j}_{m-1}}= T_{\bold{j}=\pm\bold{j}_{m}}+\hat{H}^{\#}_{\bold{j}_1,\dots,\bold{j}_{m-1}}\,.
%\end{equation}
Formally we get
\begin{eqnarray}
& &\mathscr{K}^{(N)}_{\bold{j}_1,\,\dots, \bold{j}_{m}}(w)\label{Ham-N-in}\\
&:=&\mathfrak{F}^{(N)}(\mathscr{K}^{(N-2)}_{\bold{j}_1,\,\dots, \bold{j}_{m}}(w))\\
&=&\mathscr{P}_{\psi^{\#}_{\bold{j}_1,\dots,\bold{j}_{m-1}}}(H_{\bold{j}_1,\,\dots, \bold{j}_{m}}-w)\mathscr{P}_{\psi^{\#}_{\bold{j}_1,\dots,\bold{j}_{m-1}}}-\mathscr{P}_{\psi^{\#}_{\bold{j}_1,\dots,\bold{j}_{m-1}}}\Gamma_{\bold{j}_1,\,\dots, \bold{j}_{m}\,;\,N,N}(w)\mathscr{P}_{\psi^{\#}_{\bold{j}_1,\dots,\bold{j}_{m-1}}}\quad \label{K-last-step}\\
& &-\mathscr{P}_{\psi^{\#}_{\bold{j}_1,\dots,\bold{j}_{m-1}}}(V_{\bold{j}_{m}}-\Gamma_{\bold{j}_1,\,\dots, \bold{j}_{m}\,;\,N,N}(w))\times \label{term-not-important}\\
& &\quad\quad\times \overline{\mathscr{P}_{\psi^{\#}_{\bold{j}_1,\dots,\bold{j}_{m-1}}}}\,\frac{1}{\overline{\mathscr{P}_{\psi^{\#}_{\bold{j}_1,\dots,\bold{j}_{M-1}}}}\mathscr{K}^{(N-2)}_{\bold{j}_1,\,\dots, \bold{j}_{m}}(w)\overline{\mathscr{P}_{\psi^{\#}_{\bold{j}_1,\dots,\bold{j}_{m-1}}}}}\overline{\mathscr{P}_{\psi^{\#}_{\bold{j}_1,\dots,\bold{j}_{m-1}}}}\times \nonumber\\
& &\quad\quad\quad\quad\quad\quad \times (V_{\bold{j}_{m}}-\Gamma_{\bold{j}_1,\,\dots, \bold{j}_{m}\,;\,N,N}(w))^*\mathscr{P}_{\psi^{\#}_{\bold{j}_1,\dots,\bold{j}_{m-1}}} \nonumber\\
&=:&f_{\bold{j}_1,\dots,\bold{j}_m}(w)\mathscr{P}_{\psi^{\#}_{\bold{j}_1,\dots,\bold{j}_{m-1}}}\,\label{K-last-step-bis}
\end{eqnarray}
where
\begin{eqnarray}\label{GammaBog-N}
& &\Gamma_{\bold{j}_1,\,\dots, \bold{j}_{m}\,;\,N,N}(w)\\
&:=&\check{W}_{\bold{j}_1,\,\dots, \bold{j}_{m}}\,R_{\bold{j}_1,\,\dots, \bold{j}_{m}\,;\,N-2,N-2}(w) \sum_{l_{N-2}=0}^{\infty}\Big[\Gamma_{\bold{j}_1,\,\dots, \bold{j}_{m}\,;\,N-2,N-2}(w)R_{\bold{j}_1,\,\dots, \bold{j}_{m}\,;\,N-2,N-2}(w)\Big]^{l_{N-2}}\check{W}^*_{\bold{j}_1,\,\dots, \bold{j}_{m}}\nonumber\\
&=&\check{W}_{\bold{j}_1,\,\dots, \bold{j}_{m}}\,(R_{\bold{j}_1,\,\dots, \bold{j}_{m}\,;\,N-2,N-2}(w))^{\frac{1}{2}}\times\\
& &\quad\quad\times  \sum_{l_{N-2}=0}^{\infty}\Big[(R_{\bold{j}_1,\,\dots, \bold{j}_{m}\,;\,N-2,N-2}(w))^{\frac{1}{2}}\Gamma_{\bold{j}_1,\,\dots, \bold{j}_{m}\,;\,N-2,N-2}(w)(R_{\bold{j}_1,\,\dots, \bold{j}_{m}\,;\,N-2,N-2}(w))^{\frac{1}{2}}\Big]^{l_{N-2}}\times \quad\quad\quad\\
& &\quad\quad\quad \times (R_{\bold{j}_1,\,\dots, \bold{j}_{m}\,;\,N-2,N-2}(w))^{\frac{1}{2}}\check{W}^*_{\bold{j}_1,\,\dots, \bold{j}_{m}}\label{lastGammaH}\,.
\end{eqnarray}

Then, we set $w=z+z^{\#}_{\bold{j}_1,\dots,\bold{j}_{m-1}}$ and solve the fixed point equation
 \begin{equation}f_{\bold{j}_1,\dots,\bold{j}_m}(w)=f_{\bold{j}_1,\dots,\bold{j}_m}(z+z^{\#}_{\bold{j}_1,\dots,\bold{j}_{m-1}})=0\,.\label{f-p-e}\end{equation}
The (unique) solution, $w\equiv z_{\bold{j}_1,\dots,\bold{j}_m}$, is the (non-degenerate) ground state energy of $H_{\bold{j}_1,\,\dots, \bold{j}_{m}}$. The corresponding (non-normalized) ground state vector is
\begin{eqnarray}
& &\psi_{\bold{j}_1,\dots,\bold{j}_m} \label{gs-Hm-start-bis}\\
&:=&\Big\{\charf 
%\Big[Q^{(>0)}-\frac{1}{Q^{(0)}(H^{Bog}-z_*)Q^{(0)}}Q^{(0)}(H^{Bog}-z_*)Q^{(>0)}\Big]\times\\
%& &\quad\quad\quad\times \Big\{ \prod_{i=0}^{N-2}\Big[Q^{(>i+1)}-\frac{1}{Q^{(i+1)}\mathscr{K}^{Bog\,(i)}(z_*)Q^{(i+1)}}Q^{(i+1)}\mathscr{K}^{Bog\,(i)}(z_*)Q^{(>i+1)}\Big]\Big\}\eta \quad \\
%&= &\eta \\
%& &-\frac{1}{Q^{(N-2)}\mathscr{K}^{Bog\,(N-3)}(z_*)Q^{(N-2)}}Q^{(N-2)}W^*\eta \\
%& &+\frac{1}{Q^{(N-4)}\mathscr{K}^{Bog\,(N-5)}(z_*)Q^{(N-4)}}W^*_{N-4,N-2}\frac{1}{Q^{(N-2)}\mathscr{K}^{Bog\,(N-3)}(z_*)Q^{(N-2)}}Q^{(N-2)}W^*\eta \quad \nonumber \\
%& &-\frac{1}{Q^{(N-6)}\mathscr{K}^{Bog\,(N-7)}(z_*)Q^{(N-6)}}W^*_{N-6,N-4}\frac{1}{Q^{(N-4)}\mathscr{K}^{Bog\,(N-5)}(z_*)Q^{(N-4)}}Q^{(N-4)}W^*Q^{(N-2)}\times \quad \nonumber \\
%& &\quad \times\frac{1}{Q^{(N-2)}\mathscr{K}^{Bog\,(N-3)}(z_*)Q^{(N-2)}}Q^{(N-2)}W^*\eta \\
%& &+\dots \\
-\frac{1}{\mathfrak{Q}^{(N-2, N-1)}_{\bold{j}_m}\mathscr{K}_{\bold{j}_1,\dots ,\bold{j}_m}^{(N-4)}(z_{\bold{j}_1,\dots ,\bold{j}_m})\mathfrak{Q}^{(N-2, N-1)}_{\bold{j}_m}}\mathfrak{Q}^{(N-2, N-1)}_{\bold{j}_m}\check{W}_{\bold{j}_1,\,\dots, \bold{j}_{m};N-2,N}\nonumber\\
& &\quad-\sum_{j=2}^{\frac{N-\bar{i}}{2}}\prod^{2}_{r=j}\Big[-\frac{1}{\mathfrak{Q}_{\bold{j}_m}^{(N-2r,N-2r+1)}\mathscr{K}_{\bold{j}_1,\dots ,\bold{j}_m}^{(N-2r-2)}(z_{\bold{j}_1,\dots ,\bold{j}_m})\mathfrak{Q}_{\bold{j}_m}^{(N-2r, N-2r+1)}}\check{W}_{\bold{j}_1,\dots,\bold{j}_m;N-2r,N-2r+2}\Big]\times\nonumber\\
& &\quad\quad\quad \times \frac{1}{\mathfrak{Q}_{\bold{j}_m}^{(N-2, N-1)}\mathscr{K}_{\bold{j}_1,\dots ,\bold{j}_m}^{(N-4)}(z_{\bold{j}_1,\dots, \bold{j}_m})\mathfrak{Q}^{(N-2,N-1)}_{\bold{j}_m}}\mathfrak{Q}^{(N-2, N-1)}_{\bold{j}_m}\check{W}_{\bold{j}_1,\,\dots, \bold{j}_{m};N-2,N}\Big\}\times \label{gs-vector-inter}\\
& & \times \Big[\charf -\frac{1}{\overline{\mathscr{P}_{\psi^{\#}_{\bold{j}_1,\dots,\bold{j}_{m-1}}}}\mathscr{K}^{(N-2)}_{\bold{j}_1,\dots,\bold{j}_m}(z_{\bold{j}_1,\dots,\bold{j}_m})\overline{\mathscr{P}_{\psi^{\#}_{\bold{j}_1, \dots, \bold{j}_{m-1}}}}}
\overline{\mathscr{P}_{\psi^{\#}_{\bold{j}_1\dots,\bold{j}_{m-1}}}}\mathscr{K}^{(N-2)}_{\bold{j}_1,\dots,\bold{j}_m}(z_{\bold{j}_1,\dots,\bold{j}_m})\Big]\psi^{\#}_{\bold{j}_1, \dots, \bold{j}_{m-1}}\,.\quad\quad \label{gs-Hm-fin-bis}
%&=:&T_{m}\,\psi^{Bog}_{\bold{j}_1, \dots, \bold{j}_{m-1}}
\end{eqnarray}
where $\mathscr{K}^{(\bar{i}-2)}_{\bold{j}_1,\dots,\bold{j}_m}(z_{\bold{j}_1,\dots,\bold{j}_m})=H_{\bold{j}_1,\dots,\bold{j}_m}-z_{\bold{j}_1,\dots,\bold{j}_m}$.

The ground state $\psi^{\#}_{\bold{j}_1,\dots,\bold{j}_m}$ corresponding to the ground state energy $z^{\#}_{\bold{j}_1,\dots,\bold{j}_m}$ of the Hamiltonian $H^{\#}_{\bold{j}_1,\dots,\bold{j}_m}$ has an analogous formula
\begin{eqnarray}
& &\psi^{\#}_{\bold{j}_1,\dots,\bold{j}_m} \label{gs-Hm-start-bis-diesis}\\
&:=&\Big\{\charf 
%\Big[Q^{(>0)}-\frac{1}{Q^{(0)}(H^{Bog}-z_*)Q^{(0)}}Q^{(0)}(H^{Bog}-z_*)Q^{(>0)}\Big]\times\\
%& &\quad\quad\quad\times \Big\{ \prod_{i=0}^{N-2}\Big[Q^{(>i+1)}-\frac{1}{Q^{(i+1)}\mathscr{K}^{Bog\,(i)}(z_*)Q^{(i+1)}}Q^{(i+1)}\mathscr{K}^{Bog\,(i)}(z_*)Q^{(>i+1)}\Big]\Big\}\eta \quad \\
%&= &\eta \\
%& &-\frac{1}{Q^{(N-2)}\mathscr{K}^{Bog\,(N-3)}(z_*)Q^{(N-2)}}Q^{(N-2)}W^*\eta \\
%& &+\frac{1}{Q^{(N-4)}\mathscr{K}^{Bog\,(N-5)}(z_*)Q^{(N-4)}}W^*_{N-4,N-2}\frac{1}{Q^{(N-2)}\mathscr{K}^{Bog\,(N-3)}(z_*)Q^{(N-2)}}Q^{(N-2)}W^*\eta \quad \nonumber \\
%& &-\frac{1}{Q^{(N-6)}\mathscr{K}^{Bog\,(N-7)}(z_*)Q^{(N-6)}}W^*_{N-6,N-4}\frac{1}{Q^{(N-4)}\mathscr{K}^{Bog\,(N-5)}(z_*)Q^{(N-4)}}Q^{(N-4)}W^*Q^{(N-2)}\times \quad \nonumber \\
%& &\quad \times\frac{1}{Q^{(N-2)}\mathscr{K}^{Bog\,(N-3)}(z_*)Q^{(N-2)}}Q^{(N-2)}W^*\eta \\
%& &+\dots \\
-\frac{1}{\mathfrak{Q}^{(N-2, N-1)}_{\bold{j}_m}\mathscr{K}_{\bold{j}_1,\dots ,\bold{j}_m}^{\#\,(N-4)}(z^{\#}_{\bold{j}_1,\dots ,\bold{j}_m})\mathfrak{Q}^{(N-2, N-1)}_{\bold{j}_m}}\mathfrak{Q}^{(N-2, N-1)}_{\bold{j}_m}\check{W}^{\#}_{\bold{j}_1,\,\dots, \bold{j}_{m};N-2,N}\nonumber\\
& &\quad-\sum_{j=2}^{\frac{N-\bar{i}}{2}}\prod^{2}_{r=j}\Big[-\frac{1}{\mathfrak{Q}_{\bold{j}_m}^{(N-2r,N-2r+1)}\mathscr{K}_{\bold{j}_1,\dots ,\bold{j}_m}^{\#\,(N-2r-2)}(z^{\#}_{\bold{j}_1,\dots ,\bold{j}_m})\mathfrak{Q}_{\bold{j}_m}^{(N-2r, N-2r+1)}}\check{W}^{\#}_{\bold{j}_1,\dots,\bold{j}_m;N-2r,N-2r+2}\Big]\times\nonumber\\
& &\quad\quad\quad \times \frac{1}{\mathfrak{Q}_{\bold{j}_m}^{(N-2, N-1)}\mathscr{K}_{\bold{j}_1,\dots ,\bold{j}_m}^{\#\,(N-4)}(z^{\#}_{\bold{j}_1,\dots, \bold{j}_m})\mathfrak{Q}^{(N-2,N-1)}_{\bold{j}_m}}\mathfrak{Q}^{(N-2, N-1)}_{\bold{j}_m}\check{W}^{\#}_{\bold{j}_1,\,\dots, \bold{j}_{m};N-2,N}\Big\}\times\\
& & \times \Big[\charf -\frac{1}{\overline{\mathscr{P}_{\psi^{\#}_{\bold{j}_1,\dots,\bold{j}_{m-1}}}}\mathscr{K}^{\#\,(N-2)}_{\bold{j}_1,\dots,\bold{j}_m}(z^{\#}_{\bold{j}_1,\dots,\bold{j}_m})\overline{\mathscr{P}_{\psi^{\#}_{\bold{j}_1, \dots, \bold{j}_{m-1}}}}}
\overline{\mathscr{P}_{\psi^{\#}_{\bold{j}_1\dots,\bold{j}_{m-1}}}}\mathscr{K}^{\#\,(N-2)}_{\bold{j}_1,\dots,\bold{j}_m}(z^{\#}_{\bold{j}_1,\dots,\bold{j}_m})\Big]\psi^{\#}_{\bold{j}_1, \dots, \bold{j}_{m-1}}\,\quad\quad\quad \label{gs-Hm-fin-bis-diesis}
\end{eqnarray}
where $\mathscr{K}^{\#\,(\bar{i}-2)}_{\bold{j}_1,\dots,\bold{j}_m}(z^{\#}_{\bold{j}_1,\dots,\bold{j}_m}):=H^{\#}_{\bold{j}_1,\dots,\bold{j}_m}-z^{\#}_{\bold{j}_1,\dots,\bold{j}_m}=:W^{\#}_{\bold{j}_1,\dots,\bold{j}_m}-z^{\#}_{\bold{j}_1,\dots,\bold{j}_m}$ and  $\mathscr{K}^{\#\,(i)}_{\bold{j}_1,\dots,\bold{j}_m}(z^{\#}_{\bold{j}_1,\dots,\bold{j}_m})=\mathfrak{F}^{(i)}(\mathscr{K}^{\#\,(i-2)}_{\bold{j}_1,\dots,\bold{j}_m}(z^{\#}_{\bold{j}_1,\dots,\bold{j}_m}))$ with $i\geq \bar{i}$ and even.
%&=:&T_{m}\,\psi^{Bog}_{\bold{j}_
%where  $\psi^{Bog}_{\bold{j}_1,\dots,\bold{j}_{m-1}}$ is replaced by the vector $\eta$ for $m=1$.

The main subtleties in the construction are concerned with the very first implementation (from $i=\bar{i}$ to $i=\bar{i}+2$) and the very last implementation of the Feshbach map (from $i=N-2$ to $i=N$). 
\section{Rigorous construction of the Feshbach Hamiltonians $\mathscr{K}_{\bold{j}_1,\dots,\bold{j}_m}^{(i)}(w)$}\label{rig-cos-H}
\setcounter{equation}{0}
The construction outlined in Section \ref{new-proj} must be implemented by induction in the index  $m$ ranging from $1$ to $M$. For the sake of clarity, first we show that for each $m$   the Feshbach flow can be rigorously defined from $i=\bar{i}$ up to $i=N-2$ under some conditions to be proven later in Section \ref{last-Fesh}. %First, we shall implement the Feshbach flow for $m=1$.

%We observe that because of the new projections given  for $m\geq 2$ in Section \ref{new-proj}, the Feshbach map will be implemented $\bar{i}$ times with $\bar{i}= \lfloor N^{\frac{1}{16}}\rfloor$. Starting from $i=0$ the projections $\{\mathfrak{Q}^{(i, i+1)}_{\bold{j}_m}\,\,;\,\, i=0,\dots, \bar{i}-2\}$  constrain the number of particles in the modes $\pm \bold{j}_{m}$ to values smaller than $\lfloor N^{\frac{1}{16}}\rfloor$. We also point out that there is no constraint on the occupation numbers in the modes $\{\pm \bold{j}_1, \dots, \pm \bold{j}_{m-1}\}$.
%\\

%The number of particles in the modes $\pm \bold{j}_m$ must be constrained starting from $i=0$ otherwise we cannot control the ``non-Bogoliubov" terms, $V_{\bold{j}_1,\dots,\bold{j}_m}$,  in the interaction. The price of this constraint is that for each implementation of the Feshbach map we project out a smaller subspace so that at the very last implementation of the Feshbach map a subspace of states containing ``many" particles in the nonzero modes  has to be projected out in one shot.
%\\

For the next estimates, it is important to isolate a term proportional to the kinetic energy operator $T:=\sum_{\bold{j}\in \mathbb{Z}^d}k_{\bold{j}}^2a^*_{\bold{j}}a_{\bold{j}}$ in the Hamiltonians $H_{\bold{j}_1,\dots,\bold{j}_{m}}$ and $H^{\#}_{\bold{j}_1,\dots,\bold{j}_{m}}$, by which we can dominate  some of the terms of the interaction Hamiltonian. To this purpose we introduce some definitions dependent on a suitable small ($N$-dependent) positive number $\xi$.

\noindent
For $m\geq2$
\begin{equation}\label{def-xiham-1}
(H^{\#}_{\bold{j}_1,\dots,\bold{j}_{m-1}})_{\xi}:=(1-\xi)T_{\bold{j}\notin\{\pm\bold{j}_{1};\dots; \pm\bold{j}_{m-1}\}}+(\hat{H}^{Bog}_{\bold{j}_1,\dots,\bold{j}_{m-1}})_{\xi}+V^{\#}_{\bold{j}_1,\dots,\bold{j}_{m-1}}
\end{equation}
%and
%\begin{equation}\label{def-xiham-1-bis}
%%\end{equation}
where $T_{\bold{j}\notin\{\pm\bold{j}_{1};\dots; \pm\bold{j}_{m-1}\}}$ is defined in (\ref{def-Tsenza}) and 
%\item
%\begin{equation}
%T_{\bold{j}\neq\{\pm\bold{j}_{1},\dots, \pm\bold{j}_{m}\}}:=\sum_{\bold{j}\neq\{\pm\bold{j}_{1},\dots, \pm\bold{j}_{m}\}}(k_{\bold{j}})^2a^*_{\bold{j}}a_{\bold{j}}\,;\quad\quad
%\end{equation}
%\item
\begin{equation}
(\hat{H}^{Bog}_{\bold{j}_1,\dots,\bold{j}_{m-1}})_{\xi}:=\sum_{l=1}^{m-1}(\hat{H}^{Bog}_{\bold{j}_l})_{\xi}\end{equation}
with
\begin{eqnarray}
(\hat{H}^{Bog}_{\bold{j}_l})_{\xi}
&:=&\sum_{\bold{j}=\pm \bold{j}_l} [(1-\xi)k^2_{\bold{j}}+\frac{\phi_{\bold{j}}}{N}a^*_{\bold{0}}a_{\bold{0}}]\,a_{\bold{j}}^{*}a_{\bold{j}}+\frac{1}{2}\sum_{\bold{j}=\pm \bold{j}_l}\frac{\phi_{\bold{j}}}{N}\,\Big\{a^*_{\bold{0}}a^*_{\bold{0}}a_{\bold{j}}a_{-\bold{j}}+a^*_{\bold{j}}a^*_{-\bold{j}}a_{\bold{0}}a_{\bold{0}}\Big\}\,.\quad\quad\quad\quad
\end{eqnarray}

\noindent
Thus we can write (see (\ref{def-xiham-1}))
\begin{equation}\label{rel-H-hat-H-hat-xi}
H^{\#}_{\bold{j}_1,\dots,\bold{j}_{m-1}}=(H^{\#}_{\bold{j}_1,\dots,\bold{j}_{m-1}})_{\xi}+\xi T\,,
\end{equation}
and
\begin{eqnarray}\label{xi-depend-ham}
H_{\bold{j}_1,\dots,\bold{j}_m}&=&(H^{\#}_{\bold{j}_{1},\dots,\bold{j}_{m-1}})_{\xi}-(1-\xi)T_{\bold{j}=\pm\bold{j}_m}+V_{\bold{j}_{m}}+V'_{\bold{j}_1,\dots \bold{j}_{m-1}}+(\hat{H}^{Bog}_{\bold{j}_m})_{\xi}+\xi \,T\,
%& &+\xi \sum_{ \bold{j}\in \{\pm\bold{j}_{m}\}}(k_{\bold{j}})^2a^*_{\bold{j}}a_{\bold{j}}
\end{eqnarray}
where $T_{\bold{j}=\pm\bold{j}_m}:=\sum_{\bold{j}=\pm \bold{j}_m} k^2_{\bold{j}}\,a_{\bold{j}}^{*}a_{\bold{j}}$ and  $V'_{\bold{j}_1,\dots \bold{j}_{m-1}}:=V_{\bold{j}_1,\dots \bold{j}_{m-1}}-V^{\#}_{\bold{j}_1,\dots \bold{j}_{m-1}}$.
\\

\noindent
For $m=1$ we set 
%\begin{equation}
%H^{\#}_{\bold{j}_{1},\dots,\bold{j}_{m-1}}|_{m=1}:=T\,
%\end{equation}
%and
\begin{equation}
(H^{\#}_{\bold{j}_{1},\dots,\bold{j}_{m-1}}|_{m=1})_{\xi}:=(1-\xi)T\,.
\end{equation}
%and
%\begin{equation}
%(H_{\bold{j}_{1},\dots,\bold{j}_{m-1}})_{\xi}+\xi T:=T_{\bold{j}\neq \{\pm\bold{j}_{1}\}}+\xi T_{\bold{j}=\{\pm \bold{j}_1\}}\,\label{hxim=1}
%\end{equation}
%where $T_{\bold{j}=\{\pm \bold{j}_1\}}=\sum_{\bold{j}\in\{\pm\bold{j}_{1}\}}(k_{\bold{j}})^2a^*_{\bold{j}}a_{\bold{j}}$, thus
%\begin{equation}
%(H^{\#}_{\bold{j}_{1},\dots,\bold{j}_{m-1}})_{\xi}|_{m=1}:=(1-\xi)T\,.
%_{\bold{j}\neq \pm \bold{j}_1}
%\end{equation}
We recall
%Thus, for $m=1$ the relation in (\ref{xi-depend-ham}) becomes
\begin{eqnarray}
H_{\bold{j}_1}&=&V_{\bold{j}_{1}}+\hat{H}^{Bog}_{\bold{j}_1}+T_{\bold{j}\notin \{\pm\bold{j}_{1}\}}\\
&=&V_{\bold{j}_{1}}+\hat{H}^{Bog}_{\bold{j}_1}+T-T_{\bold{j}=\pm\bold{j}_{1}}\\
%& =&(1-\xi)T_{\bold{j}\neq \{\pm\bold{j}_{1}\}}+V_{\bold{j}_1}+(\hat{H}^{Bog}_{\bold{j}_1})_{\xi}+\xi \,T\,\\
&=&(1-\xi)T-(1-\xi)T_{\bold{j}=\pm \bold{j}_1}+V_{\bold{j}_1}+(\hat{H}^{Bog}_{\bold{j}_1})_{\xi}+\xi \,T\,.
%& &+\xi \sum_{ \bold{j}\in \{\pm\bold{j}_{m}\}}(k_{\bold{j}})^2a^*_{\bold{j}}a_{\bold{j}}
\end{eqnarray}
\begin{lemma}\label{xi-cor}
Let $M\geq m\geq 1$ and assume that $\text{infspec}[H^{\#}_{\bold{j}_1,\dots, \bold{j}_{m-1}}-T_{\bold{j}=\pm \bold{j}_m}]\geq z^{\#}_{\bold{j}_1,\dots, \bold{j}_{m-1}}-\frac{(m-1)\xi^{\frac{1}{2}}}{2M}$. Then for $\xi=(\frac{1}{\ln N})^{\frac{1}{4}}$ and $N$ large enough the inequality 
\begin{equation}\label{xi-ineq}
(H^{\#}_{\bold{j}_1,\dots,\bold{j}_{m-1}})_{\xi}-(1-\xi)T_{\bold{j}=\pm\bold{j}_m}\geq z^{\#}_{\bold{j}_1,\dots\,\bold{j}_{m-1}}-\frac{(m-1)\xi^{\frac{1}{2}}}{M}
\end{equation}
holds true where $z^{\#}_{\bold{j}_1,\dots, \bold{j}_{m-1}}\equiv 0\,$  for $m=1$.
%where $z^{\#}_{\bold{j}_1,\dots\,\bold{j}_{m-1}}\equiv 0$ for $m=1$.
\end{lemma}

\noindent
\emph{Proof}

For $m=1$ the property is trivial. For $m\geq 2$, we notice that, due to the identity in (\ref{rel-H-hat-H-hat-xi}), the assumption in the statement corresponds to
\begin{equation}\label{initial}
\langle \psi \,,\,[(H^{\#}_{\bold{j}_1,\dots,\bold{j}_{m-1}})_{\xi}-T_{\bold{j}=\pm \bold{j}_m}+\xi T]\psi \rangle \geq z^{\#}_{\bold{j}_1,\dots,\bold{j}_{m-1}}-\frac{(m-1)\xi^{\frac{1}{2}}}{2M}\,,\quad \|\psi\|=1\,.
\end{equation}
Furthermore, we observe that
\begin{equation}\label{identita}
\langle \psi \,,\,[(H^{\#}_{\bold{j}_1,\dots,\bold{j}_{m-1}})_{\xi}-T_{\bold{j}=\pm \bold{j}_m}+\xi T]\psi \rangle=\langle \psi \,,\,[(H^{\#}_{\bold{j}_1,\dots,\bold{j}_{m-1}})_{\xi}-(1-\xi)T_{\bold{j}=\pm \bold{j}_m}+\xi T_{\bold{j}\notin \{\pm \bold{j}_m\}}]\psi \rangle\,.
\end{equation}
Now, we assume that $\langle \psi \,,\, T_{\bold{j}\notin \{\pm \bold{j}_m\}}\psi \rangle> \frac{1}{2M\xi^{\frac{1}{2}}}$, so that starting from the definition in (\ref{def-xiham-1}) and using a slightly modified version of Lemma \ref{control-quad} in the Appendix the inequality in (\ref{xi-ineq}) is trivially fulfilled for $N$ large enough. If $\langle \psi \,,\,T_{\bold{j}\notin \{ \pm \bold{j}_m\}} \psi \rangle\leq  \frac{1}{2M\xi^{\frac{1}{2}}}$, from (\ref{initial}) and (\ref{identita}) we readily obtain
\begin{equation}
\langle \psi \,,\,(H^{\#}_{\bold{j}_1,\dots,\bold{j}_{m-1}})_{\xi}\psi \rangle-(1-\xi)\langle \psi\,,\,T_{\bold{j}=\pm \bold{j}_m}\psi \rangle  \geq z^{\#}_{\bold{j}_1,\dots,\bold{j}_{m-1}}-\xi \langle \psi \,,\, T_{\bold{j}\notin \{\pm \bold{j}_m\}}\psi \rangle-\frac{(m-1)\xi^{\frac{1}{2}}}{2M}\geq  z^{\#}_{\bold{j}_1,\dots,\bold{j}_{m-1}}-\frac{(m-1)\xi^{\frac{1}{2}}}{M}\,.
\end{equation}
\qed

In Corollary \ref{main-lemma-H}, assuming  
\begin{equation}\label{assume}
(H^{\#}_{\bold{j}_1,\dots,\bold{j}_{m-1}})_{\xi}-(1-\xi)T_{\bold{j}=\pm\bold{j}_m}- z^{\#}_{\bold{j}_1,\dots\,\bold{j}_{m-1}}\geq -\frac{(m-1)\xi^{\frac{1}{2}}}{M}\quad 
\end{equation}
we show how  the norm bound 
\begin{equation}
\|\Big[R_{\bold{j}_1,\dots, \bold{j}_m\,;\,i-2,i-2}(w)\Big]^{\frac{1}{2}}\check{W}^*_{\bold{j}_1,\dots, \bold{j}_m\,;i-2,i}\,\Big[R_{\bold{j}_1,\dots, \bold{j}_m\,;\,i,i}(w)\Big]^{\frac{1}{2}}\|^2\leq \frac{1}{4(1+a_{\epsilon_{\bold{j}_m}}-\frac{2b_{\epsilon_{\bold{j}_m}}}{N-i+2}-\frac{1-c_{\epsilon_{\bold{j}_m}}}{(N-i+2)^2})}\label{norma}
\end{equation}
can be derived for $\bar{i}+4\leq i \leq N-2$ ($i$ even) and
\begin{equation}
w=z+z^{\#}_{\bold{j}_1,\dots,\bold{j}_{m-1}}\leq  z^{\#}_{\bold{j}_1,\dots,\bold{j}_{m-1}}+ E^{Bog}_{\bold{j}_m}+ (\delta-1)\phi_{\bold{j}_m}\sqrt{\epsilon_{\bold{j}_m}^2+2\epsilon_{\bold{j}_m}}\,, \quad \delta<2\,,
\end{equation} with $\epsilon_{\bold{j}_m}$ sufficiently small  and $N$ sufficiently large (for definitions see (\ref{a-bis}),(\ref{b-bis}) and (\ref{c-bis})).  
%The estimate in (\ref{norma})  implies
%\begin{equation}
%\|\check{\Gamma}_{\bold{j}_1,\dots,\bold{j}_{m}\,;\,i-2,i-2}(w)\|<1
%\end{equation}
%like in the proof of  \emph{\underline{Theorem 3.1} of \cite{Pi1}}.

The proof of Corollary \ref{main-lemma-H} requires some modifications with respect to \emph{\underline{Lemma 3.5} of \cite{Pi1}} and  \emph{\underline{Corollary 5.1} of \cite{Pi2}}. However, the new term, $V_{\bold{j}_1,\dots,\bold{j}_m}$, in the interaction can be controlled because it has to be evaluated only in the following expressions
\begin{equation}\label{V-terms}
\mathfrak{Q}_{\bold{j}_m}^{(i,i+1)}V_{\bold{j}_m}\mathfrak{Q}_{\bold{j}_m}^{(i,i+1)},\quad \mathfrak{Q}_{\bold{j}_m}^{(i,i+1)}V'_{\bold{j}_1,\dots, \bold{j}_m}\mathfrak{Q}_{\bold{j}_m}^{(i,i+1)},\quad,\quad \mathfrak{Q}_{\bold{j}_m}^{(i-2,i-1)}V_{\bold{j}_1,\dots, \bold{j}_m}\mathfrak{Q}_{\bold{j}_m}^{(i,i+1)}\,,
\end{equation}
%The main mechanism exploited in Corollary \ref{main-lemma-H} exploits that 
%can be %summarized as follows: (1) If $i\leq \bar{i}$ the total number of particles in the nonzero modes different %from $\pm\bold{j}_1,\dots, \pm \bold{j}_m$ is bounded by $\mathcal{O}(N^{\frac{1}{8}})$ due to the %projection $\mathfrak{Q}_{\bold{j}_m}^{(i,i+1)}$. This is enough for the control of the operators in (\ref{V-%terms})  thanks to the \emph{Non-connected Frequencies Assumption}. 
%\noindent
%(2) 
and, for  $i> \bar{i}$, $\mathfrak{Q}_{\bold{j}_m}^{(i,i+1)}$ yields a bound on the number of particles ($\leq \lfloor N^{\frac{1}{16}} \rfloor -2$) in the modes $\pm \bold{j}_m$. The details are deferred to Corollary \ref{main-lemma-H}.
%{\color{red}that are  bounded in norm by $\mathcal{O}(\frac{N^{\frac{1}{2}}N^{\frac{3}{16}}}{N})=\mathcal{O}(N^{-\frac{5}{16}})$ as the total number of particles in the nonzero modes is bounded by $\mathcal{O}(N^{\frac{1}{16}})$ due to the spectral projections $\mathfrak{Q}_{\bold{j}_1}^{(i,i+1)}, \mathfrak{Q}_{\bold{j}_1}^{(i-2,i-1)}$.} 

\begin{remark}
Corollary \ref{main-lemma-H} essentially controls all the steps of the Feshbach flow from $i=\bar{i}+2$ to $i=N-2$ provided
that  for $w=z+z^{\#}_{\bold{j}_1,\dots,\bold{j}_{m-1}}$
\begin{eqnarray}
& &\Gamma_{\bold{j}_1,\,\dots, \bold{j}_{m}\,;\,\bar{i}+2,\bar{i}+2}(w)\\
&:=&\mathfrak{Q}^{(\bar{i}+2,\bar{i}+3)}_{\bold{j}_m}\check{W}_{\bold{j}_1,\,\dots, \bold{j}_{m}}\mathfrak{Q}^{(\bar{i},\bar{i}+1)}_{\bold{j}_m}\frac{1}{\mathfrak{Q}^{(\bar{i},\bar{i}+1)}_{\bold{j}_m}(H_{\bold{j}_1,\,\dots, \bold{j}_{m}}-w)\mathfrak{Q}^{(\bar{i},\bar{i}+1)}_{\bold{j}_m}}\mathfrak{Q}^{(\bar{i},\bar{i}+1)}_{\bold{j}_m}\check{W}^*_{\bold{j}_1,\,\dots, \bold{j}_{m}}\mathfrak{Q}^{(\bar{i}+2,\bar{i}+3)}_{\bold{j}_m}\,\quad\quad \label{def-gamma-2-bis}
\end{eqnarray}
is well approximated by $\Gamma^{Bog}_{\bold{j}_{m}\,;\,\bar{i}+2,\bar{i}+2}(z)$ in a sense specified in Theorem \ref{theorem-junction} below. For this reason, the very first step, from $i=\bar{i}$ to $i=\bar{i}+2$,
requires a more careful control if compared with the Feshbach flows studied in \cite{Pi1} and \cite{Pi2}. This is due to the new choice of the very first perpendicular projection $\mathfrak{Q}^{(\bar{i},\bar{i}+1)}_{\bold{j}_m}$. In the proof of Theorem \ref{theorem-junction} we take advantage of the ``short range property of the potential in the particle states numbers" described in the introduction (see Section \ref{introduction}).
\end{remark}
\begin{thm}\label{theorem-junction}
For $M\geq m\geq 1$ assume:

\noindent
(a)  
%\begin{equation}
%(H^{Bog}_{\bold{j}_1,\dots,\bold{j}_{m-1}})_{\xi}\geq z_{\bold{j}_1,\dots\,\bold{j}_m}-\mathcal{O}(\xi)
%\end{equation}
\begin{equation}
(H^{\#}_{\bold{j}_{1},\dots,\bold{j}_{m-1}})_{\xi}-(1-\xi)T_{\bold{j}=\{\pm\bold{j}_m\}}\geq z^{\#}_{\bold{j}_1,\dots\,\bold{j}_{m-1}}-\frac{(m-1)\xi^{\frac{1}{2}}}{M}\,,\label{ass-1-main-lemma-0}
\end{equation}
where $(H^{\#}_{\bold{j}_1,\dots,\bold{j}_{m-1}})_{\xi}$ is defined in  (\ref{def-xiham-1}) for $m\geq 2$ and is equal to $(1-\xi)T$ for $m=1$, 
 and where $z^{\#}_{\bold{j}_1,\dots\,\bold{j}_{m-1}}$ is the ground state energy of $H^{\#}_{\bold{j}_1,\dots,\bold{j}_{m-1}}$.\\
 % with $z^{\#}_{\bold{j}_1,\dots\,\bold{j}_{m-1}}\equiv 0$ if $m=1$.
%
%\noindent
%b) 
%\begin{equation}
%|z^{\#}_{\bold{j}_1,\dots\,\bold{j}_{m-1}}-z_{\bold{j}_1,\dots\,\bold{j}_{m-1}}|\leq \frac{(m-1)\xi^{\frac{1}{2}}}{3M}\,.
%\end{equation}

\noindent
(b)  
\begin{equation}\label{inter-z-S}
w:=z+z^{\#}_{\bold{j}_1,\dots,\bold{j}_{m-1}}\leq  z^{\#}_{\bold{j}_1,\dots,\bold{j}_{m-1}}+ E^{Bog}_{\bold{j}_m}+ (\delta-1)\phi_{\bold{j}_m}\sqrt{\epsilon_{\bold{j}_m}^2+2\epsilon_{\bold{j}_m}}
\end{equation} with $\delta< 2$ and $\epsilon_{\bold{j}_m}$ sufficiently small.  

\noindent
Then, for $\xi=(\frac{1}{\ln N})^{\frac{1}{4}}$ and $N$ sufficiently large
\begin{eqnarray}
& &\Big\|\Big[R_{\bold{j}_1,\dots, \bold{j}_m\,;\,\bar{i}+2,\bar{i}2}(w)\Big]^{\frac{1}{2}}\Gamma_{_{\bold{j}_1,\,\dots, \bold{j}_{m}}\,;\,\bar{i}+2,\bar{i}+2}(w)\Big[R_{\bold{j}_1,\dots, \bold{j}_m\,;\,\bar{i}+2,\bar{i}+2}(w)\Big]^{\frac{1}{2}}\Big\|\\
&\leq &\frac{1}{x_{\bar{i}+2}}
\end{eqnarray}
where $x_{\bar{i}+2}$ is the $\bar{i}+2$-term of the sequence\footnote{This sequence was introduced in  \emph{\underline{Lemma 3.6} of \cite{Pi1}}}
\begin{eqnarray}
x_{2j+2}&:=&1-\frac{1}{4(1+a_{\epsilon_{\bold{j}_m}}-\frac{2b_{\epsilon_{\bold{j}_m}}}{N-2j-1}-\frac{1-c_{\epsilon_{\bold{j}_m}}}{(N-2j-1)^2})x_{2j}}\,,\quad j\in \mathbb{N},
%x_{2j+3}&:=&1-\frac{1}{4(1+a_{\epsilon}-\frac{2b_{\epsilon}}{N-2j-1}-\frac{1-c_{\epsilon}}{(N-2j-1)^2})x_{2j+1}}
\end{eqnarray}
starting from $x_{0}\equiv 1$ and defined up to $x_{N-2}$. Here,  $a_{\epsilon_{\bold{j}m}}$, $b_{\epsilon_{\bold{j}_m}}$, and $c_{\epsilon_{\bold{j}_m}}$ coincide with  those given
 in Corollary \ref{main-lemma-H}.
 % where the  corrections coming from $N$-dependent and $\xi$-dependent terms  are hidden in the term $o(\epsilon_{\bold{j}_{m}})$ which enters the definition of $a_{\epsilon_{\bold{j}_m}}$; see (\ref{a-bis}).}
\end{thm}

\noindent
\emph{Proof}

First, we observe that $\Gamma_{_{\bold{j}_1,\,\dots, \bold{j}_{m}}\,;\,\bar{i}+2,\bar{i}+2}(w)$ is well defined  thanks to the inequality in (\ref{estimate-control-quad-2}) of Lemma \ref{control-quad} that implies $$\mathfrak{Q}^{(\bar{i},\bar{i}+1)}_{\bold{j}_m}(H_{\bold{j}_1,\,\dots, \bold{j}_{m}}-w)\mathfrak{Q}^{(\bar{i},\bar{i}+1)}_{\bold{j}_m}\geq C N^{\frac{1}{16}}\mathfrak{Q}^{(\bar{i},\bar{i}+1)}_{\bold{j}_m}$$ for some $C>0$. Next, we recall that (see (\ref{xi-depend-ham}))
\begin{eqnarray}
& & \mathfrak{Q}^{(\bar{i},\bar{i}+1)}_{\bold{j}_m}[ H_{\bold{j}_1,\dots,\bold{j}_m}-w] \mathfrak{Q}^{(\bar{i},\bar{i}+1)}_{\bold{j}_m}\label{ineq-den-in}\\
&= & \mathfrak{Q}^{(\bar{i},\bar{i}+1)}_{\bold{j}_m} [(H^{\#}_{\bold{j}_1,\dots,\bold{j}_{m-1}})_{\xi}-(1-\xi)T_{\bold{j}=\pm\bold{j}_m}-z^{\#}_{\bold{j}_1,\dots,\bold{j}_{m-1}}] \mathfrak{Q}^{(\bar{i},\bar{i}+1)}_{\bold{j}_m}\\
& &+\mathfrak{Q}^{(\bar{i},\bar{i}+1)}_{\bold{j}_m}[ V_{\bold{j}_{m}}+V'_{\bold{j}_1,\dots \bold{j}_{m-1}}+(\hat{H}^{Bog}_{\bold{j}_m})_{\xi}+\xi \,T-z]\mathfrak{Q}^{(\bar{i},\bar{i}+1)}_{\bold{j}_m}\,.
%&\geq&\mathfrak{Q}^{(\bar{i},\bar{i}+1)}_{\bold{j}_m}[ V_{\bold{j}_{m}}+V'_{\bold{j}_1,\dots \bold{j}_{m-1}}+(\hat{H}^{Bog}_{\bold{j}_m})_{\xi}+\xi \,T-\frac{(m-1)\xi^{\frac{1}{2}}}{M}-z]\mathfrak{Q}^{(\bar{i},\bar{i}+1)}_{\bold{j}_m}\,\\
%&\geq &\mathfrak{Q}^{(\bar{i},\bar{i}+1)}_{\bold{j}_m}[(\hat{H}^{Bog}_{\bold{j}_m})_{\xi}-\frac{(m-1)\xi^{\frac{1}{2}}}{M}-z]\mathfrak{Q}^{(\bar{i},\bar{i}+1)}_{\bold{j}_m}\,\label{ineq-den-fin}
\end{eqnarray}
We define 
\begin{eqnarray}
V^{(3)}_{\bold{j}_m}&:= &\frac{1}{N}\sum_{\bold{j}\in \mathbb{Z}^d\setminus \{-\bold{j}_m,\bold{0}\}}a^*_{\bold{j}+\bold{j}_m}\,a^*_{\bold{0}}\,\phi_{\bold{j}_m}\,a_{\bold{j}}a_{\bold{j}_m}+h.c. \\
& &+\frac{1}{N}\sum_{\bold{j}\in \mathbb{Z}^d\setminus \{\bold{j}_m,\bold{0}\}}a^*_{\bold{j}-\bold{j}_m}\,a^*_{\bold{0}}\,\phi_{\bold{j}_m}\,a_{\bold{j}}a_{-\bold{j}_m}+h.c. \\
\end{eqnarray}
\begin{equation}\label{def-V4-m}
V^{(4)}_{\bold{j}_m}:=\frac{1}{N}\sum_{\bold{j}\in \mathbb{Z}^d\setminus\{- \bold{j}_m,\bold{0}\}}\,a^*_{\bold{j}+\bold{j}_m}a_{\bold{j}}\,\,\phi_{\bold{j}_m}\,\sum_{\bold{j}'\in \mathbb{Z}^d\setminus\{+ \bold{j}_m,\bold{0}\}}a^*_{\bold{j}'-\bold{j}_m}a_{\bold{j}'} \geq 0\,,
\end{equation}
\begin{equation}\label{def-V4-m-diesis}
V^{\#\,(4)}_{\bold{j}_m}:=\frac{1}{N}\sum_{\bold{j}\in \mathbb{Z}^d\setminus\{\pm \bold{j}_m,\bold{0}, -2\bold{j}_m\}}\,a^*_{\bold{j}+\bold{j}_m}a_{\bold{j}}\,\,\phi_{\bold{j}_m}\,\sum_{\bold{j}'\in \mathbb{Z}^d\setminus\{\pm \bold{j}_m,\bold{0}, 2\bold{j}_m\}}a^*_{\bold{j}'-\bold{j}_m}a_{\bold{j}'} \geq 0\,,
\end{equation}
\begin{equation}\label{def-V4-m-diesis}
(V^{(4)}_{\bold{j}_m})':=V^{(4)}_{\bold{j}_m}-V^{\#\,(4)}_{\bold{j}_m}=\frac{1}{N}\,a^*_{-\bold{j}_m}a_{ -2\bold{j}_m}\,\phi_{\bold{j}_m}\,a^*_{\bold{j}_m}a_{2\bold{j}_m}+h.c.
\end{equation}
so that
\begin{eqnarray}
V_{\bold{j}_m}&= &V^{(3)}_{\bold{j}_m}+\frac{1}{N}\sum_{\bold{j}\in \mathbb{Z}^d\setminus\{ -\bold{j}_m,\bold{0}\}}\,\sum_{\bold{j}'\in \mathbb{Z}^d\setminus \{+\bold{j}_m,\bold{0}\}}a^*_{\bold{j}+\bold{j}_m}\,a^*_{\bold{j}'-\bold{j}_m}\,\phi_{\bold{j}_m}\,a_{\bold{j}}a_{\bold{j}'}\\
&=&V^{(3)}_{\bold{j}_m}+(V^{(4)}_{\bold{j}_m})'+V^{\#\,(4)}_{\bold{j}_m}-\frac{1}{N}\,\phi_{\bold{j}_m}\,\sum_{\bold{j}'\in \mathbb{Z}^d\setminus\{+ \bold{j}_m,\bold{0}\}}a^*_{\bold{j}'}a_{\bold{j}'}\,. \,
%&=&V^{\#\,(4)}_{\bold{j}_m}+(V^{(4)}_{\bold{j}_m})'-\frac{1}{N}\,\phi_{\bold{j}_m}\,\sum_{\bold{j}'\in \mathbb{Z}^d\setminus\{+ \bold{j}_m,\bold{0}\}}a^*_{\bold{j}'}a_{\bold{j}'}\,.
\end{eqnarray}
We split $H_{\bold{j}_1,\dots,\bold{j}_m}-w$ into
\begin{eqnarray}
& & (H^{\#}_{\bold{j}_1,\dots,\bold{j}_{m-1}})_{\xi}-(1-\xi)T_{\bold{j}=\pm\bold{j}_m}-z^{\#}_{\bold{j}_1,\dots,\bold{j}_{m-1}}+V^{\#\,(4)}_{\bold{j}_m}+(\hat{H}^{0}_{\bold{j}_m})_{\xi}-z+\xi \,T-\frac{1}{N}\,\phi_{\bold{j}_m}\,\sum_{\bold{j}'\in \mathbb{Z}^d\setminus\{+ \bold{j}_m,\bold{0}\}}a^*_{\bold{j}'}a_{\bold{j}'}\label{free}\quad\quad\quad\\
& &+V^{(3)}_{\bold{j}_{m}}+(V^{(4)}_{\bold{j}_m})'+V'_{\bold{j}_1,\dots \bold{j}_{m-1}}+W_{\bold{j}_m}+W^*_{\bold{j}_m}\,,
%&\geq&\mathfrak{Q}^{(\bar{i},\bar{i}+1)}_{\bold{j}_m}[ V_{\bold{j}_{m}}+V'_{\bold{j}_1,\dots \bold{j}_{m-1}}+(\hat{H}^{Bog}_{\bold{j}_m})_{\xi}+\xi \,T-\frac{(m-1)\xi^{\frac{1}{2}}}{M}-z]\mathfrak{Q}^{(\bar{i},\bar{i}+1)}_{\bold{j}_m}\,\\
%&\geq &\mathfrak{Q}^{(\bar{i},\bar{i}+1)}_{\bold{j}_m}[(\hat{H}^{Bog}_{\bold{j}_m})_{\xi}-\frac{(m-1)\xi^{\frac{1}{2}}}{M}-z]\mathfrak{Q}^{(\bar{i},\bar{i}+1)}_{\bold{j}_m}\,\label{ineq-den-fin}
\end{eqnarray}
where 
 $$(\hat{H}^{0}_{\bold{j}_m})_{\xi}:=\sum_{\bold{j}=\pm \bold{j}_m} [(1-\xi)k^2_{\bold{j}}+\frac{\phi_{\bold{j}}}{N}a^*_{\bold{0}}a_{\bold{0}}]\,a_{\bold{j}}^{*}a_{\bold{j}}\,.$$
Assuming the inequality in (\ref{assume}) it is clear that for $N$ sufficiently large
\begin{equation}
\mathfrak{Q}^{(\bar{i},\bar{i}+1)}_{\bold{j}_m}(\ref{free}) \mathfrak{Q}^{(\bar{i},\bar{i}+1)}_{\bold{j}_m}>0\,
\end{equation}
because
\begin{itemize}
\item
the inequality in (\ref{ass-1-main-lemma-0}) has been assumed;
\item
 the operators $V^{\#\,(4)}_{\bold{j}_{m}}$ and $(\hat{H}^{0}_{\bold{j}_m})_{\xi}$ are non negative;
 \item $\xi \,T$ dominates $-\frac{1}{N}\,\phi_{\bold{j}_m}\,\sum_{\bold{j}'\in \mathbb{Z}^d\setminus\{+ \bold{j}_m,\bold{0}\}}a^*_{\bold{j}'}a_{\bold{j}'}$  for $\xi=(\frac{1}{\ln N})^{\frac{1}{4}}$ and $N$ large;
 \item
  $-z>0$ uniformly in $N$.
 \end{itemize} 
Hence, we define the resolvent
\begin{equation}
S_{\bold{j}_m}(z):=\mathfrak{Q}^{(\bar{i},\bar{i}+1)}_{\bold{j}_m}\frac{1}{\mathfrak{Q}^{(\bar{i},\bar{i}+1)}_{\bold{j}_m}(\ref{free}) \mathfrak{Q}^{(\bar{i},\bar{i}+1)}_{\bold{j}_m}}\mathfrak{Q}^{(\bar{i},\bar{i}+1)}_{\bold{j}_m}
\end{equation}
and implement a truncated Neumann expansion so that we obtain
\begin{eqnarray}
& &\mathfrak{Q}^{(\bar{i}+2,\bar{i}+3)}_{\bold{j}_m}\check{W}_{\bold{j}_1,\,\dots, \bold{j}_{m}}\mathfrak{Q}^{(\bar{i},\bar{i}+1)}_{\bold{j}_m}\frac{1}{\mathfrak{Q}^{(\bar{i},\bar{i}+1)}_{\bold{j}_m}(H_{\bold{j}_1,\,\dots, \bold{j}_{m}}-w)\mathfrak{Q}^{(\bar{i},\bar{i}+1)}_{\bold{j}_m}}\mathfrak{Q}^{(\bar{i},\bar{i}+1)}_{\bold{j}_m}\check{W}_{\bold{j}_1,\,\dots, \bold{j}_{m}}\mathfrak{Q}^{(\bar{i}+2,\bar{i}+3)}_{\bold{j}_m}\\
&=&\mathfrak{Q}^{(\bar{i}+2,\bar{i}+3)}_{\bold{j}_m}\check{W}_{\bold{j}_1,\,\dots, \bold{j}_{m}}\sum_{l=0}^{n'-1}S_{\bold{j}_m}(z)\Big\{(-)\,[(V^{(4)}_{\bold{j}_{m}})'+V^{(3)}_{\bold{j}_m}+V'_{\bold{j}_1,\dots \bold{j}_{m-1}}+W_{\bold{j}_m}+W^*_{\bold{j}_m}]S_{\bold{j}_m}(z)\Big\}^l\times \quad\quad\quad\label{smallterm-0}\\
& &\quad \times \check{W}_{\bold{j}_1,\,\dots, \bold{j}_{m}}\mathfrak{Q}^{(\bar{i}+2,\bar{i}+3)}_{\bold{j}_m}\nonumber\\
& &+\mathfrak{Q}^{(\bar{i}+2,\bar{i}+3)}_{\bold{j}_m}\check{W}_{\bold{j}_1,\,\dots, \bold{j}_{m}}\Big\{S_{\bold{j}_m}(z)\,(-)[(V^{(4)}_{\bold{j}_{m}})'+V^{(3)}_{\bold{j}_m}+V'_{\bold{j}_1,\dots \bold{j}_{m-1}}+W_{\bold{j}_m}+W^*_{\bold{j}_m}]\Big\}^{n'}\times \label{smallterm-1}\\
& &\quad\times \mathfrak{Q}^{(\bar{i},\bar{i}+1)}_{\bold{j}_m}\frac{1}{\mathfrak{Q}^{(\bar{i},\bar{i}+1)}_{\bold{j}_m}(H_{\bold{j}_1,\,\dots, \bold{j}_{m}}-w)\mathfrak{Q}^{(\bar{i},\bar{i}+1)}_{\bold{j}_m}}\mathfrak{Q}^{(\bar{i},\bar{i}+1)}_{\bold{j}_m}\check{W}_{\bold{j}_1,\,\dots, \bold{j}_{m}}\mathfrak{Q}^{(\bar{i}+2,\bar{i}+3)}_{\bold{j}_m}\,.\quad\quad\quad\nonumber
\end{eqnarray}
Making use of the selection rules of the operator $\check{W}_{\bold{j}_1,\,\dots, \bold{j}_{m}}$ we get
\begin{eqnarray}
& &(\ref{smallterm-0})+(\ref{smallterm-1})\\
&=&\mathfrak{Q}^{(\bar{i}+2,\bar{i}+3)}_{\bold{j}_m}\check{W}_{\bold{j}_1,\,\dots, \bold{j}_{m}}Q^{(\bar{i},\bar{i}+1)}_{\bold{j}_m}\sum_{l=0}^{n'-1}S_{\bold{j}_m}(z)\Big\{(-)\,[(V^{(4)}_{\bold{j}_{m}})'+V^{(3)}_{\bold{j}_m}+V'_{\bold{j}_1,\dots \bold{j}_{m-1}}+W_{\bold{j}_m}+W^*_{\bold{j}_m}]S_{\bold{j}_m}(z)\Big\}^l\times \label{smallterm-0-bis} \quad\quad\quad \\
& &\quad \times \check{W}_{\bold{j}_1,\,\dots, \bold{j}_{m}}\mathfrak{Q}^{(\bar{i}+2,\bar{i}+3)}_{\bold{j}_m}\nonumber\\
& &+\mathfrak{Q}^{(\bar{i}+2,\bar{i}+3)}_{\bold{j}_m}\check{W}_{\bold{j}_1,\,\dots, \bold{j}_{m}}Q^{(\bar{i},\bar{i}+1)}_{\bold{j}_m}\Big\{S_{\bold{j}_m}(z)\,(-)[(V^{(4)}_{\bold{j}_{m}})'+V^{(3)}_{\bold{j}_m}+V'_{\bold{j}_1,\dots \bold{j}_{m-1}}+W_{\bold{j}_m}+W^*_{\bold{j}_m}]\Big\}^{n'}\times \label{smallterm-1-bis}\\
& &\quad\times \mathfrak{Q}^{(\bar{i},\bar{i}+1)}_{\bold{j}_m}\frac{1}{\mathfrak{Q}^{(\bar{i},\bar{i}+1)}_{\bold{j}_m}(H_{\bold{j}_1,\,\dots, \bold{j}_{m}}-w)\mathfrak{Q}^{(\bar{i},\bar{i}+1)}_{\bold{j}_m}}\mathfrak{Q}^{(\bar{i},\bar{i}+1)}_{\bold{j}_m}\check{W}_{\bold{j}_1,\,\dots, \bold{j}_{m}}\mathfrak{Q}^{(\bar{i}+2,\bar{i}+3)}_{\bold{j}_m}\,.\quad\quad\quad\nonumber
\end{eqnarray}

\begin{remark}\label{particles-short-range-2}
We observe that in (\ref{smallterm-0-bis}), (\ref{smallterm-1-bis}), on the left of the operator $$(V^{(4)}_{\bold{j}_{m}})'+V^{(3)}_{\bold{j}_m}+V'_{\bold{j}_1,\dots \bold{j}_{m-1}}\,,$$
 we can insert the projection
\begin{equation}
\mathcal{P}^{(N^{\frac{1}{16}}+2n')}_{m}
\end{equation}
where $\mathcal{P}^{(N^{\frac{1}{16}}+2n')}_{m}$ projects onto the subspace of vectors with  at most $\lfloor N^{\frac{1}{16}}\rfloor +2n'$ particles in the modes $\pm \bold{j}_m$. We also notice that
\begin{equation}
\Big[S_{\bold{j}_m}(z)\,,\, \mathcal{P}^{(N^{\frac{1}{16}}+2n')}_{m}\Big]=0
\end{equation}
and
\begin{eqnarray}
& &\|Q^{(\bar{i},\bar{i}+1)}_{\bold{j}_m}(S_{\bold{j}_m}(z))^{\frac{1}{2}}\{[\mathcal{P}^{(N^{\frac{1}{16}}+2n')}_{m}(V^{(4)}_{\bold{j}_{m}})'+V^{(3)}_{\bold{j}_m}+\mathcal{P}^{(N^{\frac{1}{16}}+2n')}_{m}V'_{\bold{j}_1,\dots \bold{j}_{m-1}}+W_{\bold{j}_m}+W^*_{\bold{j}_m}](S_{\bold{j}_m}(z))^{\frac{1}{2}}\}^l\| \quad\quad\quad\quad\\
&\leq  &\Big\{\|(S_{\bold{j}_m}(z))^{\frac{1}{2}}[\mathcal{P}^{(N^{\frac{1}{16}}+2n')}_{m}(V^{(4)}_{\bold{j}_{m}})'+V^{(3)}_{\bold{j}_m}+\mathcal{P}^{(N^{\frac{1}{16}}+2n')}_{m}V'_{\bold{j}_1,\dots \bold{j}_{m-1}}](S_{\bold{j}_m}(z))^{\frac{1}{2}}\|+\\
& &\quad\quad 2\sup_{0\leq r \leq 2n'\,,\, r \,\, even} \|(S_{\bold{j}_m}(z))^{\frac{1}{2}}Q^{(\bar{i}-r,\bar{i}-r+1)}_{\bold{j}_m}W_{\bold{j}_m}Q^{(\bar{i}-r-2,\bar{i}-r-1)}_{\bold{j}_m}(S_{\bold{j}_m}(z))^{\frac{1}{2}}\|\Big\}^l\,.
\end{eqnarray}
\end{remark}

\noindent
Next, we observe that for each summand in $(V^{(4)}_{\bold{j}_{m}})', V^{(3)}_{\bold{j}_m}$, and $V'_{\bold{j}_1,\dots \bold{j}_{m-1}}$: 
\begin{enumerate}
\item
 at most one operator of the type $a_{\bold{0}},a^*_{\bold{0}}$ can be present; 
 \item
 at least one operator $a^*_{\bold{j}_m}$ or $a^*_{-\bold{j}_m}$ is present.
 %;due to the projections $\mathfrak{Q}^{(i,i+1)}_{\bold{j}_m}$ and $\mathfrak{Q}^{(i-2,i-1)}_{\bold{j}_m}$ on the left and on the right, respectively.
%3)  the number of particles in the modes $\pm \bold{j}_m$ is constrained by $\mathfrak{Q}^{(i,i+1)}_{\bold{j}_m}$ to values less than $\lfloor N^{\frac{1}{16}}\rfloor -5$ for $i\geq \bar{i}+4$.
  \end{enumerate}
 %Using the same argument exploited to control (\ref{V-sandw})-(\ref{def-v1...m-bis}) in Corollary \ref{main-lemma-H} we can state that
Consequently, for any $\phi \in \mathcal{F}^N$
\begin{equation}\label{ineq-form-est}
|\langle \phi\,,\,\mathcal{P}^{(N^{\frac{1}{16}}+2n')}_{m}[(V^{(4)}_{\bold{j}_{m}})'+V^{(3)}_{\bold{j}_m}+V'_{\bold{j}_1,\dots \bold{j}_{m-1}}]\phi \rangle|\leq C[\frac{N^{\frac{1}{16}}}{N}]^{\frac{1}{2}}\langle \phi\,,\,\mathcal{N}_+\phi \rangle
\end{equation}
for some $C>0$ where $\mathcal{N}_+:=\sum_{\bold{j}\in \mathbb{Z}^d\setminus\{ \bold{0}\}}a_{\bold{j}}^{*}a_{\bold{j}}$, and we derive
\begin{equation}
\|\mathcal{P}^{(N^{\frac{1}{16}}+2n')}_{m}[S_{\bold{j}_m}(z)]^{\frac{1}{2}}[(V^{(4)}_{\bold{j}_{m}})'+V^{(3)}_{\bold{j}_m}+V'_{\bold{j}_1,\dots \bold{j}_{m-1}}][S_{\bold{j}_m}(z)]^{\frac{1}{2}}\|\leq \mathcal{O}(\frac{1}{\xi}[\frac{N^{\frac{1}{16}}}{N}]^{\frac{1}{2}})\label{V-terms}
\end{equation}
and
\begin{equation}
\|\mathcal{P}^{(N^{\frac{1}{16}}+2n')}_{m}[S_{\bold{j}_m}(z)]^{\frac{1}{2}}[(V^{(4)}_{\bold{j}_{m}})'+V^{(3)}_{\bold{j}_m}+V'_{\bold{j}_1,\dots \bold{j}_{m-1}}][\mathfrak{Q}^{(\bar{i},\bar{i}+1)}_{\bold{j}_m}\frac{1}{\mathfrak{Q}^{(\bar{i},\bar{i}+1)}_{\bold{j}_m}(H_{\bold{j}_1,\,\dots, \bold{j}_{m}}-w)\mathfrak{Q}^{(\bar{i},\bar{i}+1)}_{\bold{j}_m}}\mathfrak{Q}^{(\bar{i},\bar{i}+1)}_{\bold{j}_m}]^{\frac{1}{2}}\|\leq \mathcal{O}(\frac{1}{\xi}[\frac{N^{\frac{1}{16}}}{N}]^{\frac{1}{2}})\,.\end{equation}
Next, we invoke \emph{\underline{Lemma 3.4 in \cite{Pi1}}} and estimate
\begin{eqnarray}
& &\sup_{0\leq r \leq 2n'\,,\, r \,\, even} \|(S_{\bold{j}_m}(z))^{\frac{1}{2}}Q^{(\bar{i}-r,\bar{i}-r+1)}_{\bold{j}_m}W_{\bold{j}_m}Q^{(\bar{i}-r-2,\bar{i}-r-1)}_{\bold{j}_m}(S_{\bold{j}_m}(z))^{\frac{1}{2}}\|^2 \\
&\leq &\frac{1}{4(1+a_{\epsilon_{\bold{j}_m}}-\frac{2b_{\epsilon_{\bold{j}_m}}}{N-\bar{i}-1}-\frac{1-c_{\epsilon_{\bold{j}_m}}}{(N-\bar{i}-1)^2})}
\end{eqnarray}
where  
\begin{equation}\label{setabc}
a_{\epsilon_{\bold{j}_m}}:=2\epsilon_{\bold{j}_{*}}+\mathcal{O}(\epsilon^{\nu}_{\bold{j}_m})\,,\quad
b_{\epsilon_{\bold{j}_m}}:=(1+\epsilon_{\bold{j}_m})\delta\chi_{[0,2)}(\delta)\sqrt{\epsilon_{\bold{j}_m}^2+2\epsilon_{\bold{j}_m}}, \quad 
c_{\epsilon_{\bold{j}_m}}:=-(1-\delta^2\chi_{[0,2)}(\delta))(\epsilon_{\bold{j}_m}^2+2\epsilon_{\bold{j}_m})\,
\end{equation}
 are those of  \emph{\underline{Lemma 3.6} of \cite{Pi1}} up to $N$- and $\xi$-dependent corrections that are hidden in the term $\mathcal{O}(\epsilon^{\nu}_{\bold{j}_m})$ (with $\nu>\frac{11}{8}$)  which enters the definition of $a_{\epsilon_{\bold{j}_m}}$.
% as in \emph{\underline{Lemma 3.5 of \cite{Pi1}}}.

Hence, for $N$ large enough we have derived the inequality
\begin{eqnarray}
& &\Big\|\Big[R_{\bold{j}_1,\dots, \bold{j}_m\,;\,\bar{i}+2,\bar{i}+2}(w)\Big]^{\frac{1}{2}}\check{W}_{\bold{j}_1,\,\dots, \bold{j}_{m}}\mathfrak{Q}^{(\bar{i},\bar{i}+1)}_{\bold{j}_m}\frac{1}{\mathfrak{Q}^{(\bar{i},\bar{i}+1)}_{\bold{j}_m}(H_{\bold{j}_1,\,\dots, \bold{j}_{m}}-w)\mathfrak{Q}^{(\bar{i},\bar{i}+1)}_{\bold{j}_m}}\mathfrak{Q}^{(\bar{i},\bar{i}+1)}_{\bold{j}_m}\check{W}_{\bold{j}_1,\,\dots, \bold{j}_{m}}\Big[R_{\bold{j}_1,\dots, \bold{j}_m\,;\,\bar{i}+2,\bar{i}+2}(w)\Big]^{\frac{1}{2}}\Big\|\quad\quad\quad\\
&\leq &\Big\|\Big[R_{\bold{j}_1,\dots, \bold{j}_m\,;\,\bar{i}+2,\bar{i}+2}(w)\Big]^{\frac{1}{2}}\check{W}_{\bold{j}_1,\,\dots, \bold{j}_{m}}Q^{(\bar{i},\bar{i}+1)}_{\bold{j}_m}\sum_{l=0}^{n'-1}S_{\bold{j}_m}(z)\Big\{(-)\,[W_{\bold{j}_m}+W^*_{\bold{j}_m}]S_{\bold{j}_m}(z)\Big\}^l\times \quad\quad\quad \\
& &\quad \times \check{W}_{\bold{j}_1,\,\dots, \bold{j}_{m}}\Big[R_{\bold{j}_1,\dots, \bold{j}_m\,;\,\bar{i}+2,\bar{i}+2}(w)\Big]^{\frac{1}{2}}\nonumber\Big\|\\
& &+\mathcal{O}\Big((\frac{1}{1+a_{\epsilon_{\bold{j}_m}}/4})^{n'}\Big)+\mathcal{O}\Big(\frac{1}{\xi}[\frac{N^{\frac{1}{16}}}{N}]^{\frac{1}{2}}n'\sum_{l=0}^{n'}(\frac{1}{1+a_{\epsilon_{\bold{j}_m}}/4})^{l}\Big)\\
&\leq &\Big\|\Big[R_{\bold{j}_1,\dots, \bold{j}_m\,;\,\bar{i}+2,\bar{i}+2}(w)\Big]^{\frac{1}{2}}\check{W}_{\bold{j}_1,\,\dots, \bold{j}_{m}}Q^{(\bar{i},\bar{i}+1)}_{\bold{j}_m}\sum_{l=0}^{\infty}S_{\bold{j}_m}(z)\{(-)\,[W_{\bold{j}_m}+W^*_{\bold{j}_m}]S_{\bold{j}_m}(z)\}^l\times \quad\quad\quad  \label{inter}\\
& &\quad \times Q^{(\bar{i},\bar{i}+1)}_{\bold{j}_m}\check{W}_{\bold{j}_1,\,\dots, \bold{j}_{m}}\Big[R_{\bold{j}_1,\dots, \bold{j}_m\,;\,\bar{i}+2,\bar{i}+2}(w)\Big]^{\frac{1}{2}}\nonumber\Big\|\\
& &+\mathcal{O}\Big((\frac{1}{1+a_{\epsilon_{\bold{j}_m}}/4})^{n'}\Big)+\mathcal{O}\Big(\frac{1}{\xi}[\frac{N^{\frac{1}{16}}}{N}]^{\frac{1}{2}}n'\sum_{l=0}^{n'}\frac{1}{(1+a_{\epsilon_{\bold{j}_m}}/4})^{l}\Big)\label{error}\,.
\end{eqnarray}
By setting $n'=\frac{\ln N}{\ln (1+a_{\epsilon_{\bold{j}_m}}/4)}$ and 
\begin{eqnarray}
& &\Delta H^{\#}_{\bold{j}_1,\dots,\bold{j}_m}+(\hat{H}^{Bog}_{\bold{j}_m})_{\xi}-z-\frac{(m-1)\xi^{\frac{1}{2}}}{M}+\frac{\xi \,T}{2}\label{splitting}\\
&:=&(H^{\#}_{\bold{j}_1,\dots,\bold{j}_{m-1}})_{\xi}-(1-\xi)T_{\bold{j}=\pm\bold{j}_m}-z^{\#}_{\bold{j}_1,\dots,\bold{j}_{m-1}}+V^{\#\,(4)}_{\bold{j}_{m}}+(\hat{H}^{Bog}_{\bold{j}_m})_{\xi}-z+\frac{\xi \,T}{2}-\frac{1}{N}\,\phi_{\bold{j}_m}\,\sum_{\bold{j}'\in \mathbb{Z}^d\setminus\{+ \bold{j}_m,\bold{0}\}}a^*_{\bold{j}'}a_{\bold{j}'}\quad\quad\quad\label{splitting-2}
\end{eqnarray}
where $\Delta H^{\#}_{\bold{j}_1,\dots,\bold{j}_m}\geq 0$ (recall the assumption in (\ref{ass-1-main-lemma})),
we can write
\begin{eqnarray}
&&\mathfrak{Q}^{(\bar{i},\bar{i}+1)}_{\bold{j}_m}\sum_{l=0}^{\infty}S_{\bold{j}_m}(z)\Big\{(-)\,[W_{\bold{j}_m}+W^*_{\bold{j}_m}]S_{\bold{j}_m}(z)\Big\}^l\mathfrak{Q}^{(\bar{i},\bar{i}+1)}_{\bold{j}_m}\\
&=&\mathfrak{Q}^{(\bar{i},\bar{i}+1)}_{\bold{j}_m}\frac{1}{\mathfrak{Q}^{(\bar{i},\bar{i}+1)}_{\bold{j}_m}[\Delta H^{\#}_{\bold{j}_1,\dots,\bold{j}_m}+\frac{\xi \,T}{2}+(\hat{H}^{Bog}_{\bold{j}_m})_{\xi}-z-\frac{(m-1)\xi^{\frac{1}{2}}}{M}]\mathfrak{Q}^{(\bar{i},\bar{i}+1)}_{\bold{j}_m}}\mathfrak{Q}^{(\bar{i},\bar{i}+1)}_{\bold{j}_m}
\end{eqnarray}
and estimate
\begin{eqnarray}
& &(\ref{inter})+(\ref{error})\\
&\leq &\Big\|\Big[R_{\bold{j}_1,\dots, \bold{j}_m\,;\,\bar{i}+2,\bar{i}+2}(w)\Big]^{\frac{1}{2}}\check{W}_{\bold{j}_1,\,\dots, \bold{j}_{m}}\mathfrak{Q}^{(\bar{i},\bar{i}+1)}_{\bold{j}_m}\frac{1}{\mathfrak{Q}^{(\bar{i},\bar{i}+1)}_{\bold{j}_m}[\frac{\xi \,T}{2}+(\hat{H}^{Bog}_{\bold{j}_m})_{\xi}-z-\frac{(m-1)\xi^{\frac{1}{2}}}{M}]\mathfrak{Q}^{(\bar{i},\bar{i}+1)}_{\bold{j}_m}}\times \label{step-1}\\
& &\quad\quad\quad \times \mathfrak{Q}^{(\bar{i},\bar{i}+1)}_{\bold{j}_m} \check{W}_{\bold{j}_1,\,\dots, \bold{j}_{m}}\Big[R_{\bold{j}_1,\dots, \bold{j}_m\,;\,\bar{i}+2,\bar{i}+2}(w)\Big]^{\frac{1}{2}}\nonumber\Big\|\\
& &+\mathcal{O}\Big(\frac{1}{\xi}[\frac{N^{\frac{1}{16}}}{N}]^{\frac{1}{4}}\Big)\nonumber\\
&\leq &\Big\|\Big[R_{\bold{j}_1,\dots, \bold{j}_m\,;\,\bar{i}+2,\bar{i}+2}(w)\Big]^{\frac{1}{2}}W_{\bold{j}_{m}}\times\label{step-2}\\
& &\quad\quad\quad\times \mathfrak{Q}^{(\bar{i},\bar{i}+1)}_{\bold{j}_m}\frac{1}{\mathfrak{Q}^{(\bar{i},\bar{i}+1)}_{\bold{j}_m}[\frac{\xi \,T}{2}+(\hat{H}^{Bog}_{\bold{j}_m})_{\xi}-z-\frac{(m-1)\xi^{\frac{1}{2}}}{M}]\mathfrak{Q}^{(\bar{i},\bar{i}+1)}_{\bold{j}_m}}\mathfrak{Q}^{(\bar{i},\bar{i}+1)}_{\bold{j}_m} W^*_{\bold{j}_{m}}\Big[R_{\bold{j}_1,\dots, \bold{j}_m\,;\,\bar{i}+2,\bar{i}+2}(w)\Big]^{\frac{1}{2}}\nonumber\Big\|\\
& &+\mathcal{O}\Big(\frac{1}{\xi}[\frac{N^{\frac{1}{16}}}{N}]^{\frac{1}{4}}\Big)\,.\label{rest}
\end{eqnarray}
In the step from (\ref{step-1})  to (\ref{step-2}) we make use of
\begin{eqnarray}
& &\mathfrak{Q}^{(\bar{i}+2,\bar{i}+3)}_{\bold{j}_m}\check{W}_{\bold{j}_1,\,\dots, \bold{j}_{m}}\mathfrak{Q}^{(\bar{i},\bar{i}+1)}_{\bold{j}_m}\\
&=&\mathfrak{Q}^{(\bar{i}+2,\bar{i}+3)}_{\bold{j}_m}[(V^{(4)}_{\bold{j}_m})'+V^{(3)}_{\bold{j}_m}+V'_{\bold{j}_1,\dots \bold{j}_{m-1}}+ W_{\bold{j}_{m}}+ W^*_{\bold{j}_{m}}]\mathfrak{Q}^{(\bar{i},\bar{i}+1)}_{\bold{j}_m}
\end{eqnarray}
and we estimate the term proportional to $\mathfrak{Q}^{(\bar{i}+2,\bar{i}+3)}_{\bold{j}_m}[(V^{(4)}_{\bold{j}_m})'+V^{(3)}_{\bold{j}_m}+V'_{\bold{j}_1,\dots \bold{j}_{m-1}}]\mathfrak{Q}^{(\bar{i},\bar{i}+1)}_{\bold{j}_m}$ of order $\frac{1}{\xi}[\frac{N^{\frac{1}{16}}}{N}]^{\frac{1}{2}}$; see an analogous estimate in (\ref{V-terms}).
%This estimate relies on the fact that: 1) {\color{red}at most one operator of the type $a_{\bold{0}},a^*_{\bold{0}}$ can be present in each summand; 2)  at least one operator $a^*_{\bold{j}_m}$ or $a^*_{-\bold{j}_m}$ must be present due to the projections $\mathfrak{Q}^{(i,i+1)}_{\bold{j}_m}$ and $\mathfrak{Q}^{(i-2,i-1)}_{\bold{j}_m}$ on the left and on the right, respectively; 3)  the number of particles in the modes $\pm \bold{j}_m$ is constrained by $\mathfrak{Q}^{(i,i+1)}_{\bold{j}_m}$ to values less than $\lfloor N^{\frac{1}{16}}\rfloor -1$.}

We notice that because of the semigroup property of the Feshbach map the following identity holds
\begin{eqnarray}
& &Q^{(>\bar{i}+1)}_{\bold{j}_m}W_{\bold{j}_{m}}\mathfrak{Q}^{(\bar{i},\bar{i}+1)}_{\bold{j}_m}\frac{1}{\mathfrak{Q}^{(\bar{i},\bar{i}+1)}_{\bold{j}_m}[\frac{\xi \,T}{2}+(\hat{H}^{Bog}_{\bold{j}_m})_{\xi}-z-\frac{(m-1)\xi^{\frac{1}{2}}}{M}]\mathfrak{Q}^{(\bar{i},\bar{i}+1)}_{\bold{j}_m}}\mathfrak{Q}^{(\bar{i},\bar{i}+1)}_{\bold{j}_m} W^*_{\bold{j}_{m}}Q^{(>\bar{i}+1)}_{\bold{j}_m}\quad\quad\\
&=&Q^{(>\bar{i}+1)}_{\bold{j}_m}W_{\bold{j}_m}\sum_{l=0}^{\infty}(R^{Bog}_{\bold{j}_m;\,\bar{i},\bar{i}}(z))_{\xi}\,\Big[(\Gamma^{Bog\,}_{\bold{j}_m\,;\,\bar{i},\bar{i}}(z))_{\xi}(R^{Bog}_{\bold{j}_m;\,\bar{i},\bar{i}}(z))_{\xi}\Big]^{l}W_{\bold{j}_m}^*Q^{(>\bar{i}+1)}_{\bold{j}_m}
\end{eqnarray}
where $(R^{Bog}_{\bold{j}_m;\,\bar{i},\bar{i}}(z))_{\xi}$ and $(\Gamma^{Bog\,}_{\bold{j}_m\,;\,\bar{i},\bar{i}}(z))_{\xi}$ have the same definition of $R^{Bog}_{\bold{j}_m;\,\bar{i},\bar{i}}(z)$ and $\Gamma^{Bog\,}_{\bold{j}_m\,;\,\bar{i},\bar{i}}(z)$ but  are referred to the Feshbach flow associated with $\frac{\xi \,T}{2}+(\hat{H}^{Bog}_{\bold{j}_m})_{\xi}-\frac{(m-1)\xi^{\frac{1}{2}}}{M}$. Hence, we can write
\begin{eqnarray}
& &\Big[R_{\bold{j}_1,\dots, \bold{j}_m\,;\,\bar{i}+2,\bar{i}+2}(w)\Big]^{\frac{1}{2}}W_{\bold{j}_{m}}\mathfrak{Q}^{(\bar{i},\bar{i}+1)}_{\bold{j}_m}\frac{1}{\mathfrak{Q}^{(\bar{i},\bar{i}+1)}_{\bold{j}_m}[\frac{\xi \,T}{2}+(\hat{H}^{Bog}_{\bold{j}_m})_{\xi}-z-\frac{(m-1)\xi^{\frac{1}{2}}}{M}]\mathfrak{Q}^{(\bar{i},\bar{i}+1)}_{\bold{j}_m}}\times \\
& &\quad\quad\quad \times \mathfrak{Q}^{(\bar{i},\bar{i}+1)}_{\bold{j}_m}W^*_{\bold{j}_{m}}\Big[R_{\bold{j}_1,\dots, \bold{j}_m\,;\,\bar{i}+2,\bar{i}+2}(w)\Big]^{\frac{1}{2}}\nonumber \\
&=&\Big[R_{\bold{j}_1,\dots, \bold{j}_m\,;\,\bar{i}+2,\bar{i}+2}(w)\Big]^{\frac{1}{2}}[(R^{Bog}_{\bold{j}_m;\,\bar{i}+2,\bar{i}+2}(z))_{\xi}]^{-\frac{1}{2}}\times\\
& &\quad\times[(R^{Bog}_{\bold{j}_m;\,\bar{i}+2,\bar{i}+2}(z))_{\xi}]^{\frac{1}{2}}W_{\bold{j}_{m}}\sum_{l=0}^{\infty}(R^{Bog}_{\bold{j}_m;\,\bar{i},\bar{i}}(z))_{\xi}\,\Big[(\Gamma^{Bog\,}_{\bold{j}_m\,;\,\bar{i},\bar{i}}(z))_{\xi}(R^{Bog}_{\bold{j}_m;\,\bar{i},\bar{i}}(z))_{\xi}\Big]^{l}W_{\bold{j}_m}^*[(R^{Bog}_{\bold{j}_m;\,\bar{i}+2,\bar{i}+2}(z))_{\xi}]^{\frac{1}{2}}\times\quad\quad\quad\quad \nonumber \\
& &\quad\quad\quad \times [(R^{Bog}_{\bold{j}_m;\,\bar{i}+2,\bar{i}+2}(z))_{\xi}]^{-\frac{1}{2}}\Big[R_{\bold{j}_1,\dots, \bold{j}_m\,;\,\bar{i}+2,\bar{i}+2}(w)\Big]^{\frac{1}{2}}\,.\nonumber
\end{eqnarray}
Using the splitting in (\ref{splitting})-(\ref{splitting-2}) we can estimate
\begin{equation}
\|(R_{\bold{j}_1,\,\dots, \bold{j}_{m}\,;\,\bar{i}+2,\bar{i}+2}(w))^{\frac{1}{2}}[(R^{Bog}_{\bold{j}_m;\,\bar{i}+2,\bar{i}+2}(z))_{\xi}]^{-\frac{1}{2}}\,\charf_{\mathfrak{Q}^{(\bar{i}+2,\bar{i}+3)}_{\bold{j}_m}\mathcal{F}^N}\|\leq 1
\end{equation}
(see a similar argument in Corollary \ref{main-lemma-H}). Finally, we use {\emph{\underline{Theorem 3.1} of \cite{Pi1}} to estimate 
\begin{equation}
\|\,[(R^{Bog}_{\bold{j}_m;\,\bar{i},\bar{i}}(z))_{\xi}]^{\frac{1}{2}}W_{\bold{j}_{m}}\sum_{l=0}^{\infty}(R^{Bog}_{\bold{j}_m;\,\bar{i},\bar{i}}(z))_{\xi}\,\Big[(\Gamma^{Bog\,}_{\bold{j}_m\,;\,\bar{i},\bar{i}}(z))_{\xi}(R^{Bog}_{\bold{j}_m;\,\bar{i},\bar{i}}(z))_{\xi}\Big]^{l}W_{\bold{j}_m}^*[(R^{Bog}_{\bold{j}_m;\,\bar{i},\bar{i}}(z))_{\xi}]^{\frac{1}{2}}\,\|\,,
\end{equation}
and get
\begin{eqnarray}
& &\frac{1}{\|(R_{\bold{j}_1,\,\dots, \bold{j}_{m}\,;\,\bar{i}+2,\bar{i}+2}(w))^{\frac{1}{2}}\Gamma_{_{\bold{j}_1,\,\dots, \bold{j}_{m}}\,;\,\bar{i}+2,\bar{i}+2}(w)(R_{\bold{j}_1,\,\dots, \bold{j}_{m}\,;\,\bar{i}+2,\bar{i}+2}(w))^{\frac{1}{2}}\|}\\
& \geq &\frac{1}{\|\,[(R^{Bog}_{\bold{j}_m;\,\bar{i},\bar{i}}(z))_{\xi}]^{\frac{1}{2}}W_{\bold{j}_{m}}\sum_{l=0}^{\infty}(R^{Bog}_{\bold{j}_m;\,\bar{i},\bar{i}}(z))_{\xi}\,\Big[(\Gamma^{Bog\,}_{\bold{j}_m\,;\,\bar{i},\bar{i}}(z))_{\xi}(R^{Bog}_{\bold{j}_m;\,\bar{i},\bar{i}}(z))_{\xi}\Big]^{l}W_{\bold{j}_m}^*[(R^{Bog}_{\bold{j}_m;\,\bar{i},\bar{i}}(z))_{\xi}]^{\frac{1}{2}}\,\|+\mathcal{O}\Big(\frac{1}{\xi}[\frac{N^{\frac{1}{16}}}{N}]^{\frac{1}{2}}\Big)}\quad\quad\quad\\
&\geq& x_{\bar{i}+2} \label{Gamma-ineq}
\end{eqnarray}
where $x_{\bar{i}+2}$ is the $\bar{i}+2$-term of the sequence 
\begin{eqnarray}\label{def-x}
x_{2j+2}&:=&1-\frac{1}{4(1+a_{\epsilon_{\bold{j}_m}}-\frac{2b_{\epsilon_{\bold{j}_m}}}{N-2j-1}-\frac{1-c_{\epsilon_{\bold{j}_m}}}{(N-2j-1)^2})x_{2j}}\,.
%x_{2j+3}&:=&1-\frac{1}{4(1+a_{\epsilon}-\frac{2b_{\epsilon}}{N-2j-1}-\frac{1-c_{\epsilon}}{(N-2j-1)^2})x_{2j+1}}
\end{eqnarray}
Here,  $a_{\epsilon_{\bold{j}_m}}$, $b_{\epsilon_{\bold{j}m}}$, and $c_{\epsilon_{\bold{j}_m}}$ are those in  \emph{\underline{Lemma 3.4} of \cite{Pi1}} and in (\ref{setabc}) up  to $N$- and $\xi$-dependent corrections  that are hidden in the term $\mathcal{O}(\epsilon^{\nu}_{\bold{j}_{m}})$ (with $\nu>\frac{11}{8})$ which enters the definition of $a_{\epsilon_{\bold{j}_m}}$. We use the same notation to avoid a new symbol. Moreover, the value of $a_{\epsilon_{\bold{j}_m}}$ set here will coincide with the one in Corollary \ref{main-lemma-H}.  
%In fact these corrections vanish as $N\to \infty$. (See a similar reasoning in Corollary \ref{main-lemma-H}.)}
\qed
\\

Corollary \ref{main-lemma-H} and Theorem \ref{theorem-junction} enable us  to define the Neumann expansion in (\ref{ass-nue-1})-(\ref{ass-nue-2}) rigorously, and a result analogous to \emph{\underline{Theorem 3.1} of \cite{Pi1}} can be proven with the help of Lemma \ref{formal-fesh}:
\begin{thm}\label{Feshbach-H}
Assume condition a) of Corollary \ref{main-lemma-H}. Then, for \begin{equation}
z\leq E^{Bog}_{\bold{j}_m}+ (\delta-1)\phi_{\bold{j}_m}\sqrt{\epsilon_{\bold{j}_m}^2+2\epsilon_{\bold{j}_m}}
\end{equation} with $\delta=  1+\sqrt{\epsilon_{\bold{j}_m}}$,  $\epsilon_{\bold{j}_m}$ sufficiently small and $N$ sufficiently large, the
% and \begin{equation}
%z\leq E^{Bog}_{\bold{j}_m}+ (\delta-1)\phi_{\bold{j}_m}\sqrt{\epsilon_{\bold{j}_m}^2+2\epsilon_{\bold{j}_m}}
%\end{equation} with $\delta\leq  1+\sqrt{\epsilon_{\bold{j}_m}}$ and $\epsilon_{\bold{j}_m}$ sufficiently small. 
operators $\mathscr{K}_{\bold{j}_1,\dots,\bold{j}_m}^{(i)}(z+z^{\#}_{\bold{j}_1,\dots,\bold{j}_{m-1}})$, $\bar{i}+2\leq i\leq N-2$ and even,  are well defined \footnote{$\mathscr{K}^{(i)}_{\bold{j}_1,\dots,\bold{j}_m}(z+z^{\#}_{\bold{j}_1,\dots,\bold{j}_{m-1}})$ is  self-adjoint on the domain of the Hamiltonian $\mathfrak{Q}^{(>i+1)}_{\bold{j}_m}(H^{Bog}_{\bold{j}_1,\dots,\bold{j}_m}-z-z^{\#}_{\bold{j}_1,\dots,\bold{j}_{m-1}})\mathfrak{Q}^{(>i+1)}_{\bold{j}_m}$.}. For $i=\bar{i}+2,4,6,\dots,N-2$ they correspond to
\begin{eqnarray}
& &\mathscr{K}^{(i)}_{\bold{j}_1,\,\dots, \bold{j}_{m}}(w)\label{fesh-ham-i-in-bis}\\
&:=&\mathfrak{Q}^{(>i+1)}_{\bold{j}_m}(H_{\bold{j}_1,\dots,\bold{j}_{m}}-w)\mathfrak{Q}^{(>i+1)}_{\bold{j}_m}\\
%& &-\mathfrak{Q}^{(>i+1)}_{\bold{j}_M}\check{W}\mathfrak{Q}^{(i,i+1)}_{\bold{j}_M}\,R_{i,i}(w)\sum_{l_{i}=0}^{\infty}[\Gamma_{i,i}(z) R_{i,i}(w)]^{l_{i}}\, \mathfrak{Q}^{(i,i+1)}_{\bold{j}_M}\check{W}^*\mathfrak{Q}^{(>i+1)}_{\bold{j}_M}\\
& &-\mathfrak{Q}^{(>i+1)}_{\bold{j}_m}\check{W}_{\bold{j}_1,\,\dots, \bold{j}_{m}}\,\sum_{l_i=0}^{\infty}R_{\bold{j}_1,\,\dots, \bold{j}_{m}\,;\,i,i}(w)\Big[\Gamma_{\bold{j}_1,\,\dots, \bold{j}_{m}\,;\,i,i}(w)\,R_{\bold{j}_1,\,\dots, \bold{j}_{m}\,;\,i,i}(w)\Big]^{l_i}\times \quad\quad\quad\quad  \label{fesh-ham-i-fin-bis}
\\
& &\quad\quad\quad\times\check{W}^*_{\bold{j}_1,\,\dots, \bold{j}_{m}}\mathfrak{Q}^{(>i+1)}_{\bold{j}_m} \nonumber\,.%& &-\mathscr{V}^{(i)}(w)\,R_{i,i}(w)\sum_{l_i=0}^{\infty}[\Gamma_{i,i}(w) R_{i,i}(w)]^{l_i}\,(\mathscr{V}^{(i)}(w) )^*
\end{eqnarray}
where $w=z+z^{\#}_{\bold{j}_1,\dots,\bold{j}_{m-1}}$. The operators $R_{\bold{j}_1,\dots,\bold{j}_m\,;\,i,i}(w)$,  $\Gamma_{\bold{j}_1,\dots,\bold{j}_m\,;\,i,i}(w)$ are defined in (\ref{def-Rii}) and (\ref{def-gamma-2})-(\ref{GammaH-i}), respectively.

\noindent
The following estimates hold true for $\bar{i}+2\leq i \leq N-2$ and even:
\begin{equation}
\|\check{\Gamma}_{\bold{j}_1,\dots,\bold{j}_m\,;\,i,i}(w)\|\leq \frac{1}{x_{i}}\label{Gamma-ineq}
\end{equation}
where 
\begin{equation}
\check{\Gamma}_{\bold{j}_1,\dots,\bold{j}_m\,;\,i,i}(w):=\sum_{l_i=0}^{\infty}\Big[(R_{\bold{j}_1,\dots,\bold{j}_m\,;\,i,i}(w))^{\frac{1}{2}}\Gamma_{\bold{j}_1,\dots,\bold{j}_m\,;\,i,i}(w)(R_{\bold{j}_1,\dots,\bold{j}_m\,;\,i,i}(w)^{\frac{1}{2}}\Big]^{l_i}
\end{equation}
and $x_i$ defined in (\ref{def-x}) fulfills the bound (see \underline{Lemma 3.6} of \cite{Pi1}) 
\begin{equation}\label{x-ineq}
x_{i}\geq\frac{1}{2}\Big[1+\sqrt{\eta a_{\epsilon}}-\frac{b_{\epsilon}/\sqrt{\eta a_{\epsilon}}}{N-2j-\epsilon^{\Theta}}\Big]\,,\quad \eta=1-\epsilon^{\frac{1}{2}},
\end{equation}
for $\epsilon\equiv \epsilon_{\bold{j}_m}$ small enough and $0<\Theta\leq \frac{1}{4}$.
\end{thm}

\noindent
\emph{Proof}

\noindent
The proof follows the arguments of \emph{\underline{Theorem 3.1} of \cite{Pi1}} starting from the result in Theorem \ref{theorem-junction}. \qed

Furthermore, similarly to what seen in the companion papers \cite{Pi1}, \cite{Pi2}, Corollary \ref{main-lemma-H} implies the expansion of the operators $\Gamma_{\bold{j}_1,\dots,\bold{j}_{m}\,;\,i-2,i-2}(w)$ in terms of finite sums of products of the resolvents $R_{\bold{j}_1,\dots, \bold{j}_m\,;\,j,j}(w)$, $\bar{i}\leq j\leq i-4$ with $j$ even, of the operator $\check{W}_{\bold{j}_1,\dots, \bold{j}_m}$, and of $\Gamma_{_{\bold{j}_1,\,\dots, \bold{j}_{m}}\,;\,\bar{i}+2,\bar{i}+2}(w)$. This is the content of Proposition \ref{lemma-expansion-proof-0}.

\noindent
To streamline formulae, in Definition \ref{def-sums} and in Proposition \ref{lemma-expansion-proof-0} we write $\check{W}_{j,j-2}$, $\check{W}^*_{j-2,j}$, $R_{j-2,j-2}(w)$, and $\Gamma_{j,j}(w)$ instead of $\check{W}_{\bold{j}_1,\dots, \bold{j}_m\,;j,j-2}$, $\check{W}^*_{\bold{j}_1,\dots, \bold{j}_m\,;j-2,j}$, $R_{\bold{j}_1,\dots,\bold{j}_m\,;\,j-2,j-2}(w)$, and $\Gamma_{\bold{j}_1,\dots,\bold{j}_m\,;\,j,j}(w)$, respectively.
\begin{definition} \label{def-sums}

\noindent
Let $h\in \mathbb{N}$, $h\geq 2$, and 
\begin{equation}
w\leq z^{\#}_{\bold{j}_1,\dots,\bold{j}_{m-1}}+E^{Bog}_{\bold{j}_m}+ (\delta-1)\phi_{\bold{j}_m}\sqrt{\epsilon_{\bold{j}_m}^2+2\epsilon_{\bold{j}_m}}
\end{equation} with $\delta\leq 1+\sqrt{\epsilon_{\bold{j}_m}}$ and $\epsilon_{\bold{j}_m}$ sufficiently small. Let $N$ be sufficiently large. We define:
\begin{enumerate}
\item For $N-2\geq  j \geq \bar{i}+4$ and even
\begin{equation}
[\Gamma_{j,j}(w)]_{(j-2, h_-)}=[\Gamma_{j,j}(w)]^{(0)}_{(j-2, h_-)}+[\Gamma_{j,j}(w)]^{(>0)}_{(j-2, h_-)}
\end{equation}
where
\begin{eqnarray}
& &[\Gamma_{j,j}(w)]^{(0)}_{(j-2, h_-)}:=\check{W}_{j,j-2}R_{j-2,j-2}(w) \check{W}^*_{j-2,j}\,,
\end{eqnarray}
\begin{eqnarray}
& &[\Gamma_{j,j}(w)]^{(>0)}_{(j-2, h_-)}\\
&:=&\check{W}_{j,j-2}\,(R_{j-2,j-2}(w))^{\frac{1}{2}}\times \\
& &\quad \times  \sum_{l_{j-2}=1}^{h-1}\Big[(R_{j-2,j-2}(w))^{\frac{1}{2}}\check{W}_{j-2,j-4}\,R_{j-4,j-4}(w)\check{W}_{j-4,j-2}^*(R_{j-2,j-2}(w))^{\frac{1}{2}} \Big]^{l_{j-2}}(R_{j-2,j-2}(w))^{\frac{1}{2}} \check{W}^*_{j-2,j}\nonumber \\
&=&\check{W}_{j,j-2}\,(R_{j-2,j-2}(w))^{\frac{1}{2}}\times \\
& &\quad \times  \sum_{l_{j-2}=1}^{h-1}\Big[(R_{j-2,j-2}(w))^{\frac{1}{2}}[\Gamma_{j-2,j-2}(w)]^{(0)}_{(j-4, h_-)}(R_{j-2,j-2}(w))^{\frac{1}{2}} \Big]^{l_{j-2}}(R_{j-2,j-2}(w))^{\frac{1}{2}} \check{W}^*_{j-2,j}\,,\nonumber 
\end{eqnarray}
with  $[\Gamma_{\bar{i}+2,\bar{i}+2}(w)]^{(0)}_{(\bar{i}, h_-)}:=(\ref{def-gamma-2})$;
\\

for $N-2\geq  j \geq \bar{i}+4$ and even
\begin{eqnarray}
[\Gamma_{j,j}(w)]_{(j-2, h_+)}&:=&\check{W}_{j,j-2}\,(R_{j-2,j-2}(w))^{\frac{1}{2}}\times \\
& &\quad \times  \sum_{l_{j-2}=h}^{\infty}\Big[(R_{j-2,j-2}(w))^{\frac{1}{2}}\Gamma_{j-2,j-2}(w)(R_{j-2,j-2}(w))^{\frac{1}{2}} \Big]^{l_{j-2}}\times \nonumber \\
& &\quad\quad\quad \times (R_{j-2,j-2}(w))^{\frac{1}{2}} \check{W}^*_{j-2,j}\,.\nonumber
\end{eqnarray}
\item
For $N-2\geq  j \geq \bar{i}+6$,\,  $\bar{i}+2\leq l \leq j-4$ and even
\begin{eqnarray}
& &[\Gamma_{j,j}(w)]_{(l,h_-; l+2,h_-;\dots;j-4,h_-;j-2,h_-)}\\
&:= &\check{W}_{j,j-2}\,(R_{j-2,j-2}(w))^{\frac{1}{2}} \check{\sum}_{l_{j-2}=1}^{h-1}\Big[(R_{j-2,j-2}(w))^{\frac{1}{2}}[\Gamma_{j-2,j-2}(w)]_{(l,h_-;l+2,h_-;\dots;j-4,h_-)}(R_{j-2,j-2}(w))^{\frac{1}{2}} \Big]^{l_{j-2}}\times \nonumber\\
& &\quad\quad\quad\quad \times(R_{j-2,j-2}(w))^{\frac{1}{2}} \check{W}^*_{j-2,j} \label{collection}
\end{eqnarray}
Here, the  symbol $\check{\sum}^{h-1}_{l_{j-2}=1}$ stands for a sum of terms resulting from operations $\mathcal{A}1$ and $\mathcal{A}2$ below:
\begin{itemize}
\item[$\mathcal{A}1)$]
At fixed $1\leq l_{j-2} \leq h-1$ summing all the products
\begin{equation}
\Big[(R_{j-2,j-2}(w))^{\frac{1}{2}}\mathcal{X}(R_{j-2,j-2}(w))^{\frac{1}{2}} \Big]^{l_{j-2}}
\end{equation}
that are obtained by replacing  $\mathcal{X}$ for each factor with the operators (defined by iteration) of the type $[\Gamma_{j-2,j-2}(w)]_{(s,h_-;s+2,h_-;\dots;j-4,h_-)}$ with $l\leq s\leq j-4$ and even, with the constraint that  if $l\leq j-6$ then $\mathcal{X}$ is replaced with $[\Gamma_{j-2,j-2}(w)]_{(l,h_-;4,h_-;\dots;j-4,h_-)}$ in one factor at least, whereas if $l= j-4$ then  $\mathcal{X}$ is replaced with $[\Gamma_{j-2,j-2}(w)]^{(>0)}_{(j-4,h_-)}$ in one factor at least;
\item[$\mathcal{A}2)$]
Summing from $l_{j-2}=1$ up to $l_{j-2}=h-1$.
\end{itemize}
\item
For $N-2\geq  j \geq \bar{i}+6$,  $\bar{i}+2\leq l \leq j-4$ and even
\begin{eqnarray}
& &[\Gamma_{j,j}(w)]_{(l,h_+; l+2,h_-;\dots;j-4,h_-;j-2,h_-)}\\
&:= &\check{W}_{j,j-2}\,(R_{j-2,j-2}(w))^{\frac{1}{2}} \check{\sum}_{l_{j-2}=1}^{h-1}\Big[(R_{j-2,j-2}(w))^{\frac{1}{2}}[\Gamma_{j-2,j-2}(w)]_{(l,h_+;l+2,h_-;\dots;j-4,h_-)}\times \nonumber\\
& &\quad\quad \quad\quad\quad\quad\quad\quad\quad\quad\quad\quad\times (R_{j-2,j-2}(w))^{\frac{1}{2}} \Big]^{l_{2}}(R_{j-2,j-2}(w))^{\frac{1}{2}} \check{W}^*_{j-2,j}\label{collection-bis}
\end{eqnarray}
Here, the  symbol $\check{\sum}^{h-1}_{l_{j-2}=1}$ stands for a sum of terms resulting from operations $\mathcal{B}1$ and $\mathcal{B}2$ below:
\begin{itemize}
\item[$\mathcal{B}1)$]
At fixed $1\leq l_{j-2} \leq h-1$, summing all the products
\begin{equation}
\Big[(R_{j-2,j-2}(w))^{\frac{1}{2}}\mathcal{X}(R_{j-2,j-2}(w))^{\frac{1}{2}} \Big]^{l_{j-2}}
\end{equation}
that are obtained by replacing  $\mathcal{X}$ for each factor with the operators (defined by iteration) of the type  $[\Gamma_{j-2,j-2}(w)]_{(l,h_+;l+2,h_-;\dots;j-4,h_-)}$ and $[\Gamma_{j-2,j-2}(w)]_{(s,h_-;s+2,h_-;\dots;j-4,h_-)}$  with $l\leq s\leq j-4$ and even, with the constraint   that $\mathcal{X}$ is replaced with $[\Gamma_{j-2,j-2}(w)]_{(l,h_+;l+2,h_-;\dots;j-4,h_-)}$ in one factor at least.
\item[$\mathcal{B}2)$]
Summing from $l_{j-2}=1$ up to $h-1$.
\end{itemize}
\end{enumerate}
\end{definition}
\begin{prop}\label{lemma-expansion-proof-0} Let $\epsilon_{\bold{j}_m}\equiv \epsilon$ be sufficiently small and $N$ sufficiently large.
For any fixed $2\leq h \in \mathbb{N}$ and for $N-2\geq i\geq \bar{i}+4$ and even,  the splitting
\begin{eqnarray}
\Gamma_{i,i}(w)&=&\sum_{l=\bar{i}+2,\,l\, even}^{i-2}[\Gamma_{i,i}(w)]_{(l,h_-;l+2,h_-;\dots ; i-2,h_-)}+\sum_{l=\bar{i}+2\,,\,l\, even}^{i-2}[\Gamma_{i,i}(w)]_{(l,h_+; l+2,h_-;\dots; i-2,h_-)}\quad\quad\label{decomposition}
\end{eqnarray}
holds true for $w\leq z^{\#}_{\bold{j}_1,\dots,\bold{j}_{m-1}}+E^{Bog}_{\bold{j}_m}+ (\delta-1)\phi_{\bold{j}_m}\sqrt{\epsilon_{\bold{j}_m}^2+2\epsilon_{\bold{j}_m}}$ and $\delta\leq 1+\sqrt{\epsilon_{\bold{j}_m}}$. Moreover, for $\bar{i}+2\leq l \leq i-2$ and even,  the estimates
\begin{eqnarray}\label{gamma-exp-1}
& &\Big\|(R^{Bog}_{i,i}(z))^{\frac{1}{2}}[\Gamma^{Bog\,}_{i,i}(z)]_{(l,h_-;l+2, h_-;\, \dots \,; i-2,h_-)}(R^{Bog}_{i,i}(z))^{\frac{1}{2}}\Big\|\\
& &\leq  \prod_{f=l+2\,,\, f-l\,\text{even}}^{i}\frac{K_{f,\epsilon}}{(1-Z_{f-2,\epsilon})^2}\nonumber
%\Big(\frac{2}{3}+\mathcal{O}(\sqrt{\epsilon})\Big)^{\frac{i-l}{2}}\prod_{f=l+2\,,\, f-l\,\text{even}}^{i-2}(1+a_{\epsilon}-\frac{2b_{\epsilon}}{N-f-1}-\frac{1-c_{\epsilon}}{(N-f-1)^2})^{-1}\nonumber
\end{eqnarray}
and
%\begin{equation}
%\Big\|(R^{Bog}_{i,i}(z))^{\frac{1}{2}}[\Gamma^{Bog\,}_{i,i}(z)]_{(l,h_-;\dots ; j,h_-;j-4,h_+)}
%(R^{Bog}_{i,i}(z))^{\frac{1}{2}}\Big\|\leq {\color{red}...}
%\end{equation}
\begin{eqnarray}\label{gamma-exp-2}
& &\|(R^{Bog}_{i,i}(z))^{\frac{1}{2}}[\Gamma^{Bog\,}_{i,i}(z)]_{(l,h_+; 4,h_-;\dots;i-2,h_-)}(R^{Bog}_{i,i}(z))^{\frac{1}{2}}\|\\
&\leq& (Z_{l,\epsilon})^{h}\,\prod_{f=l+2\,,\, f-l\,\text{even}}^{i}\frac{K_{f,\epsilon}}{(1-Z_{f-2,\epsilon})^2} \nonumber
%& &\quad\times \prod_{f=l+2\,,\, f-l\,\text{even}}^{i-2}(1+a_{\epsilon}-\frac{2b_{\epsilon}}{N-f-1}-\frac{1-c_{\epsilon}}{(N-f-1)^2})^{-1}\nonumber
 \end{eqnarray}
hold true, where 
\begin{equation}
K_{i,\epsilon}:=\frac{1}{4(1+a_{\epsilon}-\frac{2b_{\epsilon}}{N-i+1}-\frac{1-c_{\epsilon}}{(N-i+1)^2})}\quad,\quad Z_{i-2,\epsilon}:=\frac{1}{4(1+a_{\epsilon}-\frac{2b_{\epsilon}}{N-i+3}-\frac{1-c_{\epsilon}}{(N-i+3)^2})}\frac{2}{\Big[1+\sqrt{\eta a_{\epsilon}}-\frac{b_{\epsilon}/\sqrt{\eta a_{\epsilon}}}{N-i+4-\epsilon^{\Theta}}\Big]}
\end{equation}
%\begin{eqnarray}\label{gamma-exp-1}
%& &\Big\|(R_{i,i}(w))^{\frac{1}{2}}[\Gamma_{i,i}(w)]_{(l,h_-;l+2, h_-;\, \dots \,; i-2,h_-)}(R_{i,i}(w))^{\frac{1}{2}}\Big\|\\
%& &\leq  (\frac{2}{3}+\mathcal{O}(\sqrt{\epsilon}))^{\frac{i-l}{2}}\prod_{f=l+2\,,\, f-l\,\text{even}}^{i-2}(1+a_{\epsilon}-\frac{2b_{\epsilon}}{N-f+1}-\frac{1-c_{\epsilon}}{(N-f+1)^2})^{-1}\nonumber
%\end{eqnarray}
%and
%\begin{equation}
%\Big\|(R^{Bog}_{i,i}(z))^{\frac{1}{2}}[\Gamma^{Bog\,}_{i,i}(z)]_{(l,h_-;\dots ; j,h_-;j-4,h_+)}
%(R^{Bog}_{i,i}(z))^{\frac{1}{2}}\Big\|\leq {\color{red}...}
%\end{equation}
%\begin{eqnarray}\label{gamma-exp-2}
%& &\|(R_{i,i}(w))^{\frac{1}{2}}[\Gamma_{i,i}(w)]_{(l,h_+; 4,h_-;\dots;i-2,h_-)}(R_{i,i}(w))^{\frac{1}{2}}\|\\
%&\leq& (\frac{2}{3}+\mathcal{O}(\sqrt{\epsilon}))^{\frac{i-l}{2}}\Big[\frac{1}{4(1+a_{\epsilon}-\frac{2b_{\epsilon}}{N-i+1}-\frac{1-c_{\epsilon}}{(N-i+1)^2})}\Big]^{h}\times \\
%& &\quad\times \prod_{f=l+2\,,\, f-l\,\text{even}}^{i-2}(1+a_{\epsilon}-\frac{2b_{\epsilon}}{N-f+1}-\frac{1-c_{\epsilon}}{(N-f+1)^2})^{-1}\nonumber
 %\end{eqnarray}
%hold true, where $\prod_{f=l+2\,,\, f-l\,\text{even}}^{l}(...)\equiv 1$.
where $a_{\epsilon}, b_{\epsilon}, c_{\epsilon}$, and $0<\Theta\leq \frac{1}{4}$ are those defined in Corollary \ref{main-lemma-H} and \underline{Lemma 3.6} of \cite{Pi1}. 
\end{prop}

\noindent
\emph{Proof}

Using the results of Theorem \ref{Feshbach-H} the proof is like in  \emph{\underline{Proposition 4.10} of \cite{Pi1}}. \qed

\subsection{Last implementation of the Feshbach map}
For the last implementation of the Feshbach map, i.e., from $i=N-2$ to $i=N$, we have to make sure that
\begin{equation}\label{invers}
\frac{1}{\overline{\mathscr{P}_{\psi^{\#}_{\bold{j}_1,\dots,\bold{j}_{m-1}}}}\mathscr{K}^{(N-2)}_{\bold{j}_1,\,\dots, \bold{j}_{m}}(z+z^{\#}_{\bold{j}_1,\dots,\bold{j}_{m-1}})\overline{\mathscr{P}_{\psi^{\#}_{\bold{j}_1,\dots,\bold{j}_{m-1}}}}}
\end{equation}
is well defined in $\overline{\mathscr{P}_{\psi^{\#}_{\bold{j}_1,\dots,\bold{j}_{m-1}}}}\mathcal{F}^N$ and estimate its operator norm.  This is the content of next Proposition \ref{invertibility-bis}. 
$\overline{\mathscr{P}_{\psi^{\#}_{\bold{j}_1,\dots,\bold{j}_{m-1}}}}$ is the projection onto the subspace of vectors without particles in the modes $\pm \bold{j}_m$ and orthogonal to $\psi^{\#}_{\bold{j}_1,\dots,\bold{j}_{m-1}}$.
%two orthogonal subspaces $\mathcal{H}_{m,I}$ and $\mathcal{H}_{m,II}$ described below:
%\begin{itemize}
%\item
%$\mathcal{H}_{m,I}$ is the subspace of vectors in $\mathcal{F}^N$ with no particles in the modes $\pm \bold{j}_m$ and orthogonal to $\mathscr{P}_{\psi^{\#}_{\bold{j}_1,\dots,\bold{j}_{m-1}}}\mathcal{F}^N$;
%\item 
%$\mathcal{H}_{m,II}$ is the subspace spanned by the vectors with at least $\lfloor N^{\frac{1}{16}} \rfloor-1$ particles in the modes $\pm \bold{j}_m$  and where the operator {\color{red} $T-T_{\bold{j}=\pm \bold{j}_m}$} takes values larger than $N^{\frac{1}{8}}$.
%the subspace $\mathcal{H}_{II}$ is the linear span of the product state vectors with:

In the case $m=1$ we shall use the notation
\begin{equation}
\mathscr{P}^{\#}_{\eta}:=|\eta \rangle \langle \eta |\quad,\quad \overline{\mathscr{P}^{\#}_{\eta}}:=\mathfrak{Q}^{(>N-1)}_{\bold{j}_1} -\mathscr{P}^{\#}_{\eta}\,.
\end{equation}
%to avoid any confusion with $\overline{\mathscr{P}_{\eta}}:=Q^{(>N-1)}_{\bold{j}_1}-|\eta \rangle \langle \eta |$ defined for the Bogoliubov Hamiltonian.
\begin{proposition}\label{invertibility-bis}
Let  $1\leq m\leq  M$, $\epsilon_{\bold{j}_m}$ be sufficiently small and $N$ sufficiently large such that
 the Feshbach flow (see (\ref{fesh-ham-i-in-bis})) is well defined for $z\leq E^{Bog}_{\bold{j}_m}+ \sqrt{\epsilon_{\bold{j}_m}}\phi_{\bold{j}_m}\sqrt{\epsilon_{\bold{j}_m}^2+2\epsilon_{\bold{j}_m}}$ up to $i=N-2$.
 Assume that
 \begin{enumerate}
\item 
the Hamiltonian $H^{\#}_{\bold{j}_1,\dots,\bold{j}_{m-1}}$ has ground state energy $z^{\#}_{\bold{j}_1,\dots,\bold{j}_{m-1}}$ with ground state vector $\psi^{\#}_{\bold{j}_1,\dots,\bold{j}_{m-1}}\in \mathcal{F}^N\ominus \mathcal{F}^N_{\pm\bold{j}_m}$  where $\mathcal{F}^N_{\pm\bold{j}_m}$ is the subspace of vectors containing at least one particle in the modes $\pm \bold{j}_m$,   and 
\item
\begin{equation}\label{assumption-gap}
\text{infspec}\,\Big[H^{\#}_{\bold{j}_1,\dots,\bold{j}_{m-1}}\upharpoonright_{(\mathcal{F}^N\ominus \mathcal{F}^N_{\pm\bold{j}_m})\ominus \{\mathbb{C}\psi^{\#}_{\bold{j}_1, \dots,\bold{j}_{m-1}}\}}\Big]-z^{\#}_{\bold{j}_1,\dots,\bold{j}_{m-1}}\geq \Delta^{\#}_{m-1}
%\text{Gap}\,\Big[\Big(\hat{H}^{Bog}_{\bold{j}_1, \dots,\bold{j}_{m-1}}+\sum_{\bold{j}\neq \{\pm\bold{j}_1, \dots, \pm\bold{j}_{m}\}}(k_{\bold{j}})^2a^*_{\bold{j}}a_{\bold{j}}\Big)\upharpoonright_{\mathcal{F}\ominus \mathcal{F}_{\pm\bold{j}_m}}\Big]\geq \Delta_{m-1}
\end{equation}
for some $\Delta^{\#}_{m-1}>0$.
%(Notice that $$\hat{H}^{Bog}_{\bold{j}_1, \dots,\bold{j}_{m-1}}+\sum_{\bold{j}\neq \{\pm\bold{j}_1, \dots, \pm\bold{j}_{m}\}}(k_{\bold{j}})^2a^*_{\bold{j}}a_{\bold{j}}=H^{Bog}_{\bold{j}_1, \dots,\bold{j}_{m-1}}-(k_{\bold{j}_m})^2(a^*_{\bold{j}_m}a_{\bold{j}_m}+a^*_{-\bold{j}_m}a_{-\bold{j}_m})\,,$$ i.e., the kinetic energy associated with $\pm \bold{j}_m$ is absent).
%\item
%\begin{equation}
%|z_{\bold{j}_1,\dots\,\bold{j}_{m-1}}-z^{\#}_{\bold{j}_1,\dots\,\bold{j}_{m-1}}|\leq \frac{(m-1)\xi}{3M}\label{diff-z-zbog}
%\end{equation}
%where $z_{\bold{j}_1,\dots\,\bold{j}_{m-1}},\, z^{\#}_{\bold{j}_1,\dots\,\bold{j}_{m-1}}\equiv 0$ if $m=1$, and $\xi=(\frac{1}{\ln N})^{\frac{1}{4}}$.
\end{enumerate}
\item

\noindent
Then, there exists $C^{\#\,\perp}>0$ such that 
\begin{equation}\label{denominator}
\Big\|\overline{\mathscr{P}_{\psi^{\#}_{\bold{j}_1,\dots,\bold{j}_{m-1}}}}\frac{1}{\overline{\mathscr{P}_{\psi^{\#}_{\bold{j}_1,\dots,\bold{j}_{m-1}}}}\mathscr{K}_{\bold{j}_1,\dots,\bold{j}_m}^{\,(N-2)}(z+z^{\#}_{\bold{j}_1,\dots,\bold{j}_{m-1}})\overline{\mathscr{P}_{\psi^{\#}_{\bold{j}_1,\dots,\bold{j}_{m-1}}}}}\overline{\mathscr{P}_{\psi^{\#}_{\bold{j}_1,\dots,\bold{j}_{m-1}}}}\Big\|\leq \frac{1}{(1-\gamma)\Delta^{\#}_{m-1}}
\end{equation}
%\begin{eqnarray}
%& &\overline{\mathscr{P}_{\psi^{Bog}_{\bold{j}_1,\dots,\bold{j}_{M-1}}}}\Gamma^{Bog}_{\bold{j}_1,\dots,\bold{j}_M;\,N,N}(z+z_{\bold{j}_1,\dots,\bold{j}_{M-1}})\,\overline{\mathscr{P}_{\psi^{Bog}_{\bold{j}_1,\dots,\bold{j}_{M-1}}}}\leq -z_M\overline{\mathscr{P}_{\psi^{Bog}_{\bold{j}_1,\dots,\bold{j}_{M-1}}}}\label{ineq-z_M}
%\end{eqnarray} 
%and 
%\begin{equation}
%\overline{\mathscr{P}^{\#}_{\psi^{Bog}_{\bold{j}_1,\dots,\bold{j}_{m-1}}}}\mathscr{K}_{\bold{j}_1,\dots,\bold{j}_m}^{\,(N-2)}(z+z_{\bold{j}_1,\dots,\bold{j}_{m-1}})\overline{\mathscr{P}_{\psi^{Bog}_{\bold{j}_1,\dots,\bold{j}_{m-1}}}}\geq (1-\gamma)\Delta_{m-1}\overline{\mathscr{P}^{\#}_{\psi^{Bog}_{\bold{j}_1,\dots,\bold{j}_{m-1}}}}\label{inequa-m}
%\end{equation}
 for $N$ sufficiently large and 
\begin{equation}
z\leq \min\,\Big\{ z_{m}+\gamma \Delta^{\#}_{m-1}-\frac{m\xi^{\frac{1}{2}}}{M}-\frac{C^{\#\,\perp}}{(\ln N)^{\frac{1}{4}}}\,;\,E^{Bog}_{\bold{j}_m}+ \sqrt{\epsilon_{\bold{j}_m}}\phi_{\bold{j}_m}\sqrt{\epsilon_{\bold{j}_m}^2+2\epsilon_{\bold{j}_m}}\Big\}\label{bound-z-0-bis-bis}
\end{equation}
 where $\xi=\frac{1}{(\ln N)^{\frac{1}{4}}}$,  $z_{m}$ is the ground state energy of $H^{Bog}_{\bold{j}_m}$,  $\gamma = \frac{1}{2}$, and $\Delta_0^{\#}\equiv \Delta_0:=\min\, \Big\{k_{\bold{j}}^2\,|\,\bold{j}\in \mathbb{Z}^d \setminus \{\bold{0}\}\Big\}$.
 % fulfills the constraint in  (\ref{constraint-xi-2}) and equals zero if $m=1$. 
 %which implies
%\begin{equation}
% \xi \Delta_0\geq \mathcal{O}(\frac{1}{N^{\frac{1}{2}}})\,. \label{constraint-xi}
%\end{equation}
%\noindent
%Therefore the Hamiltonian $$\overline{\mathscr{P}^{\#}_{\psi^{Bog}_{\bold{j}_1,\dots,\bold{j}_{m-1}}}}\mathscr{K}_{\bold{j}_1,\dots,\bold{j}_m}^{\,(N-2)}(z+z_{\bold{j}_1,\dots,\bold{j}_{m-1}})\overline{\mathscr{P}^{\#}_{\psi^{Bog}_{\bold{j}_1,\dots,\bold{j}_{m-1}}}}$$ is invertible in $\overline{\mathscr{P}^{\#}_{\psi^{Bog}_{\bold{j}_1,\dots,\bold{j}_{m-1}}}}\mathcal{F}^{N}$ 
%in the range (of $z$) given in (\ref{bound-z-0-bis}).
%\begin{equation}
%\Big\|\overline{\mathscr{P}^{\#}_{\psi^{Bog}_{\bold{j}_1,\dots,\bold{j}_{m-1}}}}\frac{1}{\overline{\mathscr{P}^{\#}_{\psi^{Bog}_{\bold{j}_1,\dots,\bold{j}_{m-1}}}}\mathscr{K}_{\bold{j}_1,\dots,\bold{j}_m}^{\,(N-2)}(z+z_{\bold{j}_1,\dots,\bold{j}_{m-1}})\overline{\mathscr{P}^{\#}_{\psi^{Bog}_{\bold{j}_1,\dots,\bold{j}_{m-1}}}}}\overline{\mathscr{P}^{\#}_{\psi^{Bog}_{\bold{j}_1,\dots,\bold{j}_{m-1}}}}\Big\|\leq \frac{1}{(1-\gamma)\Delta_{m-1}}
%\end{equation}
\end{proposition}

\noindent
\emph{Proof}

We set $w\equiv z+z^{\#}_{\bold{j}_1,\dots,\bold{j}_{m-1}}$ and write
\begin{eqnarray}
& &\overline{\mathscr{P}_{\psi^{\#}_{\bold{j}_1,\dots,\bold{j}_{m-1}}}}\mathscr{K}_{\bold{j}_1,\dots,\bold{j}_m}^{(N-2)}(z+z^{\#}_{\bold{j}_1,\dots,\bold{j}_{m-1}})\overline{\mathscr{P}_{\psi^{\#}_{\bold{j}_1,\dots,\bold{j}_{m-1}}}}\\
%& =&\overline{\mathscr{P}_{\psi^{Bog}_{\bold{j}_1,\dots,\bold{j}_{m-1}}}}\mathscr{K}_{\bold{j}_1,\dots,\bold{j}_m}^{Bog\,(N-2)}(z+z_{\bold{j}_1,\dots,\bold{j}_{m-1}})\overline{\mathscr{P}_{\psi^{Bog}_{\bold{j}_1,\dots,\bold{j}_{m-1}}}}\\
&=&\overline{\mathscr{P}_{\psi^{\#}_{\bold{j}_1,\dots,\bold{j}_{m-1}}}}(H_{\bold{j}_1,\dots,\bold{j}_m}-w)\overline{\mathscr{P}_{\psi^{\#}_{\bold{j}_1,\dots,\bold{j}_{m-1}}}}\quad\quad\quad \label{rel-in}\\
%& &+\Big(\overline{\mathscr{P}_{\psi^{Bog}_{\bold{j}_1,\dots,\bold{j}_{M-1}}}^{\#}}\Big)_1\,H\,\Big(\overline{\mathscr{P}_{\psi^{Bog}_{\bold{j}_1,\dots,\bold{j}_{M-1}}}^{\#}}\Big)_2+
%\Big(\overline{\mathscr{P}_{\psi^{Bog}_{\bold{j}_1,\dots,\bold{j}_{M-1}}}^{\#}}\Big)_2\,H\,\Big(\overline{\mathscr{P}_{\psi^{Bog}_{\bold{j}_1,\dots,\bold{j}_{M-1}}}^{\#}}\Big)_1\\
%& &+\Big(\overline{\mathscr{P}_{\psi^{Bog}_{\bold{j}_1,\dots,\bold{j}_{M-1}}}^{\#}}\Big)_2(H-z-z_{\bold{j}_1,\dots,\bold{j}_{M-1}})\Big(\overline{\mathscr{P}_{\psi^{Bog}_{\bold{j}_1,\dots,\bold{j}_{M-1}}}^{\#}}\Big)_2\quad\quad\quad\\
& &-\overline{\mathscr{P}_{\psi^{\#}_{\bold{j}_1,\dots,\bold{j}_{m-1}}}}\,\Gamma_{\bold{j}_1,\dots,\bold{j}_m\,;\,N,N}(w)\,\overline{\mathscr{P}_{\psi^{\#}_{\bold{j}_1,\dots,\bold{j}_{m-1}}}}\,.\label{nonH}
% &-\Big\{\Big(\overline{\mathscr{P}_{\psi^{Bog}_{\bold{j}_1,\dots,\bold{j}_{M-1}}}^{\#}}\Big)_I\,[\Gamma_{\bold{j}_1,\dots,\bold{j}_M\,,;\,\bar{i},\bar{i}}(w)]\Big(\overline{\mathscr{P}_{\psi^{Bog}_{\bold{j}_1,\dots,\bold{j}_{M-1}}}^{\#}}\Big)_{II}+h.c.\Big\}\\
%& &-\Big(\overline{\mathscr{P}_{\psi^{Bog}_{\bold{j}_1,\dots,\bold{j}_{M-1}}}^{\#}}\Big)_2[\Gamma_{\bold{j}_1,\dots,\bold{j}_M\,,;\,\bar{i},\bar{i}}]^{(0)}(z_{\bold{j}_1,\dots,\bold{j}_{M-1}}+z)\Big(\overline{\mathscr{P}_{\psi^{Bog}_{\bold{j}_1,\dots,\bold{j}_{M-1}}}^{\#}}\Big)_2 \\
%& & -\Big(\overline{\mathscr{P}_{\psi^{Bog}_{\bold{j}_1,\dots,\bold{j}_{M-1}}}^{\#}}\Big)_{II}[\mathscr{V}^{(\bar{i}-2)}(w)]^{(0)}\Big(\overline{\mathscr{P}_{\psi^{Bog}_{\bold{j}_1,\dots,\bold{j}_{M-1}}}^{\#}}\Big)_{II} \label{rel-fin}
\end{eqnarray}

\noindent
With reference to the definition in (\ref{def-xiham-1}), 
%notice that the following identity also holds
%\begin{eqnarray}
%H_{\bold{j}_1,\dots,\bold{j}_m}&=&(H_{\bold{j}_{1},\dots,\bold{j}_{m-1}})_{\xi}+V_{\bold{j}_{m}}+(\hat{H}^{Bog}_{\bold{j}_m})_{2\xi}+\xi \,T_{\bold{j}\neq\{\pm\bold{j}_{1},\dots, \pm%\bold{j}_{m}\}}\\
%& &+\frac{1}{2}T_{\bold{j}\neq\{\pm\bold{j}_{1},\dots, \pm\bold{j}_{m}\}}+\xi \sum_{\bold{j}\in\{\pm\bold{j}_{1},\dots, \pm\bold{j}_{m}\}}(k_{\bold{j}})^2a^*_{\bold{j}}a_{\bold{j}}+\xi %\sum_{\bold{j}\in\{\pm\bold{j}_{m}\}}(k_{\bold{j}})^2a^*_{\bold{j}}a_{\bold{j}}
%\end{eqnarray}
%We shall make use of it for the estimate of 
%\begin{equation}
%(\overline{\mathscr{P}_{\psi^{Bog}_{\bold{j}_1,\dots,\bold{j}_{m-1}}}^{\#}})_{II}\,\Big[H_{\bold{j}_{1},\dots,\bold{j}_{m}}-w\Big]\,(\overline{\mathscr{P}_{\psi^{Bog}_{\bold{j}_1,\dots,%\bold{j}_{m-1}}}^{\#}})_{II}
%\end{equation}
we proceed with the identity
\begin{eqnarray}
& &(\ref{rel-in}) \label{H-terms-in-primero}\\
&=  &\overline{\mathscr{P}_{\psi^{\#}_{\bold{j}_1,\dots,\bold{j}_{m-1}}}}\,[(H^{\#}_{\bold{j}_{1},\dots,\bold{j}_{m-1}})_{\xi}-(1-\xi)T_{\bold{j}=\pm\bold{j}_m}-z^{\#}_{\bold{j}_1,\dots,\bold{j}_{m-1}}]\overline{\mathscr{P}_{\psi^{\#}_{\bold{j}_1,\dots,\bold{j}_{m-1}}}}\label{H-terms-fin-1}\\
& &+\overline{\mathscr{P}_{\psi^{\#}_{\bold{j}_1,\dots,\bold{j}_{m-1}}}}(\hat{H}^{Bog}_{\bold{j}_{m}})_{\xi}+\xi \,T+V_{\bold{j}_{m}}+V'_{\bold{j}_1,\dots \bold{j}_{m-1}}-z\Big\}\,\overline{\mathscr{P}_{\psi^{\#}_{\bold{j}_1,\dots,\bold{j}_{m-1}}}}\,.\quad\quad\quad\quad \label{H-terms-fin}
\end{eqnarray}
We recall that 
%\begin{equation}\label{def-V4-m}
%V^{(4)}_{\bold{j}_m}:=\frac{1}{N}\sum_{\bold{j}\in \mathbb{Z}^d\setminus\{- \bold{j}_m,\bold{0}\}}\,a^*_{\bold{j}+\bold{j}_m}a_{\bold{j}}\,\,\phi_{\bold{j}_m}\,\sum_{\bold{j}'\in \mathbb{Z}^d\setminus\{+ \bold{j}_m,\bold{0}\}}a^*_{\bold{j}'-\bold{j}_m}a_{\bold{j}'} \geq 0\,
%\end{equation}
%so that
\begin{eqnarray}
& &\frac{1}{N}\sum_{\bold{j}\in \mathbb{Z}^d\setminus\{ -\bold{j}_m,\bold{0}\}}\,\sum_{\bold{j}'\in \mathbb{Z}^d\setminus \{+\bold{j}_m,\bold{0}\}}a^*_{\bold{j}+\bold{j}_m}\,a^*_{\bold{j}'-\bold{j}_m}\,\phi_{\bold{j}_m}\,a_{\bold{j}}a_{\bold{j}'}\\
&=&V^{(4)}_{\bold{j}_m}-\frac{1}{N}\,\phi_{\bold{j}_m}\,\sum_{\bold{j}'\in \mathbb{Z}^d\setminus\{+ \bold{j}_m,\bold{0}\}}a^*_{\bold{j}'}a_{\bold{j}'} \,
\end{eqnarray}
where $V^{(4)}_{\bold{j}_m}\geq 0$ has been defined in (\ref{def-V4-m}). 

For $m=1$, we recall 
\begin{eqnarray}
& &H_{\bold{j}_1}-z\\
&=&(1-\xi)T-T_{\bold{j}=\pm \bold{j}_1}+V_{\bold{j}_1}+\hat{H}^{Bog}_{\bold{j}_1}+\xi \,T-z\\
&=&T_{\bold{j}\notin \{ \pm \bold{j}_1\}}+V_{\bold{j}_1}+\hat{H}^{Bog}_{\bold{j}_1}-z
%&=&(H_{\bold{j}_{1}})_{\xi}+V_{\bold{j}_{1}}+(\hat{H}^{Bog}_{\bold{j}_1})_{\xi}+\xi T-z\\
%&=& (1-\xi)T_{\bold{j}\neq \{\pm\bold{j}_{1}\}}+V_{\bold{j}_{1}}+(\hat{H}^{Bog}_{\bold{j}_1})_{\xi}+\xi T-z
\end{eqnarray}
and make use of the inequality
\begin{eqnarray}
& &\overline{\mathscr{P}_{\eta}^{\#}}\,(H_{\bold{j}_1}-z)\,\overline{\mathscr{P}_{\eta}^{\#}}\label{ineq-V4-m=1}\\
&\geq  &\overline{\mathscr{P}_{\eta}^{\#}}\,\Big[\sum_{\bold{j}\in \mathbb{Z}^d\setminus \{\bold{0}\}}\,((k_{\bold{j}})^2-\frac{\phi_{\bold{j}_1}}{N})\,a^*_{\bold{j}}a_{\bold{j}}+V^{(4)}_{\bold{j}_1}-z\Big]\overline{\mathscr{P}_{\eta}^{\#}}\,
%& &\langle (\overline{\mathscr{P}_{\psi^{Bog}_{\bold{j}_1,\dots,\bold{j}_{M-1}}}^{\#}})_{I}\psi\,,\,\charf^c_{\sigma}(\sum_{\bold{j}=\pm\bold{j}_{M}}(k_{\bold{j}})^2a^*_{\bold{j}}a_{\bold{j}}-z)\charf^c_{\sigma}\,(\overline{\mathscr{P}_{\psi^{Bog}_{\bold{j}_1,\dots,\bold{j}_{M-1}}}^{\#}})_{I}\psi \rangle \\
\end{eqnarray}
that holds because $\overline{\mathscr{P}_{\eta}^{\#}}$ projects onto a subspace with no particles in the modes $\pm \bold{j}_1$ so that $\overline{\mathscr{P}_{\eta}^{\#}}V^{(3)}_{\bold{j}_1}\overline{\mathscr{P}_{\eta}^{\#}}=\overline{\mathscr{P}_{\eta}^{\#}}\hat{H}^{Bog}_{\bold{j}_1}\overline{\mathscr{P}_{\eta}^{\#}}=0$.

\noindent
Thus, for $N$ large enough we can estimate
\begin{eqnarray}
& &\overline{\mathscr{P}_{\eta}^{\#}}\,(H_{\bold{j}_1}-z)\,\overline{\mathscr{P}_{\eta}^{\#}}
-\overline{\mathscr{P}_{\eta}^{\#}}\,\Gamma_{\bold{j}_1\,;\,N,N}(z)\,\overline{\mathscr{P}_{\eta}^{\#}}\\
&\geq &(\Delta^{\#}_0-\frac{\xi^{\frac{1}{2}}}{M}-z)\overline{\mathscr{P}_{\eta}^{\#}}-\overline{\mathscr{P}_{\eta}^{\#}}\,\Gamma_{\bold{j}_1\,;\,N,N}(z)\,\overline{\mathscr{P}_{\eta}^{\#}}\,.
\end{eqnarray}
%$\Delta_{0}:=\min\, \Big\{(k_{\bold{j}})^2\,|\,\bold{j}\in \mathbb{Z}^d \setminus \{\bold{0}\}\Big\}$.
From Lemma \ref{main-relations-H} there exists a constant $C^{\#\,\perp}$ such that
\begin{eqnarray}
& &\overline{\mathscr{P}_{\eta}^{\#}}\,\Gamma_{\bold{j}_1\,;\,N,N}(z)\,\overline{\mathscr{P}_{\eta}^{\#}}\\
%& \leq &\langle\Big(\overline{\mathscr{P}_{\eta}^{\#}}\Big)_I\psi\,,\,[\Gamma^{Bog}_{\bold{j}_1\,;\,\bar{i},\bar{i}}(z)]\,\Big(\overline{\mathscr{P}_{\eta}^{\#}}\Big)_I\psi\rangle\\
%& &+\mathcal{O}(...)\\
& \leq &\frac{\phi_{\bold{j}_{1}}}{2\epsilon_{\bold{j}_1}+2-\frac{z-\Delta^{\#}_0(1-\frac{\phi_{\bold{j}_1} \lfloor (\ln N)^{\frac{1}{2}} \rfloor }{N\Delta^{\#}_0})}{\phi_{\bold{j}_{1}}}}\check{G}_{\bold{j}_1\,;\,N-2,N-2}(z-\Delta^{\#}_0(1-\frac{\phi_{\bold{j}_1} \lfloor (\ln N)^{\frac{1}{2}} \rfloor }{N\Delta^{\#}_0}))\overline{\mathscr{P}_{\eta}^{\#}}\\
& &+\frac{C^{\#\,\perp}}{(\ln N)^{\frac{1}{4}}}\overline{\mathscr{P}_{\eta}^{\#}}
%& &+\mathcal{O}({\color{red}ESTIMATE})
\end{eqnarray}
where $\check{G}_{\bold{j}_1\,;\,N-2,N-2}(z)$ has been defined for a three modes system by recursion 
starting from 
\begin{equation}\label{def-G}
\check{\mathcal{G}}_{\bold{j}_1\,;\,i,i}(z):=\sum_{l_{i}=0}^{\infty}[\mathcal{W}_{\bold{j}_1\,;i,i-2}(z)\mathcal{W}^*_{\bold{j}_1\,;i-2,i}(z)\check{\mathcal{G}}_{\bold{j}_1\,;\,i-2,i-2}(z)]^{l_i}\quad,\quad \check{\mathcal{G}}_{\bold{j}_1\,;\,0,0}(z)\equiv 1\,.
\end{equation}
with
\begin{eqnarray}
& &\mathcal{W}_{\bold{j}_1\,;\,i,i-2}(z)\mathcal{W}^*_{\bold{j}_1\,;\,i-2,i}(z)\\
&:= &\frac{(n_{\bold{j}_0}-1)n_{\bold{j}_0}}{N^2}\,\phi^2_{\bold{j}_{1}}\,\frac{ (n_{\bold{j}_{1}}+1)(n_{-\bold{j}_{1}}+1)}{\Big[(\frac{n_{\bold{j}_0}}{N}\phi_{\bold{j}_{1}}+k_{\bold{j}_{1}}^2)(n_{\bold{j}_{1}}+n_{-\bold{j}_{1}})-z\Big]}\label{def-Wcal-1}\\
& &\quad\quad\quad \times\frac{1}{\Big[(\frac{(n_{\bold{j}_0}-2)}{N}\phi_{\bold{j}_{1}}+k_{\bold{j}_{1}}^2)(n_{\bold{j}_{1}}+n_{-\bold{j}_{1}})+2(\frac{(n_{\bold{j}_0}-2)}{N}\phi_{\bold{j}_{1}}+k_{\bold{j}_{1}}^2)-z\Big]}
%&:=& \frac{(n_{\bold{j}_0}+2)(n_{\bold{j}_0}+1)}{N^2}\,\phi^2_{\bold{j}_{*}}\,\frac{ (n_{\bold{j}_{*}}+1)(n_{-\bold{j}_{*}}+1)}{\Big[(\frac{n_{\bold{j}_0}}{N}\phi_{\bold{j}_{*}}+(k_{\bold{j}_{*}}^2))(n_{\bold{j}_{*}}+n_{-\bold{j}_{*}})-z\Big]}\times\\
%& &\times \frac{1}{\Big[(\frac{(n_{\bold{j}_0}+2)}{N}\phi_{\bold{j}_{*}}+(k_{\bold{j}_{*}}^2))(n_{\bold{j}_{*}}+n_{-\bold{j}_{*}})-2(\frac{(n_{\bold{j}_0}+2)}{N}\phi_{\bold{j}_{*}}+(k_{\bold{j}_{*}}^2))-z\Big]}\quad\quad\quad
\end{eqnarray} 
where 
\begin{equation}
n_{\bold{j}_{1}}+n_{-\bold{j}_{1}}=N-i\quad \text{with}\,\,i\,\,\text{even}\quad;\quad n_{\bold{j}_{1}}=n_{-\bold{j}_{1}}\quad;\quad n_{\bold{j}_0}=n_{\bold{0}}=i\,.\label{def-Wcal-2}
\end{equation}
$\check{\mathcal{G}}_{\bold{j}_1\,;\,N-2,N-2}(z)$  enters the fixed point equation for the Bogoliubov Hamiltonian associated with a three-modes system:
\begin{equation}\label{fp-equation-2}
z=-\frac{\phi_{\bold{j}_{1}}}{2\epsilon_{\bold{j}_1}+2-\frac{z}{\phi_{\bold{j}_{1}}}}\check{\mathcal{G}}_{\bold{j}_1\,;\,N-2,N-2}(z)\,.
\end{equation}
The additional inputs
\begin{enumerate}
\item 
$\check{G}_{\bold{j}_1\,;\,N-2,N-2}(z)$ is nondecreasing in $z$ (see \emph{\underline{Remark 4.1}} in  \cite{Pi1});
% thus (for $N$ large)  $$\check{G}_{\bold{j}_1\,;\,N-2,N-2}(z)\geq \check{G}_{\bold{j}_1\,;\,N-2,N-2}(z-\Delta_0(1-\frac{\phi_{\bold{j}_m}}{N\Delta_0}));$$
\item
the existence of the fixed point $z_1$ of (\ref{fp-equation-2}) with $|z_1-E^{Bog}_{\bold{j}_1}|=\mathcal{O}(\frac{1}{N^{\beta}})$ for any $0<\beta<1$
%+  (\frac{2\sqrt{2}+3}{6})\sqrt{\epsilon_{\bold{j}_1}} \phi_{\bold{j}^*}\sqrt{\epsilon_{\bold{j}_1}^2+2\epsilon_{\bold{j}_1}}$
 (see \emph{\underline{Lemma 5.5} of \cite{Pi1}});
\end{enumerate}
imply that for $N$ large enough and for
\begin{equation}
z\leq \min\,\Big\{ z_{1}+\gamma\Delta^{\#}_{0}-\frac{\xi^{\frac{1}{2}}}{M}-\frac{C^{\#\,\perp}}{(\ln N)^{\frac{1}{4}}}\,;\,E^{Bog}_{\bold{j}_1}+ \sqrt{\epsilon_{\bold{j}_1}}\phi_{\bold{j}_1}\sqrt{\epsilon_{\bold{j}_m}^2+2\epsilon_{\bold{j}_1}}\Big\}\label{bound-z-0-first}
\end{equation}
we can estimate
\begin{equation}
\check{G}_{\bold{j}_1\,;\,N-2,N-2}(z-\Delta^{\#}_0(1-\frac{\phi_{\bold{j}_1} \lfloor (\ln N)^{\frac{1}{2}} \rfloor }{N\Delta_0}))\leq \check{G}_{\bold{j}_1\,;\,N-2,N-2}(z_1)
\end{equation}
and, consequently,
\begin{eqnarray}
& &\overline{\mathscr{P}_{\eta}^{\#}}\,(H_{\bold{j}_1}-z)\,\overline{\mathscr{P}_{\eta}^{\#}}-\overline{\mathscr{P}_{\eta}^{\#}}\,\Gamma_{\bold{j}_1\,;\,N,N}(z)\,\overline{\mathscr{P}_{\eta}^{\#}} \label{first-bound-in}\\
&\geq &(\Delta^{\#}_0-z-\frac{\xi^{\frac{1}{2}}}{M})\overline{\mathscr{P}_{\eta}^{\#}}-\frac{\phi_{\bold{j}_{1}}}{2\epsilon_{\bold{j}_1}+2-\frac{z_1}{\phi_{\bold{j}_{1}}}}\check{G}_{\bold{j}_1\,;\,N-2,N-2}(z_1)\overline{\mathscr{P}_{\eta}^{\#}}\\
& &-\frac{C^{\#\,\perp}}{(\ln N)^{\frac{1}{4}}}\overline{\mathscr{P}_{\eta}^{\#}}\\
&= &(\Delta^{\#}_0-z-\frac{\xi^{\frac{1}{2}}}{M}+z_1-\frac{C^{\#\,\perp}}{(\ln N)^{\frac{1}{4}}})\overline{\mathscr{P}_{\eta}^{\#}}\\
&\geq&(1-\gamma)\Delta^{\#}_0\overline{\mathscr{P}_{\eta}^{\#}}\,.\label{first-bound-fin}
\end{eqnarray}

For $2\leq m \leq M$, we recall that $\overline{\mathscr{P}_{\psi^{\#}_{\bold{j}_1,\dots,\bold{j}_{m-1}}}}\mathcal{F}^N\subset (\mathcal{F}^N\ominus \mathcal{F}^N_{\pm\bold{j}_m}) $ by definition. Hence, due to the assumption in (\ref{assumption-gap}) the inequality 
\begin{eqnarray}
& &\overline{\mathscr{P}_{\psi^{\#}_{\bold{j}_1,\dots,\bold{j}_{m-1}}}}\,\Big[(H^{\#}_{\bold{j}_{1},\dots,\bold{j}_{m-1}})_{\xi}-(1-\xi)T_{\bold{j}=\pm\bold{j}_m}-z^{\#}_{\bold{j}_1,\dots,\bold{j}_{m-1}}\Big]\,\overline{\mathscr{P}_{\psi^{\#}_{\bold{j}_1,\dots,\bold{j}_{m-1}}}}\\
&=&\overline{\mathscr{P}_{\psi^{\#}_{\bold{j}_1,\dots,\bold{j}_{m-1}}}}\,\Big[(H^{\#}_{\bold{j}_{1},\dots,\bold{j}_{m-1}})_{\xi}-z^{\#}_{\bold{j}_1,\dots,\bold{j}_{m-1}}\Big]\,\overline{\mathscr{P}_{\psi^{\#}_{\bold{j}_1,\dots,\bold{j}_{m-1}}}}\\
&\geq &(\Delta^{\#}_{m-1}-\frac{(m-1)\xi^{\frac{1}{2}}}{M})\overline{\mathscr{P}_{\psi^{\#}_{\bold{j}_1,\dots,\bold{j}_{m-1}}}}
\end{eqnarray} can be proven with an argument similar to Lemma \ref{xi-cor}. Using this ingredient,
\begin{eqnarray}
& &\overline{\mathscr{P}_{\psi^{\#}_{\bold{j}_1,\dots,\bold{j}_{m-1}}}}\,\Big[(H^{\#}_{\bold{j}_{1},\dots,\bold{j}_{m-1}})_{\xi}-(1-\xi)T_{\bold{j}=\pm\bold{j}_m}-z^{\#}_{\bold{j}_1,\dots,\bold{j}_{m-1}}\Big]\overline{\mathscr{P}_{\psi^{\#}_{\bold{j}_1,\dots,\bold{j}_{m-1}}}}\\
& &+\overline{\mathscr{P}_{\psi^{\#}_{\bold{j}_1,\dots,\bold{j}_{m-1}}}}\,\Big[(\hat{H}^{Bog}_{\bold{j}_m})_{\xi}+\xi T+V_{\bold{j}_{m}}+V'_{\bold{j}_1,\dots \bold{j}_{m-1}}-z\Big]\,\overline{\mathscr{P}_{\psi^{\#}_{\bold{j}_1,\dots,\bold{j}_{m-1}}}}\quad\quad\quad\label{PIineq}\\
%& &-\Big(\overline{\mathscr{P}_{\psi^{Bog}_{\bold{j}_1,\dots,\bold{j}_{m-1}}}^{\#}}\Big)_I\,[\Gamma_{\bold{j}_1,\dots,\bold{j}_m\,,;\,\bar{i},\bar{i}}(w)]\Big(\overline{\mathscr{P}_{\psi^{Bog}_{\bold{j}_1,\dots,\bold{j}_{m-1}}}^{\#}}\Big)_I\\
&\geq &(\Delta^{\#}_{m-1}-\frac{(m-1)\xi^{\frac{1}{2}}}{M})\overline{\mathscr{P}_{\psi^{\#}_{\bold{j}_1,\dots,\bold{j}_{m-1}}}}\\
& &+\overline{\mathscr{P}_{\psi^{\#}_{\bold{j}_1,\dots,\bold{j}_{m-1}}}}\,(\sum_{\bold{j}\in \mathbb{Z}^d\setminus \{\bold{0}\}}\,[ \xi(k_{\bold{j}})^2-\frac{\phi_{\bold{j}_m}}{N}]\,a^*_{\bold{j}}a_{\bold{j}}+V^{(4)}_{\bold{j}_m}-z)\,\overline{\mathscr{P}_{\psi^{\#}_{\bold{j}_1,\dots,\bold{j}_{m-1}}}}\,\quad\quad
\end{eqnarray}
where we have exploited that $V'_{\bold{j}_1,\dots,\bold{j}_{m-1}}, V^{(3)}_{\bold{j}_m}$, and $(\hat{H}^{Bog}_{\bold{j}_m})_{\xi}$  are normal ordered and contain at least one operator $a_{\pm\bold{j}_m},a^{*}_{\pm\bold{j}_m}$ by definition.
Then, for $z$ as in (\ref{bound-z-0-bis-bis}) and $\frac{1}{N}$ small enough
\begin{eqnarray}
& &  (\ref{rel-in}) + (\ref{nonH})\label{normal-esti-m-in}\\
%& &\langle (\overline{\mathscr{P}_{\psi^{Bog}_{\bold{j}_1,\dots,\bold{j}_{M-1}}}^{\#}})_{I}\psi\,,\,\charf^c_{\sigma}(\sum_{\bold{j}=\pm\bold{j}_{M}}(k_{\bold{j}})^2a^*_{\bold{j}}a_{\bold{j}}-z)\charf^c_{\sigma}\,(\overline{\mathscr{P}_{\psi^{Bog}_{\bold{j}_1,\dots,\bold{j}_{M-1}}}^{\#}})_{I}\psi \rangle \\
&\geq &(\Delta^{\#}_{m-1}-z-\frac{(m-1)\xi^{\frac{1}{2}}}{M} )\overline{\mathscr{P}_{\psi^{\#}_{\bold{j}_1,\dots,\bold{j}_{m-1}}}} \label{invert-PI-1}\\
& &-\overline{\mathscr{P}_{\psi^{\#}_{\bold{j}_1,\dots,\bold{j}_{m-1}}}}\,\Gamma_{\bold{j}_1,\dots,\bold{j}_m\,,;\,N,N}(w)\,\Big(\overline{\mathscr{P}_{\psi^{\#}_{\bold{j}_1,\dots,\bold{j}_{m-1}}}} \label{invert-PI-2}\\
%&\geq  &{\color{red}(1-\gamma)\Delta_{m-1}}-\frac{C_m^{\#\,\perp}}{(\ln N)^{\frac{1}{4}}}\label{invert-PI-3}\\
&\geq&(\Delta^{\#}_{m-1}-z)\overline{\mathscr{P}_{\psi^{\#}_{\bold{j}_1,\dots,\bold{j}_{m-1}}}}+(z_m-\frac{C^{\#\,\perp}}{(\ln N)^{\frac{1}{4}}}-\frac{m\xi^{\frac{1}{2}}}{M})\overline{\mathscr{P}_{\psi^{\#}_{\bold{j}_1,\dots,\bold{j}_{m-1}}}}\label{-2}\\
&\geq &(1-\gamma)\Delta^{\#}_{m-1}\overline{\mathscr{P}_{\psi^{\#}_{\bold{j}_1,\dots,\bold{j}_{m-1}}}}\label{normal-esti-m-fin}
\end{eqnarray}
where for the step from (\ref{invert-PI-1})-(\ref{invert-PI-2}) to (\ref{-2}) we invoke (\ref{estimate-gammaperp-0}) in Lemma \ref{main-relations-H} and repeat the argument already used for the analogous quantity in the case $m=1$.
\\

%\noindent
%\emph{Conclusion}
%
%Making use of the previous estimates  we derive  (for $N$ sufficiently large)
%\begin{equation}
%\overline{\mathscr{P}_{\psi^{\#}_{\bold{j}_1,\dots,\bold{j}_{m-1}}}}\mathscr{K}_{\bold{j}_1,\dots,\bold{j}_m}^{\,(N-2)}(z+z_{\bold{j}_1,\dots,\bold{j}_{m-1}})\overline{\mathscr{P}_{\psi^{\#}_%{\bold{j}_1,\dots,\bold{j}_{m-1}}}}\geq (1-\gamma)\Delta^{\#}_{m}\overline{\mathscr{P}_{\psi^{\#}_{\bold{j}_1,\dots,\bold{j}_{m-1}}}}
%\end{equation}
%which concludes the proof.
 %Assuming the constraint in (\ref{constraint-xi}) and considering that by construction
%$$\|(\mathcal{N}_{+})^{\frac{1}{2}}(\overline{\mathscr{P}_{\psi^{Bog}_{\bold{j}_1,\dots,\bold{j}_{m-1}}}^{\#}})_{II}\,\psi\|^2\geq \lfloor N^{\frac{1}{16}} \rfloor \|(\overline{\mathscr{P}_{\psi^{Bog}_{\bold{j}_1,\dots,\bold{j}_{m-1}}}^{\#}})_{II}\,\psi\|^2$$
% the inequality in (\ref{inequa-m}) is proven for $N$ sufficiently large.
  \qed

\subsection{Proof by induction in the index $m$}\label{last-Fesh}

%We  assume to have constructed the ground state $\psi_{\bold{j}_1,\dots,\bold{j}_{m-1}}$ of the Hamiltonian $H_{\bold{j}_1,\dots,\bold{j}_{m-1}}$ corresponding to the ground state energy $z_{\bold{j}_1,\dots,\bold{j}_{m-1}}$, and show how to construct the ground state of $H_{\bold{j}_1,\dots,\bold{j}_{m}}$. 

We have now all the tools to state the main result of this section contained in Theorem \ref{induction-many-modes}. This theorem concerns five properties proven by induction. For the convenience of the reader we outline the structure of the proof. 

\noindent
Property 1. ensures the construction of the Feshbach flow $\{\mathscr{K}^{(i)}_{\bold{j}_1,\,\dots, \bold{j}_{m}}(z+z^{\#}_{\bold{j}_1,\dots,\bold{j}_{m-1}})\,|\,\bar{i}\leq  i\,\, \text{and even}\}$ up to the last step ($i=N$). 

\noindent
Property 2. provides the existence of the unique solution of the fixed point equation associated with the Feshbach Hamiltonian $\mathscr{K}^{(N)}_{\bold{j}_1,\,\dots, \bold{j}_{m}}(z+z^{\#}_{\bold{j}_1,\dots,\bold{j}_{m-1}})$  (defined with Property 1.) and the construction of its ground state.

\noindent
Property 3. is concerned with Property 1. and 2. but for the auxiliary Hamiltonian $H^{\#}_{\bold{j}_1,\dots,\bold{j}_{m}}$. In addition to the construction of the ground state vector,  Property 3. provides the gap condition at step $m$ that must be used to get Property 1., 2., and 3. at step $m+1$. 

\noindent
Property 4. provides the information on $$\text{infspec}[H^{\#}_{\bold{j}_1,\dots, \bold{j}_{m}}-T_{\bold{j}\in \{\pm \bold{j}_{m+1}\}}]$$ that is used at step $m+1$ in Corollary \ref{main-lemma-H}. Thanks to this input, the operator norm estimate (\ref{estimate-main-lemma-H}) in Corollary \ref{main-lemma-H} can be derived as if the modes $\bold{j}_1,\dots,\bold{j}_{m-1}$ were absent.

\noindent
Property 5. provides a bound on the expectation value of $\mathcal{N}^2_+$ in the ground state of $H^{\#}_{\bold{j}_1,\dots,\bold{j}_{m}}$. This information is needed to control the fixed point equation at step $m+1$ both for $H_{\bold{j}_1,\dots,\bold{j}_{m+1}}$ and for $H^{\#}_{\bold{j}_1,\dots,\bold{j}_{m+1}}$.

\begin{thm}\label{induction-many-modes}
Let $max_{1\leq m \leq M}\epsilon_{\bold{j}_m}$ be sufficiently small and $N$ sufficiently large. 
%Set $\Delta_0\equiv \min\, \Big\{\epsilon_{\bold{j}}\,|\,\bold{j}\in \mathbb{Z}^3 \Big\}\,$.  
Then the following properties hold true for all $1\leq m \leq M$:
\begin{enumerate}
\item There exists a constant $C^{\#\,\perp}$ such that the Feshbach Hamiltonian in (\ref{K-last-step})-(\ref{K-last-step-bis}) is well defined for \begin{equation}
 z\leq\min\,\Big\{ z_{m}+\gamma\Delta^{\#}_{m-1}-\frac{m\xi^{\frac{1}{2}}}{M}-\frac{C^{\#\,\perp}}{(\ln N)^{\frac{1}{4}}}\,;\,E^{Bog}_{\bold{j}_m}+ \sqrt{\epsilon_{\bold{j}_m}}\phi_{\bold{j}_m}\sqrt{\epsilon_{\bold{j}_m}^2+2\epsilon_{\bold{j}_m}}\Big\}\label{inter-def-finalth}
 \end{equation}
 where:
 \begin{itemize}
 \item
   $z_m$ is the ground state energy of $H^{Bog}_{\bold{j}_m}$ (see \underline{Corollary 4.6}  and \underline{Theorem 4.1} of \cite{Pi1});
 \item
 $\gamma = \frac{1}{2}$;
 \item  $\Delta^{\#}_{m-1}$ (for $m\geq 1$)  is defined iteratively from $\Delta^{\#}_{0}\equiv \Delta_0:=\min\, \Big\{(k_{\bold{j}})^2\,|\,\bold{j}\in \mathbb{Z}^d \setminus \{\bold{0}\}\Big\}$ and for $N$ large enough
  \begin{eqnarray}
\Delta^{\#}_m &:= &\min\,\Big\{ \gamma\Delta^{\#}_{m-1}-\frac{C^{\#\,\perp}}{(\ln N)^{\frac{1}{4}}}\,;\,\frac{1}{2}\sqrt{\epsilon_{\bold{j}_m}}\phi_{\bold{j}_m}\sqrt{\epsilon_{\bold{j}_m}^2+2\epsilon_{\bold{j}_m}}\Big\}-\frac{2m\xi^{\frac{1}{2}}}{M} \label{Deltagap-bis}\\
&= &\gamma\Delta^{\#}_{m-1}-\frac{C^{\#\,\perp}}{(\ln N)^{\frac{1}{4}}}-\frac{2m\xi^{\frac{1}{2}}}{M}
 \end{eqnarray}
where  $\xi=\frac{1}{(\ln N)^{\frac{1}{4}}}$.
 \end{itemize}
 \item
For $z$ as in (\ref{inter-def-finalth}), there exists a unique value $z^{\$\,(m)}$ such that (see (\ref{f-p-e})) $$f_{\bold{j}_1,\dots,\bold{j}_m}(z+z^{\#}_{\bold{j}_1,\dots,\bold{j}_{m-1}})|_{z=z^{\$\,(m)}}=0.$$ The inequality \begin{equation}\label{dist-zdollar}
|z^{\$\,(m)}-z_m|\leq (\frac{2}{\gamma})^m\frac{C^{\#}_{III}}{(\ln N)^{\frac{1}{4}}}+\frac{\tilde{C}^{\#}}{N}
\end{equation}
 holds true  with $C^{\#}_{III}:=C^{\#}_{I}+\frac{(C^{\#}_{II})^2}{(1-\gamma)\Delta_{0}}$, where $\tilde{C}^{\#}$ is defined in point 5. below, $C^{\#}_{I}$ and $C^{\#}_{II}$ are defined in Lemma \ref{main-relations-H}.
 
 \noindent
 The Hamiltonian $H_{\bold{j}_1,\dots,\bold{j}_{m}}$ has (non-degenerate) ground state energy $z_{\bold{j}_1,\dots,\bold{j}_{m}}:=z^{\#}_{\bold{j}_1,\dots,\bold{j}_{m-1}}+z^{\$\,(m)}$ where $z^{\#}_{\bold{j}_1,\dots,\bold{j}_{m-1}}|_{m=1}\equiv 0$. The corresponding eigenvector, $\psi_{\bold{j}_1,\dots,\bold{j}_m}$, is given in (\ref{gs-Hm-start-bis})-(\ref{gs-Hm-fin-bis}).
 
%\item 
% The spectral gap of the two operators $H_{\bold{j}_1,\dots,\bold{j}_{m}}$ above the ground state energy $z_{\bold{j}_1,\dots,\bold{j}_{m}}\equiv z_{\bold{j}_1,\dots,\bold{j}_{m-1}}+z^{(m)}$  is larger or equal to $\Delta_{m}$.

%\item The ground state $\psi^{Bog}_{\bold{j}_1,\bold{j}_2,\dots,\bold{j}_{m}}$ fulfills the expansion:
%\begin{equation}
%\psi^{Bog}_{\bold{j}_1,\bold{j}_2,\dots,\bold{j}_{m}}=\sum_{l=2}^{m} T_{m}\dots T_{l+1}S_{l}\,\psi^{Bog}_{\bold{j}_1,\bold{j}_2,\dots,\bold{j}_{l-1}}+T_{m}\dots T_0\eta
%\end{equation}
%\begin{equation}
%S_n:=\quad \|S_n\|\leq \frac{C^n}{N}
%\end{equation}
\item 
(a)  The operator $H^{\#}_{\bold{j}_1, \dots,\bold{j}_{m}}$ has (non-degenerate) ground state energy $z^{\#}_{\bold{j}_1,\dots,\bold{j}_{m}}$ determined via a fixed point equation analogous to (\ref{f-p-e}) and ground state vector $\psi^{\#}_{\bold{j}_1,\dots,\bold{j}_{m}}$ given by the formula in (\ref{gs-Hm-start-bis-diesis})-(\ref{gs-Hm-fin-bis-diesis}).

%\item The ground state $\psi^{Bog}_{\bold{j}_1,\bold{j}_2,\dots,\bold{j}_{m}}$ fulfills the expansion:
%\begin{equation}
%\psi^{Bog}_{\bold{j}_1,\bold{j}_2,\dots,\bold{j}_{m}}=\sum_{l=2}^{m} T_{m}\dots T_{l+1}S_{l}\,\psi^{Bog}_{\bold{j}_1,\bold{j}_2,\dots,\bold{j}_{l-1}}+T_{m}\dots T_0\eta
%\end{equation}
%\begin{equation}
%S_n:=\quad \|S_n\|\leq \frac{C^n}{N}
%\end{equation}
%
%\noindent
%(b) The inequality
%\begin{equation}
%|z_{\bold{j}_1,\dots,\bold{j}_{m}}-z^{\#}_{\bold{j}_1,\dots,\bold{j}_{m}}|\leq \frac{m\xi^{\frac{1}{2}}}{3M}
%\end{equation}
%holds true.

\noindent
(b) The operator $H^{\#}_{\bold{j}_1, \dots,\bold{j}_{m}}-T_{\bold{j}=\pm \bold{j}_{m+1}}\upharpoonright_{(\mathcal{F}^N\ominus \mathcal{F}^N_{\pm\bold{j}_{m+1}})}$ has ground state vector $\psi^{\#}_{\bold{j}_1,\dots,\bold{j}_m}$ and ground state energy $z^{\#}_{\bold{j}_1,\dots,\bold{j}_m}$, and the gap estimate
\begin{equation}
\text{infspec}[(H^{\#}_{\bold{j}_1, \dots,\bold{j}_{m}}-T_{\bold{j}=\pm \bold{j}_{m+1}})\upharpoonright_{(\mathcal{F}^N\ominus \mathcal{F}^N_{\pm\bold{j}_{m+1}})\ominus \{\mathbb{C}\psi^{\#}_{\bold{j}_1,\dots,\bold{j}_m}\}}]-z^{\#}_{\bold{j}_1,\dots,\bold{j}_{m}}\geq \Delta^{\#}_m\,
\end{equation}
holds true.
\item
The bound from below
\begin{equation}
\text{infspec}[H^{\#\,(l)}_{\bold{j}_1,\dots, \bold{j}_{m}}-T_{\bold{j}\in \{\pm \bold{j}_{m+1},\dots,\pm \bold{j}_{m+l}\}}]\geq z^{\#}_{\bold{j}_1,\dots, \bold{j}_{m}}-\frac{m\xi^{\frac{1}{2}}}{2M}\quad,\quad 1\leq l \leq M-m,
%\text{infspec}[\hat{H}^{\#}_{\bold{j}_1, \dots,\bold{j}_{m}}]\geq z^{\#}_{\bold{j}_1,\dots,\bold{j}_{m}}- ...
\end{equation}
holds true where 
\begin{equation}
H^{\#\,(l)}_{\bold{j}_1,\dots,\bold{j}_{m}}:=T_{\bold{j}\notin\{\pm\bold{j}_{1},\dots, \pm\bold{j}_{m}\}}+\hat{H}^{Bog}_{\bold{j}_1,\dots,\bold{j}_{m}}+V^{\#\,(l)}_{\bold{j}_1,\dots,\bold{j}_{m}}
\end{equation}
with 
\begin{itemize}
\item
$V^{\#\,(l)}_{\bold{j}_1,\dots,\bold{j}_{m}}$ corresponding to $V_{\bold{j}_1,\dots,\bold{j}_{m}}$ (see (\ref{pair-1})-(\ref{pair-2})) minus all the summands containing at least one of the operators $\{a_{\pm \bold{j}_{m+l'}},a^*_{\pm \bold{j}_{m+l'}}\,,\,l'=1,\dots,l\}$; consequently, $V^{\#\,(1)}_{\bold{j}_1,\dots,\bold{j}_{m}}\equiv V^{\#}_{\bold{j}_1,\dots,\bold{j}_{m}}$ and $H^{\#\,(1)}_{\bold{j}_1,\dots, \bold{j}_{m}}\equiv H^{\#}_{\bold{j}_1,\dots, \bold{j}_{m}}$;
\item $T_{\bold{j}\in\{\pm\bold{j}_{m+1},\dots, \pm\bold{j}_{m+l}\}}:=\sum_{\bold{j}\in \{\pm\bold{j}_{m+1};\dots; \pm\bold{j}_{m+l}\}}k_{\bold{j}}^2a^*_{\bold{j}}a_{\bold{j}}$.
\end{itemize}
 \item The upper bound
 \begin{equation}
\langle \frac{\psi^{\#}_{\bold{j}_1,\dots,\bold{j}_{m}}}{\|\psi^{\#}_{\bold{j}_1,\dots,\bold{j}_{m}}\|}\,,\,(\sum_{\bold{j}\in\mathbb{Z}^d\setminus \{\bold{0}\}} a_{\bold{j}}^{*}a_{\bold{j}})^2\,\frac{\psi^{\#}_{\bold{j}_1,\dots,\bold{j}_{m}}}{\|\psi^{\#}_{\bold{j}_1,\dots,\bold{j}_{m}}\|}\rangle \leq \mathcal{O}(1)\,
\end{equation}
holds true. This implies that for some  $\tilde{C}^{\#}<\infty$
\begin{equation}\|\mathscr{P}_{\psi^{\#}_{\bold{j}_1,\dots,\bold{j}_{m}}}V_{\bold{j}_{m+1}}\mathscr{P}_{\psi^{\#}_{\bold{j}_1,\dots,\bold{j}_{m}}}\|\leq \frac{\tilde{C}^{\#}}{N}\,. \end{equation}
% \item There exists $(1<)\tilde{C}_m<\infty$ such that  
% \begin{equation}
%\langle \frac{\psi^{Bog}_{\bold{j}_1,\dots,\bold{j}_{m}}}{\|\psi^{Bog}_{\bold{j}_1,\dots,\bold{j}_{m}}\|}\,,\,\sum_{\bold{j}\in\mathbb{Z}^3\setminus \{\bold{0}\}} a_{\bold{j}}^{*}a_{\bold{j}}\,\frac{\psi^{Bog}_{\bold{j}_1,\dots,\bold{j}_{m}}}{\|\psi^{Bog}_{\bold{j}_1,\dots,\bold{j}_{m}}\|}\rangle \leq \tilde{C}_m
%\end{equation}
 \end{enumerate}
%where $\tilde{C}_m=2\frac{\sum_{l=1}^{m}\phi_{\bold{j}_l}}{\Delta_0}$.
\end{thm}

\noindent
\emph{Proof}
\\

\noindent
\emph{Case $m=1$}

We observe that the  Feshbach Hamiltonian $\mathscr{K}^{(\bar{i})}_{\bold{j}_1,\,\dots, \bold{j}_{m}}(w)$, with $m=1$, in (\ref{first-fesh-ham-0})-(\ref{first-fesh-ham-1}) is well defined because of the estimate in (\ref{estimate-control-quad-2}) of Lemma \ref{control-quad} that ensures the invertibility of $\mathfrak{Q}^{(\bar{i},\bar{i}+1)}_{\bold{j}_1}(H_{\bold{j}_1}-w)\mathfrak{Q}^{(\bar{i},\bar{i}+1)}_{\bold{j}_1}$ in $\mathfrak{Q}^{(\bar{i},\bar{i}+1)}_{\bold{j}_1}\mathcal{F}^N$. Then, Property 1. follows from Corollary \ref{main-lemma-H}, Theorem \ref{theorem-junction}, Theorem \ref{Feshbach-H}, and Proposition \ref{invertibility-bis} (through Lemma \ref{main-relations-H}). Notice that assumptions 1. and 2. in Proposition \ref{invertibility-bis} are satisfied for $\psi^{\#}_{\bold{j}_1,\dots,\bold{j}_{m-1}}|_{m=1}\equiv \eta$, $z^{\#}_{\bold{j}_1,\dots,\bold{j}_{m-1}}|_{m=1}\equiv 0$ and $\Delta^{\#}_0=\min\, \Big\{(k_{\bold{j}})^2\,|\,\bold{j}\in \mathbb{Z}^d \setminus \{\bold{0}\}\Big\}$.

As far as Property 2. is concerned, at first we point out that
\begin{equation}
\mathscr{P}^{\#}_{\eta}H_{\bold{j}_1} \mathscr{P}^{\#}_{\eta}=0\,.
\end{equation}
Similarly to the fixed point problem for the intermediate Bogoliubov Hamiltonians $H^{Bog}_{\bold{j}_1,\dots,\bold{j}_m}$ with $m\geq 2$ (see \emph{\underline{Theorem 4.3} of \cite{Pi2}}), the term in (\ref{term-not-important}) is not zero but vanishes as $N\to \infty$. More precisely,
\begin{itemize}
\item the identity
\begin{equation}
\mathscr{P}^{\#}_{\eta}V_{\bold{j}_1} \overline{\mathscr{P}^{\#}_{\eta}}=0 
\end{equation}
holds true because the state $\eta$ contains only particles in the zero mode, and $V_{\bold{j}_1}$ is normal ordered and contains only ``cubic'' and ``quartic'' terms in the nonzero modes;
% and  the subspace $(\overline{\mathscr{P}^{\#}_{\eta}})_{I}\mathcal{F}^N$ does not contain particles in the modes $\pm \bold{j}_1$, and for a state in $(\overline{\mathscr{P}^{\#}_{\eta}})_{II}\mathcal{F}^N$ the number of particles in the modes $\pm \bold{j}_1$ is larger than $\lfloor N^{\frac{1}{8}} \rfloor-1$.
%; 2) the bound 
%\begin{equation}
%\langle \frac{\psi^{Bog}_{\bold{j}_1,\dots,\bold{j}_{m}}}{\|\psi^{Bog}_{\bold{j}_1,\dots,\bold{j}_{m}}\|}\,,\,\sum_{\bold{j}\in\mathbb{Z}^3\setminus \{\bold{0}\}} a_{\bold{j}}^{*}a_{\bold{j}}\,\frac{\psi^{Bog}_{\bold{j}_1,\dots,\bold{j}_{m}}}{\|\psi^{Bog}_{\bold{j}_1,\dots,\bold{j}_{m}}\|}\rangle \leq \tilde{C}_m
%\end{equation}
%holds true  where $\tilde{C}_m=2\frac{\sum_{l=1}^{m}\phi_{\bold{j}_l}}{\Delta_0}$ (see \emph{\underline{Theorem 4.3} of \cite{Pi2}});
\item
the estimate 
\begin{eqnarray}
& &\|\mathscr{P}^{\#}_{\eta}\Gamma_{\bold{j}_1;\,N,N}(z)\,\overline{\mathscr{P}^{\#}_{\eta}}\|\leq \frac{C^{\#}_{II}}{(\ln N)^{\frac{1}{4}}}\,\label{primero-3}
\end{eqnarray}
%\begin{eqnarray}
%& &\Big\|\Big(\overline{\mathscr{P}_{\psi^{Bog}_{\bold{j}_1,\dots,\bold{j}_{m-1}}}^{\#}}\Big)_I\psi\,,\,[\Gamma_{\bold{j}_1,\dots,\bold{j}_m\,,;\,\bar{i},\bar{i}}(w)]\,\Big(\overline{\mathscr{P}_{\psi^{Bog}_{\bold{j}_1,\dots,\bold{j}_{m-1}}}^{\#}}\Big)_I\Big\|\label{estimate-gammaperp-0}\\
%&\leq& \frac{\phi_{\bold{j}_{m}}}{\Delta_{m-1}+2\epsilon_{\bold{j}_m}+2-\frac{z}{\phi_{\bold{j}_{m}}}}\check{\mathcal{G}}_{\bold{j}_m\,;\,N-2,N-2}(z-\Delta_{m-1})+\mathcal{O}({\color{red}ESTIMATE2})\nonumber
%\end{eqnarray}
is derived in Lemma \ref{main-relations-H}  by means of a procedure already employed for the intermediate Bogoliubov Hamiltonians (see {\emph{\underline{Lemma 4.4} of \cite{Pi2} }}).
\end{itemize}

Finally, we solve the fixed point equation
\begin{eqnarray}\label{fp-1}
%0&=&-z-\langle \eta \,,\,\Gamma_{\bold{j}_1\,;\,N,N}(z)\,\eta \rangle \\
%& &-\langle \eta \,,\,(V_{\bold{j}_1}-\Gamma_{\bold{j}_1\,;\,N,N}(z)) \overline{\mathscr{P}^{\#}_{\eta}}\,\frac{1}{\overline{\mathscr{P}^{\#}_{\eta}}\mathscr{K}^{(\bar{i}-2)}_{\bold{j}_1}(z)\overline{\mathscr{P}^{\#}_{\eta}}}\overline{\mathscr{P}^{\#}_{\eta}}\times \\
%& &\quad\quad\times (\check{W}_{\bold{j}_1}-\Gamma_{\bold{j}_1\,;\,N,N}(z))^* \eta \rangle\, \nonumber \\
0&= &-z-\langle \eta \,,\,\Gamma_{\bold{j}_1\,;\,N,N}(z)\,\eta \rangle \\
& &-\langle \eta \,,\,\Gamma_{\bold{j}_1\,;\,N,N}(z) \overline{\mathscr{P}^{\#}_{\eta}}\,\frac{1}{\overline{\mathscr{P}^{\#}_{\eta}}\mathscr{K}^{(N-2)}_{\bold{j}_1}(z)\overline{\mathscr{P}^{\#}_{\eta}}}\overline{\mathscr{P}^{\#}_{\eta}}\,\Gamma_{\bold{j}_1\,;\,N,N}(z)^* \eta \rangle\nonumber
\end{eqnarray}
which is well defined thanks to Proposition \ref{invertibility-bis}.
We claim that there is a unique solution, $z=z^{\$\,(1)}\equiv z_{\bold{j}_1}$, to the equation in (\ref{fp-1}). Using the isospectrality of the Feshbach map, 
%{\color{red}(see the comment after {\emph{\underline{Theorem 3.1}} of \cite{Pi1}})}, 
$z_{\bold{j}_1}$ is the (nondegenerate) ground state energy of $H_{\bold{j}_1}$. Concerning the existence of $z_{\bold{j}_1}$, due to the estimates in (\ref{primero-1}), (\ref{primero-3}) of Lemma \ref{main-relations-H}  and (\ref{denominator}) of Proposition \ref{invertibility-bis}, the fixed point equation  is equivalent to
\begin{equation}\label{eq-dollar-bis}
z=-\langle \eta \,,\,\Gamma^{Bog}_{\bold{j}_1 ;N,N}(z)\,\eta \rangle +\mathcal{Y}_1(z)
\end{equation}
with $$|\mathcal{Y}_1(z)|\leq \frac{C^{\#}_I}{(\ln N)^{\frac{1}{4}}}+(\frac{2}{\gamma})\frac{(C^{\#}_{II})^2}{(\ln N)^{\frac{1}{4}}(1-\gamma)\Delta_{0}}\quad,\quad \gamma = \frac{1}{2}.$$
Then, the same argument of \emph{\underline{Theorem 4.1} of \cite{Pi1}}   implies that for $N$ sufficiently large there exists a $z^{\$\,(1)}$ that solves the equation in (\ref{eq-dollar-bis}). Furthermore, the inequality 
\begin{equation}\label{diff-z-zbog-bis}
 |z^{\$\,(1)}-z_1|\leq (\frac{2}{\gamma})\frac{C^{\#}_{III}}{(\ln N)^{\frac{1}{4}}}\,\quad,\quad C^{\#}_{III}:=C^{\#}_I+\frac{(C^{\#}_{II})^2}{(1-\gamma) \Delta^{\#}_{0}}\,,
\end{equation}
holds true for $z_1$ such that $z_1+\langle \eta \,,\,\Gamma^{Bog}_{\bold{j}_1 ;N,N}(z_1)\,\eta \rangle =0$ because the derivative w.r.t. $z$ of $$z+\langle \eta \,,\,\Gamma^{Bog}_{\bold{j}_1 ;N,N}(z)\eta \rangle$$ is not smaller than $1$; see \emph{\underline{Remark 4.1}} of \cite{Pi1}}.
The eigenvector $\psi_{\bold{j}_1}$ of $H_{\bold{j}_1}$ corresponding to $z^{\$\,(1)}$ is given in expression  (\ref{gs-Hm-start-bis})-(\ref{gs-Hm-fin-bis}) with $m=1$.

\noindent 
The uniqueness of $z^{\$\,(1)}$ follows from the fact that for any other value, $(z^{\$\,(1)})'$, that solves the fixed point problem an inequality analogous to $(\ref{diff-z-zbog-bis})$ holds, hence
\begin{equation}
 |(z^{\$\,(1)})'-z^{\$\,(1)}|\leq \mathcal{O}(\frac{1}{(\ln N)^{\frac{1}{4}}})\,.\label{closeness}
 \end{equation}   Then, using the same argument of  \emph{\underline{Theorem 4.3} of \cite{Pi2}} the closeness (see (\ref{closeness})) of the two eigenvalues $z^{\#\,(1)}$ and $(z^{\$\,(1)})'$ implies
\begin{equation}
\|\psi_{\bold{j}_1}-(\psi_{\bold{j}_1})'\|\leq  \mathcal{O}(\frac{1}{[\ln (\ln N)]^{\frac{1}{2}}})\,.
\end{equation}
where $(\psi_{\bold{j}_1})'$ is the eigenvector corresponding to $(z^{\$\,(1)})'$.
Thus for $N$ large enough the two eigenvalues must coincide. 

\noindent
Since in the interval (\ref{inter-def-finalth}) $\mathscr{K}_{\bold{j}_1}^{(N)}(z)$ is well defined  and $z^{\$\,(1)}$ is the unique fixed point of the equation in (\ref{fp-1}), we can conclude that (in the given interval) $\mathscr{K}_{\bold{j}_1}^{(N)}(z)$  is bounded invertible except for $z=z^{\$\,(1)}$ and, consequently,  $z^{\$\,(1)}$ is the ground state energy of $H_{\bold{j}_1}$. The isospectrality of the Feshbach map implies that the eigenvalue $z^{\$\,(1)}$ of $H_{\bold{j}_1}$ is nondegenerate.
\\

Regarding Property 3., as long as
 \begin{equation}
 z\leq\min\,\Big\{ z_{1}+\gamma\Delta^{\#}_{0}-\frac{\xi^{\frac{1}{2}}}{M}-\frac{C^{\#\,\perp}}{(\ln N)^{\frac{1}{4}}}\,;\,E^{Bog}_{\bold{j}_1}+ \sqrt{\epsilon_{\bold{j}_1}}\phi_{\bold{j}_1}\sqrt{\epsilon_{\bold{j}_1}^2+2\epsilon_{\bold{j}_1}}\Big\}
 \end{equation}
the Feshbach Hamiltonian $\mathscr{K}_{\bold{j}_1}^{\#\,(\bar{i})}(z)$ is well defined thanks to the bound in (\ref{estimate-control-quad}) (see Lemma \ref{control-quad}). Furthermore, we can adapt Theorem \ref{theorem-junction} and Corollary \ref{main-lemma-H} to the Hamiltonian $H^{\#}_{\bold{j}_1}$ in order to implement the Feshbach flow and define the Feshbach Hamiltonians $\mathscr{K}_{\bold{j}_1}^{\#\,(i)}(z)$ up to $i=N-2$ in the same way we proceeded for $\mathscr{K}^{(i)}_{\bold{j}_1}(z)$. To understand this, it must be noticed  that $H^{\#}_{\bold{j}_1}$  contains an interaction term less ($V'_{\bold{j}_1}$) with respect to the Hamiltonian $H_{\bold{j}_1}$ and this does not affect the proof. Next, we adapt Lemma \ref{main-relations-H} and Proposition \ref{invertibility-bis} to  the Hamiltonian  $\mathscr{K}_{\bold{j}_1}^{\#\,(N-2)}(z)$ obtaining analogous results with the same  constants.  Finally, we can define $z^{\#}_{\bold{j}_1}:=z^{\#\,(1)}$ by determining the
solution, $z^{\#\,(1)}$, of the fixed point equation associated with $\mathscr{K}^{\#\,(N)}_{\bold{j}_1}(z)$. The eigenvalue $z^{\#\,(1)}$ fulfills the bound
\begin{equation} |z^{\#\,(1)}-z_1|\leq (\frac{2}{\gamma})\frac{C^{\#}_{III}}{(\ln N)^{\frac{1}{4}}}\,.\label{diff-zdiesis-0}
\end{equation}
%and is unique (due to the same argument used for $z^{\$}_1$).
By Feshbach theory we construct the eigenvector $\psi^{\#}_{\bold{j}_1}$ as in (\ref{gs-Hm-start-bis-diesis})-(\ref{gs-Hm-fin-bis-diesis}) and similarly to our discussion on $z^{\$\,(1)}$ we conclude that $z^{\#\,(1)}$ is unique and is the (nondegenerate) ground state energy of $H^{\#}_{\bold{j}_1}$.

\noindent
We can also estimate the gap above the ground state energy.  
%Indeed, from  Proposition \ref{invertibility-bis} adapted  to $\mathscr{K}_{\bold{j}_1}^{\#\,(N-2)}(z)$ and the uniqueness of the fixed point solution $z^{\#\,(1)}$ 
Indeed, for
 \begin{equation}
z\leq \min\,\Big\{ z_{1}-\frac{C^{\#\,\perp}}{(\ln N)^{\frac{1}{4}}}-\frac{\xi^{\frac{1}{2}}}{M}+\gamma \Delta^{\#}_{0}\,;\,E^{Bog}_{\bold{j}_1}+ \sqrt{\epsilon_{\bold{j}_1}}\phi_{\bold{j}_1}\sqrt{\epsilon_{\bold{j}_1}^2+2\epsilon_{\bold{j}_1}}\Big\}\quad,\quad \gamma = \frac{1}{2}\,,\label{bound-z-0-bis}
\end{equation}
the Hamiltonian $\mathscr{K}_{\bold{j}_1}^{\#\,(N)}(z)$ is bounded invertible in $\mathscr{P}^{\#}_{\eta}\mathcal{F}^N$ except for $z\equiv z^{\#\,(1)}$. In addition, from \emph{\underline{Lemma 5.5} of \cite{Pi1}} we know that $|z_1-E^{Bog}_{\bold{j}_1}|=\mathcal{O}(\frac{1}{N^{\beta}})$ for any $0<\beta<1$.
%+  (\frac{2\sqrt{2}+3}{6})\sqrt{\epsilon_{\bold{j}_1}} \phi_{\bold{j}^*}\sqrt{\epsilon_{\bold{j}_1}^2+2\epsilon_{\bold{j}_1}}$
 This estimate combined with (\ref{diff-zdiesis-0}) imply (for $N$ large)
\begin{eqnarray}
& &\text{infspec}\,\Big[ H^{\#}_{\bold{j}_1}\upharpoonright_{\mathcal{F}^N\ominus \{\mathbb{C}\psi^{\#}_{\bold{j}_1}\}}\Big]-z^{\#}_{\bold{j}_1}\\
&\geq  & \min\,\Big\{ \gamma\Delta^{\#}_{0}-\frac{C^{\#\,\perp}}{(\ln N)^{\frac{1}{4}}}\,;\,\frac{1}{2}\sqrt{\epsilon_{\bold{j}_1}}\phi_{\bold{j}_1}\sqrt{\epsilon_{\bold{j}_1}^2+2\epsilon_{\bold{j}_1}}\Big\}-\frac{2\xi^{\frac{1}{2}}}{M}\,\quad\quad\quad\\
&=& \gamma\Delta^{\#}_{0}-\frac{C^{\#\,\perp}}{(\ln N)^{\frac{1}{4}}}-\frac{2\xi^{\frac{1}{2}}}{M}\\
& =&\Delta^{\#}_1.
 \end{eqnarray}
As for Property 3. (b), by construction $\psi^{\#}_{\bold{j}_1}$ is eigenvector with eigenvalue $z^{\#}_{\bold{j}_1}$. From Property 3. (a) we derive
\begin{equation}
\text{infspec}\,\Big[ (H^{\#}_{\bold{j}_1}-T_{\bold{j}=\pm \bold{j}_{2}})\upharpoonright_{(\mathcal{F}^N\ominus \mathcal{F}^N_{\pm\bold{j}_{2}})\ominus \{\mathbb{C}\psi^{\#}_{\bold{j}_1}\}}\Big]-z^{\#}_{\bold{j}_1}\geq  \Delta^{\#}_1\,.
 \end{equation}

Concerning Property 4., we show the procedure for $H^{\#}_{{\bold{j}_1}}-T_{\bold{j}=\pm\bold{j}_2}$. (For the cases corresponding to  $2\leq l\leq M-1$ the proof is very similar.) We can restrict $$H^{\#}_{{\bold{j}_1}}-T_{\bold{j}=\pm\bold{j}_2}-z-z^{\#}_{\bold{j}_1}$$ to any subspace $[\mathcal{F}^N]_j$ of $\mathcal{F}^N$ with fixed number of particles, $j$,  in the modes $\pm\bold{j}_2$. For simplicity, assume that  $j$ is even; the same result (Property 4.) holds if $j$ is odd. By adapting\footnote{Notice for example  that the assumption in (\ref{ass-1-main-lemma}) in Corollary \ref{main-lemma-H} can be replaced with $(1-\xi)T-(1-\xi)T_{\bold{j}\in\{\pm\bold{j}_1;\pm\bold{j}_2\}}\geq 0 $\,.} Lemma \ref{xi-cor}, Theorem \ref{theorem-junction}, and Corollary \ref{main-lemma-H}, the Feshbach flow can be implemented in the same way with minor modifications. More precisely:
\\

1)  If $j< N-\bar{i}=\lfloor N^{\frac{1}{16}} \rfloor $, we start from }$$\overline{\hat{\mathscr{P}}}^{(\bar{i})}:=\hat{\mathfrak{Q}}_{\bold{j}_1}^{(\bar{i},\bar{i}+1)}:=\mathfrak{Q}_{\bold{j}_1}^{(\bar{i},\bar{i}+1)}\charf_{[\mathcal{F}^N]_j}\quad\text{and}\quad\hat{\mathscr{P}}^{(\bar{i})}:=\hat{\mathfrak{Q}}_{\bold{j}_1}^{(>\bar{i}+1)}:=\charf_{[\mathcal{F}^N]_j}-\hat{\mathfrak{Q}}^{(\bar{i},\bar{i}+1)}_{\bold{j}_1}\,$$ 
and proceed for $i>\bar{i}$ (and even) up to the step $i=N-2$ with the definitions
$$\overline{\hat{\mathscr{P}}}^{(i)}:=\hat{\mathfrak{Q}}_{\bold{j}_1}^{(i,i+1)}:=\mathfrak{Q}_{\bold{j}_1}^{(i,i+1)}\charf_{[\mathcal{F}^N]_j}\quad\text{and}\quad\hat{\mathscr{P}}^{(i)}:=\hat{\mathfrak{Q}}_{\bold{j}_1}^{(>i+1)}:=\hat{\mathfrak{Q}}_{\bold{j}_1}^{(>i-1)}-\hat{\mathfrak{Q}}^{(i,i+1)}_{\bold{j}_1}\,.$$ 

2)  If $j\geq  N-\bar{i}$, we start from 
$$\overline{\hat{\mathscr{P}}}^{(j)}:=\hat{\mathfrak{Q}}_{\bold{j}_1}^{(j,j+1)}:=\mathfrak{Q}_{\bold{j}_1}^{(j,j+1)}\charf_{[\mathcal{F}^N]_j}\quad\text{and}\quad\hat{\mathscr{P}}^{(j)}:=\hat{\mathfrak{Q}}_{\bold{j}_1}^{(>j+1)}:=\charf_{[\mathcal{F}^N]_j}-\hat{\mathfrak{Q}}^{(j,j+1)}_{\bold{j}_1}$$ and proceed for $i>j$ (and even) up to the step $i=N-2$ with the definitions
$$\overline{\hat{\mathscr{P}}}^{(i)}:=\hat{\mathfrak{Q}}_{\bold{j}_1}^{(i,i+1)}:=\mathfrak{Q}_{\bold{j}_1}^{(i,i+1)}\charf_{[\mathcal{F}^N]_j}\quad\text{and}\quad\hat{\mathscr{P}}^{(i)}:=\hat{\mathfrak{Q}}_{\bold{j}_1}^{(>i+1)}:=\hat{\mathfrak{Q}}_{\bold{j}_1}^{(>i-1)}-\hat{\mathfrak{Q}}^{(i,i+1)}_{\bold{j}_1}\,.$$ 

 %whereas if $j\geq \lfloor N-N^{\frac{1}{16}}\rfloor+2$ then $\mathfrak{Q}^{(i,i+1)}_{\bold{j}_1}\equiv Q^{(i,i+1)}_{\bold{j}_1}\charf_{[\mathcal{F}^N]_j}$.

We call $\hat{\mathscr{K}}^{\#\,(i)}_{\bold{j}_1}(z+z^{\#}_{\bold{j}_1})$ the Feshbach Hamiltonians so defined. One can observe that $\hat{\mathfrak{Q}}_{\bold{j}_1}^{(>N-1)}$  projects  onto the subspace of states with no particles in the modes $\pm \bold{j}_1$ and $j$ particles in the modes $\pm\bold{j}_2$.
%; 2) the subspace spanned by the vectors with at least $\lfloor N^{\frac{1}{16}} \rfloor-1$ particles in the modes $\pm \bold{j}_1$, $j$ particles in the modes $\pm \bold{j}_2$,  and where the (total) kinetic energy operator for all the other modes takes values larger than $N^{\frac{1}{8}}$. The latter subspace is trivial if $j\geq  N-\lfloor N^{\frac{1}{16}} \rfloor+2$. 
We set $w=z+z^{\#}_{\bold{j}_1}$ and obtain 
\begin{eqnarray}
& &\hat{\mathscr{K}}^{\#\,(N-2)}_{\bold{j}_1}(w)\\
%&=&\mathscr{P}_{\eta}(H^{Bog}_{\bold{j}_*}-z)\mathscr{P}_{\eta}\\
%& &-\mathscr{P}_{\eta}W_{\bold{j}_*}\,Q^{(N-2)}_{\bold{j}_*}\,R^{Bog}_{\bold{j}_*\,;\,N-2,N-2}(z)\sum_{l_{N-2}=0}^{\infty}[\Gamma^{Bog}_{\bold{j}_*\,;\,N-2,N-2}(z) R^{Bog}_{\bold{j}_*\,;\,N-2,N-2}(z)]^{l_{N-2}}\, Q^{(N-2)}_{\bold{j}_*}W^*_{\bold{j}_*}\mathscr{P}_{\eta}\nonumber\\
&=&\hat{\mathfrak{Q}}_{\bold{j}_1}^{(>N-1)}(H^{\#}_{{\bold{j}_1}}-T_{\bold{j}=\pm\bold{j}_2}-w)\hat{\mathfrak{Q}}_{\bold{j}_1}^{(>N-1)}\\
& &-\hat{\mathfrak{Q}}_{\bold{j}_1}^{(>N-1)}\check{W}_{\bold{j}_1}\,\hat{R}_{\bold{j}_1\,;\,N-2,N-2}(w)\sum_{l_{N-2}=0}^{\infty}[\hat{\Gamma}_{\bold{j}_1\,;\,N-2,N-2}(w) \hat{R}_{\bold{j}_1\,;\,N-2,N-2}(w)]^{l_{N-2}}\, \check{W}^*_{\bold{j}_1}\hat{\mathfrak{Q}}_{\bold{j}_1}^{(>N-1)}\nonumber\,
%& &-\overline{\mathscr{P}_{\psi^{\#}_{\bold{j}_1,\dots,\bold{j}_{m-1}}}}\,\Gamma_{\bold{j}_1,\,\dots, \bold{j}_{m}\,;\,N,N}(w)\,\overline{\mathscr{P}_{\psi^{\#}_{\bold{j}_1,\dots,\bold{j}_{m-1}}}}\\
%& &-\hat{\mathfrak{Q}}_{\bold{j}_1}^{(>N-1)}\hat{\mathscr{V}}^{(N-2)}_{1}(w)\hat{\mathfrak{Q}}_{\bold{j}_1}^{(>N-1)}\\
%& &+\Big\{\hat{\mathfrak{Q}}_{\bold{j}_1}^{(>N-1)}\,\hat{\mathscr{V}}^{(N-2)}_{1}(w)\,\times \\
%& &\quad \quad\quad \times \sum_{l_{N-2}=0}^{\infty}\hat{R}_{\bold{j}_1\,;\,N-2,N-2}(w)\Big[\hat{\Gamma}_{\bold{j}_1\,;\,N-2,N-2}(w)R_{\bold{j}_1,\,\dots, \bold{j}_{m}\,;\,N-2,N-2}(w)\Big]^{l_{N-2}}\check{W}^*_{\bold{j}_1}\hat{\mathfrak{Q}}_{\bold{j}_1}^{(>N-1)}\,\nonumber\\
%& &\quad\quad +h.c. \Big\}\nonumber \\
%& &-\hat{\mathfrak{Q}}_{\bold{j}_1}^{(>N-1)}\,\hat{\mathscr{V}}^{(N-2)}_{1}(w)\,\times \label{irrev-fin-bis}\\
%& &\quad \times \sum_{l_{N-2}=0}^{\infty}\hat{R}_{\bold{j}_1\,;\,N-2,N-2}(w)\Big[\hat{\Gamma}_{\bold{j}_1\,;\,N-2,N-2}(w)\hat{R}_{\bold{j}_1\,;\,N-2,N-2}(w)\Big]^{l_{N-2}}(\hat{\mathscr{V}}^{(N-2)}_{1}(w))^*\,\hat{\mathfrak{Q}}_{\bold{j}_1}^{(>N-1)}\,\nonumber
\end{eqnarray}
%& &-Q^{(>N-1)}W\,Q^{(N-2)}\,R^{Bog}_{N-2,N-2}(z)\sum_{l_{N-2}=0}^{\infty}[\Gamma^{Bog}_{N-2,N-2}(z) R^{Bog}_{N-2,N-2}(z)]^{l_{N-2}}\, Q^{(N-2)}W^*Q^{(>N-1)}\quad \quad\quad\quad\label{first-term-last-Bog}\\
%& &-\mathscr{V}^{(N-1)}(z)\,R_{N-1,N-1}(z)\sum_{l_{N-1}=0}^{\infty}[\Gamma_{N-1,N-1}(z) R_{N-1,N-1}(z)]^{l_{N-1}}\,(\mathscr{V}^{(N-1)}(z) )^*\quad\quad \label{last-term-last-Bog}
where:

1) if $j<N-\bar{i}$ the operators $\hat{R}_{\bold{j}_1\,;\,i,i}(w)$,  $\hat{\Gamma}_{\bold{j}_1\,;\,i,i}(w)$  have the same definition of $R_{\bold{j}_1\,;\,i,i}(w)$, $\Gamma_{\bold{j}_1\,;\,i,i}(w)$ but in terms of $H^{\#}_{{\bold{j}_1}}-T_{\bold{j}=\pm\bold{j}_2}$ and of the new projections;  
\\

2) if $j\geq N-\bar{i}$ the operators $\hat{R}_{\bold{j}_1\,;\,i,i}(w)$,  $\hat{\Gamma}_{\bold{j}_1\,;\,i,i}(w)$ have the same definition of $R_{\bold{j}_1\,;\,i,i}(w)$, $\Gamma_{\bold{j}_1\,;\,i,i}(w)$  but in terms of $H^{\#}_{{\bold{j}_1}}-T_{\bold{j}=\pm\bold{j}_2}$ and of the new projections, and $\Gamma_{\bold{j}_1\,;\,i,i}(w)$ starts from
$$\hat{\Gamma}_{\bold{j}_1\,;\,j+2,j+2}(w):=\mathfrak{Q}^{(j+2,j+3)}_{\bold{j}_1}\check{W}_{\bold{j}_1}\mathfrak{Q}^{(j,j+1)}_{\bold{j}_1}\frac{1}{\mathfrak{Q}^{(j,j+1)}_{\bold{j}_1}(H^{\#}_{{\bold{j}_1}}-T_{\bold{j}=\pm\bold{j}_2}-w)\mathfrak{Q}^{(j,j+1)}_{\bold{j}_1}}\mathfrak{Q}^{(j,j+1)}_{\bold{j}_1}\check{W}^*_{\bold{j}_1}\mathfrak{Q}^{(j+2,j+3)}_{\bold{j}_1}\,.$$
%and
%$$\hat{\mathscr{V}}^{(j+2)}_{1}(w):=\check{W}_{\bold{j}_1}\mathfrak{Q}^{(j,j+1)}_{\bold{j}_1}\frac{1}{\mathfrak{Q}^{(j,j+1)}_{\bold{j}_1}(H^{\#}_{{\bold{j}_1}}-T_{\bold{j}=\pm\bold{j}_2}-w)\mathfrak{Q}^{(j,j+1)}_{\bold{j}_1}}\mathfrak{Q}^{(j,j+1)}_{\bold{j}_1}\check{W}^*_{\bold{j}_1}\,.$$

%As last couple of projections we consider $\hat{\mathscr{P}}^{(N-j)}:=|\eta_j\rangle \langle \eta_j|$ and $\overline{\hat{\mathscr{P}}}^{(N-j)}:=\hat{Q}_{\bold{j}_1}^{(>N-j-1)}-|\eta_j\rangle \langle \eta_j|$.  

\noindent
We want to prove that  for $N$ large enough the operator 
\begin{equation}\label{inf-flow-final}
\hat{\mathscr{K}}^{\#\,(N-2)}_{\bold{j}_1}(w)\equiv \hat{\mathfrak{Q}}_{\bold{j}_1}^{(>N-1)}\hat{\mathscr{K}}_{\bold{j}_1}^{\#\,(N-2)}(w)\hat{\mathfrak{Q}}_{\bold{j}_1}^{(>N-1)}
\end{equation} is bounded invertible as long as  $z$ is less than $-\frac{\xi^{\frac{1}{2}}}{2M}$. To this purpose we exploit the analogy with the estimate of a lower bound to the spectrum of $$\overline{\mathscr{P}_{\psi^{\#}_{\bold{j}_1}}}\mathscr{K}_{\bold{j}_1}^{\,(N-2)}(z+z^{\#}_{\bold{j}_1})\overline{\mathscr{P}_{\psi^{\#}_{\bold{j}_1}}}\,.$$ More precisely, we replace $\overline{\mathscr{P}_{\psi^{\#}_{\bold{j}_1}}}$ with $\hat{\mathfrak{Q}}_{\bold{j}_1}^{(>N-1)}$ and proceed like for \emph{Property 4.} in \emph{\underline{Theorem 4.3} of \cite{Pi2}} by adapting\footnote{We point out that a modified Corollary \ref{main-lemma-H} where condition a) is replaced with $(1-\xi)T-(1-\xi)T_{\bold{j}\in\{\pm\bold{j}_1;\pm\bold{j}_2\}}\geq 0 $ implies Lemma \ref{main-relations-H} where the Hamiltonian $H^{\#}_{{\bold{j}_1}}$ is replaced with $H^{\#}_{{\bold{j}_1}}-T_{\bold{j}=\pm\bold{j}_2}$} (to $H^{\#}_{{\bold{j}_1}}-T_{\bold{j}=\pm\bold{j}_2}$) estimate (\ref{estimate-gammaperp-0}) of Lemma \ref{main-relations-H}. 

\noindent
Finally, we conclude that  
$\hat{\mathscr{K}}_{\bold{j}_1}^{\#\,(N-2)}(w)$
is strictly positive if $z\leq-\frac{\xi^{\frac{1}{2}}}{2M}$. Then, for $z$ in the same interval, the operator $H^{\#}_{{\bold{j}_1}}-T_{\bold{j}=\pm\bold{j}_2}-z-z^{\#}_{\bold{j}_1}$ is bounded invertible by isospectrality of the Feshbach map.\\

Property 5. is a straightforward consequence of (\ref{estimate-control-quad-3}) in Lemma \ref{control-quad}.
\\

\noindent
\emph{Case $m>1$}

We assume that Properties 1., 2., 3., 4., and 5. hold for $1\leq m-1<M$ and prove that they hold for $m$. 
\\

\noindent
\emph{Property 1.}   
Since Properties 3., 4., 5.  hold for $m-1$ we can apply Lemma \ref{xi-cor},  Corollary \ref{main-lemma-H}, Theorem \ref{theorem-junction}, Theorem \ref{Feshbach-H}, Lemma \ref{main-relations-H} and Proposition \ref{invertibility-bis} and get Property 1. for $m$. 
\\

\noindent
\emph{Property 2.}   
We recall the definition of $f_{\bold{j}_1,\dots,\bold{j}_m}(z+z^{\#}_{\bold{j}_1,\dots,\bold{j}_{m-1}})$  in (\ref{K-last-step})-(\ref{K-last-step-bis}) and observe that
\begin{eqnarray}
& &\mathscr{P}_{\psi^{\#}_{\bold{j}_1,\dots,\bold{j}_{m-1}}}(H_{\bold{j}_1,\dots,\bold{j}_m}-z^{\#}_{\bold{j}_1,\dots,\bold{j}_{m-1}}) \mathscr{P}_{\psi^{\#}_{\bold{j}_1,\dots,\bold{j}_{m-1}}}\\
&=&\mathscr{P}_{\psi^{\#}_{\bold{j}_1,\dots,\bold{j}_{m-1}}}(H^{\#}_{\bold{j}_{1},\dots,\bold{j}_{m-1}}+V_{\bold{j}_{m}}+V'_{\bold{j}_1,\dots \bold{j}_{m-1}}+\hat{H}^{Bog}_{\bold{j}_m}-z^{\#}_{\bold{j}_1,\dots,\bold{j}_{m-1}})\mathscr{P}_{\psi^{\#}_{\bold{j}_1,\dots,\bold{j}_{m-1}}}\\
&=&\mathscr{P}_{\psi^{\#}_{\bold{j}_1,\dots,\bold{j}_{m-1}}}(V_{\bold{j}_{m}}+V'_{\bold{j}_1,\dots \bold{j}_{m-1}}+\hat{H}^{Bog}_{\bold{j}_m})\mathscr{P}_{\psi^{\#}_{\bold{j}_1,\dots,\bold{j}_{m-1}}}\\
%& &+\mathscr{P}_{\psi^{\#}_{\bold{j}_1,\dots,\bold{j}_{m-1}}}(z^{\#}_{\bold{j}_1,\dots,\bold{j}_{m-1}}-z_{\bold{j}_1,\dots,\bold{j}_{m-1}})\mathscr{P}_{\psi^{\#}_{\bold{j}_1,\dots,\bold{j}_{m-1}}}\\
&=&\mathscr{P}_{\psi^{\#}_{\bold{j}_1,\dots,\bold{j}_{m-1}}}V_{\bold{j}_{m}}\mathscr{P}_{\psi^{\#}_{\bold{j}_1,\dots,\bold{j}_{m-1}}}\,.
\end{eqnarray}
Furthermore,  from Property 5. at step $m-1$, we have
%$$\|\mathscr{P}_{\psi^{\#}_{\bold{j}_1,\dots,\bold{j}_{m-1}}}(z^{\#}_{\bold{j}_1,\dots,\bold{j}_{m-1}}-z_{\bold{j}_1,\dots,\bold{j}_{m-1}})\mathscr{P}_{\psi^{\#}_{\bold{j}_1,\dots,\bold{j}_{m-1}}}\|\leq (m-1)\frac{\xi^{\frac{1}{2}}}{3M}$$
%and
$$\|\mathscr{P}_{\psi^{\#}_{\bold{j}_1,\dots,\bold{j}_{m-1}}}V_{\bold{j}_{m}}\mathscr{P}_{\psi^{\#}_{\bold{j}_1,\dots,\bold{j}_{m-1}}}\|\leq \frac{\tilde{C}^{\#}}{N}\,.$$
 The rest of the proof is analogous to the case $m=1$ using (\ref{primero-1}) and (\ref{primero-3}) from Lemma \ref{main-relations-H} that can be applied thanks to Property 3.(a), 4., and 5. at the step $m-1$. These estimates yield
 \begin{equation}\label{eq-dollar-bis-bis}
z=-\langle \eta \,,\,\Gamma^{Bog}_{\bold{j}_m ;N,N}(z)\,\eta \rangle +\mathcal{Y}_m(z)
\end{equation}
with 
\begin{eqnarray}
|\mathcal{Y}_m(z)|&\leq &\frac{C^{\#}_I}{(\ln N)^{\frac{1}{4}}}+\frac{(C^{\#}_{II})^2}{(\ln N)^{\frac{1}{4}}(1-\gamma)\Delta^{\#}_{m-1}}+\frac{\tilde{C}^{\#}}{N}\quad,\quad \gamma = \frac{1}{2}\\
&=&\frac{C^{\#}_I}{(\ln N)^{\frac{1}{4}}}+(\frac{2}{\gamma})^m\frac{(C^{\#}_{II})^2}{(\ln N)^{\frac{1}{4}}(1-\gamma)\Delta_{0}}+\frac{\tilde{C}^{\#}}{N}\quad,\quad \gamma = \frac{1}{2}
\end{eqnarray}
where we have used that $\Delta^{\#}_{m-1}\geq (\frac{\gamma}{2})^m\Delta_0$ for $N$ sufficiently large.
In a similar way, for $N$ large enough one can also derive the existence of the unique solution $z^{\$\,(m)}$ with the property in (\ref{dist-zdollar}). Using the isospectrality of the Feshbach map, we deduce that $H_{\bold{j}_1,\dots,\bold{j}_{m}}$ has the nondegenerate eigenvalue $z^{\$\,(m)}+z_{\bold{j}_1,\dots,\bold{j}_{m-1}}=:z_{\bold{j}_1,\dots,\bold{j}_{m}}$
The corresponding eigenvector is given in  (\ref{gs-Hm-start})-(\ref{gs-Hm-fin}). We observe that in the interval
\begin{equation}
z\leq \min\,\Big\{ z_{m}-\frac{C^{\#\,\perp}}{(\ln N)^{\frac{1}{4}}}-\frac{m\xi^{\frac{1}{2}}}{M}+\gamma \Delta^{\#}_{m-1}\,;\,E^{Bog}_{\bold{j}_m}+ \sqrt{\epsilon_{\bold{j}_m}}\phi_{\bold{j}_m}\sqrt{\epsilon_{\bold{j}_m}^2+2\epsilon_{\bold{j}_m}}\Big\}\quad,\quad \gamma = \frac{1}{2}\,,\label{bound-z-0-bis}
\end{equation}
the Hamiltonian $\mathscr{K}_{\bold{j}_1,\dots,\bold{j}_m}^{(N)}(z+z_{\bold{j}_1,\dots,\bold{j}_{m-1}})$ is bounded invertible except for $z\equiv z^{\$\,(m)}$. Then, by the isospectrality  of the Feshbach map we deduce that $$z^{\$\,(m)}+z_{\bold{j}_1,\dots,\bold{j}_{m-1}}=:z_{\bold{j}_1,\dots,\bold{j}_{m}}$$ is the (nondegenerate) ground state energy of $H_{\bold{j}_1,\dots,\bold{j}_m}$.\\

\noindent
\emph{Property 3.}  %Indeed, the vector $\psi^{Bog}_{\bold{j}_1,\dots,\bold{j}_{m}} \in \mathcal{F}\ominus \mathcal{F}_{\pm\bold{j}_{m+1}}$ and it is also eigenvector of $\Big(\hat{H}^{Bog}_{\bold{j}_1, \dots,\bold{j}_{m}}+\sum_{\bold{j}\neq \{\pm\bold{j}_1, \dots, \pm\bold{j}_{m+1}\}}(k_{\bold{j}})^2a^*_{\bold{j}}a_{\bold{j}}\Big)\upharpoonright_{\mathcal{F}\ominus \mathcal{F}_{\pm\bold{j}_{m+1}}}$. Now if $\psi\in \mathcal{F}\ominus \mathcal{F}_{\pm\bold{j}_{m+1}}$ is perpendicular to $\psi^{Bog}_{\bold{j}_1,\dots,\bold{j}_{m}}$ then we have
%\begin{eqnarray}
%& &\langle \psi\,,\,\Big(\hat{H}^{Bog}_{\bold{j}_1, \dots,\bold{j}_{m}}+\sum_{\bold{j}\neq \{\pm\bold{j}_1, \dots, \pm\bold{j}_{m+1}\}}(k_{\bold{j}})^2a^*_{\bold{j}}a_{\bold{j}}\Big)\psi \rangle\\
%&=&\langle \psi\,,\,\Big(\hat{H}^{Bog}_{\bold{j}_1, \dots,\bold{j}_{m}}+\sum_{\bold{j}\neq \{\pm\bold{j}_1, \dots, \pm\bold{j}_{m}\}}(k_{\bold{j}})^2a^*_{\bold{j}}a_{\bold{j}}\Big)\psi \rangle\\
%&\geq &z_{\bold{j_1},\dots, \bold{j}_m}+\Delta_m\,.
%\end{eqnarray}
% because  the Feshbach Hamiltonian $\mathscr{K}_{\bold{j}_1,\dots,\bold{j}_m}^{\#\,Bog\,(N)}(z+z_{\bold{j}_1,\dots,\bold{j}_{m-1}})$ associated with  $$\Big(\hat{H}^{Bog}_{\bold{j}_1, \dots,\bold{j}_{m}}+\sum_{\bold{j}\neq \{\pm\bold{j}_1, \dots, \pm\bold{j}_{m+1}\}}(k_{\bold{j}})^2a^*_{\bold{j}}a_{\bold{j}}\Big)\upharpoonright_{\mathcal{F}\ominus \mathcal{F}_{\pm\bold{j}_{m+1}}}$$ can be defined for the same $z$ where we have defined $\mathscr{K}_{\bold{j}_1,\dots,\bold{j}_m}^{Bog\,(N)}(z+z_{\bold{j}_1,\dots,\bold{j}_{m-1}})$,  and for those values of $z$ the two Feshbach Hamiltonians coincide. 
Assuming Properties 3., 4., 5.  at  step $m-1$ we can adapt Theorem \ref{theorem-junction}  and Corollary \ref{main-lemma-H} to the Hamiltonian $H^{\#}_{\bold{j}_1,\dots,\bold{j}_{m}}$ in order to implement the Feshbach flow and define the Feshbach Hamiltonians $\mathscr{K}_{\bold{j}_1,\dots,\bold{j}_m}^{\#\,(i)}(z+z^{\#}_{\bold{j}_1,\dots,\bold{j}_{m-1}})$ up to $i=N-2$ in the same way we have proceeded for $\mathscr{K}_{\bold{j}_1}^{\#\,(i)}(z)$. We can also adapt Lemma \ref{main-relations-H} and Proposition \ref{invertibility-bis} to  the Hamiltonian  $\mathscr{K}_{\bold{j}_1,\dots,\bold{j}_m}^{\#\,(N-2)}(z+z_{\bold{j}_1,\dots,\bold{j}_{m-1}})$ obtaining analogous estimates with the same  constants (recall that $H^{\#}_{\bold{j}_1,\dots,\bold{j}_m}$  contains an interaction term less ($V'_{\bold{j}_1,\dots,\bold{j}_{m}}$) with respect to the Hamiltonian $H_{\bold{j}_1,\dots,\bold{j}_{m}}$). Finally, we can define the eigenvalue $z^{\#}_{\bold{j}_1, \dots,\bold{j}_{m}}:=z^{\#\,(m)}+z^{\#}_{\bold{j}_1, \dots,\bold{j}_{m-1}}$ of $H^{\#}_{\bold{j}_1, \dots,\bold{j}_{m}}$ by determining the 
unique solution, $z^{\#\,(m)}$, of the fixed point equation corresponding to $\mathscr{K}_{\bold{j}_1,\dots,\bold{j}_m}^{\#\,(N)}(z+z_{\bold{j}_1,\dots,\bold{j}_{m-1}})$. This solution fulfills the bound
\begin{equation} |z^{\#\,(m)}-z_m|\leq (\frac{2}{\gamma})^m\frac{C^{\#}_{III}}{(\ln N)^{\frac{1}{4}}}+\frac{\tilde{C}^{\#}}{N}\,.\label{diff-zdiesis}
\end{equation}
By Feshbach theory we then construct the eigenvector of $H^{\#}_{\bold{j}_1,\dots,\bold{j}_m}$, $\psi^{\#}_{\bold{j}_1, \dots,\bold{j}_{m}}$,  as in (\ref{gs-Hm-start-bis-diesis})-(\ref{gs-Hm-fin-bis-diesis}).  Similarly to the case $m=1$ we can conclude that $z^{\#\,(m)}+z^{\#}_{\bold{j}_1, \dots,\bold{j}_{m-1}}$ and $\psi^{\#}_{\bold{j}_1, \dots,\bold{j}_{m}}$ are the (non-degenerate) ground state energy and ground state vector, respectively.

\noindent
We observe that in the interval
\begin{equation}
z\leq \min\,\Big\{ z_{m}-\frac{C^{\#\,\perp}}{(\ln N)^{\frac{1}{4}}}-\frac{m\xi^{\frac{1}{2}}}{M}+\gamma \Delta^{\#}_{m-1}\,;\,E^{Bog}_{\bold{j}_m}+ \sqrt{\epsilon_{\bold{j}_m}}\phi_{\bold{j}_m}\sqrt{\epsilon_{\bold{j}_m}^2+2\epsilon_{\bold{j}_m}}\Big\}\quad,\quad \gamma = \frac{1}{2}\,,\label{bound-z-0-bis}
\end{equation}
the Hamiltonian $\mathscr{K}_{\bold{j}_1,\dots,\bold{j}_m}^{\#\,(N)}(z+z_{\bold{j}_1,\dots,\bold{j}_{m-1}})$ is bounded invertible except for $z\equiv z^{\#\,(m)}$. 
From \emph{\underline{Corollary 4.6} of \cite{Pi1}} we know that  $|z_m-E^{Bog}_{\bold{j}_m}|=\mathcal{O}(\frac{1}{N^{\beta}})$ for any $0<\beta<1$. This estimate combined with (\ref{diff-zdiesis}) imply
\begin{eqnarray}
& &\text{infspec}\,\Big[ H^{\#}_{\bold{j}_1, \dots,\bold{j}_{m}}\upharpoonright_{\mathcal{F}^N\ominus \{\mathbb{C}\psi^{\#}_{\bold{j}_1, \dots,\bold{j}_{m}}\}}\Big]-z^{\#}_{\bold{j}_1, \dots,\bold{j}_{m}}\\
&\geq  &\Delta^{\#}_m:=  \gamma\Delta^{\#}_{m-1}-\frac{C^{\#\,\perp}}{(\ln N)^{\frac{1}{4}}}-\frac{2m\xi^{\frac{1}{2}}}{M}\,,\quad\quad\quad
 \end{eqnarray}
 and, consequently, Property 3. (b):
\begin{eqnarray}
& &\text{infspec}\,\Big[ (H^{\#}_{\bold{j}_1, \dots,\bold{j}_{m}}-T_{\bold{j}=\pm \bold{j}_{m+1}})\upharpoonright_{(\mathcal{F}^N\ominus \mathcal{F}^N_{\pm\bold{j}_{m+1}})\ominus \{\mathbb{C}\psi^{\#}_{\bold{j}_1, \dots,\bold{j}_{m}}\}}\Big]-z^{\#}_{\bold{j}_1, \dots,\bold{j}_{m}}\\
%& \geq &\text{infspec}\,\Big[ H^{\#}_{\bold{j}_1, \dots,\bold{j}_{m}}\upharpoonright_{(\mathcal{F}^N\ominus \mathcal{F}^N_{\pm\bold{j}_{m+1}})\ominus \{\mathbb{C}\psi^{\#}_{\bold{j}_1, \dots,\bold{j}_{m}}\}}\Big]-z^{\#}_{\bold{j}_1, \dots,\bold{j}_{m}}\\
&\geq  &\Delta^{\#}_m\,.
 \end{eqnarray}

\noindent
\emph{Property 4.}   The argument is analogous to the case $m=1$ given Properties 1.-5. at step $m-1$. In particular, notice that in order to apply (a suitably adapted version of) Theorem \ref{theorem-junction} and Corollary \ref{main-lemma-H} to $$H^{\#\,(\bar{l})}_{\bold{j}_1,\dots, \bold{j}_{m}}-T_{\bold{j}\in\{\pm \bold{j}_{m+1};\dots;\pm \bold{j}_{m+\bar{l}}\}}\quad,\quad 1\leq \bar{l} \leq M-m,$$
Property 4. for $$H^{\#\,(\bar{l}+1)}_{\bold{j}_1,\dots, \bold{j}_{m-1}}-T_{\bold{j}\in\{\pm \bold{j}_{m};\dots;\pm \bold{j}_{m+\bar{l}}\}}=H^{\#\,(\bar{l}+1)}_{\bold{j}_1,\dots, \bold{j}_{m-1}}-T_{\bold{j}=\{\pm \bold{j}_{m-1+1},\dots,\pm \bold{j}_{m-1+(\bar{l}+1)}\}}$$ is needed.

\noindent
\emph{Property 5.}  
This is a straightforward consequence of (\ref{estimate-control-quad-3}) in Lemma \ref{control-quad}.
 
\qed

The very last result of this section concerns the expansion of the ground state vector $\psi_{\bold{j}_1, \dots, \bold{j}_{M}}$ (and of $\psi^{\#}_{\bold{j}_1, \dots, \bold{j}_{m}}$,  $1\leq m\leq M-1$) in terms of the bare quantities.
 
\begin{corollary}\label{expansion}
Assume the hypotheses of Theorem \ref {induction-many-modes}. Then, for any arbitrarily small $\zeta>0$,  there exists $N_{\zeta}<\infty$ and a vector $(\psi_{\bold{j}_1, \dots, \bold{j}_{m}})_{\zeta}$,  corresponding to a ($\xi$-dependent) finite sum of ($\xi$-dependent)  finite products of the interaction terms $W^*_{\bold{j}_l}+W_{\bold{j}_l}$ and of the resolvents $\frac{1}{\hat{H}^0_{\bold{j}_l}-E^{Bog}_{\bold{j}_{l}}}$ (see (\ref{H0j})), $1\leq l \leq M$, applied to $\eta$,  such that 
$$\| \psi_{\bold{j}_1, \dots, \bold{j}_{m}}-(\psi_{\bold{j}_1, \dots, \bold{j}_{m}})_{\zeta}\|\leq \zeta$$  
for $N>N_{\zeta}$.
\end{corollary}

\noindent
\emph{Proof}

The proof is very similar to the analogous result for the ground state of the Bogoliubov Hamiltonian $H^{Bog}_{\bold{j}_{1},\dots,\bold{j}_{M}}$ derived in {\emph{\underline{Corollary 4.6} of \cite{Pi2}}.} However, it is important to notice that the re-expansion of the  factors
\begin{equation}
\frac{1}{\mathfrak{Q}_{\bold{j}_m}^{(N-2r,N-2r+1)}\mathscr{K}_{\bold{j}_1,\dots ,\bold{j}_m}^{(N-2r-2)}(z_{\bold{j}_1,\dots ,\bold{j}_m})\mathfrak{Q}_{\bold{j}_m}^{(N-2r, N-2r+1)}}
\end{equation}
in (\ref{gs-vector-inter}) produces terms containing $\Gamma_{\bold{j}_1,\dots,\bold{j}_m\,;\,\bar{i}+2,\bar{i}+2}(z_{\bold{j}_1,\dots ,\bold{j}_m})$  that is an operator that cannot be re-expanded. This is not a problem because the norm of the sum of the contributions proportional to $\Gamma_{\bold{j}_1,\dots,\bold{j}_m\,;\,\bar{i}+2,\bar{i}+2}(z_{\bold{j}_1,\dots ,\bold{j}_m})$ is arbitrarily small for $N$ sufficiently large.
\qed
% where $S$ is the subset of $\mathbb{Z}^{d}\setminus \{\bold{0},\pm \bold{j}_m\}$ such that all $\rho_{\bold{j}}$ are positive.

\section{Appendix}
\setcounter{equation}{0}
\begin{corollary}\label{main-lemma-H}
For $M\geq m\geq 1$ assume:

\noindent
(a)  
%\begin{equation}
%(H^{Bog}_{\bold{j}_1,\dots,\bold{j}_{m-1}})_{\xi}\geq z_{\bold{j}_1,\dots\,\bold{j}_m}-\mathcal{O}(\xi)
%\end{equation}
\begin{equation}
(H^{\#}_{\bold{j}_{1},\dots,\bold{j}_{m-1}})_{\xi}-(1-\xi)T_{\bold{j}=\{\pm\bold{j}_m\}}\geq z^{\#}_{\bold{j}_1,\dots\,\bold{j}_{m-1}}-\frac{(m-1)\xi^{\frac{1}{2}}}{M}\,,\label{ass-1-main-lemma}
\end{equation}
where $(H^{\#}_{\bold{j}_1,\dots,\bold{j}_{m-1}})_{\xi}$ is defined in  (\ref{def-xiham-1}) for $m\geq 2$ and is equal to $(1-\xi)T$ for $m=1$, 
 and where $z^{\#}_{\bold{j}_1,\dots\,\bold{j}_{m-1}}$ is the ground state energy of $H^{\#}_{\bold{j}_1,\dots,\bold{j}_{m-1}}$.\\
 % with $z^{\#}_{\bold{j}_1,\dots\,\bold{j}_{m-1}}\equiv 0$ if $m=1$.
%
%\noindent
%b) 
%\begin{equation}
%|z^{\#}_{\bold{j}_1,\dots\,\bold{j}_{m-1}}-z_{\bold{j}_1,\dots\,\bold{j}_{m-1}}|\leq \frac{(m-1)\xi^{\frac{1}{2}}}{3M}\,.
%\end{equation}

\noindent
(b)  
\begin{equation}\label{inter-z-S}
w:=z+z^{\#}_{\bold{j}_1,\dots,\bold{j}_{m-1}}\leq  z^{\#}_{\bold{j}_1,\dots,\bold{j}_{m-1}}+ E^{Bog}_{\bold{j}_m}+ (\delta-1)\phi_{\bold{j}_m}\sqrt{\epsilon_{\bold{j}_m}^2+2\epsilon_{\bold{j}_m}}
\end{equation} with $\delta< 2$ and $\epsilon_{\bold{j}_m}$ sufficiently small.  

\noindent
Then, for $\xi=(\frac{1}{\ln N})^{\frac{1}{4}}$ and $N$ sufficiently large
\begin{eqnarray}\label{estimate-main-lemma-H}
%& &\frac{\sqrt{\frac{(N-l-1)}{(N-l-2)}}}{\Big[1+\frac{N}{N-l-2}(\epsilon+\frac{1+\epsilon-\sqrt{\epsilon^2+2\epsilon}}{l})\Big]^{\frac{1}{2}} \Big[1+\frac{N}{N-l}(\epsilon-\frac{1+\epsilon+\sqrt{\epsilon^2+2\epsilon}}{l})\Big]^{\frac{1}{2}}}\\
& &\| \Big[R_{\bold{j}_1,\dots,\bold{j}_m\,;\,i,i}(w)\Big]^{\frac{1}{2}}\,\check{W}_{\bold{j}_1\,\dots,\bold{j}_m\,;i,i-2}\,\Big[R_{\bold{j}_1,\dots,\bold{j}_m\,;\,i-2,i-2}(w)\Big]^{\frac{1}{2}}\|^2\quad \quad\quad\quad \\
%%%%
%& &\|\Big[R^{Bog}_{i-2,i-2}(z)\Big]^{\frac{1}{2}}\,W_{i-2,i}\Big[R^{Bog}_{i,i}(z)\Big]^{\frac{1}{2}}\|\,\|\Big[R^{Bog}_{i,i}(z)\Big]^{\frac{1}{2}}W^*_{i,i-2}\,\Big[R^{Bog}_{i-2,i-2}(z)\Big]^{\frac{1}{2}}\|\\
%&\leq&\frac{1}{4\Big[1+\epsilon+\frac{\epsilon+1-\sqrt{\epsilon^2+2\epsilon}}{(N-i+2)}\Big]}\frac{1}{ \Big[1+\epsilon-\frac{\epsilon+1+\sqrt{\epsilon^2+2\epsilon}}{(N-i+2)}\Big]}+CN^{-\eta}\\
%%%%
%&\leq&\frac{1}{4\Big[1+\epsilon_{\bold{j}_m}+o(\epsilon_{\bold{j}_{m}})+\frac{\epsilon_{\bold{j}_m}+1-\delta\sqrt{\epsilon_{\bold{j}_m}^2+2\epsilon_{\bold{j}_m}}}{(N-i+1)}\Big]}\frac{1}{ \Big[1+\epsilon_{\bold{j}_m}+o(\epsilon_{\bold{j}_{m}})-\frac{\epsilon_{\bold{j}_m}+1+\delta\sqrt{\epsilon_{\bold{j}_m}^2+2\epsilon_{\bold{j}_m}}}{(N-i+1)}\Big]}\quad\quad\label{main-estimate-intermediate}\\
%%%
%\frac{1}{2\Big[1+(\epsilon+\frac{1+\epsilon-\sqrt{\epsilon^2+2\epsilon}}{N-i})-\mathcal{O}(\epsilon^2)\Big]^{\frac{1}{2}} \Big[1+(\epsilon-\frac{1+\epsilon+\sqrt{\epsilon^2+2\epsilon}}{N-i})-\mathcal{O}(\epsilon^2)\Big]^{\frac{1}{2}}}\\
%%%
&\leq &\frac{1}{4(1+a_{\epsilon_{\bold{j}_m}}-\frac{2b_{\epsilon_{\bold{j}_m}}}{N-i+1}-\frac{1-c_{\epsilon_{\bold{j}_m}}}{(N-i+1)^2})}\label{def-deltabog}
\end{eqnarray}
holds for $\bar{i}+4\leq i\leq N-2$. Here,  \begin{equation}\label{a-bis}
a_{\epsilon_{\bold{j}_{m}}}:=2\epsilon_{\bold{j}_{m}}+\mathcal{O}(\epsilon^{\nu}_{\bold{j}_m})\,, \quad \nu>\frac{11}{8}\,, 
\end{equation}
\begin{equation}\label{b-bis}
b_{\epsilon_{\bold{j}_{m}}}:=(1+\epsilon_{\bold{j}_{m}})\delta\chi_{[0,2)}(\delta)\sqrt{\epsilon_{\bold{j}_{m}}^2+2\epsilon_{\bold{j}_{m}}}\,,
\end{equation}
and
\begin{equation}\label{c-bis}
c_{\epsilon_{\bold{j}_{m}}}:=-(1-\delta^2\chi_{[0,2)}(\delta))(\epsilon_{\bold{j}_{m}}^2+2\epsilon_{\bold{j}_{m}})
\end{equation}
with $\chi_{[0,2)}(\delta)$ the characteristic function of the interval $[0,2)$.
%(Notice that for $\delta<2$$$\frac{\epsilon_{\bold{j}_{m}}+1+\delta\sqrt{\epsilon^2_{\bold{j}_{m}}+2\epsilon_{\bold{j}_{m}}}}{(N-i+2)}<1+\epsilon_{\bold{j}_{m}}$$ because $N-i\geq 1$ and $\epsilon_{\bold{j}_{m}}>0$. Therefore, the denominator in (\ref{main-estimate-intermediate}) is strictly positive.)
\end{corollary}

\noindent
\emph{Proof}

\noindent
%We start with our main estimate
%\begin{equation}
%\|\Big[R^{Bog}_n_{\bold{j}_0}n_{\bold{j}_0}{i-2,i-2}(z)\Big]^{\frac{1}{2}}\,W^*_{i-2,i}\,\Big[R^{Bog}_{i,i}(z)\Big]^{\frac{1}{2}}\|\leq \mathcal{O}(\rho^{\frac{1}{2}})
%\end{equation}
\noindent
%We show the result for $z\leq  E^{Bog}$. Then, it will be clear that an analogous result holds for $z\leq \xi^{Bog} E^{Bog}$ for some $\xi^{Bog}(<1)$ sufficiently close to $1$.

For $\bar{i}+2\leq i\leq N-2$, consider the operator 
\begin{equation}\label{S-operator}
S_{\bold{j}_m\,;\,i,i}(z):=\mathfrak{Q}^{(i,i+1)}_{\bold{j}_m}\frac{1}{\mathfrak{Q}^{(i,i+1)}_{\bold{j}_m}[\, V_{\bold{j}_{m}}+V'_{\bold{j}_1,\dots \bold{j}_{m-1}}+(\hat{H}^{Bog}_{\bold{j}_m})_{\xi}+\xi \,T-\frac{(m-1)\xi^{\frac{1}{2}}}{M}-z\,]\mathfrak{Q}^{(i,i+1)}_{\bold{j}_m}}\mathfrak{Q}^{(i,i+1)}_{\bold{j}_m}\,.
\end{equation}
If \begin{equation}\label{ineq-S}
 \mathfrak{Q}^{(i,i+1)}_{\bold{j}_m}[ V_{\bold{j}_{m}}+V'_{\bold{j}_1,\dots \bold{j}_{m-1}}+(\hat{H}^{Bog}_{\bold{j}_m})_{\xi}+\xi \,T-\frac{(m-1)\xi^{\frac{1}{2}}}{M}-z ] \mathfrak{Q}^{(i,i+1)}_{\bold{j}_m}> 0
\end{equation}
then we can estimate
\begin{eqnarray}
& &\| \Big[R_{\bold{j}_1,\dots,\bold{j}_m\,;\,i,i}(w)\Big]^{\frac{1}{2}}\,\check{W}_{\bold{j}_1,\dots,\bold{j}_m\,;\,i,i-2}\,\Big[R_{\bold{j}_1,\dots,\bold{j}_m\,;\,i-2,i-2}(w)\Big]^{\frac{1}{2}}\|\nonumber 
\end{eqnarray}
by inserting (for $N$ sufficiently large)
\begin{equation}
\charf_{\mathfrak{Q}^{(i-2,i-1)}_{\bold{j}_m}\mathcal{F}^N}=\Big[S_{\bold{j}_m\,;\,i-2,i-2}(w)\Big]^{\frac{1}{2}}\frac{1}{\Big[S_{\bold{j}_m\,;\,i-2,i-2}(w)\Big]^{\frac{1}{2}}}\quad,\quad i-2\geq \bar{i}+2\,,
\end{equation} 
and
\begin{equation}
\charf_{\mathfrak{Q}^{(i,i+1)}_{\bold{j}_m}\mathcal{F}^N}=\frac{1}{\Big[S_{\bold{j}_m\,;\,i,i}(w)\Big]^{\frac{1}{2}}}\Big[S_{\bold{j}_m\,;\,i,i}(w)\Big]^{\frac{1}{2}}
\end{equation} 
on the right and on the left of $\check{W}_{\bold{j}_1,\dots,\bold{j}_m\,;\,i,i-2}$, respectively, i.e.,
\begin{eqnarray}
& &\| \Big[R_{\bold{j}_1,\dots,\bold{j}_m\,;\,i,i}(z)\Big]^{\frac{1}{2}}\,\check{W}_{\bold{j}_1,\dots,\bold{j}_m\,;\,i,i-2}\,\Big[R_{\bold{j}_1,\dots,\bold{j}_m\,;\,i-2,i-2}(z)\Big]^{\frac{1}{2}}\| \nonumber \\
&= &\|\Big[R_{\bold{j}_1,\dots,\bold{j}_m\,;\,i,i}(z)\Big]^{\frac{1}{2}}\,\frac{1}{\Big[S_{\bold{j}_m\,;\,i,i}(z)\Big]^{\frac{1}{2}}}\Big[S_{\bold{j}_m\,;\,i,i}(z)\Big]^{\frac{1}{2}}\,\check{W}_{\bold{j}_1,\dots,\bold{j}_m\,;\,i,i-2}\,\Big[S_{\bold{j}_m\,;\,i-2,i-2}(z)\Big]^{\frac{1}{2}}\times\\
& &\quad\quad\quad \times \frac{1}{\Big[S_{\bold{j}_m\,;\,i-2,i-2}(z)\Big]^{\frac{1}{2}}}\,\Big[R_{\bold{j}_1,\dots,\bold{j}_m\,;\,i-2,i-2}(z)\Big]^{\frac{1}{2}}\|\,. \nonumber
\end{eqnarray}
Next, we show that the inequality in (\ref{ineq-S}) holds
for $\xi=(\frac{1}{\ln N})^{\frac{1}{4}}$, $N$ sufficiently large and $z$ in the interval (\ref{inter-z-S}). To this purpose it is helpful to recall that \begin{eqnarray}
H_{\bold{j}_1,\dots,\bold{j}_m}&=&(H^{\#}_{\bold{j}_{1},\dots,\bold{j}_{m-1}})_{\xi}-(1-\xi)T_{\bold{j}=\pm\bold{j}_m}+V_{\bold{j}_{m}}+V'_{\bold{j}_1,\dots \bold{j}_{m-1}}+(\hat{H}^{Bog}_{\bold{j}_m})_{\xi}+\xi \,T\,.
%H_{\bold{j}_1,\dots,\bold{j}_m}&=&(H_{\bold{j}_{1},\dots,\bold{j}_{m-1}})_{\xi}+V_{\bold{j}_{m}}+(\hat{H}^{Bog}_{\bold{j}_m})_{\xi}+\xi \,T
\end{eqnarray}

%Hence, due to the assumption
%\begin{equation}
%(H^{\#}_{\bold{j}_{1},\dots,\bold{j}_{m-1}})_{\xi}-(1-\xi)T_{\bold{j}=\{\pm\bold{j}_m\}}\geq z^{\#}_{\bold{j}_1,\dots\,\bold{j}_{m-1}}-\frac{2(m-1)\xi^{\frac{1}{2}}}{3M}\,
%\end{equation}
%and
%\begin{equation}
%|z_{\bold{j}_1,\dots\,\bold{j}_{m-1}}-z^{\#}_{\bold{j}_1,\dots\,\bold{j}_{m-1}}|\leq \frac{(m-1)\xi^{\frac{1}{2}}}{3M}\,,
%\end{equation}
%we conclude 
%\begin{equation}
%(H^{\#}_{\bold{j}_{1},\dots,\bold{j}_{m-1}})_{\xi}-(1-\xi)T_{\bold{j}=\{\pm\bold{j}_m\}}-z_{\bold{j}_1,\dots\,\bold{j}_{m-1}}\geq -\frac{(m-1)\xi^{\frac{1}{2}}}{M}\,.
%\end{equation}

\noindent
We point out that for $i\geq \bar{i}+2$:

\noindent
(1) $ \mathfrak{Q}^{(i,i+1)}_{\bold{j}_m}(\hat{H}^{Bog}_{\bold{j}_m})_{\xi}  \mathfrak{Q}^{(i,i+1)}_{\bold{j}_m}\geq 0$ ;

\noindent
(2) the kinetic term $\mathfrak{Q}^{(i,i+1)}_{\bold{j}_m}\xi T\mathfrak{Q}^{(i,i+1)}_{\bold{j}_m}$ dominates the non-positive part of $\mathfrak{Q}^{(i,i+1)}_{\bold{j}_m} (V_{\bold{j}_{m}}+V'_{\bold{j}_1,\dots \bold{j}_{m-1}})\mathfrak{Q}^{(i,i+1)}_{\bold{j}_m} $ as it is explained below.

\noindent
Regarding $\mathfrak{Q}^{(i,i+1)}_{\bold{j}_m} V_{\bold{j}_{m}}\mathfrak{Q}^{(i,i+1)}_{\bold{j}_m} $, we split it into 
\begin{eqnarray}
&&\mathfrak{Q}^{(i,i+1)}_{\bold{j}_m}V_{\bold{j}_m}\mathfrak{Q}^{(i,i+1)}_{\bold{j}_m}\\
&=&\mathfrak{Q}^{(i,i+1)}_{\bold{j}_m}\Big\{\frac{1}{N}\sum_{\bold{j}\in \mathbb{Z}^d\setminus \{ -\bold{j}_m,\bold{0}\}}a^*_{\bold{j}+\bold{j}_m}\,a^*_{\bold{0}}\,\phi_{\bold{j}_m}\,a_{\bold{j}}a_{\bold{j}_m}+h.c.\Big\}\mathfrak{Q}^{(i,i+1)}_{\bold{j}_m} \label{cubic-1}\\
& &+\mathfrak{Q}^{(i,i+1)}_{\bold{j}_m}\Big\{\frac{1}{N}\sum_{\bold{j}\in \mathbb{Z}^d\setminus \{ \bold{j}_m,\bold{0}\}}a^*_{\bold{j}-\bold{j}_m}\,a^*_{\bold{0}}\,\phi_{\bold{j}_m}\,a_{\bold{j}}a_{-\bold{j}_m}+h.c.\Big\}\mathfrak{Q}^{(i,i+1)}_{\bold{j}_m} \label{cubic-2}\\
& &+\mathfrak{Q}^{(i,i+1)}_{\bold{j}_m}\Big\{\frac{1}{N}\sum_{\bold{j}\in \mathbb{Z}^d\setminus\{ -\bold{j}_m,\bold{0}\}}\,\sum_{\bold{j}'\in \mathbb{Z}^d\,\setminus\{\bold{j}_m,\bold{0}\}}a^*_{\bold{j}+\bold{j}_m}\,a^*_{\bold{j}'-\bold{j}_m}\,\phi_{\bold{j}_m}\,a_{\bold{j}}a_{\bold{j}'}\Big\}\mathfrak{Q}^{(i,i+1)}_{\bold{j}_m} \label{quartic}
\end{eqnarray}
and proceed with two observations:
\noindent
\begin{itemize}
\item
For the control of (\ref{cubic-1})-(\ref{cubic-2}) we point out that  in each summand there is at most one of the operators $a_{\bold{0}}$, $a^*_{\bold{0}}$ and at least  one of the operators  $a_{\pm\bold{j}_m}$, $a^*_{\pm\bold{j}_m}$. Thus, we can exploit that  the number of particles in the modes $\pm \bold{j}_m$ is constrained by $\mathfrak{Q}^{(i,i+1)}_{\bold{j}_m}$ to the value $N-i$ or $N-i-1$ that are smaller than $\lfloor N^{\frac{1}{16}}\rfloor -1$, and use an estimate analogous to (\ref{ineq-form-est}).
\item For the  control of (\ref{quartic}) we exploit 
\begin{eqnarray}
%& &(\overline{\mathscr{P}_{\eta}^{\#}})_{I}\,(\frac{1}{3}H_{\bold{j}_1})\,(\overline{\mathscr{P}_{\eta}^{\#}})_{I}\\
%&& \mathfrak{Q}^{(i,i+1)}_{\bold{j}_m}V_{\bold{j}_{m}} \mathfrak{Q}^{(i,i+1)}_{\bold{j}_m}\\
(\ref{quartic})&\geq & \mathfrak{Q}^{(i,i+1)}_{\bold{j}_m}\Big\{\sum_{\bold{j}\in \mathbb{Z}^d\setminus \{\bold{0}\}}(-)\frac{\phi_{\bold{j}_m}}{N}\,a^*_{\bold{j}}a_{\bold{j}}+V^{(4)}_{\bold{j}_m}\Big\} \mathfrak{Q}^{(i,i+1)}_{\bold{j}_m}
%& &\langle (\overline{\mathscr{P}_{\psi^{Bog}_{\bold{j}_1,\dots,\bold{j}_{M-1}}}^{\#}})_{I}\psi\,,\,\charf^c_{\sigma}(\sum_{\bold{j}=\pm\bold{j}_{M}}(k_{\bold{j}})^2a^*_{\bold{j}}a_{\bold{j}}-z)\charf^c_{\sigma}\,(\overline{\mathscr{P}_{\psi^{Bog}_{\bold{j}_1,\dots,\bold{j}_{M-1}}}^{\#}})_{I}\psi \rangle \\
\end{eqnarray}
where
\begin{equation}\label{def-V4}
V^{(4)}_{\bold{j}_m}=\frac{1}{N}\sum_{\bold{j}\in \mathbb{Z}^d\setminus\{- \bold{j}_m,\bold{0}\}}\,a^*_{\bold{j}+\bold{j}_m}a_{\bold{j}}\,\,\phi_{\bold{j}_m}\,\sum_{\bold{j}'\in \mathbb{Z}^d\setminus\{+ \bold{j}_m,\bold{0}\}}a^*_{\bold{j}'-\bold{j}_m}a_{\bold{j}'} \geq 0\,.
\end{equation}

\end{itemize}

\noindent
Regarding $\mathfrak{Q}^{(i,i+1)}_{\bold{j}_m} V'_{\bold{j}_1,\dots, \bold{j}_{m}}\mathfrak{Q}^{(i,i+1)}_{\bold{j}_m} $, we  can repeat the strategy used to control (\ref{cubic-1}) and (\ref{cubic-2}).
%$\mathfrak{Q}^{(i,i+1)}_{\bold{j}_m}V_{\bold{j}_m}\mathfrak{Q}^{(i,i+1)}_{\bold{j}_m}$.

Due to the assumption in (\ref{ass-1-main-lemma}) and being $z<0$ uniformly in $N$, we deduce that for $N$ large enough
\begin{eqnarray}
& &  \mathfrak{Q}^{(i,i+1)}_{\bold{j}_m}[ H_{\bold{j}_1,\dots,\bold{j}_m}-w]  \mathfrak{Q}^{(i,i+1)}_{\bold{j}_m}\\
&= &  \mathfrak{Q}^{(i,i+1)}_{\bold{j}_m} [(H^{\#}_{\bold{j}_1,\dots,\bold{j}_{m-1}})_{\xi}-(1-\xi)T-z^{\#}_{\bold{j}_1,\dots,\bold{j}_{m-1}}+\frac{(m-1)\xi^{\frac{1}{2}}}{M}] \mathfrak{Q}^{(i,i+1)}_{\bold{j}_m} \label{pos-1}\\
& &+\mathfrak{Q}^{(i,i+1)}_{\bold{j}_m}[ V_{\bold{j}_{m}}+V'_{\bold{j}_1,\dots \bold{j}_{m-1}}+(\hat{H}^{Bog}_{\bold{j}_m})_{\xi}+\xi \,T-\frac{(m-1)\xi^{\frac{1}{2}}}{M}-z]\mathfrak{Q}^{(i,i+1)}_{\bold{j}_m}\label{pos-2}\\
&\geq&\mathfrak{Q}^{(i,i+1)}_{\bold{j}_m}[ V_{\bold{j}_{m}}+V'_{\bold{j}_1,\dots \bold{j}_{m-1}}+(\hat{H}^{Bog}_{\bold{j}_m})_{\xi}+\xi \,T-\frac{(m-1)\xi^{\frac{1}{2}}}{M}-z]\mathfrak{Q}^{(i,i+1)}_{\bold{j}_m}\,\\
&>&0\,.
\end{eqnarray}
%where $(\ref{pos-1})\geq0$ due to the assumption in  (\ref{ass-1-main-lemma}), and  (\ref{pos-2}) is strictly positive in $\mathfrak{Q}^{(i,i+1)}_{\bold{j}_m}\mathcal{F}^N$. 
Consequently, 
we can conclude that for $i\geq \bar{i}+2$
\begin{equation}
\|\Big[R_{\bold{j}_1,\dots,\bold{j}_m\,;\,i,i}(w)\Big]^{\frac{1}{2}}\,\frac{1}{\Big[S_{\bold{j}_m\,;\,i,i}(z)\Big]^{\frac{1}{2}}}\,\charf_{\mathfrak{Q}^{(i,i+1)}_{\bold{j}_m}\mathcal{F}^N}\|\leq 1
\end{equation}
which implies that for $i\geq \bar{i}+4$
\begin{eqnarray}
& &\| \Big[R_{\bold{j}_1,\dots,\bold{j}_m\,;\,i,i}(w)\Big]^{\frac{1}{2}}\,\check{W}_{\bold{j}_1,\dots,\bold{j}_m\,;\,i,i-2}\,\Big[R_{\bold{j}_1,\dots,\bold{j}_m\,;\,i-2,i-2}(w)\Big]^{\frac{1}{2}}\| \\
&\leq & \|\Big[S_{\bold{j}_m\,;\,i,i}(z)\Big]^{\frac{1}{2}}\,\check{W}_{\bold{j}_1, \dots, \bold{j}_m\,;\,i,i-2}\,\Big[S_{\bold{j}_m\,;\,i-2,i-2}(z)\Big]^{\frac{1}{2}}\|\,.
\end{eqnarray}

The next step is showing that for $i\geq \bar{i}+4$
\begin{equation}
\|\Big[S_{\bold{j}_m\,;\,i,i}(z)\Big]^{\frac{1}{2}}\,V_{\bold{j}_1, \dots, \bold{j}_m\,;\,i,i-2}\,\Big[S_{\bold{j}_m\,;\,i-2,i-2}(z)\Big]^{\frac{1}{2}}\|\leq \mathcal{O}(\frac{(N^{\frac{1}{16}})^{\frac{1}{2}}}{\xi N^{\frac{1}{2}}})\,.\label{estimate-V}
\end{equation}
where, for convenience, we recall
%If $1<m\leq M$, we use that in
%\begin{eqnarray}
%&&\mathfrak{Q}^{(i,i+1)}_{\bold{j}_m}V_{\bold{j}_1,\dots \bold{j}_m}\mathfrak{Q}^{(i-2,i-1)}_{\bold{j}_m}\\
%&=&\mathfrak{Q}^{(i,i+1)}_{\bold{j}_m}\frac{1}{N}\sum_{l=1}^{m}\sum_{\bold{j}\in \mathbb{Z}^d\,,\,\bold{j}\neq -\bold{j}_l}a^*_{\bold{j}+\bold{j}_l}\,a^*_%{\bold{0}}\,\phi_{\bold{j}_l}\,a_{\bold{j}}a_{\bold{j}_l}\mathfrak{Q}^{(i-2,i-1)}_{\bold{j}_m}+h.c. \label{def-v1...m}\\
%& &+\mathfrak{Q}^{(i,i+1)}_{\bold{j}_m}\frac{1}{N}\sum_{l=1}^{m}\sum_{\bold{j}\in \mathbb{Z}^d\,,\,\bold{j}\neq -\bold{j}_l}\,\sum_{\bold{j}'\in \mathbb{Z}^d%\,,\,\bold{j}'\neq \bold{j}_l}a^*_{\bold{j}+\bold{j}_l}\,a^*_{\bold{j}'-\bold{j}_l}\,\phi_{\bold{j}_l}\,a_{\bold{j}}a_{\bold{j}'}\mathfrak{Q}^{(i-2,i-1)}_{\bold{j}_m} \label%{def-v1...m-bis}
%\end{eqnarray}
%at least one of the operators must be associated with $\pm \bold{j}_m$ and we have a constraint on the number of particles in the modes $\pm \bold{j}_m%$. The other operators are easily controlled by the kinetic energy in the denominator and by the power $\frac{1}{N^{\frac{1}{2}}}$.

\begin{eqnarray}
&&\mathfrak{Q}^{(i,i+1)}_{\bold{j}_m}V_{\bold{j}_1,\dots \bold{j}_m}\mathfrak{Q}^{(i-2,i-1)}_{\bold{j}_m}\label{V-sandw}\\
&=&\mathfrak{Q}^{(i,i+1)}_{\bold{j}_m}\Big\{\frac{1}{N}\sum_{l=1}^{m}\sum_{\bold{j}\in \mathbb{Z}^d\setminus \{ -\bold{j}_l,\bold{0}\}}a^*_{\bold{j}+\bold{j}_l}\,a^*_{\bold{0}}\,\phi_{\bold{j}_l}\,a_{\bold{j}}a_{\bold{j}_l}+h.c.\Big\}\mathfrak{Q}^{(i-2,i-1)}_{\bold{j}_m} \label{def-v1...m}\\
& &+\mathfrak{Q}^{(i,i+1)}_{\bold{j}_m}\Big\{\frac{1}{N}\sum_{l=1}^{m}\sum_{\bold{j}\in \mathbb{Z}^d\setminus \{ \bold{j}_l,\bold{0}\}}a^*_{\bold{j}-\bold{j}_l}\,a^*_{\bold{0}}\,\phi_{\bold{j}_l}\,a_{\bold{j}}a_{-\bold{j}_l}+h.c.\Big\}\mathfrak{Q}^{(i-2,i-1)}_{\bold{j}_m} \label{def-v1...m-bis}\\
& &+\mathfrak{Q}^{(i,i+1)}_{\bold{j}_m}\Big\{\frac{1}{N}\sum_{l=1}^{m}\sum_{\bold{j}\in \mathbb{Z}^d\setminus\{ -\bold{j}_l,\bold{0}\}}\,\sum_{\bold{j}'\in \mathbb{Z}^d\,\setminus\{\bold{j}_l,\bold{0}\}}a^*_{\bold{j}+\bold{j}_l}\,a^*_{\bold{j}'-\bold{j}_l}\,\phi_{\bold{j}_l}\,a_{\bold{j}}a_{\bold{j}'}\Big\}\mathfrak{Q}^{(i-2,i-1)}_{\bold{j}_m} \label{def-v1...m-bis}\,.
\end{eqnarray}
Our strategy to control (\ref{V-sandw}) and provide the estimate in (\ref{estimate-V}) relies on the fact that in expressions (\ref{def-v1...m})-(\ref{def-v1...m-bis}): 
1) at most one operator of the type $a_{\bold{0}},a^*_{\bold{0}}$ can be present in each summand; 2)  at least one operator $a^*_{\bold{j}_m}$ or $a^*_{-\bold{j}_m}$ must be present due to the projections $\mathfrak{Q}^{(i,i+1)}_{\bold{j}_m}$ and $\mathfrak{Q}^{(i-2,i-1)}_{\bold{j}_m}$ on the left and on the right, respectively; 3)  the number of particles in the modes $\pm \bold{j}_m$ is constrained by $\mathfrak{Q}^{(i,i+1)}_{\bold{j}_m}$ to values less than $\lfloor N^{\frac{1}{16}}\rfloor -5$ for $i\geq \bar{i}+4$.

The leading term that remains is
\begin{equation}
[S_{\bold{j}_m\,;\,i,i}(z)\Big]^{\frac{1}{2}}\,W_{\bold{j}_m\,;\,i,i-2}\,\Big[S_{\bold{j}_m\,;\,i-2,i-2}(z)\Big]^{\frac{1}{2}}
\end{equation}
which can be estimated like in \emph{\underline{Lemma 3.4} in \cite{Pi1}}. Due to the choice of $\xi$, we arrive at the estimate in (\ref{def-deltabog}) where the corrections coming from (\ref{estimate-V}) and the $\xi$-dependent terms in (\ref{S-operator}) are hidden in the term $o(\epsilon_{\bold{j}_{m}})$ which enters the definition of $a_{\epsilon_{\bold{j}_m}}$; see (\ref{a-bis}). In fact these corrections vanish as $N\to \infty$.
\qed

\begin{lemma}\label{main-relations-H}
For $M\geq m\geq 1$ assume:

\noindent
(i)  
%\begin{equation}
%(H^{Bog}_{\bold{j}_1,\dots,\bold{j}_{m-1}})_{\xi}\geq z_{\bold{j}_1,\dots\,\bold{j}_m}-\mathcal{O}(\xi)
%\end{equation}
\begin{equation}
(H^{\#}_{\bold{j}_{1},\dots,\bold{j}_{m-1}})_{\xi}-(1-\xi)T_{\bold{j}=\{\pm\bold{j}_m\}}\geq z^{\#}_{\bold{j}_1,\dots\,\bold{j}_{m-1}}-\frac{(m-1)\xi^{\frac{1}{2}}}{M}\,,
\end{equation}
where $(H^{\#}_{\bold{j}_1,\dots,\bold{j}_{m-1}})_{\xi}$ is defined in  (\ref{def-xiham-1}) for $m\geq 2$ and is equal to $(1-\xi)T$ for $m=1$, 
 and where $z^{\#}_{\bold{j}_1,\dots\,\bold{j}_{m-1}}$ is the ground state energy of $H^{\#}_{\bold{j}_1,\dots,\bold{j}_{m-1}}$.  \\
 
%
%\noindent
%(ii) 
%\begin{equation}
%|z^{\#}_{\bold{j}_1,\dots\,\bold{j}_{m-1}}-z_{\bold{j}_1,\dots\,\bold{j}_{m-1}}|\leq \frac{(m-1)\xi^{\frac{1}{2}}}{3M}\,.
%\end{equation}

%\noindent
%(iii)

%\begin{equation}
%\text{infspec}\,\Big[ (H^{\#}_{\bold{j}_1, \dots,\bold{j}_{m}}-T_{\bold{j}=\pm \bold{j}_{m+1}})\upharpoonright_{(\mathcal{F}^N\ominus \mathcal{F}^N_{\pm\bold{j}_{m+1}})\ominus \{\mathbb{C}\psi^{\#}_{\bold{j}_1, \dots,\bold{j}_{m}}\}}\Big]-z^{\#}_{\bold{j}_1, \dots,\bold{j}_{m}}\\
%& \geq &\text{infspec}\,\Big[ H^{\#}_{\bold{j}_1, \dots,\bold{j}_{m}}\upharpoonright_{(\mathcal{F}^N\ominus \mathcal{F}^N_{\pm\bold{j}_{m+1}})\ominus \{\mathbb{C}\psi^{\#}_{\bold{j}_1, \dots,\bold{j}_{m}}\}}\Big]-z^{\#}_{\bold{j}_1, \dots,\bold{j}_{m}}\\
%\geq  \Delta^{\#}_m\,.
% \end{equation}

\noindent
(ii) 
The upper bound
 \begin{equation}\label{bound-number-squared}
\langle \frac{\psi^{\#}_{\bold{j}_1,\dots,\bold{j}_{m-1}}}{\|\psi^{\#}_{\bold{j}_1,\dots,\bold{j}_{m-1}}\|}\,,\,(\sum_{\bold{j}\in\mathbb{Z}^d\setminus \{\bold{0}\}} a_{\bold{j}}^{*}a_{\bold{j}})^2\,\frac{\psi^{\#}_{\bold{j}_1,\dots,\bold{j}_{m-1}}}{\|\psi^{\#}_{\bold{j}_1,\dots,\bold{j}_{m-1}}\|}\rangle \leq \mathcal{O}(1)\,
\end{equation}
holds true where $\psi^{\#}_{\bold{j}_1,\dots,\bold{j}_{m-1}}$ is the ground state vector of $H^{\#}_{\bold{j}_1,\dots,\bold{j}_{m-1}}$.
\\

\noindent
Let $\epsilon_{\bold{j}_m}$ be sufficiently small and $N$ sufficiently large. Then, for $z\leq E^{Bog}_{\bold{j}_m}+ \sqrt{\epsilon_{\bold{j}_m}}\phi_{\bold{j}_m}\sqrt{\epsilon_{\bold{j}_m}^2+2\epsilon_{\bold{j}_m}}$ there are constants $0<C^{\#}_{I}, C^{\#}_{II},C^{\#\,\perp}<\infty $ such that
\begin{eqnarray}
& &\Big|\langle  \frac{\psi^{\#}_{\bold{j}_1,\dots,\bold{j}_{m-1}}}{\|\psi^{\#}_{\bold{j}_1,\dots,\bold{j}_{m-1}}\|}\,,\,\Gamma_{\bold{j}_1,\dots,\bold{j}_m ;N,N}(z+z^{\#}_{\bold{j}_1,\dots,\bold{j}_{m-1}}) \frac{\psi^{\#}_{\bold{j}_1,\dots,\bold{j}_{m-1}}}{\|\psi^{\#}_{\bold{j}_1,\dots,\bold{j}_{m-1}}\|}\rangle- \langle  \eta\,,\,\Gamma^{Bog}_{\bold{j}_m ;N,N}(z) \eta \rangle \Big| \leq \frac{C^{\#}_I}{(\ln N)^{\frac{1}{4}}}\,,\quad \quad\quad \label{primero-1}
%&= &\langle  \frac{\psi^{Bog}_{\bold{j}_1,\dots,\bold{j}_{m-1}}}{\|\psi^{Bog}_{\bold{j}_1,\dots,\bold{j}_{m-1}}\|}\,,\,W_{\bold{j}_2}\,R^{Bog}_{\bold{j}_1,\,\bold{j}_2\,;\,N-2,N-2}(z_{\bold{j}_1}+z)\times\\
%& &\quad\quad \times \sum_{l_{N-2}=0}^{\infty}\Big[\Gamma^{Bog}_{\bold{j}_1,\,\bold{j}_2\,;\,N-2,N-2}(z_{\bold{j}_1}+z)R^{Bog}_{\bold{j}_1,\,\bold{j}_2\,;\,\,N-2,N-2}(z_{\bold{j}_1}+z)\Big]^{l_{N-2}}W^*_{\bold{j}_2}\, \frac{\psi^{Bog}_{\bold{j}_1,\dots,\bold{j}_{m-1}}}{\|\psi^{Bog}_{\bold{j}_1,\dots,\bold{j}_{m-1}}\|}\rangle \quad\quad\quad\\
%&= &\langle  \eta\,,\,\tilde{\Gamma}^{Bog}_{\bold{j}_1,\dots,\bold{j}_m ;N,N}(z) \eta \rangle +{\color{red}\frac{C_I}{N^{\frac{1}{16}}}}\label{A-remainder}
\end{eqnarray}
\begin{eqnarray}
& &\|\mathscr{P}^{\#}_{\psi^{\#}_{\bold{j}_1,\dots,\bold{j}_{m-1}}}(V_{\bold{j}_m}-\Gamma_{\bold{j}_1,\dots,\bold{j}_m;\,N,N}(z+z^{\#}_{\bold{j}_1,\dots,\bold{j}_{m-1}}))\,\overline{\mathscr{P}^{\#}_{\psi^{\#}_{\bold{j}_1,\dots,\bold{j}_{m-1}}}}\|\leq \frac{C^{\#}_{II}}{(\ln N)^{\frac{1}{4}}}\,,\label{primero-3}
\end{eqnarray}
and
\begin{eqnarray}
& &\Big\|\Big(\overline{\mathscr{P}_{\psi^{\#}_{\bold{j}_1,\dots,\bold{j}_{m-1}}}}\Big)_I\,(\Gamma_{\bold{j}_1,\dots,\bold{j}_m\,,;\,N,N}(z+z^{\#}_{\bold{j}_1,\dots,\bold{j}_{m-1}})\,\Big(\overline{\mathscr{P}_{\psi^{\#}_{\bold{j}_1,\dots,\bold{j}_{m-1}}}}\Big)_I\Big\|\label{estimate-gammaperp-0}\\
&\leq& \frac{\phi_{\bold{j}_{m}}}{2\epsilon_{\bold{j}_m}+2-\frac{z-\Delta^{\#}_{m-1}(1-\frac{\phi_{\bold{j}_m}\lfloor (\ln N)^{\frac{1}{2}} \rfloor}{N\Delta^{\#}_{m-1}})}{\phi_{\bold{j}_{m}}}}\check{\mathcal{G}}_{\bold{j}_m\,;\,N-2,N-2}(z-\Delta^{\#}_{m-1}(1-\frac{\phi_{\bold{j}_m}\lfloor (\ln N)^{\frac{1}{2}} \rfloor}{N\Delta^{\#}_{m-1}}))+\frac{C^{\#\,\perp}}{{(\ln N)^{\frac{1}{4}}}}\,.\nonumber
\end{eqnarray}
%for $m\geq 2$, whereas for $m=1$
%\begin{eqnarray}
%& &\Big\|\Big(\overline{\mathscr{P}_{\eta}^{\#}}\Big)_I\,\Gamma_{\bold{j}_1\,,;\,N,N}(w)\,\Big(\overline{\mathscr{P}_{\eta}^{\#}}\Big)_I\Big\|\label{estimate-gammaperp}\\
%&\leq& \frac{\phi_{\bold{j}_{m}}}{2\epsilon_{\bold{j}_1}+2-\frac{z-\Delta^{\#}_0(1-\frac{\phi_{\bold{j}_1}}{N\Delta^{\#}_0})}{\phi_{\bold{j}_{1}}}}\check{\mathcal{G}}_{\bold{j}_1\,;\,N-2,N-2}(z-\Delta^{\#}_0(1-\frac{\phi_{\bold{j}_1}}{N\Delta^{\#}_0}))+\frac{C^{\#\,\perp}}{{(\ln N)^{\frac{1}{4}}}}\,.\nonumber
%\end{eqnarray}
\end{lemma}

\emph{Proof}

The proof is very similar to the proof of the analogous inequalities in \emph{\underline{Lemma 4.3} and \underline{Lemma 4.4} of \cite{Pi2}}.   As far as (\ref{primero-1}) and (\ref{primero-3}) are concerned,  the role of the Hamiltonian $\hat{H}^{Bog}_{\bold{j}_1,\dots,\bold{j}_{m-1}}$ in the analogous estimate of  \emph{\underline{Lemma 4.3} of \cite{Pi2}} is played by the operator
$$H^{\#}_{\bold{j}_{1},\dots,\bold{j}_{m-1}}-T_{\bold{j}=\pm\bold{j}_m}+V_{\bold{j}_{m}}+V'_{\bold{j}_1,\dots \bold{j}_{m-1}}\,$$
with the help of the assumption in (\ref{bound-number-squared}).
Likewise,  in (\ref{primero-3}) the term proportional to $V_{\bold{j}_m}$ is estimated using the assumption in (\ref{bound-number-squared}).
\qed

\begin{lemma}\label{control-quad}
Let $1\leq m\leq M<\infty$. Then, assuming that the operators below are restricted to $\mathcal{F}^N$,  the following inequalities hold true, 
\begin{equation}\label{estimate-control-quad}
H^{\#}_{\bold{j}_1, \dots,\bold{j}_{m-1}}\geq T-\sum_{l=1}^{m-1}\phi_{\bold{j}_l}\quad \text{with}\quad T:=\sum_{\bold{j}\in \mathbb{Z}^d}k^2_{\bold{j}}a_{\bold{j}}^{*}a_{\bold{j}}\,,
\end{equation}
\begin{equation}\label{estimate-control-quad-2}
H_{\bold{j}_1, \dots,\bold{j}_{m-1}}\geq T-\sum_{l=1}^{m-1}\phi_{\bold{j}_l}\,,\,
\end{equation}
and
\begin{equation}\label{estimate-control-quad-3}
(H^{\#}_{\bold{j}_1, \dots,\bold{j}_{m-1}})^2\geq C_1\mathcal{N}_+^2-C_2, \quad \mathcal{N}_+:=\sum_{\bold{j}\in \mathbb{Z}^d \setminus \{\bold{0}\}}a_{\bold{j}}^{*}a_{\bold{j}}\,,
\end{equation}for some  $C_1, C_2>0$.
\end{lemma}

\noindent
\emph{Proof}

\noindent
Starting from the identity 
\begin{eqnarray}
H^{\#}_{\bold{j}_1, \dots,\bold{j}_{m-1}}
& =&\sum_{\bold{j}\in \mathbb{Z}^d}k^2_{\bold{j}}a_{\bold{j}}^{*}a_{\bold{j}}\\
& &+\sum_{l=1}^{m-1}\frac{\phi_{\bold{j}_l}}{N}(a^*_{\bold{0}}a_{\bold{j}_l}+a_{\bold{0}}a^*_{-\bold{j}_l}\,+\sum_{\bold{j}'\in \mathbb{Z}^d\setminus\{ \bold{j}_l\,;\,\bold{0}\,;\,\pm\bold{j}_m\,;\,\pm\bold{j}_m+\bold{j}_l\}}\,a^*_{\bold{j}'-\bold{j}_l}\,a_{\bold{j}'})\times\\
& &\quad\quad\quad\quad\times (a_{\bold{0}}a^*_{\bold{j}_l}+a^*_{\bold{0}}a_{-\bold{j}_l} +\sum_{\bold{j}\in \mathbb{Z}^d\setminus\{ -\bold{j}_l\,;\,\bold{0}\,;\,\pm\bold{j}_m\,;\,\pm\bold{j}_m-\bold{j}_l\}}a^*_{\bold{j}+\bold{j}_l}\,a_{\bold{j}} )\\
& &-\sum_{l=1}^{m-1}\frac{\phi_{\bold{j}_l}}{N}\Big[a^*_{-\bold{j}_l}a_{-\bold{j}_l}+a^*_{\bold{0}}a_{\bold{0}}+\sum_{\bold{j}\in \mathbb{Z}^d\setminus\{ -\bold{j}_l\,;\,\bold{0}\,;\,\pm\bold{j}_m\,;\,\pm\bold{j}_m-\bold{j}_l\}}a^*_{\bold{j}}\,a_{\bold{j}}\Big]\,.\quad \nonumber\\
%& &\frac{1}{N}\sum_{l=1}^{m-1}\sum_{\bold{j}\in \mathbb{Z}^d\setminus\{ -\bold{j}_l\,;\,\bold{0}\,;\,\pm\bold{j}_m\,;\,\pm\bold{j}_m-\bold{j}_l\}}\,\sum_{\bold{j}'\in \mathbb{Z}^d\setminus\{ \bold{j}_l\,;\,\bold{0}\,;\,\pm\bold{j}_m\,;\,\pm\bold{j}_m+\bold{j}_l\}}a^*_{\bold{j}+\bold{j}_l}\,a^*_{\bold{j}'-\bold{j}_l}\,\phi_{\bold{j}_l}\,a_{\bold{j}}a_{\bold{j}'} 
\end{eqnarray}
it is convenient to set
\begin{equation}
\mathcal{A}_l:=(a_{\bold{0}}a^*_{\bold{j}_l}+a^*_{\bold{0}}a_{-\bold{j}_l} +\sum_{\bold{j}\in \mathbb{Z}^d\setminus\{ -\bold{j}_l\,;\,\bold{0}\,;\,\pm\bold{j}_m\,;\,\pm\bold{j}_m-\bold{j}_l\}}a^*_{\bold{j}+\bold{j}_l}\,a_{\bold{j}} )\,,
\end{equation}
\begin{equation} \mathcal{B}_l:=a^*_{-\bold{j}_l}a_{-\bold{j}_l}+a^*_{\bold{0}}a_{\bold{0}}+\sum_{\bold{j}\in \mathbb{Z}^d\setminus\{ -\bold{j}_l\,;\,\bold{0}\,;\,\pm\bold{j}_m\,;\,\pm\bold{j}_m-\bold{j}_l\}}a^*_{\bold{j}}\,a_{\bold{j}}\,.
\end{equation}
Then, the inequality in (\ref{estimate-control-quad}) is obvious
since $\phi_{\bold{j}_l}>0$, $\mathcal{A}_l^*\mathcal{A}_l\geq 0$, and
\begin{equation}
a^*_{-\bold{j}_l}a_{-\bold{j}_l}+a^*_{\bold{0}}a_{\bold{0}}+\sum_{\bold{j}\in \mathbb{Z}^d\setminus\{ -\bold{j}_l\,;\,\bold{0}\,;\,\pm\bold{j}_m\,;\,\pm\bold{j}_m-\bold{j}_l\}}a^*_{\bold{j}}\,a_{\bold{j}}\leq N\,.
\end{equation}
The proof of inequality (\ref{estimate-control-quad-2}) is essentially the same.

Regarding the third inequality, we can write
\begin{eqnarray}
&&(H^{\#}_{\bold{j}_1, \dots,\bold{j}_{m-1}})^2\\
& =&T^2+(\sum_{l=1}^{m-1}\frac{\phi_{\bold{j}_l}}{N}\mathcal{A}_l^*\mathcal{A}_l)^2+(\sum_{l=1}^{m-1}\frac{\phi_{\bold{j}_l}}{N}\mathcal{B}_l)^2-2T(\sum_{l=1}^{m-1}\frac{\phi_{\bold{j}_l}}{N}\mathcal{B}_l)\\
& &+\Big\{T-\sum_{l=1}^{m-1}\frac{\phi_{\bold{j}_l}}{N}\mathcal{B}_l \Big\}(\sum_{l=1}^{m-1}\frac{\phi_{\bold{j}_l}}{N}\mathcal{A}_l^*\mathcal{A}_l)+(\sum_{l=1}^{m-1}\frac{\phi_{\bold{j}_l}}{N}\mathcal{A}_l^*\mathcal{A}_l)\Big\{ T-\sum_{l=1}^{m-1}\frac{\phi_{\bold{j}_l}}{N}\mathcal{B}_l \Big\}\quad
%& &\frac{1}{N}\sum_{l=1}^{m-1}\sum_{\bold{j}\in \mathbb{Z}^d\setminus\{ -\bold{j}_l\,;\,\bold{0}\,;\,\pm\bold{j}_m\,;\,\pm\bold{j}_m-\bold{j}_l\}}\,\sum_{\bold{j}'\in \mathbb{Z}^d\setminus\{ \bold{j}_l\,;\,\bold{0}\,;\,\pm\bold{j}_m\,;\,\pm\bold{j}_m+\bold{j}_l\}}a^*_{\bold{j}+\bold{j}_l}\,a^*_{\bold{j}'-\bold{j}_l}\,\phi_{\bold{j}_l}\,a_{\bold{j}}a_{\bold{j}'} 
\end{eqnarray}
and compute
\begin{equation}
T(\frac{\phi_{\bold{j}_l}}{N}\mathcal{A}_l^*\mathcal{A}_l)=\sum_{\bold{j}\in \mathbb{Z}^d}k^2_{\bold{j}}\,a_{\bold{j}}^{*}\frac{\phi_{\bold{j}_l}}{N}\mathcal{A}_l^*\mathcal{A}_la_{\bold{j}}+\sum_{\bold{j}\in \mathbb{Z}^d}k^2_{\bold{j}}\,a_{\bold{j}}^{*}[a_{\bold{j}}
\,,\,\frac{\phi_{\bold{j}_l}}{N}\mathcal{A}_l^*\mathcal{A}_l]\,.
\end{equation}
We also observe that
\begin{itemize}
\item
 the expression 
\begin{equation}
\sum_{\bold{j}\in \mathbb{Z}^d}k^2_{\bold{j}}\,a_{\bold{j}}^{*}[a_{\bold{j}}
\,,\,\frac{\phi_{\bold{j}_l}}{N}\mathcal{A}_l^*\mathcal{A}_l]=\sum_{\bold{j}\in \mathbb{Z}^d}k^2_{\bold{j}}\,a_{\bold{j}}^{*}\frac{\phi_{\bold{j}_l}}{N}\Big\{[a_{\bold{j}}\,,\,\mathcal{A}_l^*]\mathcal{A}_l+\mathcal{A}^*_l[a_{\bold{j}}\,,\,\mathcal{A}_l]\Big\}
\end{equation}
is dominated by a  constant times 
\begin{equation}
\mathcal{N}_++\frac{\phi_{\bold{j}_l}}{N}\mathcal{A}_l^*\mathcal{A}_l
\end{equation}
\item
the expressions
\begin{equation}
\frac{\phi_{\bold{j}_l}}{N}\mathcal{B}_l\frac{\phi_{\bold{j}_l}}{N}\mathcal{A}_l^*\mathcal{A}_l\quad,\quad-2T\frac{\phi_{\bold{j}_l}}{N}\mathcal{B}_l
\end{equation}
are dominated by a constant times $T$.
\end{itemize}
Using  $T^2\geq \Delta^2_0\mathcal{N}_+^2$ we can determine two positive constants $C_1,C_2$ 
%(dependent on the Fourier components $\phi_{\bold{j}_l}$, on $\Delta_0$, and on $M$)
 such that
\begin{equation}
(H^{\#}_{\bold{j}_1, \dots,\bold{j}_{m-1}})^2\geq C_1\mathcal{N}_+^2-C_2\,.
\end{equation}
%and
%\begin{equation}
%U\geq -\frac{\mathcal{N}_+}{2N}\phi(0) 
%\end{equation}
\qed
\\

\noindent
{\bf{Acknowledgements}}

I would like to thank D.-A. Deckert, J. Fr\"ohlich, and P. Pickl for many stimulating discussions on the contents of this paper.  I am indebted to D.-A. Deckert for his contribution  to this project in its earliest stage and for helping with the implementation of crucial numerical simulations.

\end{document}